\documentclass[12pt]{article}
\usepackage{amsfonts}

\let\ssection=\section
\renewcommand{\section}{\setcounter{equation}{0}\ssection}

\newcommand\mathC{\mkern1mu\raise2.2pt\hbox{$\scriptscriptstyle|$}
        {\mkern-7mu\rm C}} 
\newcommand{\mathR}{{\rm I\! R}}         

\newcommand\mapup[1]{\Big\uparrow
                        \rlap{$\vcenter{\hbox{$\scriptstyle#1$}}$}}
\newcommand\mapright[1]{\smash{
        \mathop{\mbox{\large{$\longrightarrow$}}}\limits^{#1}}}
\newcommand\bundle[3]{\begin{array}[t]{c}
        {#3}\\ \mapup{#2}\\ {#1}\end{array}}
\newcommand\bundlemap[2]{\begin{array}[t]{c}
\mapright{#2}\\
\phantom{\mapup{}}\\\mapright{#1}\\\end
{array}}

\newcommand{\R}{\ensuremath{{\mathR}}}
\renewcommand{\o}{\ensuremath{{\omega}}}
\renewcommand{\O}{\ensuremath{{\Omega}}}

\newcommand\bi{\begin{itemize}}
\newcommand\ei{\end{itemize}}
\newcommand\be{\begin{equation}}
\newcommand\ee{\end{equation}}

\newcommand{\al}{\ensuremath{{\alpha}}}
\newcommand{\bb}{\ensuremath{{\beta}}}
\newcommand{\pl}{\ensuremath{{\partial}}}

\newcommand{\fl}{\ensuremath{{\flat}}}
\newcommand{\ra}{\ensuremath{{\rightarrow}}}
\newcommand{\sh}{\ensuremath{{\sharp}}}

\newcommand{\la}{\ensuremath{{\lambda}}}
\newcommand{\mg}{\ensuremath{{\mathfrak{g}}}}
\newcommand{\mh}{\ensuremath{{\mathfrak{h}}}}
\newcommand{\mgl}{\ensuremath{{\mathfrak{gl}}}}
\newcommand{\mso}{\ensuremath{{\mathfrak{so}}}}

\begin{document}
\begin{titlepage}

\begin{center}

{\large\bf On Symplectic Reduction in Classical Mechanics}
\end{center}

\vspace{0.3 truecm}

\begin{center}
        J.~Butterfield\footnote{email: jb56@cus.cam.ac.uk;
            jeremy.butterfield@all-souls.oxford.ac.uk}\\
            [10pt] All Souls College\\ Oxford OX1 4AL
\end{center}

\begin{center}
     Thursday 21 July 2005; a Chapter of {\em The North Holland Handbook of 
Philosophy of Physics}
\end{center}

\begin{abstract}
This Chapter expounds the modern theory of symplectic reduction in 
finite-dimensional Hamiltonian mechanics. This theory generalizes the 
well-known connection between continuous symmetries and conserved quantities, 
i.e. Noether's theorem. It also illustrates one of mechanics' grand themes: 
exploiting a  symmetry so as to reduce the number of variables needed to treat 
a problem. The exposition emphasises how the theory provides insights about the 
rotation group and the rigid body. The theory's device of quotienting a state 
space also casts light on philosophical issues about whether two  apparently 
distinct but utterly indiscernible possibilities should be ruled to be one and 
the same. These issues are illustrated using ``relationist'' mechanics.
\end{abstract}

\begin{center}
         {\bf Mottoes}
\end{center}

\noindent The current vitality of mechanics, including the investigation of 
fundamental questions, is quite remarkable, given its long history and 
development. This vitality comes about through rich interactions with pure 
mathematics (from topology and geometry to group representation theory), and 
through  new and exciting applications to areas like control theory. It is 
perhaps even more remarkable that absolutely fundamental points, such as a 
clear and unambiguous linking of Lie's work on the Lie-Poisson bracket on the 
dual of a Lie algebra ... with the most basic of examples in mechanics, such as 
the rigid body and the motion of ideal fluids, took nearly a century to 
complete.\\
\indent Marsden and Ratiu (1999, pp. 431-432). 

\noindent In the ordinary theory of the rigid body, six different 
three-dimensional  spaces $\R^3, \R^{3*}, \mg,\\ \mg^*, TG_g, T^*G_g$ are 
identified.\\
\indent Arnold (1989, p. 324).

\end{titlepage}

\tableofcontents

\newpage
\section{Introduction}\label{intro}
\subsection{Why classical mechanics?}\label{whyclassmech}
All hail the rise of modern physics! Between 1890 and 1930, the quantum and 
relativity revolutions and the consolidation of statistical physics through the 
discovery of atoms, utterly transformed our understanding of nature; and had an 
enormous influence on philosophy; (e.g. Kragh 1999; Ryckman 2005). Accordingly, 
this Handbook concentrates on those three pillars of modern physics---quantum 
theories, spacetime theories and thermal physics. So some initial explanation  
of the inclusion of a Chapter on classical mechanics, indeed the classical 
mechanics of finite-dimensional systems, is in order.    

The first point to make is that the various fields of classical physics, such 
as mechanics and optics, are wonderfully rich and deep, not only in their 
technicalities, but also in their implications for the philosophy and 
foundations of physics. From Newton's time onwards, classical mechanics and 
optics have engendered an enormous amount of philosophical reflection. As 
regards mechanics, the central philosophical topics are usually taken  (and 
have traditionally been taken) to be space, time, determinism and the 
action-at-a-distance nature of Newtonian gravity. Despite their importance, I 
will not discuss these topics; but some other Chapters will do so (at least in 
part, and sometimes in connection with theories other than classical 
mechanics). I will instead focus on the theory of symplectic reduction, which 
develops the well-known connection between continuous symmetries and conserved 
quantities, summed up in Noether's ``first theorem''. I choose this focus 
partly by way of preparation for parts of some other Chapters; and partly 
because, as we will see in a moment, symplectic reduction  plays a central role 
in the current renaissance of classical mechanics, and in its relation to 
quantum physics. 

I said that classical physics engendered a lot of philosophical reflection. It 
is worth stressing two, mutually related, reasons for this: reasons which 
today's philosophical emphasis on the quantum and relativity revolutions tends 
to make us forget.

\indent First: in the two centuries following Newton, these fields of classical 
physics were transformed out of all recognition, so that the framework for 
philosophical reflection about them also changed. Think of how in the 
nineteenth century, classical mechanics and optics gave rise  to classical 
field theories, especially electromagnetism. And within this Chapter's specific 
field, the classical mechanics of finite-dimensional systems, think of how even 
its central theoretical principles were successively recast, in fundamental 
ways, by figures such Euler, Lagrange, Hamilton and Jacobi.

\indent Second, various difficult problems beset the attempt to rigorously 
formulate classical mechanics and optics; some of which have considerable 
philosophical aspects. It is {\em not} true that once we set aside the familiar 
war-horse topics---space, time, determinism and  action-at-a-distance---the 
world-picture of classical mechanics   is straightforward: just ``matter in 
motion''. On the contrary. Even if we consider only finite-dimensional systems, 
we can ask, for example:\\
\indent\indent (i) For point-particles (material points): can they have 
different masses, and if so how? What happens when they collide? Indeed, for 
point-particles interacting only by Newtonian gravity, a collision involves 
infinite kinetic energy.  \\
\indent\indent (ii) For extended bodies treated as finite-dimensional because 
rigid: what happens when they collide? Rigidity implies that forces, and 
displacements, are transmitted ``infinitely fast'' through the body. Surely 
that should not be taken literally? But if so, what justifies this 
idealization; and what are its scope and limits?\\
\indent As to infinite-dimensional systems (elastic solids, fluids and fields), 
many parts of their theories remain active research areas, especially  as 
regards rigorous formulations and results. For contemporary work on elastic 
solids, for example, cf. Marsden and Hughes (1982). As to fluids, the existence 
and uniqueness of rigorous solutions of the main governing equations, the 
Navier-Stokes equations, is still an open problem. This problem not only has an 
obvious bearing on determinism; it is regarded as scientifically significant 
enough that its solution would secure a million-dollar Clay Millennium prize.

\indent  These two reasons---the successive reformulations of classical 
mechanics, and its philosophical problems---are of course related. The 
monumental figures of classical mechanics recognized and debated the  problems, 
and much of their technical work was aimed at solving them. As a result, there 
was a rich debate about the  foundations of classical physics, in particular 
mechanics, for the two centuries after Newton's {\em Principia} (1687). A 
well-known example is Duhem's instrumentalist philosophy of science, which 
arose in large measure from his realization how hard it was to secure rigorous 
foundations at the microscopic level for classical mechanics. A similar example 
is Hilbert's being prompted by his contemporaries' continuing controversies 
about the foundations of mechanics, to choose as the sixth of his famous list 
of outstanding mathematical problems, the axiomatization of mechanics and 
probability; (but for some history of this list, cf. Grattan-Guinness (2000)). 
A third example, spanning both centuries, concerns variational principles: the 
various principles of least action  formulated first by Maupertuis, then by 
Euler and later figures---first for finite classical mechanical systems, then 
for infinite ones---prompted much discussion of teleology. Indeed, this 
discussion  ensnared the logical empiricists (St\"{o}ltzner 2003); it also 
bears on contemporary philosophy of modality (Butterfield 2004).   

\indent In the first half of the twentieth century, the quantum and relativity 
revolutions tended to distract physicists, and  thereby philosophers, from 
these and similar problems. The excitement of developing the new theories, and 
of debating their implications for natural philosophy, made it understandable, 
even inevitable, that the foundational problems of classical mechanics were 
ignored.

\indent Besides, this tendency was strengthened by the demands of pedagogy: the 
necessity of including the new theories in physics undergraduate degrees.  By 
mid-century, the constraints of time on the physics curriculum had led many 
physics undergraduates' education in classical mechanics to finish with the 
elementary parts of analytical mechanics, especially of finite-dimensional 
systems: for example, with the material in Goldstein's well-known textbook 
(1950). Such a restriction is understandable, not least because: (i) the 
elementary theory of Lagrange's and Hamilton's equations requires knowledge of 
ordinary differential equations, and (ii) elementary Hamiltonian mechanics 
forms a springboard to learning elementary canonical quantization (as does 
Hamilton-Jacobi theory, from another perspective). Besides, as I mentioned: 
even this restricted body of theory provides plenty of material for 
philosophical analysis---witness my examples above, and the discussions of the 
great figures such Euler, Lagrange, Hamilton and Jacobi.

However, the second half of the twentieth century saw a renaissance in research 
in classical mechanics: hence my first motto. There are four  obvious reasons 
for this: the first two ``academic'', and the second two ``practical''. \\
\indent (i): Thanks partly to developments in mathematics in the decades after 
Hilbert's list of problems, the foundational questions were addressed afresh, 
as much by mathematicians and mathematically-minded engineers as by physicists. 
The most relevant developments lay in such fields as topology, differential 
geometry, measure theory and functional analysis. In this revival, the 
contributions of the Soviet school, always strong in mechanics and probability, 
were second to none. And relatedly:--- \\
\indent (ii): The quest to deepen the formulation of quantum theory, especially 
quantum field theory, prompted investigation of (a) the structure of classical 
mechanics and (b) quantization. For both (a) and (b), special interest attaches 
to the generally much harder case of  infinite systems.\\
\indent (iii): The coming of spaceflight, which spurred the development of 
celestial mechanics. And relatedly:---  \\
\indent   (iv): The study of non-linear dynamics (``chaos theory''), which was 
spurred by the invention of computers.   

With these diverse causes and aspects, this renaissance continues to 
flourish---and accordingly, I shall duck out of trying to further adumbrate it! 
I shall even duck out of trying to survey the philosophical questions that 
arise from the various formulations of mechanics from Newton to Jacobi and 
Poincar{\'{e}}. Suffice it to say here that to the various topics mentioned 
above, one could add, for example, the following two: the first broadly 
ontological, the second broadly epistemological.\\
\indent (a): The analysis of  notions such as mass and force (including how 
they change over time). For this topic, older books include Jammer (1957, 1961) 
and McMullin (1978);  recent books include Boudri (2002), Jammer (2000), Lutzen 
(2005) and Slovik (2002); and Grattan-Guinness (2006) is a fine recent synopsis 
of the history, with many references.\\
\indent (b): The analysis of what it is to have an explicit solution of a 
mechanical problem (including how the notion of explicit solution was gradually 
generalized). This topic is multi-faceted. It not only relates to the gradual 
generalization of the notion of  function (a grand theme in the history of 
mathematics---well surveyed by Lutzen 2003), and to modern non-linear dynamics 
(cf. (iv) above). It also relates to the simplification of problems by 
exploiting a symmetry so as to reduce the number of variables one needs---and 
{\em this} is the core idea of symplectic reduction. I turn  to introducing it.

\subsection{Prospectus}\label{introprospectus}
The strategy of simplifying a mechanical problem by exploiting a symmetry so as 
to reduce the number of variables is  one of classical mechanics' grand themes. 
It is theoretically deep, practically important, and recurrent in the history 
of the subject. The best-known general theorem about the strategy is 
undoubtedly Noether's theorem, which describes a correspondence  between 
continuous symmetries and conserved quantities. There is both a Lagrangian and 
a Hamiltonian version of this theorem, though for historical reasons the name 
`Noether's theorem' is more strongly attached to the Lagrangian version. 
However, we shall only need the Hamiltonian version of the theorem: it will be 
the ``springboard'' for our exposition of symplectic reduction.\footnote{For 
discussion of the Lagrangian version, cf. e.g. Brading and Castellani (this 
vol., ch. 13) or (restricted to finite-dimensional systems) Butterfield (2004a: 
Section 4.7). For an exposition of both versions that is complementary to this 
paper (and restricted to finite-dimensional systems), cf. Butterfield (2006). 
Brading and Castellani also bring out that, even apart from Noether's theorems 
in other branches of mathematics, there are other `Noether's theorems' about 
symmetries in classical dynamics; so the present theorem is sometimes called 
Noether's ``first theorem''. Note also (though I shall not develop this point)  
that symplectic structure can be seen in the classical solution space of 
Lagrange's equations, so that symplectic reduction can be developed in the 
Lagrangian framework; cf. e.g. Marsden and Ratiu (1999: p. 10, Sections 
7.2-7.5, and 13.5).} 

So I shall begin by briefly reviewing the Hamiltonian version in Section 
\ref{july05}. For the moment, suffice it to make four comments (in ascending 
order of importance for what follows):\\
\indent (i): Both versions are underpinned by the  theorems in elementary  
Lagrangian and Hamiltonian mechanics about cyclic (ignorable) coordinates and 
their corresponding conserved momenta.\footnote{Here we glimpse the long  
history of our subject: these theorems were of course clear to these subjects' 
founders. Indeed the strategy of exploiting a symmetry to reduce the number of 
variables occurs already in 1687, in Newton's solution  of the Kepler problem; 
(or more generally, the problem of two bodies exerting equal and opposite 
forces along the line between them). The symmetries are translations and 
rotations, and the corresponding conserved quantities are the linear and 
angular momenta. In what follows, these symmetries and quantities will provide 
us with several examples.}

\indent (ii): In fact, the Hamiltonian version of the theorem is stronger. This  
reflects the fact that the canonical transformations form a ``larger'' group 
than the point transformations. A bit more precisely: though the point 
transformations $q \rightarrow q'$ on the configuration space $Q$ induce 
canonical transformations on the phase space $\Gamma$ of the $q$s and $p$s, $q 
\rightarrow q', p \rightarrow p'$ , there are yet other canonical 
transformations which ``mix'' the $q$s and $p$s in ways that transformations 
induced by point transformations do not.

(iii): I shall limit our discussion to (a) time-independent Hamiltonians and 
(b) time-independent transformations. Agreed, analytical mechanics can be 
developed, in both Lagrangian and Hamiltonian frameworks, while allowing 
time-dependent dynamics and transformations. For example, in the Lagrangian 
framework, allowing velocity-dependent potentials and-or time-dependent 
constraints would  prompt one to use what is often called the  `extended 
configuration space' $Q \times \mathR$. And in the Hamiltonian framework, 
time-dependence prompts one to use an `extended phase space' $\Gamma \times 
\mathR$. Besides, from a philosophical viewpoint, it is important to consider 
time-dependent transformations: for they include boosts, which are central to 
the philosophical discussion of  spacetime symmetry groups, and especially of  
relativity principles. But beware: rough-and-ready statements about symmetry, 
e.g. that the Hamiltonian must be invariant under a symmetry transformation, 
are liable to stumble on these transformations.  To give the simplest example: 
the Hamiltonian of a free particle is just its kinetic energy, which can be 
made zero by transforming to the particle's rest frame; i.e. it is not 
invariant under boosts.\\
\indent So a full treatment of symmetry in Hamiltonian mechanics, and thereby 
of symplectic reduction, needs to treat time-dependent transformations---and to 
beware!  But I will set aside all these complications. Here it must suffice to 
assert, without any details, that the modern theory of symplectic reduction 
{\em does} cope with boosts; and more generally, with time-dependent dynamics 
and transformations.

\indent (iv): As we shall see in detail, there are three main ways in which the 
theory of symplectic reduction generalizes Noether's theorem. As one might 
expect, these three ways are intimately related to one another.\\
\indent \indent (a): Noether's theorem is ``one-dimensional'' in the sense that 
for each symmetry (a vector field of a special kind on the phase space), it 
provides a conserved quantity, i.e. a real-valued function on the phase space, 
whose value stays constant over time. So in particular, different components of 
a conserved vector quantity, such as total linear momentum, are treated 
separately; (in this example, the corresponding  vector fields generate 
translations in three different spatial directions). But in symplectic 
reduction, the notion of a {\em momentum map} provides a ``unified'' 
description of these different components.\\
\indent \indent (b): Given a symmetry, Noether's theorem enables us to confine 
our attention to the level surface of the conserved quantity, i.e. the 
sub-manifold of phase space on which the quantity takes its initial value: for 
the system's time-evolution is confined to that surface. In that sense, the 
number of variables we need to consider is reduced. But in symplectic 
reduction, we go further and form a {\em quotient space} from the phase space. 
That is, in the jargon of logic: we define on phase space an equivalence 
relation  (not in general  so simple as having a common value for a conserved 
quantity) and form the set of equivalence classes. In the jargon of group 
actions: we form the set of orbits. Passage to this quotient space can have 
various good technical, and even philosophical, motivations. And under good 
conditions, this set is itself a manifold with lower dimension.\\
\indent \indent (c): Hamiltonian mechanics, and so Noether's theorem, is 
usually formulated in terms of symplectic manifolds, in particular the 
cotangent bundle $T^*Q$ of the configuration space $Q$. (Section \ref{july05} 
will give details.) But in symplectic reduction, we often need a  (mild) 
generalization of the idea of a symplectic manifold, called a {\em Poisson 
manifold}, in which a bracket, with some of the properties of the Poisson 
bracket, is taken as the primitive notion. Besides, this is related to (b) in 
that we are often led to a Poisson manifold, and dynamics on it, by taking the 
quotient of a symplectic manifold (i.e. a phase space of the usual kind) by the 
action of a symmetry  group.

As comment (iv) hints, symplectic reduction is a large subject. So there are 
several motivations for expounding it. As regards physics, many of the ideas 
and results can be developed for finite-dimensional classical   systems (to 
which I will confine myself), but then generalized to infinite-dimensional 
systems. And in either the original or the generalized form, they underpin 
developments in quantum theories. So these ideas and results have formed part 
of the contemporary renaissance in classical mechanics; cf. (i) and (ii) at the 
end of Section \ref{whyclassmech}. 

As regards philosophy, symmetry is both a long-established focus for  
philosophical  discussion, and a currently active one: cf. Brading and 
Castellani (2003). But philosophical discussion of symplectic reduction  seems 
to have begun only recently, especially in some papers of Belot and Earman. 
This delay is presumably because the technical material is more sophisticated: 
indeed, the theory of symplectic reduction was cast in its current general form 
only in the 1970s. But as Belot and Earman emphasise, the philosophical 
benefits are worth the price of learning the technicalities. The most obvious 
issue is that symplectic reduction's device of quotienting a state space casts 
light on philosophical issues about whether two  apparently distinct but 
utterly indiscernible possibilities should be ruled to be one and the same. In 
Section \ref{rednintro}, I will follow Belot in illustrating this issue with 
``relationist'' mechanics. Indeed, I have  selected the topics for my 
exposition with an eye to giving philosophical readers the background needed 
for  some of Belot's discussions. His papers (which I will cite in Section 
\ref{rednintro}) make many judicious philosophical points, without burdening 
the reader with an exposition of technicalities: excellent stuff---but to fully 
appreciate the issues, one of course has to slog through the details.

Finally, in the context of this volume, symplectic reduction  provides some 
background for the Chapters on the representation of time in mechanics (Belot, 
this vol., ch. 2), and on the relations between classical and quantum physics 
(Landsman, this vol., ch. 5, especially Sections 4.3-4.5 and 6.5; Dickson, this 
vol., ch. 4).

The plan of the Chapter is as follows. I first review Noether's theorem in 
Hamiltonian mechanics as usually formulated, in Section \ref{july05}. Then I 
introduce the themes mentioned in (b) and (c) above, of quotienting a phase 
space, and Poisson manifolds (Section \ref{rednprospectus}); and illustrate 
these themes with ``relationist'' mechanics (Section \ref{relnalmechsexample}).

 Thereafter, I expound the basics of symplectic reduction: (confining myself to 
finite-dimensional  Hamiltonian mechanics). Section by Section, the plan will 
be as follows. Sections \ref{tool} and \ref{actionlg} review the modern 
geometry that will be needed. Section \ref{tool} is mostly about Frobenius' 
theorem, Lie algebras and Lie groups.\footnote{Its first two Subsections also 
provide some pre-requisites for Malament (this vol.).} Section \ref{actionlg} 
expounds Lie group actions. It ends with the central idea of the co-adjoint  
representation of a Lie group $G$ on the dual $\mg^*$ of its Lie algebra. This 
review enables us to better understand the motivations for Poisson manifolds 
(\ref{pmspreamble}); and then to exhibit examples, and prove some main 
properties (Section \ref{basics} onwards). Section \ref{symyredn} applies this 
material to symmetry and conservation in mechanical systems. In particular, it 
expresses conserved quantities as momentum maps, and proves Noether's theorem 
for Hamiltonian mechanics on Poisson manifolds. Finally, in Section \ref{redn}, 
we prove one of the several main theorems about symplectic reduction. It 
concerns the case where the natural configuration space for a system is itself 
a Lie group $G$: this occurs both for the rigid body and ideal fluids.  In this 
case, quotienting the natural phase space (the cotangent bundle on $G$) gives a 
Poisson manifold that ``is'' the dual $\mg^*$ of $G$'s Lie algebra.\footnote{In 
this endeavour, my sources are four books by masters of the subject: Abraham 
and Marsden (1978), Arnold (1989), Marsden and Ratiu (1999) and Olver (2000). 
But again, be warned: my selection is severe, as anyone acquainted with these 
or similar books will recognize.}

To sum up:--- The overall effect of this exposition is, I hope, to illustrate 
this Chapter's mottoes: that classical mechanics is alive and kicking, not 
least through deepening our understanding of time-honoured systems such as the 
rigid body---whose analysis in traditional textbooks can be all too confusing!

\section{Symplectic reduction: an overview}\label{rednintro}

We begin by briefly reviewing Hamiltonian mechanics and Noether's theorem, in 
Section \ref{july05}.\footnote{For more details about differential geometry, 
cf. Sections \ref{prelimtool} and \ref{lgla}. For more details about the 
geometric formulation of mechanics, cf. Arnold (1989) or Marsden and Ratiu 
(1999); or Singer (2001) (more elementary than this exposition) or Abraham and 
Marsden (1978) (more advanced); or Butterfield (2006) (at the same level). Of  
many good textbooks of mechanics, I admire especially Desloge (1982) and Johns 
(2005).} This prepares us for the idea of symplectic reduction, Section 
\ref{rednprospectus}: which we then illustrate using ``relationist'' mechanics, 
Section \ref{relnalmechsexample}.

\subsection{Hamiltonian mechanics and Noether's theorem: a 
review}\label{july05}
\subsubsection{Symplectic manifolds; the cotangent bundle as a symplectic 
manifold}\label{cotgtblesymp}
A {\em symplectic structure} or {\em symplectic form} on a manifold $M$ is 
defined to be a differential 2-form $\o$ on $M$ that is closed (i.e. its 
exterior derivative ${\bf d} \o$ vanishes) and is non-degenerate. That is: for 
any $x \in M$, and any two tangent  vectors at $x$, $\sigma, \tau \in T_x$:
\be
{\bf d}\o = 0 \;\; \mbox{ and } \;\; \forall \; \tau \neq 0, \;\; \exists 
\sigma: \;\;\; \o(\tau,\sigma) \neq 0 \;\; .
\label{definesympstruc} 
\ee 
Such a pair $(M, \o)$ is called a {\em symplectic  manifold}. There is a rich 
theory of symplectic manifolds; but we shall only need a small fragment of it. 
(In particular, the fact that we mostly avoid the theory of canonical 
transformations means we will not need the theory of Lagrangian sub-manifolds.)

\indent First, it follows from the non-degeneracy of $\o$ that $M$ is 
even-dimensional. The reason lies in a theorem of linear algebra, which one 
then applies to the tangent space at each point of $M$. Namely, for any 
bilinear form $\o: V \times V \rightarrow \mathR$: if $\o$ is antisymmetric of 
rank $r \leq m \equiv {\rm{dim}}(V)$, then $r$ is even. That is: $r = 2n$ for 
some integer $n$, and there is a basis $e_1,...,e_i,...,e_m$ of $V$ for which 
$\o$ has a simple expansion as wedge-products 
\be
\o \; = \; \Sigma^n_{i= 1} \; e^i \wedge e^{i+n} \;\; ;
\label{omegawiths}
\ee  
equivalently, $\o$ has the $m \times m$ matrix 
\be
\o \;  = \; \left(\begin{array}{ccc}
{\bf 0} & {\bf 1} & {\bf 0} \\
{\bf {-1}} & {\bf 0} & {\bf 0} \\
{\bf {0}} & {\bf 0} & {\bf 0} \\
\end{array}\right) \;\; .
\label{eq;gramschmidt}
\ee
where ${\bf 1}$ is the $n \times n$ identity matrix, and similarly for the zero 
matrices of various sizes. This {\em normal form} of antisymmetric bilinear 
forms is an analogue of the Gram-Schmidt theorem that an inner product space 
has an orthonormal basis, and is proved by an analogous argument.\\
\indent So if an antisymmetric  bilinear form is non-degenerate, then $r \equiv 
2n = m$. That is: eq. \ref{eq;gramschmidt} loses its bottom row and right 
column consisting of zero matrices, and  reduces to the $2n \times 2n$ {\em 
symplectic matrix} ${\o}$ given by
\be
\o : = \left(\begin{array}{cc}
{\bf 0} & {\bf 1} \\
{\bf {-1}} & {\bf 0} \\
\end{array}\right) \;\; .
\label{eq;defineOmega}
\ee

\indent Second, the non-degeneracy of $\o$ implies that at any $x \in M$, there 
is a basis-independent isomorphism $\o^{\fl}$ from the tangent space $T_x$ to 
its dual $T^*_x$.  Namely: for any $x \in M$ and $\tau \in T_x$, the value of 
the 1-form $\o^{\fl}(\tau) \in T^*_x$ is defined by
\be
\o^{\fl}(\tau)(\sigma) := \o(\sigma,\tau) \;\;\; \forall \sigma \in T_x \; .
\label{symimpliescanlisomm}
\ee
This also means that a symplectic structure enables a covector field, i.e. a 
differential one-form, to determine a vector field. Thus for any function $H: M 
\rightarrow \mathR$, so that $dH$ is a differential 1-form on $M$, the inverse 
of $\o^{\fl}$ (which we might write as $\o^{\sh}$), carries $dH$ to a vector 
field on $M$, written $X_H$. This is the key idea whereby in Hamiltonian 
mechanics,  a scalar function $H$ determines a dynamics; cf. Section 
\ref{geomHamEq}. 

So far, we have noted some implications of $\o$ being non-degenerate. The other 
part of the definition of a symplectic form (for a manifold), viz. $\o$ being 
closed, ${\bf d} \o = 0$, is also important. We shall see in Section 
\ref{noetcomplete} that it implies that a vector field $X$ on a symplectic 
manifold $M$ preserves the symplectic form $\o$ (i.e.  in more physical jargon: 
generates (a one-parameter family of) canonical transformations) iff $X$ is 
Hamiltonian in the sense that there is a scalar function $f$ such that  $X = 
X_f \equiv \o^{\sh}(df)$. Or in terms of the Poisson bracket, with $\cdot$ 
representing the argument place for a scalar function: $ X(\cdot) = X_f(\cdot) 
\equiv \{\cdot, f\}$.

So much by way of introducing symplectic manifolds. I turn to showing that any 
cotangent bundle $T^*Q$ is such a manifold. That is: it has, independently  of 
a choice of coordinates or bases, a symplectic structure.

Given a manifold $Q$ (dim($Q$)=$n$) which we think of as the system's 
configuration space, choose any local coordinate system $q$ on $Q$ , and  the 
natural local coordinates $q,p$ thereby induced on $T^*Q$. We define the 2-form 
\be
dp \wedge dq := dp_i \wedge dq^i := \Sigma^n_{i=1} dp_i \wedge dq^i \; .
\label{defineomega}
\ee
In fact, eq. \ref{defineomega} defines the same 2-form, whatever choice we make 
of the chart $q$ on $Q$. For $dp \wedge dq$ is the exterior derivative of a 
1-form on $T^*Q$ which is  defined naturally (i.e. independently of coordinates 
or bases) from the derivative (also known as: tangent) map of the projection
\be
\pi: (q,p) \in T^*Q  \mapsto q \in Q .
\ee 
Thus consider a tangent vector $\tau$ (not to $Q$, but) to the cotangent bundle 
$T^*Q$  at a point $\eta = (q,p) \in T^*Q,$ i.e. $q \in Q$ and $p \in T^*_q$. 
Let us write this as: $\tau \in T_{\eta}(T^*Q) \equiv T_{(q,p)}(T^*Q)$. The 
derivative map, $D \pi$ say,  of the natural projection $\pi$ applies to 
$\tau$:
\be
D\pi: \tau \in T_{(q,p)}(T^*Q) \mapsto (D\pi(\tau)) \in T_q \; \; .
\label{derivprojectionpi}
\ee
Now define a 1-form $\theta_H$ on $T^*Q$ by
\be
\theta_H: \tau \in  T_{(q,p)}(T^*Q) \mapsto p(D\pi(\tau)) \in \mathR \; ;
\label{definethetaH}
\ee 
where in this definition  of $\theta_H$, $p$ is defined to be the second 
component of $\tau$'s base-point $(q,p) \in T^*Q$; i.e. $\tau \in 
T_{(q,p)}(T^*Q)$ and $p \in T^*_q$.

This 1-form is called the {\em canonical 1-form} on $T^*Q$. One now checks that 
in any natural local coordinates $q,p$, $\theta_H$ is given by 
\be
\theta_H = p_i dq^i.
\label{definethetaHagain}
\ee
Finally, we define a 2-form by taking the exterior derivative of $\theta_H$:
\be
 {\bf d}(\theta_H) := {\bf d}(p_i dq^i) \equiv dp_i \wedge dq^i \; .
\label{defineDthetaH}
\ee
One checks that this 2-form is closed (since ${\bf d}^2 = 0$) and 
non-degenerate. So $(T^*Q, {\bf d}(\theta_H))$ is a symplectic manifold. 
Accordingly,  ${\bf d}(\theta_H)$, or its negative $-{\bf d}(\theta_H)$, is 
called the {\em canonical symplectic form}, or {\em canonical 2-form}.

There is a theorem (Darboux's theorem) to the effect that locally, any 
symplectic manifold ``looks like'' a cotangent bundle: or in other words,   a 
cotangent bundle is locally a ``universal'' example of  symplectic structure. 
We will not go into details; but in Section \ref{DarbouxPoiss}, we will  
discuss the generalization of this theorem for Poisson manifolds. But first we 
review, in the next two Subsections, Hamilton's equations, and Noether's 
theorem.  

\subsubsection{Geometric formulations of Hamilton's equations}\label{geomHamEq}
As we already emphasised, the main geometric idea behind Hamilton's equations 
is that a gradient, i.e. covector, field $dH$ determines a vector field $X_H$. 
So to give a geometric formulation of Hamilton's equations at a point $x = 
(q,p)$ in a cotangent bundle $T^*Q$, let us write $\o^{\sh}$ for the 
(basis-independent) isomorphism from the cotangent space to the tangent space, 
$T^*_x \rightarrow T_x$, induced by $\o := - {\bf d}(\theta_H) = dq^i \wedge 
dp_i$ (cf. eq.  \ref{symimpliescanlisomm}). Then Hamilton's equations may be 
written as:
\be
{\dot x} = X_H (x) = \o^{\sh} ({\bf d}H(x)) = \o^{\sh} (dH(x)) \;\; .
\label{HEgeomic1}
\ee
There are various other formulations. Applying $\o^{\fl}$, the inverse 
isomorphism $T_x \rightarrow T^*_x$, to  both sides, we get
\be
\o^{\fl} X_H (x) = dH(x) \;\; .
\label{HEgeomic2}
\ee
In terms of the symplectic form $\o$ at $x$, this is: for all vectors $\tau \in 
T_x$
\be
\o(X_H(x), \tau) = dH(x) \cdot \tau \;\; ;
\label{HEgeomic3}
\ee
or in terms of the contraction (also known as: interior product) ${\bf i}_X 
\al$ of a differential form $\al$ with a vector field $X$, with $\cdot$ marking 
the argument place of $\tau \in T_x$:
\be
{\bf i}_{X_H} \o :=  \o(X_H(x), \cdot) = dH(x)(\cdot) \;\; .
\label{HEgeomic4}
\ee
More briefly, and now written for any function $f$, it is:
\be
{\bf i}_{X_f} \o = df \; .
\label{HEgeomic4forf}
\ee
Finally, recall the relation between the Poisson bracket and the directional 
derivative  (or the Lie derivative  $\cal L$) of a function: viz.
\be
{\cal L}_{X_f} g = dg(X_f) = X_f(g) = \{ g, f \} \; .
\label{Liepbbasic}
\ee
Combining this with eq. \ref{HEgeomic4forf}, we can state the relation  between 
the symplectic form and Poisson bracket in the form: 
\be
\{g, f \} = dg(X_f) = {\bf i}_{X_f} dg = {\bf i}_{X_f} ({\bf i}_{X_g} \o) = 
\o(X_g,X_f) \; .
\label{sympPBgeomic}
\ee

\subsubsection{Noether's theorem}\label{noetcomplete}
The core idea of Noether's theorem, in both the Lagrangian and Hamiltonian 
frameworks, is that to every continuous symmetry of the system there 
corresponds a conserved quantity (a first integral, a constant of the motion). 
The idea of a continuous symmetry is made precise along the following lines: a 
symmetry is a vector field on the state-space that (i) preserves the Lagrangian 
(respectively, Hamiltonian) and (ii) ``respects'' the structure of the 
state-space.    

In the Hamiltonian framework, the heart of the proof is a ``one-liner'' based 
on the fact that the Poisson bracket is  antisymmetric. Thus for any scalar 
functions $f$ and $H$ on a symplectic manifold $(M, \o)$ (and so with a Poisson 
bracket given by eq. \ref{sympPBgeomic}), we have that at any point $x \in M$ 
\be
X_f(H)(x)  \equiv \{H, f\}(x) = 0 \;\;\;\; {\mbox{ iff }} \;\; \;\;
0 = \{f,H\}(x) \equiv X_H(f)(x) \; \; .
\label{naivenoetham} 
\ee
In words: around $x$, $H$ is constant under the flow of the vector field $X_f$ 
(i.e. under what the evolution would be if $f$ was the Hamiltonian) iff $f$ is 
constant under the flow $X_H$. Thinking of $H$ as the physical Hamiltonian, so 
that $X_H$ represents the real time-evolution (sometimes called: the dynamical 
flow), this means: around $x$, $X_f$ preserves the Hamiltonian iff $f$ is 
constant under time-evolution, i.e. $f$ is a conserved quantity (a constant of 
the motion).

But we need to be careful about clause (ii) above: the idea that a vector field 
respects'' the structure of the state-space. In the Hamiltonian framework, this 
is made precise as preserving the symplectic form. Thus we define a vector 
field $X$ on a symplectic manifold $(M,\o)$ to be {\em symplectic} (also known 
as: {\em canonical}) iff  the Lie-derivative along $X$ of  the symplectic form 
vanishes, i.e. ${\cal L}_X \o = 0$. (This definition is equivalent to $X$'s 
generating (active) canonical transformations, and to its preserving the 
Poisson bracket. But I will not go into details about these equivalences: for 
they belong to the theory of canonical transformations, which, as mentioned, I 
will not need to develop.)

We also define a {\em Hamilton system} to be a triple $(M, \o, H)$ where 
$(M,\o)$ is a symplectic manifold and $H: M \rightarrow \mathR$, i.e. $M \in 
{\cal F}(M)$. And then we define a (continuous) {\em symmetry} of a Hamiltonian  
system to be a vector field $X$ on $M$ that:\\
\indent (i) preserves the Hamiltonian function, ${\cal L}_X H = 0$; and \\
\indent (ii) preserves the symplectic form, ${\cal L}_X \o = 0$. 

These definitions mean that to prove Noether's theorem from eq. 
\ref{naivenoetham}, it will suffice to prove that a vector field $X$ is 
symplectic iff it is locally of the form $X_f$. Such a vector field is called 
{\em locally Hamiltonian}. (And a vector field is called {\em Hamiltonian} if 
there is a global scalar $f: M \rightarrow \mathR$ such that $X = X_f$.) In 
fact, two results from the theory of differential forms, the Poincar\'{e} Lemma 
and  Cartan's magic formula, make it easy to prove this; (for a vector field on 
any symplectic manifold $(M,\o)$, i.e. $(M,\o)$ does not need to be a cotangent 
bundle). 

Again writing $\bf d$ for the exterior derivative, we recall that a $k$-form 
$\al$ is called:\\
\indent (i):  {\em exact} if there is a $(k-1)$-form $\bb$ such that $\al = 
{\bf d}\bb$; (cf. the elementary definition of an exact differential);\\
\indent (ii): {\em  closed} if ${\bf d}\al = 0$.\\
The {\em Poincar\'{e} Lemma} states that every closed form is locally exact.  
To be precise:  for any open set $U$ of $M$, we define the vector space 
$\O^k(U)$ of $k$-form fields on $U$. Then the  Poincar\'{e} Lemma states that 
if $\al \in \O^k(M)$ is closed, then at every $x \in M$ there is a 
neighbourhood $U$ such that $\al\mid_U \; \in \; \O^k(U)$ is exact.

{\em Cartan's magic formula} is a useful formula (proved by straightforward 
calculation) relating the Lie derivative, contraction and the exterior 
derivative. It says that if $X$ is a vector field and $\al$ a $k$-form on a 
manifold $M$, then the Lie derivative of $\al$ with respect to $X$ (i.e. along 
the flow of $X$) is
\be
{\cal L}_X \al = {\bf d}{\bf i}_X \al + {\bf i}_X{\bf d}\al \;\; .
\label{magic}
\ee

We now argue as follows. Since $\o$ is closed, i.e. ${\bf d}\o = 0$, Cartan's 
magic formula, eq. \ref{magic}, applied to $\o$ becomes
\be
{\cal L}_X \o \equiv {\bf d}{\bf i}_X \o + {\bf i}_X{\bf d}\o  =  {\bf d}{\bf 
i}_X \o \;\; .
\label{magicforomega}
\ee
So for $X$ to be symplectic is for ${\bf i}_X \o$ to be closed. But by the 
Poincar\'{e} Lemma, if ${\bf i}_X \o$ closed, it is locally exact. That is: 
there locally exists a scalar function $f: M \rightarrow \mathR$ such that 
\be
{\bf i}_X \o = df \;\; {\rm{i.e.}} \;\; X = X_f \; .
\ee
So for $X$ to be symplectic is equivalent to $X$ being locally Hamiltonian. 

Thus we have
\begin{quote}
{\bf {Noether's theorem for a Hamilton system}} If $X$ is a symmetry of a 
Hamiltonian system $(M, \o, H)$, then locally $X = X_f$; so by the 
anti-symmetry of the Poisson bracket, eq. \ref{naivenoetham}, $f$ is a constant 
of the motion. And conversely: if $f: M \rightarrow \mathR$ is a constant of 
the motion, then $X_f$ is a symmetry.
\end{quote}

We will see in Section \ref{mommmapsnoet} that most of this approach to 
Noether's theorem, in particular the ``one-liner'' appeal to the anti-symmetry 
of the Poisson bracket, eq. \ref{naivenoetham}, carries over to the more 
general framework of Poisson manifolds. For the moment, we mention an example 
(which we will also return to). 

For most Hamiltonian systems in euclidean space $\mathR^3$, spatial 
translations and rotations are (continuous) symmetries.  Let us consider in 
particular a system we will discuss in more detail in Section 
\ref{relnalmechsexample}: $N$ point-particles interacting by Newtonian gravity. 
The Hamiltonian is a sum of two terms, which are each individually invariant 
under translations and rotations:\\
\indent  (i) a kinetic energy term $K$; though I will not go into details, it 
is in fact defined by the euclidean metric of $\mathR^3$, and is thereby 
invariant; and\\
\indent  (ii) a potential energy term $V$; it depends only on the particles' 
relative distances, and is thereby invariant.\\
\indent The corresponding conserved quantities are the total linear and angular 
momentum.\footnote{By the way, this Hamiltonian is {\em not} invariant under 
boosts. But as I said in (iii) of Section \ref{introprospectus}, I restrict 
myself to time-independent transformations; the treatment of symmetries that 
``represent the relativity of motion'' needs separate discussion.}

\subsection{The road ahead}\label{rednprospectus}
In this Subsection, four comments will expand on the introductory comment (iv) 
of Section \ref{introprospectus}, and also give some information about the 
history of symplectic reduction and about some crucial examples.

(1): {\em Generalizing from Noether's theorem; Poisson manifolds}:---\\
Noether's theorem tells us that a continuous symmetry, i.e. a one-parameter 
group of symmetries, determines a first integral (i.e. a constant of the 
motion). So a larger group of symmetries, i.e. a group with several parameters, 
implies several first integrals. The phase flow is therefore confined to the 
intersection of the level surfaces of these integrals: an intersection which is 
in general a manifold. In other words: the simultaneous level manifold of these 
integrals is an invariant manifold of the phase flow. 

It turns out that, in many useful cases,  this manifold is also invariant under 
an appropriately chosen subgroup of the group of symmetries; and that the 
quotient space, i.e. the set of orbits under the action of this subgroup, is a 
manifold with a natural structure induced by the original Hamiltonian system 
that is sufficient to do mechanics in Hamiltonian style. The quotient space is 
therefore called the `reduced phase space'. 

But in some cases, this natural structure is  not a symplectic form, but a 
(mild) generalization in which the the form is allowed to be degenerate; i.e. 
like eq. \ref{eq;gramschmidt} rather than eq. \ref{eq;defineOmega}. A manifold 
equipped with such a structure need not be a quotient manifold. It can instead 
be defined in terms of a generalization of the usual Poisson bracket, as 
defined in terms of the symplectic form by eq. \ref{sympPBgeomic}.\\
\indent The key idea is to postulate a bracket, acting on the scalar functions 
$F: M \rightarrow \mathR$ on {\em any} manifold $M$, and possessing four 
properties enjoyed by the usual Poisson bracket. One of the properties is 
anti-symmetry, emphasised in Section \ref{noetcomplete}'s proof of Noether's 
theorem. The other three are that the postulated bracket, again written $\{ , 
\}$, is: to be bilinear; to obey the Jacobi identity  for any real functions 
$F,G,H$ on $M$, i.e.
\be
\{\{F, H \}, G \} + \{\{G, F \}, H \} + \{\{H, G \}, F \}= 0 \;\; ;
\label{Jacobiidentity2nd}
\ee
and to obey Leibniz' rule for products, i.e.
\be
\{F, H \cdot G\} = \{F, H\}\cdot G + H \cdot \{F, G\} \;\; . 
\label{Pbofproduct2nd}
\ee
We will see in Section \ref{pms} that such a bracket, again called `Poisson 
bracket',  provides a sufficient framework for mechanics in Hamiltonian style. 
In particular, it induces an anti-symmetric bilinear  form that may be 
degenerate, as in eq. \ref{eq;gramschmidt}. A manifold $M$ equipped with such a 
bracket is called a {\em Poisson manifold}.

The allowance of degeneracy means that a Poisson manifold  can have {\em odd} 
dimension; while we saw in Section \ref{cotgtblesymp} that any symplectic 
manifold  is even-dimensional. On the other hand, this generalized Hamiltonian 
mechanics  will have clear connections with the usual formulation of Section 
\ref{july05}.  The main connection will be the result that any Poisson  
manifold $M$  is a disjoint union of even-dimensional  manifolds, on which 
$M$'s degenerate antisymmetric bilinear form restricts to be 
non-degenerate.\footnote{Because of these clear connections, it is natural to 
still call the more general framework `Hamiltonian'; as is usually done. But of 
course this is just a verbal matter.}

(2): {\em Historical roots}:---\\
 The theory of symplectic reduction has deep historical roots in the  work of 
classical mechanics' monumental figures. In part, this is no surprise. As 
mentioned in (i) of Section \ref{introprospectus}, cyclic coordinates underpin 
the role of symmetry in mechanics, and in particular Noether's theorem. And  
Newton's solution of the Kepler problem provides an example: witness textbooks' 
expositions of the transition to centre-of-mass coordinates, and of polar 
coordinates with the angle being cyclic (yielding angular momentum as the 
conserved quantity).  So it is unsurprising that various results and ideas of 
symplectic reduction  can be seen in the work of such masters as Euler, 
Lagrange, Hamilton, Jacobi, Lie and Poincar\'{e}; for example (as we will see), 
in Euler's theory of the rigid body.\\
\indent But the history also holds a surprise. It turns out that Lie's 
epoch-making work on Lie groups already contained a detailed development of 
much of the general, modern theory.\footnote{The main source is his (1890). 
Besides, Arnold (1989: 456) reports that the prototype example of a Poisson 
manifold, viz. the dual of a finite-dimensional Lie algebra, was already 
understood by Jacobi.} The sad irony is that most of Lie's insights were not 
taken up---and were then repeatedly re-discovered. So this is yet another 
example (of which the history of mathematics has so many!)  of the saying that 
he who does not learn from history is doomed to repeat it. The consolation is 
of course  that it is often easier, and more fun, to re-discover something than 
to learn it from another...  \\
\indent Thus it was only from the mid-1960s that the theory, in essentially the 
form Lie had, was recovered and cast in the  geometric language adopted by 
modern mechanics; namely, by contemporary masters such as Arnold, Kostant, 
Marsden, Meyer, Smale, Souriau and Weinstein; (cf. this Chapter's first motto). 
Happily, several of these modern authors are scholars of the history, and even 
their textbooks give some historical details: cf. Marsden and Ratiu (1999, pp. 
336-8, 369-370, 430-432), and the notes to each Chapter of Olver (2000: 
especially  p.427-428). (Hawkins (2000) is a full history of Lie groups from 
1869 to 1926; for Lie, cf. especially its Sections 1.3, 2.5 and Chapter 3, 
especially 3.2.)\\
\indent In any case, setting history aside: symplectic reduction has continued 
since the 1970s to be an active research area  in contemporary mechanics, and 
allied fields such as symplectic geometry. So it has now taken its rightful 
place as a major part of the contemporary renaissance of classical mechanics: 
as shown by ... 

(3): {\em Two examples: the rigid body and the ideal fluid}:---\\
Two examples  illustrate vividly how symplectic reduction can give new physical 
understanding, even of time-honoured examples: the rigid body and the ideal 
fluid---as attested by this Chapter's mottoes. (Section 
\ref{relnalmechsexample} will develop a third example, more closely related  to 
philosophy.)

\indent As to the rigid body: we will see (especially  in Section \ref{pms}) 
that symplectic reduction considerably clarifies  the elementary theory of the 
rigid body (Euler's equations, Euler angles etc.): which, notoriously, can be 
all too confusing! For simplicity, I shall take the rigid body to be pivoted, 
so as to set aside translational motion. This will mean that the group of 
symmetries defining the quotienting procedure will be the rotation group. It 
will also mean that the rigid body's configuration space is given by the 
rotation group, since any configuration can be labelled by the rotation that 
obtains it from some reference-configuration. So in this application of 
symplectic reduction, the symmetry group (viz. the rotation group) will act on 
{\em itself} as the configuration space. This example will also give us our 
prototype example of a Poisson manifold. 

\indent As to the ideal fluid, i.e. a fluid that is incompressible and inviscid 
(with zero viscosity): this is of course an infinite-dimensional system, and so 
(as I announced in Section \ref{introprospectus}) outside the scope of this 
Chapter. So I will not go into any details, but just report the main idea.\\
\indent The equations of motion of an ideal fluid, Euler's equations, are 
usually derived either by applying Newton's second law ${\bf F} = m{\bf a}$ to 
a small fluid element; or by a heuristic use of the Lagrangian or Hamiltonian 
approach (as in heuristic classical field theories). But in the mid-1960s, 
Arnold showed how the latter derivations could be understood in terms  of a 
striking, even beautiful, analogy with the above treatment of the rigid body. 
Namely, the analogy shows that the configuration space of the fluid is an 
infinite-dimensional group; as follows. The configuration of an ideal fluid 
confined to some container occupying a volume $V \subset \mathR^3$ is an 
assignment to each spatial   position $x \in V$ of an infinitesimal fluid 
element. Given such an assignment as a reference-configuration, any other 
configuration can be labelled by the volume-preserving diffeomorphism $d$ from 
$V$ to $V$ that carries the reference-configuration to the given one, by 
dragging each fluid element along by $d$. So given a choice of 
reference-configuration, the fluid's  configuration space is given by the 
infinite-dimensional group $\cal D$ of diffeomorphisms $d: V \rightarrow V$: 
just as the rotation group is the configuration space of a (pivoted) rigid 
body. $\cal D$ then forms the basis for rigorous Lagrangian and Hamiltonian 
theories of an ideal fluid.

\indent These theories turn out to have considerable analogies with the 
Lagrangian and Hamiltonian theories of the rigid body, thanks to the fact that 
in both cases the symmetry group forms the configuration space. In particular, 
Euler's equations for ideal fluids are the analogues of Euler's equations for a 
rigid body. Besides, these rigorous theories of fluids (and symplectic 
reduction applied to them) are scientifically important: they have yielded 
various general theorems, and solved previously intractable problems. (For  
more details, cf. Abraham and Marsden (1978: Sections 4.4 and 4.6 for the rigid 
body, and 5.5.8 for the ideal fluid), Arnold (1989: Appendix 2:C to 2:F for the 
rigid body, and 2:G to 2:L for the ideal fluid), and  Marsden and Ratiu (1999: 
Chapters 1.4 and 15 for the rigid body, and 1.5, p. 266, for the ideal fluid).)

(4): {\em Philosophical importance}:---\\
Symplectic reduction is also, I submit, philosophically important; in at least 
two ways. The first way is specific: it illustrates some methodological morals 
about how classical mechanics analyses problems. I develop this theme in 
(Butterfield 2005). The second way is more general: the theory, or rather 
various applications of it, is directly relevant to disputes in the philosophy  
of space and time, and of mechanics. This relevance is recognized in 
contemporary philosophy of physics. So far as I know, the authors who develop 
these connections in most detail are Belot and Earman. They discuss symplectic 
reduction in connection with such topics as:\\
\indent (i) the treatment of symmetries, including gauge symmetries;\\
\indent  (ii) the dispute between absolute and relationist conceptions of space 
and time; and\\
\indent  (iii) the interpretation  of classical general relativity  (a topic 
which  connects (i) and (ii), and bears on heuristics for quantum gravity).\\
Thus Belot (1999, 2000, 2001, 2003, 2003a) and Earman (2003) discuss mainly (i) 
and-or (ii); Belot and Earman (2001) discusses (iii). For (i) and (ii), I also 
recommend Wallace (2003).

But these papers have  a demanding pre-requisite: they invoke, but do not 
expound, the theory of symplectic reduction. They also discuss 
infinite-dimensional systems (especially classical electromagnetism and general 
relativity), without developing finite-dimensional examples like the rigid 
body.  Indeed,  there is, so far as I know, no  article-length exposition of 
the theory which is not unduly forbidding for philosophers. So I aim to give 
such an exposition, to help  readers of papers such as those cited.\footnote{As 
I said in Section \ref{introprospectus}, my material is drawn from the books by 
Abraham and Marsden, Arnold, Marsden and Ratiu, and Olver. More precisely, I 
will mostly draw on: Abraham and Marsden (1978: Sections 3.1-3.3, 4.1-4.3), 
Arnold (1989: Appendices 2, 5 and 14), Marsden and Ratiu (1999: Chapters 9-13) 
and Olver (2000: Chapter 6). And much of what follows---in spirit, and even in 
letter---is already in Lie (1890)! As a (non-philosophical) introduction to 
symplectic reduction, I also recommend Singer (2001). It is at a yet more 
elementary level than what follows; e.g. it omits Poisson manifolds and  
co-adjoint representations.}

As an appetizer for this exposition, I will first (in Section 
\ref{relnalmechsexample}) follow Belot in presenting the general features of a 
finite-dimensional symplectic reduction which has vivid philosophical 
connections, viz. to the absolute vs. relationist   debate. This example 
concerns a system of point-particles in Euclidean space, either moving freely 
or interacting by a force such as Newtonian gravity. (The symmetries defining 
the quotienting procedure are given by the Euclidean group of translations and 
rotations.) For philosophers, this will be a good appetizer for symplectic 
reduction, since it sheds considerable light on relationism about space of the 
sort advocated by Leibniz and Mach. 

\subsection{Appetizer: Belot on relationist 
mechanics}\label{relnalmechsexample}
\subsubsection{Comparing two quotienting procedures}\label{compare2quots}
In several papers, Belot discusses how symplectic reduction bears on the 
absolute-vs.-relational debate about space. I shall pick out one main theme of 
his discussions: the comparison of a relational classical mechanical theory  
with what one gets by quotienting the orthodox absolutist (also called a 
`substantivalist') classical mechanics, by an appropriate symmetry group. His 
main contention---which I endorse---is that this comparison sheds considerable 
light on relationism: on both its motivation, and its advantages and 
disadvantages.\footnote{The main references are Belot (1999, 2001, 2003: 
Sections 3.5, 5). Cf. also his (2000: Sections 4 to 5.3), (2003a: Section 6). 
Though I recommend all these papers, the closest template for what follows is 
(2001: Section VI et seq.).} 

Belot's overall idea is as follows. Where the relationist admits one possible 
configuration, as (roughly) a specification of all the distances (and thereby 
angles) between all the parts of matter, the absolutist (or substantivalist) 
sees an infinity of possibilities: one for each way the relationist's 
configuration (a {\em relative configuration}) can be embedded in the absolute 
space. 
This makes it natural to take the relationist to be envisaging a mechanics 
which is some sort of ``quotient'' of the absolutist's mechanics.\\
\indent In particular, on the traditional conception of space as Euclidean 
(modelled by $\mathR^3$), each of the relationist's relative configurations 
corresponds to an equivalence class of absolutist configurations (i.e. 
embeddings of arrangements of matter into $\mathR^3$), with the members of the 
class related by spatial translations and rotations, i.e. elements of the 
Euclidean group. In the jargon of group actions, to be developed in Section 
\ref{actionlg}: the Euclidean group acts on the set of all  absolutist 
configurations, and a relative configuration corresponds to an orbit of this 
action. So it is natural to take the relationist to be envisaging a mechanics 
which quotients this action of the Euclidean group, to get a {\em relative 
configuration space}. A relationist mechanics, of Lagrangian or Hamiltonian 
type, is then to be built up on this space of relative configurations.  \\
\indent But as Belot emphasises, one can instead consider quotienting the 
absolutist's state-space---i.e. in a Hamiltonian framework, the phase  
space---rather than their  configuration space. Indeed, this is exactly what 
one does in symplectic reduction. In particular, the Euclidean group's action 
on the absolutist's configuration space, $Q$ say, can be lifted to give an 
action on the cotangent bundle $T^*Q$; which is accordingly called the 
`cotangent lift'. One can then take the quotient, i.e. consider the orbits into 
which $T^*Q$ is partitioned by the cotangent  lift.\\
\indent We thus have two kinds of theories to compare: (i) the relationist 
theories, built up from the relative configuration space; which for the sake of 
comparison with symplectic reduction we take to be Hamiltonian, rather than 
Lagrangian; (ii) theories obtained by quotienting ``later'', i.e. quotienting 
the absolutist's cotangent bundle.

I will now spell out this comparison. But I will not try to summarize Belot's 
more detailed conclusions, about what such a comparison reveals about the 
advantages and disadvantages of relationism. They are admirably subtle, and so 
defy summary: they can mainly be found at his (2000: p. 573-574, 582; 2001: 
Sections VIII to X). (Rovelli (this volume) also discusses relationism.)
    
\indent As befits an appetizer, I will also (like Belot) concentrate on as 
simple a case  as possible: a mechanics of $N$ point-particles, which is to 
assume a Euclidean spatial  geometry. Of course, the absolutist make this 
assumption by postulating a Euclidean space; but for the relationist, the 
assumption is encoded in constraints relating the various inter-particle 
distances. The main current example of a relationist mechanics of such a system 
is due to Barbour and Bertotti (1982), though they develop it in the Lagrangian 
framework; (to be precise, in terms of Jacobi's principle). Belot also 
discusses other relational theories, including field theories, i.e. theories  
of infinite systems; some of them also due to Barbour, and in a Lagrangian 
framework. But in this Section I only consider $N$ point-particles.\\
\indent Also, I will also not discuss boosts, though of course the relationist 
traditionally proposes to identify any two absolutist states of motion related 
by a boost. In terms of group actions, this means I will consider quotienting 
by an action of the euclidean group, but not the Galilei group. (Cf. how I set 
aside time-dependent transformations already in (iii) of Section 
\ref{introprospectus}.) I will also postpone  to later Sections technical 
details, even when our previous discussion makes them accessible.\\
\indent Finally, a warning to avoid later disappointment! The later Sections 
will not include a full analysis of the euclidean group's actions on 
configuration space and phase space, and their quotients. That would involve  
technicalities going beyond an appetizer. Instead (as mentioned at the end of 
Section \ref{introprospectus}), the material in later Sections is chosen so as 
to lead up to Section \ref{redn}'s theorem, the Lie-Poisson reduction theorem, 
about quotienting the phase space of a system whose configuration space is a 
Lie group. Further reasons for presenting the material for this theorem will be 
given in Section \ref{pmspreamble}. 

\subsubsection{The spaces and group actions introduced}\label{spacesactions}
Let us begin by formulating the orthodox absolutist mechanics of $N$ 
point-particles interacting by Newtonian gravity, together with the action of 
the Euclidean group.

 Each point-particle occupies a point of $\mathR^3$, so that the configuration 
space $Q$ is $\mathR^{3N}$: dim($Q$) = 3$N$. So the phase space for Hamiltonian 
mechanics will be the cotangent bundle $T^*Q \ni (q,p)$: dim($T^*Q$) = 6$N$.\\
\indent The Hamiltonian is a sum of kinetic and potential  terms, $K$ and $V$. 
$K$ depends only on the $p$s, and $V$ only on the $q$s. In cartesian 
coordinates, with $i$ now labelling particles $i= 1,...,N$ rather than degrees 
of freedom, we have the familiar expressions:
\be
H(q,p) = K(p) + V(q) \;\; {\rm{with}} \;\; K \; = \; \Sigma_i \frac{{\bf 
p}^2_i}{2m_i} \;\; , \;\;
V(q) \; = \; G \; \Sigma_{i < j} \frac{m_i m_j}{ \parallel {\bf q}_i - {\bf 
q}_j \parallel} 
\label{HNNewtonianpps}
\ee
where $m_i$ are the masses and $G$ is the gravitational constant.\footnote{From 
the broader philosophical perspective, the most significant feature of eq. 
\ref{HNNewtonianpps} is no doubt the fact that the potential  is a sum of all 
the two-body potential energies for the configuration $q \in Q$: there are no 
many-body interactions.}$^,$\footnote{Incidental remark. In fact, the kinetic 
energy can be represented by a metric $g$ on the configuration space. For 
Hamiltonian mechanics,  this means that the kinetic energy scalar $K$ on the 
cotangent bundle $T^*Q$ can be defined by applying $Q$'s metric $g$ to the 
projections of the momenta $p$, where at each point $(q,p) \in T^*Q$ the 
projection is made with the preferred isomorphism  $\o^{\sh}:T^*_q \rightarrow 
T_q$; (cf. eq. \ref{HEgeomic1}). That is:---
\be
K:(q,p) \in T^*Q \mapsto {g}_q(\o^{\sh}(p), \o^{\sh}(p)) \; .
\label{TasscalargeomHam}
\ee}

The euclidean group $E$ (aka: $E(3)$) is the group (under composition) of  
translations, rotations and reflections on $\mathR^3$. But since we will be 
interested in continuous symmetries, we will ignore reflections, and so 
consider the subgroup of orientation-preserving translations and rotations; 
i.e. the component of the group connected to the identity transformation (which 
I will also write as $E$). This is a Lie group, i.e. a group which is also a 
manifold, with the group operations smooth with respect to the manifold 
structure. Section \ref{tool} will give formal details. Here we just note that 
we need three real numbers to specify a translation (${\bf x} = (x,y,z)$), and 
three to specify a rotation (two for an axis, and one for the angle through 
which to rotate); and accordingly, it is unsurprising that as a manifold, the 
dimension of $E$ is 6: dim($E$) = 6.

$E$ {\em acts} in the obvious sense on $\mathR^3$. For example, if $g \in E$ is 
translation by ${\bf x} \in \mathR^3$, $g$ induces the map ${\bf q} \in 
\mathR^3 \mapsto {\bf q} + {\bf x}$. Similarly for a rotation induces: again, 
Section \ref{tool} will give a formal definition.

Now let $E$ act in this way on each of the $N$ factor spaces $\mathR^3$ of our 
system's configuration manifold $Q = \mathR^{3N}$. This defines an action 
$\Phi$ on $Q$: i.e. for all $g \in E$, there is  a map $\Phi_g: Q \rightarrow 
Q$. For example, for $g =$ a translation by ${\bf x} \in \mathR^3$, we have
\be
\Phi_g: ({\bf q}_j)=({\bf q}_1,...,{\bf q}_{N}) \in Q \mapsto ({\bf q}_1 + {\bf 
x},...,{\bf q}_{N} + {\bf x}) \in Q \;\; ;
\ee  
and similarly for rotations. Since the potential function $V: Q \rightarrow 
\mathR$ of eq. \ref{HNNewtonianpps} depends only on inter-particle distances, 
each map $\Phi_g: Q \rightarrow Q$ is a symmetry of the potential; i.e. we have 
$V(\Phi_g(q)) = V(q)$. 

The action $\Phi$ (i.e. the assignment $g \in E \mapsto \Phi_g$) induces an 
action of $E$ on $T^*Q = T^*\mathR^{3N}$, called the {\em  cotangent lift}  of 
$\Phi$ to $T^*Q$, and usually written as $\Phi^*$; so that we have for each $g 
\in E$ a lifted map $\Phi^*_g: T^*Q \rightarrow T^*Q$. Again, the details can 
wait till later (Section \ref{actionlg}). But the idea is that each map 
$\Phi_g$ on $Q$ is smooth, and so maps curves to curves, and so vectors to 
vectors, and so covectors to covectors, and so on.  

\indent Unsurprisingly, each of the lifted maps $\Phi^*_g: T^*Q \rightarrow 
T^*Q$ leaves the potential $V$, now considered as a scalar on $T^*Q$, 
invariant: i.e. we have $V(\Phi^*_g(q,p)) = V(q,p) \equiv V(q)$. But 
furthermore, each of the lifted maps $\Phi^*_g$ is a symmetry of the Hamilton 
system, in our previous sense (Section \ref{noetcomplete}). That is: $\Phi^*_g$ 
preserves the Hamiltonian (indeed the kinetic and potential terms are 
separately invariant); and it preserves the symplectic structure. This means 
the dynamics is invariant under the action of all $g \in G$: the dynamical 
histories of the system through $(q,p)$ and through  $\Phi^*_g(q,p)$ match 
exactly at each time. They are qualitatively indistinguishable: in contemporary 
metaphysical jargon, they are {\em duplicates}. 

At this point, of course, we meet the absolute-vs.-relational debate about 
space. The absolutist asserts, and the relationist denies, that there being two 
such indistinguishable possibilities makes sense.\footnote{The {\em locus 
classicus} for this debate is of course the Leibniz-Clarke correspondence, 
though the protagonists' argumentation is of course  sometimes theological. 
Clarke the absolutist maintains that there are many possible arrangements of 
bits of matter in space consistent with a specification of all relative 
distances, saying `if [the mere will of God] could in no case act without a 
pre-determining cause ... this would tend to take away all power of choosing, 
and to introduce fatality.' Leibniz claims there is only one such arrangement: 
`those two states ... would not at all differ from one another. Their 
difference therefore is only to be found in our chimerical supposition of the 
reality of space in itself.'} So the relationist, presented with the theory 
above, says we should cut down the space of possibilities. As I said in Section 
\ref{compare2quots}, it is natural to make this precise in terms of quotienting 
the action of the euclidean group: a set of absolutist possibilities related 
one to another by elements of the euclidean group form an equivalence class (an 
orbit) which is to represent {\em one} relationist possibility.

But here we need to distinguish two different quotienting procedures. I will 
call them {\em Relationism} and {\em Reductionism} (with capital R's), since 
the former is close to both traditional and contemporary relationist proposals, 
and the latter is an example of the orthodox idea of symplectic reduction. As I 
said in Section \ref{compare2quots}, the main difference will be that:\\
\indent (i): Relationism performs the quotient on $E$'s action on the 
configuration space $Q$; the set of orbits form a {\em relative configuration 
space}, on which the relationist proposes to build a dynamics, whether 
Lagrangian or Hamiltonian---yielding in the latter case, a {\em relative phase 
space}; whereas\\
\indent (ii):  Reductionism performs the quotient on $E$'s action on the usual 
phase space $T^*Q$, the set of orbits forming a {\em reduced phase space}.

Since our discussion adopts the Hamiltonian framework,  it will not matter for 
what follows, that Relationism, as defined, can adopt the Lagrangian framework, 
while Reductionism is committed to the Hamiltonian one. What {\em will} matter 
is that (i) and (ii) make for phase spaces of different dimensions; the reduced 
phase space has six more dimensions than the relative phase space. The  
``dimension gap'' is six.\\
\indent We will see that four of the six variables that describe these 
dimensions are constants of the motion; the other two vary with time. And for 
certain choices of values of the constants of the motion (roughly: no 
rotation), the time-varying variables drop out, and  the dynamics according to 
the Reductionist theory simplifies so as to coincide with that of the 
Relationist theory. In other words: if we impose no rotation, then the 
heterodox Relationist dynamics matches the conventional Reduced dynamics. 

\subsubsection{The Relationist procedure}\label{relationproced} 
The Relationist  seeks a mechanics based on the {\em relative configuration 
space} (RCS). An element of the RCS  is to be a pattern of inter-particle 
distances and angles that is geometrically possible, i.e. compatible with the 
$N$ particles being embedded in $\mathR^3$. So, roughly speaking, an element of 
the RCS is a euclidean configuration, modulo isometries; and the RCS will be 
the set of orbits $\mathR^{3N}/E$.\\
\indent Even before giving a more precise statement, we can state the 
``punchline'' about dimensions, as follows. Since dim$(E) = 6$, quotienting by 
$E$ subtracts six dimensions: that is, the dimension of the RCS will be 3$N$-6.

But we need to be more precise about the RCS. For the orbits and quotient 
spaces to be manifolds, and for dimensions to add or subtract in this simple 
way, we need to excise two classes of ``special'' points from $\mathR^{3N}$, 
before we quotient. (But I postpone till Section \ref{actionlg} the technical 
rationale for these excisions.)\\
\indent Let $\delta_Q \subset \mathR^{3N}$ be the set of configurations which 
are symmetric: i.e. each is fixed by some element of $E$ (other than the 
identity element!). Any configuration in which all the point-particles are 
collinear provides an example: the configuration is fixed by any rotation about 
the line as axis. Let $\Delta_Q$ be the set of collision configurations; i.e. 
configurations in which two or more particles are coincident in the usual 
configuration space $\mathR^{3N}$. (The $Q$ subscripts will later serve as a 
reminder that these sets are sets of configurations.) $\delta_Q$ and $\Delta_Q$ 
are both of measure zero in $\mathR^{3N}$. Excise both of them, and call the 
resulting space, which is again of dimension $3N$: $Q := \mathR^{3N} - 
(\delta_Q \cup \Delta_Q)$.

$\delta_Q$ and $\Delta_Q$ are each closed under the action of $E$. That is, 
each is a union of orbits: a euclidean transformation of a symmetric 
(collision) configuration is also symmetric (collision). So $E$ acts on $Q$. 
Now quotient $Q$ by $E$. $Q/E$ is the Relationist's RCS. Since dim$(E) = 6$, we 
have: dim($Q/E$) = 3$N$-6.\\
\indent  These $3N - 6$ variables encode all of a (relative) configuration's 
particle-pair relative distances, $r_{ij} \in \mathR$ (with $i,j$ labelling 
particles).  Note that there are $N(N-1)/2$ such relative distances; and for $N 
> 4$, this is greater than $3N - 6$: (for $N >> 4$, it is much greater). So the 
relative distances, though physically intuitive, give an {\em over-complete} 
set of coordinates  on $Q/E$. (So they cannot be freely chosen: there are  {\em 
constraints} between them.)

So the Relationist seeks a mechanics that uses this RCS. Newton's second law 
being second-order in time means that she will also need quantities  like 
velocities (in a Lagrangian framework) or like momenta (in a Hamiltonian 
framework). For the former, she will naturally consider the $N(N-1)/2$  
relative velocities ${\dot r}_{ij} := \frac{d}{dt}r_{ij}$; and for the latter, 
the corresponding momenta $p_{ij} := \frac{\pl L}{\pl {\dot r}_{ij}}$. Again, 
she must beware of constraints. The tangent and cotangent bundles built on her 
RCS $Q/E$ will each have dimension  $2(3N - 6) = 6N - 12$.  So again, for $N > 
4$, the number $N(N-1)/2$ of relative velocities  ${\dot r}_{ij}$, or of 
relative momenta  $p_{ij}$, is greater than the number of degrees of freedom 
concerned; and for $N >> 4$, it is much greater. So again, the relative 
velocities or relative momenta are over-complete: there are constraints.    

On the other hand, if the Relationist uses only these relative quantities, 
$r_{ij}$ and either ${\dot r}_{ij}$ or  $p_{ij}$ (or ``equivalent'' coordinates 
on $T(Q/E)$ or $T^*(Q/E)$ that are not over-complete), she faces a traditional 
problem---whatever the other details of her theory. At least, she faces a 
problem if she hopes for a deterministic theory which is empirically equivalent 
to the orthodox absolutist theory. I will follow tradition and state the 
problem in terms of relative velocities rather than momenta.

\indent The problem concerns rotation; (and herein lies the strength of 
Newton's and Clarke's position in the debate against Leibniz). For according to 
the absolutist theory  two systems of point-particles could match with respect 
to all relative distances and relative  velocities, and yet have different 
future evolutions; so that a theory allowing the same possibilities as the 
absolutist one, yet using only these relative quantities (or ``equivalent'' 
variables) would have to be indeterministic.\\
\indent The simplest example is an  analogue for point-particles of Newton's  
two globes thought-experiment. Thus the systems could each consist of just two 
point-particles with zero relative velocity. One system could be non-rotating, 
so that the point-particles fall towards each other under gravity; while the 
other system could be rotating about an axis normal to the line between the 
particles, and rotating at just such a rate as to balance the attractive force 
of gravity. 

The Relationist has traditionally replied that they do {\em not} hope for a 
theory empirically equivalent to the absolutist one.  Rather, they envisage a 
mechanics in which, of the two systems mentioned, only the non-rotating 
evolution is possible: more generally, a mechanics  in which the universe as a 
whole must have zero angular momentum. Originally, in authors like Leibniz and 
Mach, this reply was a promissory note. But modern Relationist theories such as 
Barbour and Bertotti's (1982) have made good the promise; and they have been 
extended well beyond point-particles interacting by Newtonian gravity. Besides, 
since the universe seems in fact to be non-rotating, these theories can even 
claim to be empirically adequate, at least as regards this principal difference 
from absolutist theories.\footnote{An advocate of the absolutist theory might 
say that it is odd to make what seems a contingent feature of the universe, 
non-rotation, a principle of mechanics; and the Relationist might reply that 
their view has the merit of predicting that the universe does not rotate! I 
fear there are no clear criteria for settling this methodological dispute; 
anyway, I will not pursue it.}\\
\indent But it is not my brief to go into these theories' details, except by 
way of comparison with a quotiented version of the absolutist theory: cf. 
Section \ref{reductionproced}.

\subsubsection{The Reductionist procedure}\label{reductionproced}
The Reductionist's main idea is to quotient only after passing to the orthodox 
phase space for $N$ point-particles, i.e. the cotangent bundle $T^* 
\mathR^{3N}$ of $\mathR^{3N}$. So the idea is to consider $(T^* 
\mathR^{3N})/E$, i.e. the quotient of $T^* \mathR^{3N}$ by the cotangent-lifted  
action $\Phi^*$ of the euclidean group $E$.

More precisely, we again proceed by first excising special points that would 
give technical trouble. But now the points to be excised are in the cotangent  
bundle $T^* \mathR^{3N}$, not in $\mathR^{3N}$.  So let $\delta \subset 
T^*\mathR^{3N}$ be the set of phase space states whose configurations are 
symmetric (in the sense of Section \ref{relationproced}'s $\delta_Q$). Let 
$\Delta \subset T^*\mathR^{3N}$ be the set of collision points; i.e. states in 
which two or more particles are coincident in the configuration space 
$\mathR^{3N}$. Both $\delta$ and $\Delta$ are of measure zero. Excise both of 
them, and call the resulting phase space, which is again of dimension $6N$: $M 
: = T^*\mathR^{3N} - (\delta \cup \Delta)$.

$\delta$ and $\Delta$ are each closed under the cotangent-lifted action of $E$ 
on $T^*\mathR^{3N}$. That is, each is a union of orbits: the cotangent lift of 
a euclidean transformation acting on a phase space state with a symmetric 
(collision) configuration yields a state which also has a symmetric (collision) 
configuration. So $E$ acts on $M$. Now quotient $M$ by $E$, getting ${\bar M} 
:= M/E$. This is called {\em reduced phase space}. We have: dim(${\bar M}$) = 
dim($M$) - dim($E$) = $6N - 6$.

\indent As emphasised at the end of Section \ref{spacesactions}, ${\bar M}$ has 
six more dimensions than the corresponding Relationist phase space (whether the 
velocity phase space (tangent bundle) or the momentum phase space (cotangent 
bundle)). The dimension of {\em those} phase spaces is $2(3N - 6) = 6N - 12$. 
Indeed, we can better understand both the reduced phase space ${\bar M}$ and 
Relationist phase spaces by considering this ``dimension gap''. There are two 
extended comments to make. 

(1): {\em Obtaining the Relationist phase space}:---\\
We can  obtain the Relationist momentum phase space from our original phase 
space $M$. Thus let $M_0$ be the subspace of $M$ in which the system has total 
linear momentum  and total angular momentum both equal to zero. Since these are 
constants of the motion, $M_0$ is dynamically closed and so supports a 
Hamiltonian dynamics given just by restriction of the original dynamics. With 
linear and angular momentum each contributing three real numbers, dim($M_0$) = 
dim($M$) - 6 = $6N - 6$. Furthermore, $M_0$ is closed under (is a union of 
orbits under) the cotangent-lifted action of $E$. So let us quotient  $M_0$ by 
this action of $E$, and write  ${\bar M}_0 : = M_0/E$. Then dim(${\bar M}_0$) =  
$6N - 6 - 6 = 6N -12$.

Now recall that this is the dimension of the phase space of the envisaged 
Relationist theory built on the RCS $Q/E$. And indeed, as one would hope: 
${\bar M}_0$ is the Hamiltonian version of Barbour and Bertotti's 1982 
Relational theory; (recall that they work in a Lagrangian framework).\\
\indent That is: ${\bar M}_0$ is a symplectic manifold, and points in ${\bar 
M}_0$ are parametrized by all the particle-to-particle  relative distances and 
relative velocities.  There is a deterministic dynamics which matches that of 
the original absolutist theory, once the original dynamics is projected down to 
Section \ref{relationproced}'s relative configuration space $Q/E$.\\
\indent In short: the vanishing total linear and angular momenta mean that an 
initial state comprising only relative quantities {\em is} sufficient to 
determine all future relative quantities.

(2): {\em Decomposing the Reductionist reduced phase space}:---\\
Let us return to the reduced phase space ${\bar M}$. The first point to make is 
that since the Hamiltonian $H$ on $M$, or indeed on $T^*\mathR^{3N}$, is 
invariant under the cotangent-lifted action of $E$, the usual dynamics on $M$ 
projects down to ${\bar M} = M/E$. That is: the reduced phase space dynamics 
captures all the $E$-invariant features of the usual dynamics.

\indent In fact, ${\bar M}$ is a {\em Poisson manifold}. So it is our first 
example of the more general framework for Hamiltonian mechanics announced in 
(1) of Section \ref{rednprospectus}. Again, I postpone technical detail till 
later (especially  Sections \ref{pmspreamble} and \ref{LPso(3)}). But the idea 
is that a Poisson manifold has a {\em degenerate} antisymmetric bilinear map, 
which implies that the manifold is a disjoint union of  symplectic manifolds. 
Each symplectic manifold is called a {\em leaf} of the Poisson manifold. The 
leaves' symplectic structures ``mesh'' with one another; and within each leaf 
there is a conventional Hamiltonian dynamics. 

Even without a precise definition of a Poisson manifold, we can describe how 
$M$ is decomposed into symplectic manifolds, each with a Hamiltonian dynamics. 
Recall that we have: dim(${\bar M}$) = dim($M$) - dim($E$) = $6N - 6$. This 
breaks down as:
\be
6N - 6 = (6N - 12) + 3 + 3 = 2(3N - 6) + 3 + 3 =: \alpha + \beta + \gamma  
\ee
where the right hand side defines $\alpha, \beta, \gamma$ respectively as $2(3N 
- 6), 3$ and $3$. In terms of ${\bar M}$, this means the following.

\indent (i): $\alpha$ corresponds to (1)'s ${\bar M}_0$, i.e. to $T^*(Q/E)$. As 
discussed, $3N - 6$ variables encode all the particle-pair relative distances; 
and the other $3N - 6$ variables encode all the particle-pair relative momenta.

The six extra variables additional to these 6$N$-12 relative quantities consist 
of: four constants of the motion, and two other variables which are dynamical, 
i.e. change in time.

\indent (ii): $\beta$ stands for three of the four constants of the motion: 
viz. the three variables that encode the total linear momentum of the system, 
i.e. the momentum of the centre of mass. These constants of the motion are 
``just parameters'' in the sense that: (a) not only does specifying a value for 
all three of them fix a surface, i.e. a $(6N - 9)$-dimensional hypersurface in 
${\bar M}$, on which there is a Hamiltonian dynamics; also (b) this Hamiltonian 
and symplectic structure is independent of the values we specify.\footnote{As 
mentioned at the end of Section \ref{compare2quots}, the relationist 
traditionally proposes to identify absolutist states of motion that differ just 
by the value of the total momentum. And indeed, the proposal can be implemented 
by considering an action of the Galilean group on the absolutist phase space 
$M$, and identifying points related by Galilean boosts. For discussion and 
references, cf. Belot (2000: Section 5.3).}

\indent (iii): $\gamma$ stands for the three variables that encode the total 
angular momentum of the system. One of these is a fourth constant of the 
motion, viz. the magnitude $L$ of the total angular momentum. The other two 
time-varying  quantities fix a point on a sphere (2-sphere) of radius $L$, 
encoding the direction of the angular momentum of the system {\em in a frame 
rotating with it}. The situation is as in the elementary theory of the rigid 
body: though the total angular momentum relative to coordinates fixed in space 
is a constant of the motion (three constant real numbers), the total angular 
momentum relative to the body is constant only in magnitude (one real number 
$L$), not in direction. This will be clearer in Section \ref{pms} onwards, when 
we describe the Poisson manifold structure  in the theory of the rigid body. 
For the moment, there are two main comments to make about the $N$ particle 
system:---

\indent (a): If we specify  $L$, in addition to the momentum of the centre of 
mass of the system, we get a $(6N - 10)$-dimensional hypersurface in ${\bar 
M}$, on which (as in (ii))  there is a Hamiltonian dynamics. So we can think of 
${\bar M}$ as consisting of the four real-parameter family of these 
hypersurfaces, with each point of each hypersurface being equipped with a 
sphere of radius $L$; (subject to a qualification in (b) below).\\
\indent Note that here `each point being equipped' does not mean that the 
sphere gives the extra dimensions that would constitute ${\bar M}$ as a fibre 
bundle; (there would be two dimensions lacking). Rather: in the point's 
representation by $6N - 10$ real numbers, two of the numbers can be taken to 
represent a point on a sphere.

\indent (b): But unlike the situation for $\beta$ in (ii) above, the 
Hamiltonian dynamics on such a hypersurface {\em depends} on the value of $L$. 
In particular, if $L = 0$ the sphere representing the body angular momentum is 
degenerate: it is of radius zero, and the other two time-varying  quantities 
drop out. A point in the hypersurface is represented by $6N - 12$ real numbers; 
i.e. the hypersurface is $6N - 12$-dimensional.\\
\indent Now recall from Section \ref{relationproced} or (1) above that $6N - 
12$ is the dimension of the phase space of the envisaged Relationist theory 
built on the RCS $Q/E$.  And indeed, just as one would hope: the hypersurface 
with $L = 0$ and also with vanishing linear momentum, with its dynamics, {\em 
is} the symplectic manifold and dynamics that is the Hamiltonian version of 
Barbour and Bertotti's 1982 Relational theory of $N$ point-particles.  In terms 
of (1)'s notation, this hypersurface is ${\bar M}_0$.

\indent We can sum up this comparison as follows. On this hypersurface ${\bar 
M}_0$, the dynamics in the reduced phase space coincides  with the dynamics one 
obtains for the relative variables, if one arbitrarily embeds their initial 
values in the usual absolutist phase space $T^*\mathR^{3N}$, subject to the 
constraint that the total angular and linear momenta vanish, and then reads off 
(just by projection) their evolution from the usual evolution in 
$T^*\mathR^{3N}$.

\subsubsection{Comparing the Relationist and Reductionist 
procedures}\label{compareproced}
In comparing the Relationist and Reductionist procedures, I shall just make 
just two extended comments, and refer to Belot for further discussion. The gist 
of both comments is  that Reductionism suffices: Relationism is not needed. The 
first is a commonplace point; the second is due to Belot.

\paragraph{2.3.5.A Reductionism allows for rotation}\label{5.2.5.A 
RednmAllowsRotn}
The first comment reiterates the Reductionist's ability, and the Relationist's 
inability, to endorse Newton's globes (or bucket) thought-experiment. The 
Reductionist can work in either\\
\indent  (i) the  $(6N - 6)$-dimensional phase space  ${\bar M} = M/E$; or\\
\indent  (ii) the $(6N - 9)$-dimensional hypersurface got from (i) by 
specifying the centre of mass' linear momentum; or\\
\indent  (iii) the  $(6N - 10)$-dimensional hypersurface got from (ii) by also 
specifying a non-zero value of $L$.\\
\indent In all three cases, the Reductionist can describe  rotation in a way 
that the Relationist with their $(6N - 12)$-dimensional space cannot. For she 
has to hand the three extra non-relative variables ($L$ and two others) that 
describe the rotation of the system as a whole. (Incidentally: that they 
describe the system as a whole is suggested by there being just three of them, 
whatever the value of $N$.)  In particular, she can distinguish states of 
rotation and non-rotation ($L = 0$), in the sense of endorsing the distinctions 
advocated by the globes and bucket thought-experiments.

The Reductionist can also satisfy a traditional motivation for relationism, 
which concerns general philosophy, rather than the theory of motion. It is 
especially associated with Leibniz: namely, our theory (or our metaphysics) 
should not admit distinct but utterly indiscernible possibilities. One might 
well ask why we should endorse this ``principle of the identity of 
indiscernibles'' for possibilities rather than objects. For Leibniz himself, 
the answer lies (as Belot's (2001) brings out) in his principle of sufficient 
reason, and ultimately in theology.\\
\indent But in any case the Reductionist can satisfy the requirement. Agreed, 
the usual absolutist theory, cast in $T^*\mathR^{3N}$ (or if you prefer, $M = 
T^*\mathR^{3N} - (\delta \cup \Delta)$) has nine variables that describe (i) 
the position of the centre of mass, (ii) the orientation of the system about 
its centre of mass, and (iii) the system's total linear momentum: i.e. three 
variables, a vector in $\mathR^3$, for each of (i)-(iii). So the usual 
absolutist theory has a nine-dimensional ``profligacy'' of distinct but 
indiscernible possibilities. But as we have seen, the Reductionist quotients by 
the action of the euclidean group $E$, and so works in ${\bar M} = M/E$: which 
removes the profigacy about (i) and (ii).\\
\indent As to (iii), I agree that for all I have said, a job remains to be 
done. The foliation of ${\bar M}$ by a three real-parameter family of $(6N - 
9)$-dimensional  hypersurfaces, labelled by the system's total linear momentum, 
codifies the profligacy---but does not eliminate it. But as I mentioned above 
(cf. footnote 16), the Reductionist can in fact quotient further, by 
considering the action of Galilean boosts and identifying phase space points 
that differ by a boost; i.e. defining orbits transverse to these hypersurfaces.  

\paragraph{2.3.5.B Analogous reductions in other theories}\label{5.2.5.B 
AnalogousRednms}
I close my philosophers' appetizer for symplectic reduction by summarizing some 
general remarks of Belot's (2001: Sections VIII-IX); cf. also his (2003a, 
Sections 12, 13). They are about how our discussion of relational mechanics is 
typical of many cases; and how symplectic reduction can be physically 
important. I label them (1)-(3).

\indent (1): {\em A general contrast: when to quotient}:---\\
 The example of $N$ point-particles interacting by Newtonian gravity is typical 
of a large class of cases (infinite-dimensional, as well as 
finite-dimensional). There is a configuration space $Q$, acted on by a 
continuous group $G$ of symmetries, which lifts to the cotangent  bundle 
$T^*Q$, with the cotangent lift leaving invariant the Hamiltonian, and so the 
dynamics. So we can quotient $T^*Q$ by $G$ to give a reduced theory. (There is 
a Lagrangian  analogue; but as above, we set it aside.) But there is also some 
motivation for quotienting  $G$'s action on $Q$, irrespective of how we then go 
on the construct dynamics.  Let us adopt `relationism' as a mnemonic label for 
whatever motivates quotienting the configuration space. Then with suitable 
technical conditions assumed (recall our excision of $\delta$ and $\Delta$), we 
will have:\\
\indent (i): for the reduced Hamiltonian theory:  dim($(T^*Q)/G$) =  2 dim$Q$ - 
dim$G$;\\
\indent (ii): for the relationist theory, in a Lagrangian or Hamiltonian 
framework:\\ dim$(T(Q/G))$ = dim$(T^*(Q/G))$ = 2(dim$Q$ - dim$G$)\\
\indent So we  have in the reduced theory, dim $G$ variables that do not occur 
in the relationist theory: let us call them `non-relational variables'.

(2): {\em The non-relational variables}:---\\
 Typically, these non-relational variables represent global, i.e. collective, 
properties of the system. That is unsurprising since the number, dim $G$, of 
these variables is independent of the number of degrees of freedom of the 
system (dim $Q$, or 2dim $Q$ if you count rate of change degrees of freedom 
separately).\\
\indent Some of these variables are conserved quantities, which arise (by 
Noether's theorem) from the symmetries. Furthermore, there can be specific 
values of the conserved quantities, like the vanishing angular momentum of 
Section \ref{reductionproced}, for which the reduced theory collapses into the 
relationist theory. That is, not only are the relevant state spaces of equal 
dimension; but also their dynamics agree.  

\indent (3): {\em The reduced theory}:---\\ 
Typically, the topology and geometry of the reduced phase space $(T^*Q)/G$, and 
the Hamiltonian function on it, ${\bar H}: (T^*Q)/G \rightarrow \mathR$ say, 
are more complex than the corresponding features of the unreduced theory on 
$T^*Q$. In particular, the reduced Hamiltonian ${\bar H}$ typically has 
potential energy terms corresponding to forces that are absent from the 
unreduced theory. But this should not be taken as necessarily a defect, for two 
reasons.\\
\indent First, there are famous cases in which the reduced theory has a 
distinctive motivation. One example is  Hertz' programme in mechanics, viz. to 
``explain away'' the apparent forces of our macroscopic experience (e.g. 
gravity) as arising from reduction of a theory that has suitable symmetries. 
(The programme envisaged cyclic variables for  microscopic degrees of freedom 
that were unknown to us; cf. Lutzen (1995, 2005).) Another famous example is 
the Kaluza-Klein treatment of the force exerted on a charged particle by the 
electromagnetic field. That is: the familiar Lorentz force-law describing a 
charged particle's motion in four spacetime dimensions can be shown to arise by 
symplectic reduction from a theory postulating a spacetime with a fifth (tiny 
and closed) spatial dimension, in which the particle undergoes straight-line 
motion. Remarkably, the relevant conserved quantity, viz. momentum along the 
fifth dimension, can be identified with electric charge; so that the theory can 
claim to explain the conservation of electric charge. (This example generalizes 
to other fields: for details and references, cf. Marsden and Ratiu (1999, 
Section 7.6).) \\
\indent Second, the reduced theory need not be so complicated as to be 
impossible to work with. Indeed, these two examples prove this point, since in 
them the reduced theory is entirely tractable: for it is the familiar 
theory---that one might resist abandoning for the sake of the postulated 
unreduced theory.\footnote{And here one should resist being prejudiced because 
of familiarity. Why {\em not} have Newtonian gravity arise from a microscopic 
cyclic degree of freedom? Why {\em not} have the Lorentz force law arise from 
geodesic motion in a five-dimensional spacetime with the fifth dimension 
wrapped up, so that conservation of charge is explained, in Noether's theorem 
fashion, by a symmetry?} Besides, Belot describes how, even when the reduced 
theory seems complicated (and not just because it is unfamiliar!), the general 
theory of symplectic reduction, as developed over the last forty years, has 
shown that one can often ``do physics'' in the reduced phase space: and that, 
as in the Kaluza-Klein example, the physics in the reduced phase space can be 
heuristically, as well as interpretatively, valuable.

\section{Some geometric tools}\label{tool}
So much by way of an appetizer. The rest of the Chapter, comprising this 
Section and the next four, is the five-course banquet! 
This Section  expounds some modern differential geometry, especially about Lie 
algebras and Lie groups. Section \ref{actionlg} takes up actions by Lie groups. 
Then Section \ref{pms} describes Poisson manifolds as a generalized framework 
for Hamiltonian mechanics. As I mentioned in (2) of Section 
\ref{rednprospectus}, Lie himself developed this framework; so in effect, he 
knew everything in these two Sections---so it is a true (though painful!) pun 
to say that these three Sections give us the ``Lie of the land''. In any case, 
these two Sections will prepare us for Section \ref{symyredn}'s description of 
symmetry and conservation in terms of momentum maps. Finally, Section 
\ref{redn} will present one of the main theorems about symplectic reduction. It 
concerns the case where the natural configuration space for a system is itself 
a Lie group $G$; (cf. (3) of Section \ref{rednprospectus}).  Quotienting the 
natural phase space (the cotangent bundle on $G$) will give a Poisson manifold 
that ``is'' the dual of $G$'s Lie algebra.

In this Section, I first sketch some notions of differential geometry, and fix 
notation (Section \ref{prelimtool}). Then I introduce Lie algebras and Lie 
brackets of vector fields (Section \ref{lgla}). Though most of this Section 
(indeed this Chapter!) is about differential rather than integral notions, I 
will later need Frobenius' theorem, which I present in Section \ref{Frob}. Then 
I give some basic information about Lie groups and their Lie algebras (Section 
\ref{lg}).

\subsection{Vector fields on manifolds}\label{prelimtool}
\subsubsection{Manifolds, vectors, curves and derivatives}\label{mfdvecsetc}
By way of fixing ideas and notation, I begin by giving details about some ideas 
in differential geometry (some already used in Section \ref{july05}), and 
introducing some new notation for them. 

A manifold $M$ will be finite-dimensional, except for obvious and explicit 
exceptions such as the infinite-dimensional group of diffeomorphisms of a (as 
usual: finite-dimensional!) manifold. I will not be concerned about the degree 
of differentiability in the definition of a manifold, or of any associated 
geometric objects: `smooth' can be taken  throughout what follows to mean 
$C^{\infty}$. I will often not be concerned with global, as against local, 
structures and results; (though the reduction results we are driving towards 
{\em are} global in nature). For example, I will not be concerned about whether 
curves are inextendible, or flows are complete.
 
I shall in general write a vector at a point $x \in M$ as $X$; or in terms of 
local coordinates $x^i$, as $X = X^i \frac{\pl }{\pl x^i}$ (summation 
convention). {\em From now on}, I shall write the tangent space  at a point $x 
\in M$ as $T_xM$ (rather than just $T_x$), thus explicitly indicating the 
manifold $M$ to which it is tangent. As before, I write the tangent bundle, 
consisting of the ``meshing collection'' of these tangent spaces, as $TM$. 
Similarly, I write a  1-form (covector) at a point $x \in M$ as $\al$; and so 
the cotangent space at $x \in M$ as $T^*_xM$; and as before, the cotangent 
bundle as $T^*M$.

\indent A smooth map $f: M \rightarrow N$ between manifolds $M$ and $N$ (maybe 
$N = M$) maps smooth curves to smooth curves, and so tangent vectors to tangent 
vectors; and so on for 1-forms  and higher tensors. It is convenient to write 
$Tf$, called the {\em derivative} or {\em tangent}  of $f$ (also written as 
$f_*$ or $df$ or $Df$), for the induced map on the tangent bundle.\\ 
\indent In more detail: let us take a curve $c$ in $M$ to be a smooth map from 
an interval $I \subset \mathR$ to $M$, and a tangent vector at $x \in M$, $X 
\in T_xM$, to be an equivalence class $[c]_x$ of curves through $x$. (The 
equivalence relation is that the curves be tangent at $x$, with respect to 
every local chart at $x$; but I omit the details of this.) Then we define $Tf: 
TM \rightarrow TN$ (also written $f_*: TM \rightarrow TN$) by
\be
f_*([c]_x) \equiv Tf([c]_x) := [f \circ c]_{f(x)}, \;\; {\rm for \; all} \;\; x 
\in M.
\label{definetgt}
\ee
We sometimes write $T_x f$ for the restriction of $Tf$ to just the tangent 
space $T_xM$ at $x$; i.e. 
\be
T_x f: [c]_x \in T_xM \mapsto [f \circ c]_{f(x)} \in T_{f(x)}N.
\label{definetgtatx}
\ee  
In Section \ref{fieldtoflow}.B, we will discuss how one can instead define 
tangent vectors to be differential operators on the set of all scalar functions 
defined in some neighbourhood of the point in question, rather than equivalence 
classes of curves. One can then define the tangent map $f_* \equiv Tf$ in a way 
provably equivalent to that above.

\subsubsection{Vector fields, integral curves and flows}\label{fieldtoflow}
We will be especially concerned with vector fields defined on $M$, i.e. $X: x 
\in M \mapsto X(x) \in T_xM$, or on a subset $U \subset M$. So suppose that  
$X$ is vector field on $M$ and $f:M \rightarrow N$ is a smooth map, so that 
$T_x f: T_xM \rightarrow T_{f(x)}N$.

\paragraph{3.1.2.A Push-forwards and pullbacks}\label{612Apushpull}
 It is important to note $(T_x f)(X(x))$ does {\em not} in general define a 
vector field on $N$. For $f(M)$ may not be all of $N$, so that for $y \in (N - 
{\rm{ran}}(f))$ $(T_x f)(X(x))$ assigns no element of $T_yN$. And $f$ may not 
be injective, so that we could have $x, x' \in M$ and $f(x) = f(x')$ with $(T_x 
f)(X(x)) \neq (T_{x'} f)(X(x'))$. Thus we say that vector fields do not {\em 
push forward}.

On the other hand, suppose that $f:M \rightarrow N$ is a {\em diffeomorphism} 
onto $N$: that is, the smooth map $f$ is a bijection, and its inverse $f^{-1}$ 
is also smooth. Then for any vector field $X$ on $M$, $Tf(X)$ is a vector field 
on $N$. So in this case, the vector field does push forward. Accordingly, 
$Tf(X)$ is called the {\em push-forward} of $X$; it is often written  as 
$f_*(X)$. So for any $x \in M$, the pushed forward vector field at the image 
point $f(x)$ is given by 
\be
(f_*(X))(f(x)) := T_x f \cdot X(x) \; .
\label{definepushfwd}
\ee 
(Note the previous use of the asterisk-subscript for the derivative of $f$, in 
eq. \ref{definetgt}.)

This prompts three more general comments.\\
\indent (1): More generally: we say that two vector fields, $X$ on $M$ and $Y$ 
on $N$, are $f$-{\em related} on $M$ (respectively: on $S \subset M$) if 
$(Tf)(X) = Y$ at all $x \in M$ (respectively: $x \in S$).

\indent (2): We can generalize the idea that a diffeomorphism implies that a 
vector field can be pushed forward, in two ways. First, the diffeomorphism need 
only be defined locally, on some neighbourhood of the point $x \in M$ of 
interest. Second,  a diffeomorphism establishes a one-one correspondence, not 
just between vector fields defined on its domain and codomain, but also between 
all differential geometric objects defined on its domain and codomain: in 
particular, 1-form  fields, and higher rank tensors.

\indent (3): (This continues comment (2).) Though  vector fields do not in 
general push forward, 1-form fields {\em do} in general {\em pull back}. This 
is written with an asterisk-superscript.  That is: for any smooth $f:M 
\rightarrow N$, not necessarily a diffeomorphism (even locally), and any 1-form 
field (differential 1-form) $\al$ on $N$, we define the pullback $f^*(\al)$ to 
be the 1-form on $M$ whose action, for each $x \in M$, and each $X \in T_xM$, 
is given by:
\be
(f^*(\al))(X) := \al\mid_{f(x)}(Tf(X)) \; .
\label{definepullback}
\ee 
Similarly, of course if the map $f$ is defined only locally on a subset of $M$: 
a 1-form defined on the range of $f$ pulls back to a 1-form on the domain of 
$f$.

\paragraph{3.1.2.B The correspondence between vector fields and 
flows}\label{612Bvecfieldsflows}
 The leading idea about vector fields is that, for any manifold, the theorems 
on the local existence, uniqueness and differentiability of solutions of 
systems of ordinary differential equations (e.g. Arnold (1973: 48-49, 77-78, 
249-250), Olver (2000: Prop 1.29)) secure a one-one correspondence  between 
four notions:\\
\indent (i): Vector fields $X$ on a subset $U \subset M$, on which they are 
non-zero; $X: x \in U \mapsto X(x)  \in T_xM, X(x) \neq 0$;\\
\indent (ii): Non-zero directional derivatives at each point $x \in U$, in the 
direction of the vector   $X(x)$. In terms of coordinates ${\bf x} = 
x^1,...,x^n$, these are  first-order linear differential operators $X^1({\bf 
x})\frac{\pl}{\pl x^1} + \dots + X^n({\bf x})\frac{\pl}{\pl x^n}$, with 
$X^i({\bf x})$ the $i$-component in this coordinate system of the vector 
$X(x)$. Such an operator is often introduced abstractly as a {\em derivation}: 
a map on the set of smooth real-valued functions defined on a neighbourhood of 
$x$, that is linear and obeys the Leibniz rule.  \\
\indent (iii): Integral curves (aka: solution curves) of the fields $X$ in $U$; 
i.e. smooth maps $\phi: I \rightarrow M$ from a real open interval $I \subset 
\mathR$ to $U$, with $0 \in I$, $\phi(0) = x \in U$, and whose tangent vector 
at each $\phi(\tau), \tau \in I$ is $X(\phi(\tau)).$\\
\indent (iv): Flows $X^{\tau}$ mapping, for each field $X$ and each $x \in U$, 
some appropriate subset of $U$ to another: $X^{\tau}: U \rightarrow M$. This 
flow is guaranteed to exist only in some neighbourhood of a given point $x$, 
and for $\tau$ in some neighbourhood  of $0 \in \mathR$; but this will be 
enough  for us. Such a flow is a one-parameter subgroup of the 
``infinite-dimensional group'' of all local diffeomorphisms.

I spell out this correspondence in a bit more detail:--- In local coordinates 
$x^1,...,x^n$, any smooth curve $\phi: I \rightarrow M$ is given by $n$ smooth 
functions $\phi(\tau) = (\phi^1(\tau),...,\phi^n(\tau))$, and the tangent 
vector to $\phi$ at $\phi(\tau) \in M$ is
\be
{\dot \phi}(\tau) = {\dot \phi}^1(\tau)\frac{\pl}{\pl x^1} + \dots + {\dot 
\phi}^n(\tau) \frac{\pl}{\pl x^n}.
\ee   
So for $\phi$ to be an integral curve of $X$ requires that for all $i = 
1,...,n$ and all $\tau \in I$
\be
{\dot \phi}^i(\tau) = X^i(\tau).
\ee
The local existence and uniqueness, for a given vector field $X$ and $x \in M$, 
of the integral curve $\phi_{X,x}$ through $x$ (with $\phi(0) = x$) then 
ensures that the flow, written either as $X^{\tau}$ or as $\phi_X(\tau)$ 
\be
X^{\tau}: x \in M \mapsto X^{\tau}(x) \equiv \phi_{X,x}(\tau) \in M \; \; ,
\label{defineflow} 
\ee
is (at least locally) well-defined. The flow is a one-parameter group of 
transformations of $M$, and $X$ is called its {\em infinitesimal generator}.\\
\indent The exponential notation
\be
\exp(\tau X)(x) := X^{\tau}(x) \equiv \phi_{X,x}(\tau)
\label{expnlintrodd}
\ee
is suggestive. For example, the group operation in the flow, i.e. 
\be
X^{\tau + \sigma}(x) = X^{\tau}(X^{\sigma}(x)) \; \; ,
\ee 
is  written in the suggestive notation
\be
\exp((\tau + \sigma)X)(x) = \exp(\tau X)(\exp(\sigma X)(x)) \; \; .
\ee 
So computing the flow for a given $X$ (i.e. solving a system of $n$ first-order 
differential equations!) is called {\em exponentiation} of the vector field 
$X$.

Remark:--- The above correspondence can be related to our discussion of 
diffeomorphisms and pushing forward vector fields. In particular: if two vector 
fields, $X$ on $M$ and $Y$ on $N$, are $f$-{\em related} by $f:M \rightarrow 
N$, so that $(Tf)(X(x)) = Y(f(x))$, then $f$ induces a map from integral curves 
of $X$ to integral curves of $Y$. We can express this in terms of 
exponentiation of $X$ and $Y = (Tf)(X)$:
\be
f(\exp(\tau X)x) = \exp(\tau (Tf)(X))(f(x)).
\ee  

Remark:--- I emphasise that the above correspondence between (i), (ii), (iii) 
and (iv) is not true at a {\em single} point. More precisely:\\
\indent (a): On the one hand: the correspondence between (i) and (ii) holds at 
a point; and also holds for zero vectors. That is: a single vector $X \in T_xM$ 
corresponds to a directional derivative operator (derivation) at $x$; and $X = 
0$ corresponds to the zero derivative  operator mapping all local scalars to 0.  
(Indeed, as I mentioned: vectors are often  defined as such 
operators/derivations). But: \\
\indent (b): On the other hand: the correspondence between (i) and (iii), or 
between  (i) and (iv), requires a neighbourhood. For a single vector $X \in 
T_xM$ corresponds to a whole class of curves (and so: of flows) through $x$, 
not to a single curve. Namely, it corresponds to all the curves (flows) with 
$X$ as their tangent vector.\\
\indent However, we shall see (starting in Section \ref{lg}) that for a 
manifold with suitable extra structure, a single vector {\em does} determine a 
curve. (And we will again talk of exponentiation.)

We need to generalize one aspect of the above correspondence (i)-(iv),  namely 
the (i)-(ii) correspondence between vectors and directional derivatives. This 
generalization is the Lie derivative.

\subsubsection{The Lie derivative}\label{Lie deriv}
Some previous Sections have briefly used the Lie derivative. Since we will use 
it a lot in the sequel, we now introduce it more thoroughly.

We have seen that given a vector field $X$ on a manifold $M$, a point $x \in 
M$, and any scalar function $f$ defined on a neighbourhood of $x$, there is a 
naturally defined rate of change of $f$ along $X$ at $x$: the directional 
derivative $X(x)(f)$.\\
\indent Now we will define the Lie derivative along $X$ as an operator ${\cal 
L}_X$ that defines a rate of change along $X$: not only for locally defined 
functions (for which the definition will agree with our previous notion, i.e. 
we will have ${\cal L}_X(f) = X(f)$); but also for vector fields and 
differential 1-forms.\footnote{Indeed, the definition can be extended to all 
higher rank tensors. But I will not develop those details, since---apart from 
Section \ref{noetcomplete}'s mention of the Lie derivative of the symplectic 
form ${\cal L}_X \o$ (viz. the requirement that if $X$ is a symmetry, ${\cal 
L}_X \o =  0$)---we shall not need them.} We proceed in three stages.   

(1): We first define the Lie derivative as an operator on  scalar functions, in 
terms of the vector field $X$ on $M$. We define the {\em Lie derivative} along 
the field $X$  (aka: the derivative in the direction of $X$), ${\cal L}_X$, as 
the operator on scalar functions $f:M \rightarrow \mathR$ defined by:
\be
{\cal L}_X: f \mapsto {\cal L}_X f:M \rightarrow \mathR \; {\rm{with \;}} 
\forall x \in M: \;\;
({\cal L}_X f)(x) := \frac{d}{d \tau}\mid_{\tau=0} f(X^{\tau}(x)) \; \equiv \; 
X(x)(f).
\label{defineLiederiv}
\ee
Though this definition assumes that both $X$ and $f$ are defined globally, i.e. 
on all of $M$, it can of course be restricted to a neighbourhood. Thus defined, 
${\cal L}_X$ is linear and obeys the Leibniz rule, i.e.
\be
 {\cal L}_X(fg) = f{\cal L}_X(g) + g{\cal L}_X(f) \; \; ;
\ee
In coordinates ${\bf x} = x^1,...,x^n$, ${\cal L}_X f$ is given by
\be
{\cal L}_X f = X^1({\bf x})\frac{\pl f}{\pl x^1} + \dots + X^n({\bf 
x})\frac{\pl f}{\pl x^n},
\label{ldscalarcoords}
\ee
with $X^i({\bf x})$ the $i$-component of the vector $X(x)$. Eq. 
\ref{ldscalarcoords} means that despite eq. \ref{defineLiederiv}'s mention of 
the flow $X^{\tau}$, the Lie derivative of a scalar agrees with our previous 
notion of directional derivative: that is, for all $f$,  ${\cal L}_X(f) = 
X(f)$.

(2): In (1), the vector field  $X$ determined the operator ${\cal L}_X$: in 
terms of Section \ref{fieldtoflow}.B's correspondence, we moved from (i) to 
(ii). But we can conversely define a vector field in terms of its Lie 
derivative; and in Section \ref{lavecfields}'s discussion of the Lie bracket, 
we shall do exactly this.\\
\indent In a bit more detail:--- We note that the set ${\cal F}(M)$ of all 
scalar fields on $M$, $f: M \rightarrow \mathR$ forms an (infinite-dimensional) 
real vector space under pointwise addition. So also does the set ${\cal X}(M)$ 
of all vector fields on $M$, $X: x  \in M \mapsto X(x) \in T_xM$. Furthermore,  
${\cal X}(M)$ is isomorphic as a real vector space, and as an module over the 
scalar fields, to the collection of operators ${\cal L}_X$. The isomorphism is 
given by the map $\theta: X \mapsto {\cal L}_X$ defined in (1).

(3): We now extend the definition of ${\cal L}_X$ so as to define it on vector 
fields $Y$ and 1-forms $\al$. We can temporarily use $\theta$ as notation for 
{\em either} a vector field $Y$ {\em or} a differential 1-form $\al$. Given a 
vector field $X$ and flow $X^{\tau} \equiv \phi_X(\tau)$, we need to compare 
$\theta$ at the point $x \in M$ with $\theta$ at the nearby point $X^{\tau}(x) 
\equiv \phi_{X,x}(\tau)$, in the limit as $\tau$ tends to zero. But the value 
of $\theta$ at $X^{\tau}(x)$ is in the tangent space, or cotangent space, at 
$X^{\tau}(x)$: $T_{X^{\tau}(x)}M$ or $T^*_{X^{\tau}(x)}M$. So to make the 
comparison, we need to somehow transport back this value to $T_xM$ or $T^*_xM$.

\indent Fortunately, the vector field $X$ provides a natural way to define such 
a transport. For the vector field $Y$, we use the differential (i.e. 
push-forward) of the inverse flow, to ``get back'' from $X^{\tau}(x)$  to $x$. 
Using $\phi^*(\tau)$ for this ``pullback'' of $\phi_{X,x}(\tau)$, we define
\be
\phi^*(\tau) := T(\exp(-\tau X)) \equiv d \exp(-\tau X): T_{X^{\tau}(x)}M 
\equiv T_{\exp (\tau X)(x)}M \rightarrow T_xM \;\; .
\label{geomdefLiederivvecfield}
\ee 
For the 1-form $\al$, we define the transport by the pullback, already defined 
by eq. \ref{definepullback}:
\be
\phi^*(\tau) := (\exp(-\tau X))^* : T^*_{X^{\tau}(x)}M \equiv T^*_{\exp (\tau 
X)(x)}M \rightarrow T^*_xM \;\; .
\label{geomdefLiederiv1form}
\ee 
With these definitions of $\phi^*(\tau)$, we now define the {\em Lie 
derivative} ${\cal L}_X \theta$, where $\theta$ is a vector field $Y$ {\em or} 
a differential 1-form $\al$, as the vector field or differential 1-form 
respectively, with value at $x$ given by
\be
\lim_{\tau \rightarrow 0} \; \frac{\phi^*(\tau)(\theta\mid_{X^{\tau}(x)}) - 
\theta\mid_x}{\tau}
\;\; \; = \;\; \;\frac{d}{d \tau} \mid_{\tau = 0} \; 
\phi^*(\tau)(\theta\mid_{X^{\tau}(x)}) \;\; .
\label{Liederivgenldefn}
\ee 

Finally, an incidental result to illustrate this Chapter's ``story so far''. It 
connects Noether's theorem, from Section \ref{noetcomplete}, to this Section's 
details about the Lie derivative,  and  to the theorem stating the local 
existence and uniqueness of solutions of ordinary differential equations (cf. 
the start of  Section \ref{fieldtoflow}.B). This latter theorem implies that on 
any manifold any vector field $X$ can be ``straightened out'', in the sense 
that around any point at which $X$ is non-zero, there is a local coordinate 
system in which $X$ has all but one component vanish and the last component  
equal to 1.  Using this theorem, it is straightforward to show that on any 
even-dimensional manifold {\em any} vector field $X$ is locally Hamiltonian, 
with respect to {\em some} symplectic form, around a point where $X$ is 
non-zero. One just defines the symplectic form by Lie-dragging from a surface 
transverse to $X$'s integral curves.

\subsection{Lie  algebras and brackets}\label{lgla}
I now introduce Lie algebras and the Lie bracket of two vector fields.

\subsubsection{Lie algebras}\label{la}
A {\em Lie algebra} is a vector space $V$ equipped with a bilinear 
anti-symmetric operation, usually  denoted by square brackets (and called 
`bracket' or `commutator'),  $[,]: V \times V \rightarrow V$, that satisfies 
the Jacobi identity, i.e.
\be
[[X,Y],Z] + [[Y,Z],X] + [[Z,X],Y] = 0 \;\; .
\label{jacobi}
\ee

\paragraph{3.2.1.A Examples; rotations introduced} Here are three examples.\\
\indent (i): $n \times n$ matrices equipped with the usual commutator, i.e. 
$[X,Y] := XY - YX$. (So the matrix multiplication ``contributes'' to the 
bracket, but not to the underlying vector space structure.)\\
\indent (ii): $3 \times 3$ anti-symmetric matrices, equipped with the usual 
commutator.\\
\indent   (iii): $\mathR^3$ equipped with vector multiplication. 
\indent   In fact, example (iii) is essentially the same as example (ii); and 
this example will recur in what follows,  in connection with rotations  and the 
rigid body. (We will also see that example (ii) is in a sense more 
fundamental.)\\
\indent To explain this, we first recall that every anti-symmetric operator $A$ 
on a three-dimensional oriented euclidean space is the operator of vector 
multiplication by a fixed vector, ${\bf \omega}$ say. That is: for all ${\bf 
q}, A{\bf q} = [{\bf \omega}, {\bf q}] \equiv {\bf \omega} \wedge {\bf q}$. 
(Proof: the anti-symmetric operators on $\mathR^3$ for a 3-dimensional vector 
space, since an anti-symmetric $3 \times 3$ matrix has three independent 
components. Vector multiplication by a vector ${\bf {\omega}}$ is a linear and 
anti-symmetric operator; varying ${\bf \omega}$ we get a subspace of the space 
of all anti-symmetric operators on $\mathR^3$; but this subspace has dimension 
3; so it coincides with the space of all anti-symmetric operators.)

With this result in hand,  the following three points  are all readily 
verified.\\
\indent (1):  The matrix representation of $A$ in cartesian coordinates is then
\be
A = \left( \begin{array}{ccc}
0 & -\omega_3 & \omega_2 \\
\omega_3 & 0 & -\omega_1 \\
-\omega_2 & \omega_1 & 0 
\end{array}
\right).
\label{ansym}
\ee
We can write
\be
A \leftrightarrow {\bf \omega} \;\; {\rm{or}} \;\; 
A_{ij} =  -\epsilon_{ijk}\omega_k \;\; {\rm{or}} \;\;
\omega_i = -\frac{1}{2}\epsilon_{ijk}A_{jk}.
\label{3dcorrspce}
\ee
\indent (2): The plane $\Pi$ of vectors perpendicular to ${\bf \omega}$ is an 
invariant subspace for $A$, i.e. $A(\Pi) = \Pi$. And ${\bf \omega}$ is an 
eigenvector for $A$ with eigenvalue 0. This suggest a familiar elementary 
interpretation, which will be confirmed later (Section \ref{lg}): viz. that any 
$3 \times 3$ anti-symmetric matrix $A$ represents a infinitesimal rotation, and 
${\bf \omega}$ represents instantaneous angular velocity. That is, we will 
have, for all ${\bf q} \in \mathR^3$: ${\dot {\bf q}} = A{\bf q} = [{\bf 
\omega}, {\bf q}]$.\\
\indent (3): The commutator of any two $3 \times 3$ anti-symmetric matrices $A, 
B$, i.e. $[A,B] := AB - BA$, corresponds by eq. \ref{3dcorrspce} to vector 
multiplication of the axes of rotation. That is: writing eq. \ref{3dcorrspce}'s 
bijection from vectors  to matrices  as $\Theta: {\bf \omega} \mapsto A =: 
\Theta({\bf \omega})$, we have for vectors ${\bf q,r,s}$
\begin{eqnarray}
(\Theta({\bf q})\Theta({\bf r}) - \Theta({\bf r})\Theta({\bf q})){\bf s} = 
\Theta({\bf q})[{\bf r},{\bf s}] - \Theta({\bf r})[{\bf q},{\bf s}] \\ 
= [{\bf q}, [{\bf r},{\bf s}]] - [{\bf r}, [{\bf q},{\bf s}]] \\ 
= [[{\bf q}, {\bf r}], {\bf s}] = \Theta([{\bf q}, {\bf r}]) \cdot {\bf s}.
\label{theta3dcorrspce} 
\end{eqnarray}
where the [,] represents vector multiplication, i.e. $[{\bf q}, {\bf r}] \equiv 
{\bf q} \wedge {\bf r}$.

Eq. \ref{theta3dcorrspce} means that $\Theta$ gives a Lie algebra isomorphism; 
and so  our example (iii) is essentially the same as example (ii).\\
\indent Besides, we can already glimpse why example (ii) is in a sense more 
fundamental. For this correspondence between anti-symmetric operators (or 
matrices) and vectors, eq. \ref{3dcorrspce}, is specific to three dimensions. 
In $n$ dimensions, the number of independent  components of an anti-symmetric 
matrix is $n(n-1)/2$: only for $n=3$ is this equal to $n$. Yet we will see 
later (Section \ref{rotgrp}) that rotations on euclidean space $\mathR^n$ of 
any dimension $n$ are generated, in a precise sense, by the Lie algebra of $n 
\times n$ anti-symmetric matrices. So only for $n = 3$ is there a corresponding  
representation of rotations by vectors in $\mathR^n$.

\indent   In the next two Subsections, we shall see other examples of Lie 
algebras: whose vectors are vector fields (Section \ref{lavecfields}), or 
tangent vectors at the identity element  of a Lie group (Section \ref{lg}). The 
first example will be an infinite-dimensional Lie algebra; the second 
finite-dimensional (since we will only consider finite-dimensional Lie groups). 
Besides, the above examples (i) and (ii) (equivalently: (i) and (iii)) will 
recur: each will be the vector space of tangent vectors at the identity element  
of a Lie group.

\paragraph{3.2.1.B Structure constants} A finite-dimensional Lie algebra is 
characterized, relative to a basis, by a set of numbers, called {\em structure 
constants} that specify the bracket operation. Thus if $\{v_1,...,v_n\}$ is a 
basis of a Lie algebra $V$, we define the structure constants $c^k_{ij}, (i,j,k 
= 1,...,n)$ by expanding, in terms of this basis, the bracket of any two basis 
elements
\be
[v_i,v_j] = \Sigma_k c^k_{ij} v_k \;\; .
\label{defstrucconst}
\ee 
The bilinearity of the bracket implies that eq. \ref{defstrucconst} determines 
the bracket of all pairs of vectors $v, w \in V$. And the bracket's obeying 
anti-symmetry and the Jacobi identity implies that, for any basis, the 
structure constants obey
\be
c^k_{ij} = - c^k_{ji} \;\; ; \;\;
\Sigma_k (c^k_{ij}c^m_{kl} + c^k_{li}c^m_{kj}+ c^k_{jl}c^m_{ki}) = 0 
\label{propstrucconst}
\ee
Conversely, any set of constants $c^k_{ij}$ obeying eq. \ref{propstrucconst} 
are the structure constants of an $n$-dimensional  Lie algebra.
 
\subsubsection{The Lie bracket of two vector fields}\label{lavecfields}
Given two vector fields $X,Y$ on a manifold $M$, the corresponding flows do not 
in general commute: $X^tY^s \neq Y^sX^t$. The non-commutativity is measured  by 
the commutator of the Lie derivatives of $X$ and of $Y$, i.e. ${\cal L}_X {\cal 
L}_Y - {\cal L}_Y {\cal L}_X$. (Cf. eq. \ref{defineLiederiv} and 
\ref{Liederivgenldefn} for a definition of the Lie derivative.) Here, 
`measured' can be  made precise by considering Taylor expansions; but I shall 
not go into detail about this. 

What matters  for us is that this commutator, which is at first glance seems to 
be a second-order operator, is in fact a first-order operator. This is verified 
by calculating in a coordinate system, and seeing that the second derivatives 
occur twice with opposite signs:
\begin{eqnarray}
({\cal L}_X {\cal L}_Y  - {\cal L}_Y {\cal L}_X) f= \Sigma_i \; X^i 
\frac{\pl}{\pl x^i} \left(\Sigma_j Y^j \frac{\pl f}{\pl x^j} \right) \; - \; 
\Sigma_j \; Y^j \frac{\pl}{\pl x^j} \left(\Sigma_i X^i \frac{\pl f}{\pl 
x^i}\right) 
\\ 
= ... = \Sigma_{i,j} \; \left(X^i \frac{\pl Y^j}{\pl x^i} - Y^i \frac{\pl 
X^j}{\pl x^i} \right)\frac{\pl f}{\pl x^j} .
\label{Liebr1storder}
\end{eqnarray} 
So ${\cal L}_X {\cal L}_Y - {\cal L}_Y {\cal L}_X$ corresponds to a vector 
field: (recall (2) of Section \ref{Lie deriv}, about defining a vector field 
from its Lie derivative). We call this field $Z$ the {\em Lie bracket} (also 
known as:  Poisson bracket, commutator, and Jacobi-Lie bracket!) of the fields 
$X$ and $Y$, and write it as $[X,Y]$. It is also written as ${\cal L}_X Y$ and 
called the {\em Lie derivative of} $Y$ {\em with respect to} $X$. (Beware: some 
books use an opposite sign convention.)

Thus $Z \equiv [X,Y] \equiv {\cal L}_X Y$ is defined to be the vector field 
such that 
\be
{\cal L}_Z \equiv {\cal L}_{[X,Y]} = {\cal L}_X {\cal L}_Y - {\cal L}_Y {\cal 
L}_X \;\; .
\ee
It follows that $Z \equiv [X,Y]$'s components in a coordinate system  are given 
by eq. \ref{Liebr1storder}. This formula can be remembered by writing it  (with 
summation convention, i.e. omitting the $\Sigma$) as
\be
\left[X^i \frac{\pl }{\pl x^i}, Y^j \frac{\pl }{\pl x^J}\right] = 
X^i \frac{\pl Y^j}{\pl x^i}\frac{\pl }{\pl x^j} - Y^j \frac{\pl X^i}{\pl 
x^j}\frac{\pl }{\pl x^i}
\label{Liemnemonic}
\ee
Another way to write eq. \ref{Liebr1storder} is as:
\be
[X,Y]^j = \;\; (X \cdot \nabla)Y^j \; - \; (Y \cdot \nabla)X^j \;\; ;
\label{Lienabla}
\ee
or without coordinates, writing $\bf D$ for the derivative map given by the 
Jacobian matrix, as 
\be
[X,Y] = {\bf D}Y \cdot X - {\bf D}X \cdot Y.
\label{LieBoldD}
\ee
Again, the vector field $Z \equiv [X,Y]$ measures the non-commutation of the 
flows $X^t$ and $Y^s$: in particular, these flows commute iff $[X,Y] = 0$.

We will need three results about the Lie bracket. They concern, respectively, 
the relation to Lie algebras, to Poisson brackets, and to Frobenius' theorem.\\  
\indent (1): The Lie bracket is obviously a bilinear and anti-symmetric 
operation on the (infinite-dimensional) vector space  ${\cal X}(M)$ of all 
vector fields on $M$: $[,]: {\cal X}(M) \times {\cal X}(M) \rightarrow {\cal 
X}(M)$. One readily checks that it satisfied the Jacobi identity. (Expand 
${\cal L}_{[[X,Y],Z]} = {\cal L}_{[X,Y]} {\cal L}_Z - {\cal L}_Z {\cal 
L}_{[X,Y]}$ etc.) So: ${\cal X}(M)$ is an (infinite-dimensional) Lie algebra.

(2): Returning to Hamiltonian mechanics (Section \ref{july05}): there is a 
simple and fundamental relation between the Lie bracket and the Poisson 
bracket, via the notion of Hamiltonian vector fields (Section 
\ref{noetcomplete}).\\
\indent Namely: the Hamiltonian vector field  of the Poisson bracket of two 
scalar functions $f, g$ on the symplectic manifold $M$ is, upto a sign, the Lie 
bracket of the Hamiltonian vector fields, $X_f$ and $X_g$, of $f$ and $g$: 
\be
X_{\{f, g\}} = - [X_f, X_g] = [X_g, X_f] . 
\label{HamvfsclosedunderLie;symp}
\ee 
Proof: apply the rhs to an arbitrary scalar $h: M \rightarrow \mathR$. One 
easily obtains $X_{\{f, g\}}(h)$, by using:\\
\indent (i) the definition of a Hamiltonian vector field;\\
\indent (ii) the Lie derivative of a function equals its elementary directional 
derivative  eq. \ref{defineLiederiv}; and \\
\indent (iii) the Poisson bracket is antisymmetric and obeys the Jacobi 
identity.

This result means that the Hamiltonian vector fields on a symplectic  manifold 
$M$, equipped with the Poisson bracket, form an (infinite-dimensional) Lie 
subalgebra of the Lie algebra ${\cal X}(M)$ of all vector fields on the 
symplectic manifold $M$.  Later, it will be important that this result extends 
from symplectic manifolds to Poisson manifolds; (details in Section 
\ref{hvfs}).

(3): For Frobenius' theorem (Section \ref{Frob}), we need to relate the Lie 
bracket to Section \ref{fieldtoflow}'s idea of  vector fields being $f$-related 
by a map $f: M \rightarrow N$ between manifolds $M$ and $N$. In short: if two 
pairs of vector fields are $f$-related, so is their Lie bracket. More 
explicitly: if $X, Y$ are vector fields on $M$, and  $f: M \rightarrow N$ is a 
map such that $(Tf)(X), (Tf)(Y)$ are well-defined vector fields on $N$, then 
$Tf$ commutes with the Lie bracket:
\be
(Tf)[X,Y] = [(Tf)X,(Tf)Y] \; .
\label{TfcommLB}
\ee

\subsection{Submanifolds and Frobenius' theorem}\label{Frob}
This Subsection differs from the preceding ones in three ways. First, it 
emphasises integral, rather than differential, notions.

Second: Section \ref{fieldtoflow}.B have emphasised that the integral curves of 
a vector field correspond to integrating a system of ordinary differential 
equations. Since such curves are one-dimensional submanifolds of the given 
manifold, our present topic, viz. higher-dimensional submanifolds, naturally 
suggests partial differential equations. For their integration involves 
finding, given an assignment to each point $x$ of a manifold $M$ of a subspace 
$S_x$ (with dimension greater than one) of the tangent space $T_xM$, an 
integral surface, i.e. a submanifold $S$ of $M$ whose tangent space at each of 
its points is $S_x$.\footnote{Beware: there is no analogue for partial 
differential equations of the local existence and uniqueness theorem for 
ordinary differential equations. Even a field of two-dimensional planes in 
three-dimensional space is in general not integrable, e.g. the field of planes 
given by the equation $dz = ydx$. So integrable fields of planes, or other 
tangent subspaces on a manifold, are an exception; and accordingly, the 
integration theory for partial differential equations is less unified, and more 
complicated, than that for ordinary differential equations.}\\
\indent However, we will {\em not} be concerned with partial differential 
equations. For us, submanifolds of dimension higher than one arise when the 
span $S_x$ of the tangent vectors at $x$ to a set of {\em vector} fields fit 
together to form a submanifold. Thus Frobenius' theorem  states, roughly 
speaking, that a finite set of vector fields is integrable in this sense iff 
the vector fields are in involution. That is: iff their pairwise Lie brackets 
are expandable in terms of the fields; i.e. the vector fields  form a Lie 
subalgebra of the entire Lie algebra of vector fields. We will not need to 
prove this theorem. But we need to state it and use it---in particular, for the 
foliation of Poisson manifolds.  

Third: a warning is in order. The intuitive idea of a subset $S \subset M$ that 
is a smooth manifold ``in its own right'' can be made precise in different 
ways. So there are subtleties about the definition of `submanifold', and 
terminology varies between expositions---in a way it does not for the material 
in previous Sections. I will adopt what seems to be a widespread, if not 
majority, terminology.\footnote{My treatment is based on Marsden and Ratiu 
(1999, p. 124-127, 140) for Section \ref{323ASubmanifolds}, and  Olver (2000, 
p. 38-40) for Section \ref{323BFrobTheorem}.  As to varying terminology:  Olver 
(2000, p. 9) defines `submanifold' to be what we will call an immersed 
submanifold; (which latter, for us, does {\em not} have to be a submanifold, 
since the immersion need not be an embedding). Bishop and Goldberg (1980, p. 
40-41) provide a similar example.  For a detailed introduction to the different 
notions of submanifold, cf. Darling (1994, Chapters 3 and 5). Note that I will 
also omit some details, in particular about Frobenius' theorem providing 
regular immersions.}

\subsubsection{Submanifolds}\label{323ASubmanifolds}
The fundamental definition is:\\
\indent Given a manifold $M$ (dim($M$)=$n$), a {\em submanifold} of $M$ of 
dimension $k$ is a subset $N \subset M$ such that for every $y \in N$ there is 
an admissible local chart (i.e. a chart in $M$'s maximal atlas) $(U,\phi)$ with 
$y \in U$ and with the {\em submanifold property}, viz.
\be
{\rm{(SM)}}. \; \phi: U \rightarrow \mathR^k \times \mathR^{n-k} \;\; 
{\rm{and}} \;\; 
\phi(U \cap N) = \phi(U) \cap (\mathR^k \times \{{\bf 0}\}).
\label{defsubmfd}  
\ee
The set $N$ becomes a manifold, generated by the atlas of all charts of the 
form $(U \cap N, \phi\mid(U \cap N))$, where $(U,\phi)$ is a chart of $M$ 
having the submanifold property. (This makes the topology of $N$ the relative 
topology.)

We need to take note of two ways in which submanifolds can be specified in 
terms of smooth functions between manifolds.\\
\indent (1): A submanifold can be specified as the set on which a smooth 
function $f: M \rightarrow P$ between manifolds  takes a certain value.  In 
effect, this will be  a generalization of eq. \ref{defsubmfd}'s requirement 
that $n - k$ coordinate-components of a chart $\phi$ take the value zero. This 
will involve the idea that the tangent map $Tf$ is  surjective, in which case 
$f$ will be called a {\em submersion}. We will need this approach for quotients 
of actions of Lie groups.\\
\indent (2): A submanifold can be specified parametrically, as the set of 
values of a local parametrization: i.e. as the range of a smooth function $f$ 
with $M$ as {\em codomain}. This will involve the idea that the tangent map 
$Tf$ is injective, in which case $f$ will be called an {\em immersion}. We will 
need this approach for Frobenius' theorem.

(1): {\em Submersions}:---\\
 If $f:M \rightarrow P$ is a smooth map between manifolds, a point $x \in M$ is 
called a {\em regular point} if the tangent map $T_x f$ is surjective; 
otherwise $x$ is a {\em critical point} of $f$. If $C \subset M$ is the set of 
critical points of $M$, we say $f(C)$ is the set of {\em critical values} of 
$f$, and $P - f(C)$ is the set of {\em regular values} of $f$. So if $p \in P$ 
is a regular value of $f$, then at every $x \in M$ with $f(x)=p$, $T_x f$ is 
surjective.

\indent The {\em submersion theorem} states that if $p \in P$ is a regular 
value of $f$, then:\\
\indent (i): $f^{-1}(p)$ is a submanifold of $M$ of dimension dim($M$) - 
dim($P$); and\\
\indent (ii): the tangent space of this submanifold at any point $x \in 
f^{-1}(p)$ is the kernel of $f$'s tangent map:
\be
T_x (f^{-1}(p)) = {\rm{ker}}T_x f \;.
\ee 
If $T_x f$ is surjective for every $x \in M$, $f$ is called a {\em submersion}.

(2): {\em Immersions}:---\\
 A smooth map between manifolds $f:M \rightarrow P$ is called an {\em 
immersion} if $T_x f$ is injective at every $x \in M$. The {\em immersion 
theorem} states that $T_x f$ is injective iff there is a neighbourhood $U$ of 
$x$ in $M$ such that $f(U)$ is a submanifold of $P$ and $f \mid_U : U 
\rightarrow f(U)$ is a diffeomorphism. 

\indent NB: This does not say that $f(M)$ is a submanifold of $P$. For $f$ may 
not be injective (so that $f(M)$ has self-intersections). And even if $f$ is 
injective, $f$ can fail to be a homeomorphism between $M$ and $f(M)$, equipped 
with the relative topology induced from $P$. A standard simple example is an 
injection of an open interval of $\mathR$ into an ``almost-closed'' 
figure-of-eight in $\mathR^2$.\\ 
\indent Nevertheless, when $f:M \rightarrow P$ is an immersion, and is also 
injective, we call $f(M)$ an {\em injectively immersed submanifold} (or 
shorter: an {\em immersed submanifold}): though $f(M)$ might not be a 
submanifold.  

We also define an {\em embedding} to be an immersion that is also a 
homeomorphism (and so injective) between $M$ and $f(M)$ (where the latter has 
the relative topology induced from $P$). If $f$ is an embedding, $f(M)$ {\em 
is} a submanifold of $N$ and $f$ is a diffeomorphism $f: M \rightarrow f(M)$.

In fact, Frobenius' theorem will provide injectively immersed submanifolds that 
need not be embedded, and so need not be submanifolds. (They must also obey 
another condition, called `regularity', that I will not go into.)

\subsubsection{The theorem}\label{323BFrobTheorem}
We saw at the end of Section \ref{lavecfields}  that if two pairs of vector 
fields are $f$-related, so is their Lie bracket: cf. eq. \ref{TfcommLB}. This 
result immediately yields a necessary condition for two vector fields to be 
tangent to an embedded submanifold: namely\\
\indent If $X_1,X_2$ are vector fields on $M$ that are tangent to an embedded 
submanifold $S$ (i.e. at each $x \in S$, $X_i(x) \in T_xS < T_xM$), then their 
Lie bracket $[X_1,X_2]$ is also tangent to $S$.\\
\indent This follows by  considering the diffeomorphism $f:{\tilde {S}} 
\rightarrow S$ that gives an embedding of $S$ in $M$. One then uses  the fact 
that $Tf$ commutes with the Lie bracket, eq. \ref{TfcommLB}. That is: the Lie 
bracket of the $f$-related vector fields ${\tilde{X}}_1, {\tilde{X}}_2$ on 
${\tilde{S}}$, which is of course tangent to ${\tilde{S}}$, is carried by $Tf$ 
to the Lie bracket $[X_1,X_2]$ of $X_1$ and $X_2$. So $[X_1,X_2]$ is tangent to 
$S$.

The idea of Frobenius' theorem will be that this necessary condition of two 
vector fields being tangent to a submanifold is also sufficient. To be more 
precise, we need the following definitions.

\indent  A {\em distribution} $D$ on a manifold $M$ is a subset of the tangent 
bundle $TM$ such that at each $x \in M$, $D_x := D  \cap T_xM$ is a vector 
space. The dimension of $D_x$ is the {\em rank} of $D$ at $x$. If the rank of 
$D$ is constant on $M$, we say the distribution is {\em regular}.\\
\indent A distribution is {\em smooth} if for every $x \in M$, and every $X_0 
\in D_x$, there is a neighbourhood $U \subset M$ of $x$, and a smooth vector 
field $X$ on $U$ such that (i) $X(x) = X_0$, (ii) for all $y \in U$, $X(y) \in 
D_y$. Such a vector field $X$ is called a {\em local section} of $D$. Example: 
a set of $r$ vector fields, $X_1,...,X_r$ each defined on $M$, together define 
a smooth distribution of rank at most $r$.\\
\indent A distribution is {\em involutive} if for any pair $X_1,X_2$ of local 
sections, the Lie bracket $[X_1,X_2](y) \in D_y$ in the two sections' common 
domain of definition.

\indent We similarly say that a set of $r$ smooth vector fields, $X_1,...,X_r$, 
on a manifold $M$ is {\em in involution} if  everywhere in $M$ they span their 
Lie brackets. That is: there are smooth real functions $h^k_{ij}:M \rightarrow 
R, i,j,k = 1,...,r$ such that at each $x \in M$
\be
[X_i,X_j](x) = \Sigma_k \; h^k_{ij}(x) X_k(x).
\label{defineinvol}
\ee
(Beware: involution is used in a different  sense in connection with 
Liouville's theorem, viz. a set of real functions on phase space is said to be 
in involution when all their pairwise Poisson brackets vanish.)

\indent A distribution $D$ on $M$ is {\em integrable} if for each $x \in M$ 
there is a local submanifold $N(x)$ of $M$ whose tangent bundle equals the 
restriction of $D$ to $N(x)$. If $D$ is integrable, the various $N(x)$ can be 
extended to get, through each $x \in M$, a unique maximal connected set whose 
tangent space at each of its elements $y$ is $D_y$. Such a set is called a 
(maximal) {\em integral manifold}.\\
\indent NB: In general, each integral manifold is injectively immersed in $M$, 
but not embedded in it; and so, by the discussion in (2) of Section 
\ref{323ASubmanifolds}, an integral manifold might not be a submanifold of $M$. 
But (like most treatments), I shall ignore this point, and talk of them as 
submanifolds, {\em integral submanifolds}.\\
\indent If the rank of $D$ is constant on $M$, all the integral submanifolds 
have a common dimension: the rank of $D$. But in general the rank of $D$ varies 
across $M$, and so does the dimension of the integral submanifolds.

\indent We similarly say that a set of $r$ vector fields, $X_1,...,X_r$, is 
{\em integrable}; viz. if through every $x \in M$ there passes a local 
submanifold $N(x)$ of $M$ whose tangent  space  at each of its points  is 
spanned by $X_1,...,X_r$. (Again: we allow that at some $x$, 
$X_1(x),...,X_r(x)$ may be linearly dependent, so that the dimension of the 
submanifolds varies.)

\indent We say (both for distributions and sets of vector fields) that the 
collection of integral manifolds is a {\em foliation} of $M$, and its elements 
are {\em leaves}. Again: if  the dimension  of the leaves is constant on $M$, 
we say the foliation is {\em regular}. 

With these definitions in hand, we can now state Frobenius' theorem:  both in 
its usual form, which concerns the case of constant rank, i.e. regular 
distributions and vector fields that are everywhere linearly independent; and 
in a generalized form. The usual form is:    
\begin{quote}
{\bf {Frobenius' theorem (usual form)}} A smooth regular distribution is 
integrable iff it is involutive.\\ 
\indent Or in terms of vector fields: a set of $r$ smooth vector fields, 
$X_1,...,X_r$, on a manifold $M$, that are everywhere linearly independent, is 
integrable iff it is in involution. 
\end{quote}
The generalization comes in two stages. The first stage concerns varying rank, 
but assumes a  finite set of vector fields. It is straightforward: this very 
same statement holds. That is: 
a set of $r$ smooth vector fields, $X_1,...,X_r$, on a manifold $M$ (perhaps 
not everywhere linearly independent) is integrable iff it is in involution.

But  for the foliation of Poisson manifolds (Section \ref{Pomaps}), we need to 
consider an infinite set of vector fields, perhaps  with varying rank; and for 
such a set, this statement fails. Fortunately, there is a useful 
generalization; as follows. 

\indent Let $\cal X$ be a set of vector fields on a manifold $M$, that forms a 
vector space.  So in the above discussion of $r$ vector fields, $\cal X$ can be 
taken as all the linear combinations $\Sigma^r_{i=1} \; f_i(x)X_i(x), x \in M$, 
where the $f_i$ are arbitrary smooth functions $f:M \rightarrow \mathR$. Such 
an $\cal X$ is called {\em finitely generated}. \\
\indent For any $\cal X$ forming a vector space, we say (as before) that $\cal 
X$ is {\em in involution} if $[X,Y] \in {\cal X}$ whenever $X,Y \in {\cal X}$. 
Let ${\cal X}_x$ be the subspace of $T_xM$ spanned by the $X(x)$ for all $X \in 
{\cal X}$. As before, we define: an {\em integral manifold} of $\cal X$ is a 
submanifold $N \subset M$ such that for all $y \in N$, $T_yN = {\cal X}_y$; and 
${\cal X}$ is called {\em integrable} iff through each $x \in M$ there passes 
an integral manifold.\\
\indent As before: if $\cal X$ is integrable, it is in involution. But the 
converse fails. A further condition is needed, as follows.\\
\indent We say that $\cal X$ is {\em rank-invariant} if for any vector field $X 
\in {\cal X}$, the dimension of the subspace ${\cal X}_{\exp(\tau X)(x)}$ along 
the flow generated by $X$ is a constant, independent of $\tau$. (But it can 
depend on the point $x$.)\\
\indent Since the integral curve ${\exp(\tau X)(x)}$ through $x$ should be 
contained in any integral submanifold, rank-invariance is certainly a necessary 
condition of integrability. (It also follows from $\cal X$ being finitely 
generated.) In fact we have:
\begin{quote}
{\bf {Frobenius' theorem (generalized form)}} A system $\cal X$ of vector 
fields on $M$ is integrable iff it is rank-invariant and in involution. 
\end{quote}
The idea of the proof is to directly construct the integral submanifolds. The 
submanifold through $x$ is obtained as
\be
N = \{\exp(X_1)\exp(X_2)....\exp(X_p)(x): p \geq 1, X_i \in {\cal X} \}.
\ee 
The rank-invariance secures that for any $y \in N$, ${\cal X}_y$ has dimension 
dim($N$).

\subsection{Lie groups, and their Lie algebras}\label{lg}
I introduce Lie groups and their Lie algebras. By the last two Subsections 
(Sections \ref{examsubsub} and \ref{rotgrp}), we will have enough theory to 
compute efficiently the Lie algebra of a fundamentally important Lie group, the 
rotation group.

\subsubsection{Lie groups and matrix Lie groups}\label{lgintrodcd}
A {\em Lie group} is a group $G$ which is also a manifold, and for which the 
product and inverse operations $G \times G \rightarrow G$ and $G \rightarrow G$ 
are smooth.

Examples:----\\
\indent (i): $\mathR^n$ under addition.\\
\indent (ii): The group of linear isomorphisms of $\mathR^n$ to $\mathR^n$, 
denoted $GL(n,\mathR)$ and called the {\em general linear group}; represented 
by the real invertible $n \times n$ matrices. This is an open subset of 
$\mathR^{n^2}$, and so a manifold  of dimension $n^2$; and the formulas for the 
product and inverse of matrices are smooth in the matrix components.\\
\indent  (iii) The group of rotations about the origin of $\mathR^3$, 
represented by  $3 \times 3$ orthogonal matrices of determinant 1; denoted 
$SO(3)$, where $S$ stands for `special' (i.e. determinant 1), and $O$ for 
`orthogonal'.

In fact, all three examples can be regarded as Lie groups of matrices, with 
matrix multiplication as the operation. In example (i), consider the 
isomorphism $\theta$ between $\mathR^n$ under addition and $(n+1) \times (n+1)$ 
matrices with diagonal entries all equal to 1, other rightmost column entries 
equal to the given vector in $\mathR^n$, and all other entries zero. Thus 
consider, for the case $n = 3$:
\be
\theta: \left( \begin{array}{c}
x \\
y \\
z
\end{array}
\right) \mapsto 
\left( \begin{array}{cccc}
1 & 0 & 0 & x \\
0 & 1 & 0 & y \\
0 & 0 & 1 & z \\
0 & 0 & 0 & 1 
\end{array}
\right).
\ee
This suggests that we define a {\em matrix Lie group} to be any set of 
invertible real matrices, under matrix multiplication, that is closed under 
multiplication, inversion and taking of limits. That a matrix Lie group is a 
Lie group will then follow from $GL(n,\mathR)$ being a Lie group, and the 
theorem below (in Section \ref{examsubsub}) that any closed subgroup of a Lie 
group is itself a Lie group.

For matrix Lie groups, some of the  theory below simplifies. For example, the 
definition of exponentiation of an element of the group's Lie algebra  reduces 
to exponentiation of a matrix. But we will develop some of the general theory, 
since (as always!) it is enlightening and powerful.

\subsubsection{The Lie algebra of a Lie group}\label{LieOfLie}
The main result in this Subsection is that for any Lie group $G$, the tangent  
space $T_eG$ at the identity $e \in G$ has a natural Lie algebra structure that 
is induced by certain natural vector fields on $G$; as follows.

\paragraph{3.4.2.A Left-invariant vector fields define the Lie 
algebra}\label{livfdefinela}: \\
Let $G$ be a Lie group. Each $g \in G$ defines a diffeomorphism of $G$ onto 
itself by {\em left translation}, and similarly by {\em right translation}:
\be
L_g: h \in G \mapsto gh \in G \;\; ; \;\; 
R_g: h \in G \mapsto hg \in G .
\label{lrtransl}
\ee
Remark: In Section \ref{actionlg} we will describe this in the language of 
group actions, saying that in eq. \ref{lrtransl} $G$ {\em acts on itself} by 
left and right translation.

  Now consider the induced maps on the tangent spaces, i.e. the tangent (aka: 
derivative) maps; cf. eq.s \ref{definetgt}, \ref{definetgtatx}. They are 
$(L_g)_* =:  L_{g*}, (R_g)_* =:  R_{g*}$ where for each $h \in G$:
\be
L_{g*}:T_hG \rightarrow T_{gh}G \;\; \mbox{ and} \;\; R_{g*}:T_hG \rightarrow 
T_{hg}G .
\label{derivsleftrighttrans}
\ee
In particular: the derivative $(R_g)_*$ at $e \in G$ maps $T_eG$ to $T_gG$. 
This implies that every vector $\xi \in T_eG$ defines a vector field on $G$: 
its value at any $g \in G$ is the image $(R_g)_*\xi$ of $\xi$ under $(R_g)_*$. 
Such a vector field is called a {\em right-invariant} vector field: it is 
uniquely defined by (applying the derivative of right translation to) its value 
at the identity $e \in G$.

In more detail, and now defining {\em left-invariant} vector fields:---\\
 A vector field $X$ on $G$ is called {\em left-invariant} if for every $g \in 
G$, $(L_g)_* X = X$. More explicitly, let us write $T_h L_g$ for the tangent or 
derivative of $L_g$ at $h$, i.e. for $L_{g*}:T_hG \rightarrow T_{gh}G$. Then 
left-invariance requires that 
\be
(T_h L_g) X(h) = X(gh) \;\; {\rm{for \; every \;}} g {\rm{\; and \;}} h \in G.
\label{linvarce}
\ee
Thus every vector $\xi \in T_eG$ defines a left-invariant  vector field, 
written $X_{\xi}$, on $G$: $X_{\xi}$'s value at any $g \in G$ is the image 
$(L_g)_*\xi$ of $\xi$ under $(L_g)_*$. In other words: $X_{\xi}(g) := (T_e 
L_g)\xi$. 

Not only is a left-invariant vector field uniquely defined by its value at the 
identity $e \in G$. Also, the set ${\cal X}_L(G)$ of left-invariant vector 
fields on $G$ is isomorphic as a vector space to the tangent space $T_eG$ at 
the identity $e$. For the linear maps $\al,\bb$ defined by
\be
\al: X \in {\cal X}_L(G) \mapsto X(e) \in T_eG \;\; ; \;\; {\rm{and}} \;\; \bb: 
\xi \in T_eG \mapsto \{g \mapsto X_{\xi}(g) := (T_eL_g)\xi \} \in {\cal X}_L(G)
\label{defalbeta}
\ee
compose to give the identity maps:
\be
\bb \circ \al = id_{{\cal X}_L(G)} \;\; ; \;\; 
\al \circ \bb = id_{T_eG}.
\label{defalbetaidy}
\ee

${\cal X}_L(G)$ is a Lie subalgebra of the Lie algebra of all vector fields on 
$G$, because it is closed under the Lie bracket. That is: the Lie bracket of 
left-invariant vector fields $X$ and $Y$ is itself left-invariant, since one 
can check that for every $g \in G$ we have (with $L$ meaning `left' not `Lie'!)
\be
L_{g*}[X,Y] = [L_{g*}X, L_{g*}Y] = [X,Y].
\label{liclosedLieb}
\ee
If we now define a bracket on $T_eG$ by 
\be
[\xi,\eta] := [X_{\xi},X_{\eta}](e)
\label{defineLA'sbracket}
\ee
then $T_eG$ becomes a Lie algebra. It is called {\em the Lie algebra of} $G$, 
written $\mg$ (or, to avoid ambiguity about which Lie group is in question: 
$\mg(G)$). It follows from eq. \ref{liclosedLieb} that 
\be
[X_{\xi}, X_{\eta}] = X_{[\xi,\eta]} \;\; ;
\label{LIcommuteLiebr}
\ee
that is to say, the maps $\al, \bb$ are Lie algebra isomorphisms.

This result, that $T_eG$ has a natural Lie algebra structure, is {\em very} 
important. For, as we shall see in the rest of Section \ref{lg}: the structure 
of a Lie group is very largely determined by the structure of this Lie algebra.  
Accordingly, as we shall see in Sections \ref{actionlg} and \ref{pms} et seq.: 
this Lie algebra underpins most of the constructions made with the Lie group, 
e.g. in Lie group actions. Thus Olver writes that this result `is the 
cornerstone of Lie group theory ... almost the entire range of applications of 
Lie groups to differential equations ultimately rests on this one 
construction!' (Olver 2000: 42).\\
\indent Before turning in the next Subsection to examples, and the topic of 
subgroups and subalgebras, I end with four results, (1)-(4), which will be 
needed later; and a remark.

\paragraph{3.4.2.B Four results}\label{4results}: \\
(1): Lie group structure determines Lie algebra structure in the following 
sense. If $G, H$ are Lie groups, and $f:G \rightarrow H$ is a smooth 
homomorphism, then the derivative of $f$ at the identity $T_e f: \mg(G) 
\rightarrow \mg(H)$ is a Lie algebra homomorphism. In particular, for all $\xi, 
\eta \in \mg(G)$, $(T_e f)[\xi, \eta] = [T_e f(\xi),T_e f(\eta)]$. (Cf. eq. 
\ref{TfcommLB}.)

(2): {\em Exponentiation again; a correspondence between left-invariant vector 
fields and one-dimensional subgroups}:\\
Recall from Section \ref{prelimtool}, especially eq. \ref{expnlintrodd}, that 
each vector field $X$ on the manifold $G$ determines an integral  curve 
$\phi_X$ in $G$ passing through the identity  $e$ (with $\phi_X(0) = e$). We 
now write the points  in (the image of) this curve  as $g_{\tau}$ ($X$ and $e$ 
being understood):
\be
\exp(\tau X)(e) \equiv X^{\tau}(e) \equiv \phi_{X,e}(\tau) =: g_{\tau}.
\label{expnlintroddagain}
\ee  
It is straightforward to show that if $X$ is left-invariant, this (image of a) 
curve is a {\em one-parameter subgroup} of $G$: i.e. not just as eq. 
\ref{defineflow} et seq., a one-parameter subgroup of the group of 
diffeomorphisms of the manifold $G$. In fact:
\be
g_{\tau + \sigma} = g_{\tau}g_{\sigma} \;\;\; g_{0} = e \;\;\;
g^{-1}_{\tau} = g_{- \tau} \;\; .
\label{oneparasubgrp}
\ee 
Besides, the group is defined for all $\tau \in \mathR$; and is isomorphic to 
either $\mathR$ or the circle group $SO(2)$. Conversely, any connected 
one-parameter subgroup of $G$ is generated by a left-invariant vector field in 
this way.

\indent Accordingly, we define exponentiation of elements $\xi$ of $\mg$  by 
reference to the isomorphisms eq. \ref{defalbeta} and \ref{defalbetaidy}. It is 
also convenient to define this as a map taking values in $G$. Thus for $\xi \in 
\mg$ and its corresponding left-invariant  vector field  $X_{\xi}$ that takes 
as value at $g \in G$, $X_{\xi}(g) := (T_e L_g)(\xi)$, we write the integral  
curve of $X_{\xi}$ that passes through  $e$ (with value $e$ for argument $\tau 
= 0$) as 
\be
\phi_{\xi}: \tau \in \mathR \mapsto \exp(\tau X_{\xi})(e) \in G \;\; .
\label{intcveasexpl}
\ee 
Then we define the {\em  exponential map} of $\mg$ into $G$ to be the map
\be
\exp: \xi \in \mg \mapsto \phi_{\xi}(1) \in G \; .
\label{explofLiealg}
\ee  
Using the linearity of $\bb$ as defined by eq. \ref{defalbeta}, these two 
equations, eq. \ref{intcveasexpl} and \ref{explofLiealg}, are related very 
simply:
\be
\exp(\tau \xi) := \phi_{\tau \xi} (1) := \exp(1. X_{\tau \xi})(e) = \exp(\tau 
X_{\xi}) \; .
\label{relateexpxitoexpXsubxi}
\ee  
We write  $\exp_G$ rather than $\exp$ when the context could suggest a Lie 
group other than $G$.

\indent The map $\exp$ is a local diffeomorphism of a neighbourhood of $0 \in 
\mg$ to a neighbourhood of $e \in G$; but not in general a global 
diffeomorphism onto $G$. In modern terms, this result follows by applying the 
inverse function theorem to the discussion above. (It also represents an 
interesting example of the history of subject; cf. Hawkins (2000: 82-83) for 
Lie's version of this result, without explicit mention of its local nature.)

The map $\exp$ also has the basic property, adding to result (1) above, that 
...

(3): {\em Homomorphisms respect exponentiation}:\\
If $f: G \rightarrow H$ is a smooth homomorphism of Lie groups, then for all 
$\xi \in \mg$, 
\be
f(\exp_G \xi) = \exp_H ((T_e f)(\xi)).
\label{homorespect}
\ee

\indent (4): {\em Right-invariant vector fields as an alternative approach}:\\
We have followed the usual practice of defining $\mg$ in terms of 
left-invariant vector fields. One can instead use right-invariant vector 
fields. This produces some changes in signs, and in whether certain defined 
operations respect or reverse the order of two elements used in their 
definition. I will not go into many details about this. But some will be needed 
when we consider:\\
\indent (i): Lie group actions, and especially their infinitesimal generators 
(Section \ref{inflgenor} and \ref{example:adcoad});\\
\indent (ii): reduction on the cotangent bundle of a Lie group---as occurs in 
the theory of the rigid body (Section \ref{mommmapcotgt} and 
\ref{rednmommfn}).\\
For the moment we just note two basic results, (A) and (B); postponing others 
to Section \ref{inflgenor} et seq..   

(A): Corresponding to the vector space isomorphism between $\mg$ and the 
left-invariant vector fields, as in eq. \ref{defalbeta}. viz.
\be
\xi \in T_eG \mapsto X_{\xi} \in {\cal X}_L(G)  \;\; {\rm{with}} \;\; 
X_{\xi}(g) := (T_eL_g)\xi \; ,
\label{xiXiagain}
\ee 
there is a vector space isomorphism to the set of right-invariant vector fields
\be
\xi \in T_eG \mapsto Y_{\xi} \in {\cal X}_R(G)  \;\; {\rm{with}} \;\; 
Y_{\xi}(g) := (T_eR_g)\xi \; .
\label{isotoYxitoo}
\ee 
Besides, the Lie bracket of right-invariant vector fields is itself 
right-invariant. So corresponding to our previous definition, eq. 
\ref{defineLA'sbracket}, of a Lie bracket on $T_eG$, and its corollary eq. 
\ref{LIcommuteLiebr}, i.e. $[X_{\xi}, X_{\eta}] = X_{[\xi,\eta]}$, that makes 
$T_eG \cong {\cal X}_L(G)$ a Lie algebra isomorphism: we can also define a Lie 
bracket on $T_eG$ by 
\be
[\xi, \eta]_R := [Y_{\xi}, Y_{\eta}](e) \; ,
\label{defineLA'sbracketbyRI}
\ee 
and get a Lie algebra isomorphism $T_eG \cong {\cal X}_R(G)$.

(B): But the two Lie brackets, eq. \ref{defineLA'sbracket} and 
\ref{defineLA'sbracketbyRI}, on $T_eG$ are different. In fact one can show 
that:\\
\indent \indent (i): $X_{\xi}$ and $Y_{\xi}$ are related by
\be
I_* X_{\xi} = - Y_{\xi}
\label{XiminusYi}
\ee  
where $I:G \rightarrow G$ is the inversion map $I(g) := g^{-1}$, and $I_*$ is 
the push-forward on vector fields induced by $I$, cf. eq. \ref{definepushfwd}, 
i.e.
\be
(I_* X_{\xi})(g) := (TI \circ X_{\xi} \circ I^{-1})(g) \; .
\label{defineI_*}
\ee 
Besides, since $I$ is a diffeomorphism, eq. \ref{XiminusYi} makes $I_*$ a 
vector space isomorphism.\\
\indent \indent (ii): It follows from eq. \ref{XiminusYi} that 
\be
[X_{\xi}, X_{\eta}](e) = - [Y_{\xi}, Y_{\eta}](e) \;\;; \;\; {\rm{so}} \;\;
[\xi, \eta] = - [\xi, \eta]_R \; .
\label{RILBminusUsualLB}
\ee

Finally, a remark about physics. In applications to physics, $G$ is usually the 
group of symmetries of a physical system, and so a vector field on $G$ is the 
infinitesimal generator of a one-parameter group of symmetries. For mechanics, 
we saw this repeatedly in Section \ref{rednintro}, especially as regards the 
group of translations and rotations about the origin, in physical space 
$\mathR^3$. This Subsection's isomorphism between the Lie algebra  $\mg$ and 
left-invariant vector fields on $G$ means that we can think of $\mg$ also as 
consisting of infinitesimal symmetries of the system. (The $\xi \in \mg$ are 
also called generators of the group $G$.)

\subsubsection{Examples, subgroups and subalgebras}\label{examsubsub} 
I begin with the first two of Section \ref{lgintrodcd}'s three examples. That 
will prompt a little more theory, which will enable us to deal efficiently in 
the next Subsection with the third example, viz. the rotation group.

(1): {\em Examples}:--- \\
\indent (i): $G := \mathR^n$ under addition. $G$ is abelian so that left and 
right translation coincide. The invariant vector fields are just the constant 
vector fields, so that ${\cal X}_L(G) \equiv {\cal X}_R(G) \cong \mathR^n$. So 
the tangent space at the identity $T_eG$, i.e. the Lie algebra $\mg$, is itself 
$\mathR^n$. The bracket structure is wholly degenerate: for all invariant 
vector fields $X,Y$, $[X,Y] = 0$; and for all $\xi,\eta \in \mg$, $[\xi,\eta] = 
0$. 

 (ii): $G := GL(n,\mathR)$, the general linear group. Since $G$ is open in 
$End(\mathR^n,\mathR^n)$, the vector space of all linear maps on $\mathR^n$ 
(`$End$' for `endomorphism'), $G$'s Lie algebra, as a vector space, is 
$End(\mathR^n,\mathR^n)$; (cf. example (i)). To compute what the Lie bracket 
is, we first note that  any $\xi \in End(\mathR^n,\mathR^n)$ defines a 
corresponding vector field on $GL(n,\mathR)$  by
\be
X_{\xi}: A \in GL(n,\mathR) \mapsto A\xi \in End(\mathR^n,\mathR^n) \;\; .
\ee
Besides, $X_{\xi}$ is left-invariant, since for every $B \in GL(n,\mathR)$, the 
left translation
\be
L_B: A \in GL(n,\mathR) \mapsto BA \in GL(n,\mathR) 
\ee
is linear, and so
\be
X_{\xi}(L_B A) = BA\xi = T_AL_BX_{\xi}(A) \;\; .
\ee 
Applying now eq. \ref{LieBoldD} at the identity $I \in GL(n,\mathR)$ to the 
definition of the bracket in the Lie algebra, eq. \ref{defineLA'sbracket}, we 
have:
\be
[\xi,\eta] := [X_{\xi},X_{\eta}](I) = 
{\bf D}X_{\eta}(I) \cdot X_{\xi}(I) - {\bf D}X_{\xi}(I) \cdot X_{\eta}(I).
\ee 
But $X_{\eta}A = A\eta$ is linear  in $A$, so ${\bf D}X_{\eta}(I) \cdot B = 
B\eta$. This means that
\be
{\bf D}X_{\eta}(I) \cdot X_{\xi}(I) = \xi\eta \; ;
\ee
and similarly
\be
{\bf D}X_{\xi}(I) \cdot X_{\eta}(I) = \eta\xi .
\ee
So the Lie algebra $End(\mathR^n,\mathR^n)$ has the usual matrix commutator as 
its bracket: $[\xi,\eta] = \xi\eta - \eta\xi$. This Lie algebra is often 
written $\mgl(n,\mathR)$.

\indent Let us  apply to this example, result (2) from Section 
\ref{LieOfLie}.B. In short, the result said that left-invariant vector fields 
correspond (by exponentiation through $e \in G$) to connected one-parameter 
subgroups of $G$. To find the one-parameter subgroup $\exp(\tau X_{\xi})(e)$ of 
$GL(n,\mathR)$, we take the matrix entries $x_{ij}, (i,j = 1,...,n)$ as the 
$n^2$ coordinates on $GL(n,\mathR)$, so that the tangent  space at the identity 
matrix $I$ is the set of vectors
\be
\Sigma_{ij} \;\; \xi_{ij} \; \frac{\pl }{\pl x_{ij}}\mid_I
\ee    
with $\xi = (\xi_{ij})$ an arbitrary matrix. For given $\xi$, $\exp(\tau 
X_{\xi})e$ is found by integrating the $n^2$ ordinary differential equations 
\be
\frac{d x_{ij}}{d \tau} = \Sigma_k \xi_{ik} x_{kj} \;\; ; \;\; x_{ij}(0) = 
\delta_{ij}.
\ee
The solution is just the matrix exponential:
\be
X(\tau) = \exp(\tau \xi).
\label{justmxexpl}
\ee

More generally, let us return to Section \ref{lgintrodcd}'s  idea of a matrix 
Lie group. For a matrix Lie group $G$, the definition of its Lie algebra can be 
given as: 
\be
\mg = \{\mathrm {\; the \; set \; of \; matrices \;} \xi = \phi'(0) : \; \phi 
\; \mathrm{a \; differentiable \; map}: \mathR \rightarrow G, \phi(0) = e_G \}.
\ee 
The deduction of the structure of the Lie algebra then proceeds 
straightforwardly. In particular, we get the result that the one-parameter 
subgroup generated by $\xi \in \mg$ is given by matrix exponentials, as in eq. 
\ref{justmxexpl}: the group  is $\{\exp(\tau \xi): \tau \in \mathR \}$.

This result will help us compute our third example: finding the Lie algebra of 
the rotation group. But for that  example, it is worth first developing a 
little the  result (2)  from Section \ref{LieOfLie}.B: i.e. the correspondence 
between left-invariant vector fields and connected one-parameter subgroups of 
$G$.

(2): {\em More theory}:---\\
\indent First, a warning remark. We will later need to take notice of the fact 
that  a subgroup, even a one-parameter subgroup, of a Lie group $G$ need not be 
a submanifold of $G$. Here we recall Section \ref{323ASubmanifolds}'s 
definitions of immersion and embedding.  Accordingly, we now define a subgroup 
$H$ of a Lie group $G$ to be a {\em Lie subgroup} of $G$ if the inclusion map 
$i: H \rightarrow G$ is an injective immersion.\\
\indent Just as we saw in Section \ref{323ASubmanifolds} that not every 
injective immersion is an embedding, so also there are examples of Lie 
subgroups that are not submanifolds. Example: the torus $\mathsf{T}^2$ can be 
made into a Lie group in a natural way (exercise: do this!); the one-parameter 
subgroups on the torus $\mathsf{T}^2$ that wind densely on the torus are Lie 
subgroups that are not submanifolds. (For more details about this example, cf. 
Arnold (1973: 160-167) or Arnold (1989: 72-74) or Butterfield (2004a: Section 
2.1.3.B).)\\
\indent But it turns out that being closed is a sufficient, and necessary,  
further condition. That is:
\begin{quote}
If $H$ is a closed subgroup of a Lie group $G$, then $H$ is a submanifold of 
$G$ and in particular a Lie subgroup. And conversely, if $H$ is a Lie subgroup 
that is also a submanifold, then $H$ is closed.
\end{quote}

Result (2) from Section \ref{LieOfLie}.B, i.e. the correspondence between 
one-dimensional subgroups of $G$ and one-dimensional subspaces (and so 
subalgebras) of $\mg$, generalizes to higher-dimensional subgroups and 
subalgebras. That is to say:
\begin{quote}
If $H \subset G$ is a Lie subgroup of $G$, then its Lie algebra $\mh := \mg(H)$ 
is a subalgebra of $\mg \equiv \mg(G)$. In fact
\be
\mh = \{ \xi \in \mg : \exp (\tau X_{\xi})(e) \in H \; {\rm{, \; for \; all}}\;  
\tau \in \mathR \}.
\label{characliesubalg}
\ee
And conversely, if $\mh$ is any $m$-dimensional subalgebra of $\mg$, then there 
is a unique connected $m$-dimensional Lie subgroup $H$ of $G$ with Lie algebra 
$\mh$.
\end{quote}
The proof of the first two statements uses result (1) of Section 
\ref{LieOfLie}.B. For the third, i.e. converse, statement, the main idea is 
that $\mh$ defines  $m$ vector fields on $G$ that are linearly independent and 
in involution, so that one can apply Frobenius' theorem to infer an integral 
submanifold. One then has to prove that $H$ is a Lie subgroup: Olver (2000: 
Theorem 1.51) and Marsden and Ratiu (1999: 279-280) give details and 
references. (Historical note: to see that this result, sometimes called Lie's 
`third fundamental theorem', is close to what Lie himself called the main 
theorem of his theory of groups, cf. Hawkins (2000: 83).)

This general correspondence between Lie subgroups and Lie subalgebras prompts 
the question whether every finite-dimensional Lie algebra $\mg$ is the Lie 
algebra of a Lie group. The answer is Yes. Besides, the question reduces to the 
case of a matrix Lie group (i.e. a Lie subgroup of $GL(n,\mathR)$), in the 
sense that: every finite-dimensional Lie algebra  $\mg$ is isomorphic to a 
subalgebra of $\mgl(n, \mathR)$, for some $n$. But be warned: this does not 
imply (and it is not true) that every Lie group is realizable as a matrix Lie 
group, i.e. that every Lie group is isomorphic to a Lie subgroup of 
$GL(n,\mathR)$.

This general correspondence also simplifies greatly the computation of the Lie 
algebras of Lie groups, for example $H := SO(3)$, that are Lie subgroups of 
$GL(n,\mathR)$. We only need to combine it with  example (ii) above, that 
$\mgl(n,\mathR)$ is $End(\mathR^n,\mathR^n)$ with the usual matrix commutator 
as its bracket: $[\xi,\eta] = \xi\eta - \eta\xi$.\\
\indent  Thus we infer that the Lie algebra of $SO(3)$, written $\mso(3)$, is a 
subalgebra of $End(\mathR^n,\mathR^n)$ with the matrix commutator as bracket. 
Besides, we can identify $\mso(3)$ by looking at all the one-dimensional 
subgroups of $G$ contained in it. Combining eq. \ref{justmxexpl} and 
\ref{characliesubalg}, we have
\be
\mso(3) = \{\xi \in \mgl(n,\mathR): {\rm {the \; matrix \; exponential \;}} 
\exp(\tau \xi) \in SO(3), \; \forall \tau \in \mathR \}.
\label{computeh}
\ee
With this result in hand, we can now compute $\mso(3)$.

\subsubsection{The Lie algebra of the rotation group}\label{rotgrp}
Our first aim is to calculate the Lie algebra $\mso(3)$ (also written: $so(3)$) 
of $H := SO(3)$, the rotation group. This will lead us back to Section 
\ref{la}.A's correspondence between anti-symmetric matrices and vectors in 
$\mathR^3$.

 $SO(3)$ is represented by  $3 \times 3$ orthogonal matrices of determinant 1. 
So the requirement in eq. \ref{computeh} becomes, now writing $e$, not $\exp$:
\be
(e^{\tau \xi})(e^{\tau \xi})^T = I \;\; {\rm {and} \;\; det}(e^{\tau \xi}) = 1.
\ee
Differentiating the first equation with respect to $\tau$ and setting $\tau = 
0$ yields 
\be
\xi + \xi^T = 0.
\ee
So $\xi$ must be anti-symmetric, i.e. represented by an anti-symmetric matrix. 
Conversely, for any such anti-symmetric matrix $\xi$, we can show that 
det$(e^{\tau \xi})=1$. So, indeed:
\be
\mso(3) = \{ 3 \times 3 \; \mathrm{antisymmetric \; matrices} \}.
\ee

Notice that the argument is independent of choosing $n = 3$. It similarly 
computes $\mso(n)$ for any integer $n$:
\be
\mso(n) = \{ n \times n \; \mathrm{antisymmetric \; matrices} \}.
\ee
Thus the rotations on euclidean space $\mathR^n$ of any dimension $n$ are 
generated by the Lie algebra of $n \times n$ anti-symmetric matrices.\\
\indent This justifies our assertion at the end of  Section \ref{la}.A that the 
rotation group in three dimensions is special in being representable by vectors 
in the space on which it acts, i.e. $\mathR^3$. For as we have just seen, in 
general the infinitesimal generators of rotations are  anti-symmetric matrices, 
which in $n$ dimensions have $n(n-1)/2$ independent components. But only for $n 
= 3$ does this equal $n$.

Remark: An informal computation of $\mso(3)$, based on the idea that 
higher-order terms in $e^{\tau \xi}$ can be neglected (cf. the physical idea 
that $\xi$ represents an infinitesimal rotation), goes as follows.\\
\indent For $(I + \tau \xi)$ to be a rotation requires that 
\be
(I + \tau \xi)(I + \tau \xi)^T = I \;\; {\rm{and \;\; det}}(I + \xi \tau) = 1.
\label{informal}
\ee
Dropping higher-order terms, the first equation yields
\be
I + \tau(\xi + \xi^T) = I \;\;\; {\rm {i.e.}}\;\; \xi + \xi^T = 0.
\ee
Besides, the second equation in eq. \ref{informal} yields no further 
constraint, since for any anti-symmetric matrix $\xi$ written as (cf. eq. 
\ref{ansym})
\be
\xi =  \left( \begin{array}{ccc}
0 & -\xi_3 & \xi_2 \\
\xi_3 & 0 & -\xi_1 \\
-\xi_2 & \xi_1 & 0 
\end{array}
\right),
\label{ansymxi}
\ee
we immediately compute that det$(I + \xi \tau) = 1 + \tau^2(\xi^2_1 + \xi^2_2 + 
\xi^2_3)$. So, dropping higher-order terms, det$(I + \xi \tau) = 1$. In short, 
we again conclude that
\be
\mso(3) = \{ 3 \times 3 \; \mathrm{antisymmetric \; matrices} \}.
\ee

For later use (e.g. Sections \ref{inflgenor} and \ref{adj;was6.5.5.A}), we note 
that the three matrices   
\be
A^x = \left( \begin{array}{ccc}
0 & 0 & 0 \\
0 & 0 & -1 \\
0 & 1 & 0 
\end{array}\right), \;\;
A^y = \left( \begin{array}{ccc}
0 & 0 & 1 \\
0 & 0 & 0 \\
-1 & 0 & 0 
\end{array}
\right), \;\;
A^z = \left( \begin{array}{ccc}
0 & -1 & 0 \\
1 & 0 & 0 \\
0 & 0 & 0 
\end{array}
\right)
\label{so3genors}
\ee
span $\mso(3)$, and generate the one-parameter subgroups 
\be
R^x_{\theta} = \left( \begin{array}{ccc}
1 & 0 & 0 \\
0 & \cos \theta & -\sin \theta \\
0 & \sin \theta & \cos \theta 
\end{array}\right), \;\;
R^y_{\theta} = \left( \begin{array}{ccc}
\cos \theta & 0 & \sin \theta \\
0 & 1 & 0 \\
-\sin \theta & 0 & \cos \theta 
\end{array}
\right), \;\;
R^z_{\theta} = \left( \begin{array}{ccc}
\cos \theta & -\sin \theta & 0 \\
\sin \theta & \cos \theta & 0 \\
0 & 0 & 1 
\end{array}
\right)
\label{so3subgrps}
\ee
representing anticlockwise rotation around the respective coordinate axes in 
the physical space $\mathR^3$. 

Having computed $\mso(3)$ to consist of antisymmetric matrices, we can use 
Section \ref{la}.A's correspondence between these and vectors in $\mathR^3$ so 
as to realize $\mso(3)$ as vectors with the Lie bracket as vector 
multiplication. With these realizations in hand, we can readily obtain several 
further results about rotations. We will not need any. But a good example, 
which  uses eq. \ref{3dcorrspce}'s isomorphism  $\Theta$ from vectors  ${\bf 
\omega} \in \mathR^3$   to matrices $A \in \mso(3)$, is as follows:---\\    
\indent $\exp(\tau \; \Theta(\omega))$ is a rotation about the axis $\o$ by the 
angle $\tau\parallel \o \parallel$.

We can now begin to see the point of this Chapter's second motto (from Arnold), 
that the elementary theory of the rigid body confuses six conceptually 
different three-dimensional spaces. For our discussion has already 
distinguished three of the six spaces which Arnold lists (in a different 
notation). Namely, we have just distinguished:\\
\indent (i) $\mathR^3$, especially when taken as physical space; from (ii) 
$\mso(3) \equiv T_e(SO(3))$, the generators of rotations; though they are 
isomorphic as Lie algebras, by eq. \ref{3dcorrspce}'s bijection $\Theta$ from 
vectors  ${\bf \omega} \in \mathR^3$   to matrices $A \in \mso(3)$;\\
\indent (ii) $\mso(3) \equiv T_e(SO(3))$ from its isomorphic copy under the 
derivative of left translation by $g$ (i.e. under $(L_g)_*$), viz. 
$T_g(SO(3))$: cf. eq. \ref{linvarce}. (In the motto, Arnold writes $g$ for 
$\mso(3)$ and $G$ for $SO(3)$.)

\indent In Section \ref{LPso(3)} we will grasp (even without developing the 
theory of the rigid body!) the rest of the motto. That is, we will see why  
Arnold also mentions the three corresponding dual spaces, $\mathR^{3*}, 
\mso(3)^*$ and $T^*_g(SO(3))$. But  we can already say more about the two 
tangent spaces $\mso(3) \equiv T_e(SO(3))$ and $T_g(SO(3))$, in connection with 
the idea that for a pivoted rigid body, the configuration space can be taken as 
$SO(3)$; (cf. (3) of Section \ref{rednprospectus}). We will show that there are 
two isomorphisms from $T_g(SO(3))$ to $T_e(SO(3))$ that are natural, not only 
in the mathematical sense of being basis-independent but also in the sense of 
having a physical  interpretation. Namely, they represent the computation of 
the angular velocity  from the Lagrangian generalized velocity, i.e. $\dot q$. 
In effect, one isomorphism computes the angular velocity's components  with 
respect to  an orthonormal frame fixed in space (called {\em spatial 
coordinates}); and the other computes it with respect to a frame fixed in the 
rigid body ({\em body coordinates}). In fact, these isomorphisms are the 
derivatives of  right and left translation, respectively; (cf. eq. 
\ref{lrtransl} and \ref{derivsleftrighttrans}).

So suppose a pivoted rigid body has a right-handed orthonormal frame $\{a, b, c 
\}$ fixed in it. We can think of the three unit vectors $a, b, c$ as column 
vectors in $\mathR^3$. Arranging them in a $3 \times 3$ matrix $g := (a \; b \; 
c) \in GL(3, \mathR)$, we get a matrix that maps the unit $x$-vector $e_1$ to 
$a$, the unit $y$-vector $e_2$ to $b$, etc. That is: $g$ maps the standard 
frame $e_1, e_2, e_3$ to $a, b, c$, and $g$ is an orthogonal matrix: $g \in 
S0(3) = \{ g \in GL(3,\mathR) \mid {\tilde{g}}g = I \; \}$. Thus $g$ represents 
the configuration of the body, and the configuration space is $SO(3)$.\\
\indent By differentiating the condition ${\tilde{g}}g = I$, we deduce that the 
tangent space at a specific $g$ $T_g(SO(3))$, i.e. the space of velocities 
$\dot g$, is the 3-dimensional vector subspace of $GL(3,\mathR)$:
\be
T_g(SO(3)) = \{ {\dot g} \in GL(3,\mathR) \mid \;\; {\dot{\tilde{g}}}g + 
{\tilde{g}}{\dot g} = 0 \;\; \}
\ee
\indent Now recall examples (ii) and (iii) of Section \ref{la}.A. We saw there 
that though the angular velocity  of the body is usually taken to be the vector 
$\o$ such that, with our ``body-vectors'' $a, b, c$,
\be
{\dot a} = \o \wedge a, \;\; {\dot b} = \o \wedge b, \;\; {\dot c} = \o \wedge 
c \; :
\label{naiveangvelyrb}
\ee 
we can instead encode the angular velocity by the antisymmetric matrix $A := 
\Theta(\o) \in \mg \equiv T_e(SO(3))$. As we saw, eq. \ref{naiveangvelyrb} then 
becomes
\be
{\dot a} = \Theta(\o) a, \;\; {\dot b} = \Theta(\o) b, \;\; {\dot c} = 
\Theta(\o) c \; :
\label{matrixangvelyrb}
\ee    
or equivalently the matrix equation for the configuration $g = (a \; b \; c)$,
\be
{\dot g} \equiv ({\dot a} \; {\dot b} \; {\dot c}) = \Theta(\o) g \;\; ; \;\; 
{\rm {i.e.}} \;\; \Theta(\o) = {\dot g} g^{-1} \; .
\label{matrixangvelyrbwithg}
\ee 
Thus we see that the map from $T_g(SO(3))$ to $\mg = T_e(SO(3))$
\be
{\dot g} \in T_g(SO(3)) \mapsto {\dot g} g^{-1} \equiv {\dot g}{\tilde g} \in 
\mg
\label{callitrtrn}
\ee
maps the generalized velocity ${\dot g}$ to the angular velocity $\Theta(\o)$. 
This is the angular velocity represented in the usual elementary way, with 
respect to coordinates  fixed in space. One immediately checks that it is an 
isomorphism (exercise!).\\
\indent On the other hand, let us consider $\Theta(\o)$ as a linear 
transformation $\Theta(\o): \mathR^3 \rightarrow \mathR^3$, and express it in 
the {\em body} coordinates $a, b, c$. This gives $g^{-1}\Theta(\o)g \equiv 
g^{-1}{\dot g}$. Thus the map  
\be 
{\dot g} \in T_g(SO(3)) \mapsto g^{-1}{\dot g}  \equiv {\tilde g}{\dot g} \in 
\mg
\label{callitltrn}
\ee
maps the generalized velocity ${\dot g}$ to the angular velocity expressed in 
body coordinates. It also is clearly an isomorphism.\\
\indent Summing up: we have two natural isomorphisms that compute the angular 
velocity, in spatial and body coordinates respectively, from the generalized 
velocity $\dot g$.\\
\indent Incidentally, one can verify directly that the images ${\dot g}{\tilde 
g}$ and ${\tilde g}{\dot g}$ of the isomorphisms eq. \ref{callitrtrn} and 
\ref{callitltrn} lie in $\mg$, i.e. are antisymmetric matrices. Thus with 
$\cdot$ for the elementary dot-product, we have:
\be
g^{-1}{\dot g}  \equiv {\tilde g}{\dot g} = \left( \begin{array}{c}
{\tilde a}\\
{\tilde b} \\
{\tilde c}
\end{array}\right)
({\dot a} \; {\dot b} \; {\dot c}) = 
\left( \begin{array}{ccc}
0 & {a \cdot {\dot b}} & {a \cdot {\dot c}} \\
{b \cdot {\dot a}} & 0 & {b \cdot {\dot c}} \\
{c \cdot {\dot a}} & {c \cdot {\dot b}} & 0 
\end{array}\right) \; .
\label{calcg-1gdot}
\ee  
This is an antisymmetric matrix, since differentiating $a \cdot b = b \cdot c = 
a \cdot c = 0$ with respect to time gives $a \cdot {\dot b} + {\dot a} \cdot b 
= 0$ etc. Finally, we deduce that ${\dot g}{\tilde g}$ is antisymmetric from 
the facts that ${\dot g}{\tilde g} = g(g^{-1}{\dot g})g^{-1}$ and antisymmetry 
is preserved by conjugation by $g$.

We end this Subsection with two incidental remarks; (they will not be used in 
what follows).\\
\indent (1): In Section \ref{cotgtblesymp}, we could have specialized the 
discussion from a symplectic manifold to a symplectic vector space, i.e. a 
(real, finite-dimensional) vector space equipped with a  non-degenerate 
anti-symmetric bilinear form $\o: Z \times Z \rightarrow \mathR$. It follows 
that $Z$ is of even dimension. The  question then arises which linear maps $A:Z 
\rightarrow Z$ preserve the normal form of $\o$ given by eq. 
\ref{eq;defineOmega}. It is straightforward to show that this is equivalent to 
$A$ preserving the form of Hamilton's equations (for any Hamiltonian); so that 
these maps $A$ are called {\em symplectic} (or {\em canonical}, or {\em 
Poisson}). The set of all such maps form a Lie group, the {\em symplectic 
group}, written Sp($Z,\o$). But since this Chapter will not need the theory of 
canonical transformations, I leave the study of Sp($Z,\o$)'s structure as an 
exercise! (For details, cf. e.g. Abraham and Marsden (1978: 167-174),  Marsden 
and Ratiu (1999: 69-72, 293-299).)   

(2): Finally, a glimpse of the infinite-dimensional manifolds that this Chapter 
has foresworn. Consider the infinite-dimensional Lie group $Diff(M)$ of all 
diffeomorphisms on $M$. An element of its Lie algebra, i.e. a vector $A \in 
T_e(Diff(M))$, is a vector field, or equivalently a flow, on $M$. Besides, the 
Lie bracket in this Lie algebra  $T_e(Diff(M))$, as defined by eq. 
\ref{defineLA'sbracket} turns out to be the usual Lie bracket of the vector 
fields on $M$, as defined in Section \ref{lavecfields}.

\section{Actions of Lie groups}\label{actionlg}
We turn to actions of Lie groups on manifolds. The notions, results and 
examples in this Section will be crucial from Section \ref{sympcoad} onwards. 
Fortunately, the foregoing provides several examples of the notions and results 
we need. Section \ref{basicaction} will give basic material, including the 
crucial notion of cotangent lifts. Sections \ref{quotstruc} and 
\ref{332Aproper} describe conditions for orbits and quotient spaces to be 
manifolds. Section \ref{inflgenor} describes actions infinitesimally, i.e. in 
terms of their infinitesimal generators.  Section \ref{example:adcoad} presents 
two important representations of a Lie group, its adjoint and co-adjoint 
representations, on its Lie algebra $\mg$ and on the dual $\mg^*$ respectively. 
Finally, Section \ref{kinicslgs} gathers some threads concerning our central, 
recurring example, viz. the rotation group.

\subsection{Basic definitions and examples}\label{basicaction}
A {\em left action} of a Lie group $G$ on a manifold $M$ is a smooth map $\Phi: 
G \times M \rightarrow M$ such that:\\
\indent (i): $\Phi(e,x) = x$ for all $x \in M$\\
\indent (ii): $\Phi(g,\Phi(h,x)) = \Phi(gh,x)$ for all $g,h \in G$ and all $x 
\in M$.\\
\indent We sometimes write $g \cdot x$ for $\Phi(g,x)$.

Similarly, a {\em right action} of a Lie group $G$ on a manifold $M$ is a 
smooth map $\Psi: M \times G \rightarrow M$ satisfying (i) $\Psi(x,e) = x$ and 
(ii) $\Psi(\Psi(x,g),h) = \Psi(x,gh)$. We sometimes write $x \cdot g$ for 
$\Psi(x,g)$.

It is convenient to also have a subscript notation. For every $g \in G$, we 
define
\be
\Phi_g: M \rightarrow M \;\; : \;\; x \mapsto \Phi(g,x).
\label{definesubscripaction}
\ee 
In this notation, (i) becomes $\Phi_e = id_M$ and (ii) becomes $\Phi_{gh} = 
\Phi_g \circ \Phi_h$. For right actions, (ii) becomes $\Psi_{gh} = \Psi_h \circ 
\Psi_g$.

One immediately verifies that any left action $\Phi$ of $G$ on a manifold $M$, 
$g \mapsto \Phi_g: M \rightarrow M$, defines a right action $\Psi$ by
\be
g \mapsto \Psi_g := \Phi_{g^{-1}}: M \rightarrow M \;\; ; \;\; {\rm {i.e.}} 
\;\; 
\Psi: (x,g) \in M \times G \mapsto \Phi(g^{-1}, x) \in M \; .
\label{rightfromleftbyinverse} 
\ee
(Use the fact that in $G$, $(gh)^{-1} = h^{-1}g^{-1}$.) Similarly, a right 
action defines a left action, by taking the inverse in $G$. We will 
occasionally make use of this left-right ``flip''.

\indent The definition of left action is equivalent to saying that the map $g 
\mapsto \Phi_g$ is a homomorphism of $G$ into Diff($M$), the group of 
diffeomorphisms of $M$. In the special case where $M$ is a Banach space $V$ and 
each $\Phi_g: V \rightarrow V$ is a continuous linear transformation, the 
action of $G$ on $V$ is called a {\em representation} of $G$ on $V$.

The {\em orbit} of $x \in M$ (under the action $\Phi$) is the set
\be
{\rm {Orb}}(x) = \{\Phi_g(x) : g \in G \} \subset M.
\ee
The action is called {\em transitive} if there is just one orbit, i.e. for all 
$x,y \in M$, there is a $g \in G$ such that $g \cdot x = y$. It is called {\em 
effective} (or {\em faithful}) if $\Phi_g = {\rm{id}}_M$ implies $g = e$, i.e. 
if $g \mapsto \Phi_g$ is one-to-one. It is called {\em free} if it has no fixed 
points for any $g \neq e$: that is, $\Phi_g(x) = x$ implies $g = e$. In other 
words, it is free if for each $x \in M$, $g \mapsto \Phi_g(x)$ is one-to-one.
(So: every free action is faithful.) 

\paragraph{4.1.A  Examples; cotangent lifts} We begin with geometric examples; 
and then return to mechanics, giving first some general theory, followed by 
some examples.

(1): {\em Geometric examples}:---\\
\indent (i): $SO(3)$ acts on $\mathR^3$ by $(A,x) \mapsto Ax$. The action is 
faithful. But it is neither free (each rotation fixes the points on its axis) 
nor transitive (the orbits are the spheres centred at the origin).\\
\indent (ii): $GL(n,\mathR)$ acts on $\mathR^n$ by $(A,x) \mapsto Ax$. The 
action is faithful, not free, and ``almost transitive'': the zero subspace $\{ 
{\bf 0} \}$ is an orbit, and so is $\mathR^n - \{ {\bf 0} \}$.\\
\indent (iii): Suppose $X$ is a vector field on $M$ which is {\em complete} in 
the sense that the flow $\phi_X(\tau)$ of eq. \ref{defineflow} is defined for 
all $\tau \in \mathR$. Then this flow defines an action of $\mathR$ on $M$.  \\
\indent We turn to two examples which will be central, and recurring, in our 
discussion of symplectic reduction.

(iv): Left translation by each $g \in G$, $L_g: h \in G \mapsto gh \in G$ (cf. 
eq. \ref{lrtransl}), defines a left action of $G$ on itself. Since $G$ is a 
group, it is transitive and free (and so faithful). Similarly, right 
translation, $g \mapsto R_g$ with $R_g: h \in G \mapsto hg \in G$, defines a 
{\em right} action. And $g \mapsto R_{g^{-1}}$ defines a {\em left} action; cf. 
eq. \ref{rightfromleftbyinverse}. \\
\indent One readily proves that left translation lifts to the tangent bundle 
$TG$ as a left action. That is: one verifies by the chain rule that
\be
\Phi_g : TG \rightarrow TG \;\;\; : \;\;\; v \equiv v_h \in T_h G \; \mapsto \; 
(T_h L_g)(v) \in T_{gh}G
\ee
defines a left action on $TG$. Similarly, right translation lifts to a right 
action on $TG$. But our interest in Hamiltonian mechanics of course makes us 
more interested in {\em cotangent} lifts. See (2) below for the general 
definitions, and example (viii) in (3) below for the cotangent lift of left 
translation.

(v): $G$ acts on itself by {\em conjugation} (inner automorphism): $g \mapsto 
K_g := R_{g^{-1}} \circ L_g$. That is: $K_g: h \in G \mapsto ghg^{-1} \in G$. 
Each $K_g$ is an isomorphism of $G$. The orbits are conjugacy classes. Section 
\ref{example:adcoad} will introduce two ``differentiated versions'' of action 
by conjugation, viz. the adjoint and co-adjoint actions, which will be 
important in symplectic reduction.

(2): {\em Hamiltonian symmetries and cotangent lifts}:---\\
We turn to Hamiltonian mechanics. Following the discussion in Section  
\ref{noetcomplete}, we say: given a Hamilton  system $(M,\o,H)$ with $(M,\o)$ a 
symplectic manifold and $H:M \rightarrow \mathR$, a {\em Hamiltonian group of 
symmetries} is a Lie group $G$ acting on $M$ such that each $\Phi_g: M 
\rightarrow M$ preserves both $\o$ and $H$. Then the simplest possible examples 
are spatial translations  and-or rotations acting on the free particle. The 
details of these examples, (vi) and (vii) below, will be clearer if we first 
develop some general theory.\\
\indent This theory will illustrate the interaction between the left-right 
contrast for actions, and the tangent-cotangent contrast for bundles. Besides, 
both the general theory and the examples' details will carry over 
straightforwardly, i.e. component by component, to the case of $N$ particles 
interacting by Newtonian gravity, discussed in Section \ref{spacesactions}: the 
action defined on a single particle is just repeated for each of the $N$ 
particles.
 
So we will take $M := (\mathR^3) \times (\mathR^3)^*, \omega := dq^i \wedge 
dp^i, H := p^2/2m.$ In the first place, both translations (by ${\bf x} \in 
\mathR^3$) and rotations (by $A \in SO(3)$) act on the configuration space $Q = 
\mathR^3$. We have actions of $\mathR^3$ and $SO(3)$ on $\mathR^3$ by
\be
\Phi_{\bf x}({\bf q}) = {\bf q} + {\bf x} \;\;\; ; \;\;\; 
\Phi_{A}({\bf q}) = A{\bf q} \; .
\label{trslnrotnActonR3}
\ee

But these actions {\em lift} to the cotangent bundle $T^*Q = (\mathR^3) \times 
(\mathR^3)^* \cong \mathR^6$; (as mentioned in Section \ref{spacesactions}). 
The lift of these actions is defined using a result that does not use the 
notion of an action. Namely: 
\begin{quote}
Any diffeomorphism $f:Q_1 \rightarrow Q_2$ induces a {\em cotangent lift} 
$T^*f: T^*Q_2 \rightarrow T^*Q_1$ (i.e. in the opposite direction) which is 
symplectic, i.e. maps the canonical one-form, and so symplectic form, on 
$T^*Q_2$ to that of $T^*Q_1$.
\end{quote}
To define the lift of an action, it is worth going into detail about the 
definition of $T^*f$. (But I will not prove the result just stated; for 
details, cf. Marsden and Ratiu (1999: Section  6.3).)\\
\indent The idea is that $T^*f$ is to be the ``pointwise adjoint'' of the 
tangent map $Tf: TQ_1 \rightarrow TQ_2$ (eq. \ref{definetgt}). That is: we 
define $T^*f$ in terms of the contraction of its value, for an arbitrary 
argument $\al \in T^*_{q_2}Q_2$, with an arbitrary  tangent vector $v \in 
T_{f^{-1}(q_2)}Q_1$. (Here it will be harmless to (follow many presentations 
and) conflate a point in $T^*Q_2$, i.e. strictly speaking a pair $(q_2, \al), 
q_2 \in Q_2, \al \in T^*_{q_2}Q_2$, with its form $\al$. And similarly it will 
be harmless to conflate a point $(q_1, v)$ in $TQ_1$ with its vector $v \in 
T_{q_1}Q_1$.)\\
\indent We  recall that any finite-dimensional vector space is naturally, i.e. 
basis-independently, isomorphic to its double dual:  $(V^*)^* \cong V$; and we 
will  use angle brackets $< \; ; \; >$ for the natural pairing between $V$ and 
$V^*$. So we define $T^*f; T^*Q_2 \rightarrow T^*Q_1$ by requiring:
\be
< \; (T^*f)(\al) ; v \; > \;\; := \;\; < \; \al ; (Tf)(v)\; > \; , \;\;
\forall \;\; \al \in T^*_{q_2}Q_2, \; v \in T_{f^{-1}(q_2)}Q_1 \;\; .
\label{defineT^*f}
\ee
NB: Because $T^*f$ ``goes in the opposite direction'', the composition of lift 
with function-composition involves a reversal of the order. That is: if $Q_1 = 
Q_2 \equiv Q$ and $f, g$ are two diffeomorphisms of $Q$, then
\be
T^*(f \circ g) = T^*g \circ T^*f.
\label{cotgtliftreversal}
\ee 

With this definition of $T^*f$, a left action $\Phi$ of $G$ on the manifold $Q$ 
induces for each $g \in G$ the cotangent  lift of $\Phi_g: Q \rightarrow Q$. 
That is:  we have the map
\be
T^*\Phi_g \equiv T^*(\Phi_g): T^*Q \rightarrow T^*Q, \; {\rm{with}} \; \al \in 
T^*_qQ \mapsto (T^*\Phi_g)(\al) \in T^*_{g^{-1} \; \cdot \; q}Q \; .
\label{clarifyT^*Phig}
\ee
Now consider the map assigning to each $g \in G$, $T^*\Phi_g$:
\be
g \in G \mapsto T^*\Phi_g: T^*Q \rightarrow T^*Q \; .
\label{defineright}
\ee
To check that this is indeed an action of $G$ on $T^*Q$, we first check that 
since $\Phi_e = id_Q$, $T \Phi_e: TQ \rightarrow TQ$ is $id_{TQ}$ and 
$T^*(\Phi_e)$ is $id_{T^*Q}$. But beware: eq. \ref{cotgtliftreversal} yields
\be
T^*\Phi_{gh} = T^*(\Phi_g \circ \Phi_h) = T^*\Phi_h \circ T^*\Phi_g \; ,
\label{cotgtnaturallyyieldsright}
\ee
so that eq. \ref{defineright} defines a {\em right} action.

But here we recall that any left action defines a right action by using the 
inverse; cf. eq. \ref{rightfromleftbyinverse}. Combining this with the idea of 
the cotangent lift of an action on $Q$, we get:\\
\indent The left action $\Phi$ on $Q$ defines, not only the right action eq. 
\ref{defineright} on $T^*Q$, but also a {\em left} action on $T^*Q$, viz. by
\be
g \in G \mapsto \Psi_g := T^*(\Phi_{g^{-1}}): T^*Q \rightarrow T^*Q \; .
\label{defineleft}
\ee
For since $(gh)^{-1} = h^{-1}g^{-1}$,
\be
\Psi_{gh} \equiv T^*(\Phi_{(gh)^{-1}}) = T^*(\Phi_{h^{-1}g^{-1}}) = 
T^*(\Phi_{h^{-1}} \circ \Phi_{g^{-1}}) = T^*\Phi_{g^{-1}} \circ 
T^*\Phi_{h^{-1}} \equiv \Psi_g \circ \Psi_h \; .
\label{checkleftdefined}
\ee
In short, the two reversals of order cancel out. This sort of left-right flip 
will recur in some important contexts in the following, in particular in 
Sections \ref{mommmapcotgt} and \ref{redn}.

(3): {\em Mechanical examples}:---\\
So much by way of generalities. Now we apply them to translations  and 
rotations of a free particle, to rotations of a pivoted rigid body, and to $N$ 
point-particles.

(vi): Let the translation group $G = (\mathR^3, +)$ act on the free particle's 
configuration space $Q = \mathR^3$ by 
\be
\Phi_{\bf x}({\bf q}) = {\bf q} + {\bf x} \; .
\ee
Since $G$ is abelian, the distinction between left and right actions of $G$ 
collapses. (And if we identify $G$ with $Q$, this is left=right translation by 
$\mathR^3$ on itself, i.e. example (iv) again: and so transitive and free.) But 
of course the lifted actions we have defined, ``with $g$'' and ``with 
$g^{-1}$'', eq. \ref{defineright} and \ref{defineleft} respectively, remain 
distinct actions.\\
\indent Then, writing $\al = ({\bf q},{\bf p}) \in T^*_{\bf q}Q$, and using the 
fact that $T \Phi_{\bf x}({\bf q} - {\bf x},{\dot {\bf q}}) = ({\bf q},{\dot 
{\bf q}})$, we see that eq. \ref{defineT^*f} implies that: first,
\be
T^*(\Phi_{\bf x})({\bf q},{\bf p}) \in T^*_{{\bf q} - {\bf x}}Q \; ;
\ee
and second, that for all ${\dot {\bf q}} \in T_{{\bf q} - {\bf x}}Q$,
\be
< T^*(\Phi_{\bf x})({\bf q},{\bf p}) ; ({\bf q} - {\bf x},{\dot {\bf q}}) \; >  
\;\;
= \;\;
< \; ({\bf q},{\bf p}) ; ({\bf q},{\dot {\bf q}}) > \;\;
\equiv \;\; {\bf p}({\dot {\bf q}}) \; .
\label{rightlifttranslnelypcle}
\ee
For eq. \ref{rightlifttranslnelypcle} to hold for all ${\dot {\bf q}} \in 
T_{{\bf q} - {\bf x}}Q$ requires that $T^*(\Phi_{\bf x})({\bf q},{\bf p})$ does 
not affect ${\bf p}$, i.e.
\be
T^*(\Phi_{\bf x})({\bf q},{\bf p}) = ({\bf q} - {\bf x},{\bf p}) \; .
\label{deducedwithgaction}
\ee
So this is the lifted action ``with $g$'', corresponding to eq. 
\ref{defineright}. Similarly, the lifted action ``with $g^{-1}$'', 
corresponding to eq. \ref{defineleft}, is: $\Psi_{\bf x}({\bf q},{\bf p}) := 
T^*(\Phi_{- {\bf x}})({\bf q},{\bf p}) = ({\bf q} + {\bf x},{\bf p})$.

One readily checks that these lifted actions preserve both $\o = dq^i \wedge 
dp^i$ (an exercise in manipulating the exterior derivative) and $H := p^2/2m.$ 
So we have a Hamiltonian symmetry group. The action is not transitive: the 
orbits are labelled by their values of ${\bf p} \in (\mathR^3)^*$. But it is 
free.

(vii): Let $SO(3)$ act on the left on  $Q = \mathR^3$ by
\be
\Phi_{A}({\bf q}) = A{\bf q} \; .
\ee
(This is example (i) again.) Let us lift this action ``with $g$'', i.e. eq. 
\ref{defineright}, so as to get a right action on $T^*Q$. \\
\indent As in example (vi), we write $\al = ({\bf q},{\bf p}) \in T^*_{\bf 
q}Q$. Using the fact that $T \Phi_{A}({\bf q},{\dot {\bf q}}) = (A{\bf 
q},A{\dot {\bf q}})$, eq. \ref{defineT^*f} then implies that: first,
\be
T^*(\Phi_{A})({\bf q},{\bf p}) \in T^*_{A^{-1}{\bf q}}Q \; ;
\ee
and second, that for all ${\dot {\bf q}} \in T_{A^{-1}{\bf q}}Q$,
\be
< T^*(\Phi_{A})({\bf q},{\bf p}) ; (A^{-1}{\bf q},{\dot {\bf q}}) \; >  \;\;
= \;\;
< \; ({\bf q},{\bf p}) ; ({\bf q},A{\dot {\bf q}}) > \;\;
\equiv \;\; {\bf p}(A{\dot {\bf q}}) \;\; \equiv \;\; p_i A^i_j {\dot q}^j \; .
\label{rightliftrotnelypcle}
\ee
For eq. \ref{rightliftrotnelypcle} to hold for all ${\dot {\bf q}} \in 
T_{A^{-1}{\bf q}}Q$ requires that 
\be
T^*(\Phi_{A})({\bf q},{\bf p}) = (A^{-1}{\bf q},{\bf p}A) \; ,
\label{deducedwithgactionsecond}
\ee
where ${\bf p}A$ is a row-vector. Or if one thinks of the ${\bf p}$ components 
as a column vector, it requires:
\be
T^*(\Phi_{A})({\bf q},{\bf p}) = (A^{-1}{\bf q},{\tilde A}{\bf p}) = 
(A^{-1}{\bf q}, A^{-1}{\bf p}) \; ,
\ee
where $\; \tilde{} \;$ represents the transpose of a matrix, and the last 
equation holds because $A$ is an orthogonal matrix.\\ 
\indent So this is the lifted action ``with $g$'', corresponding to eq. 
\ref{defineright}. Similarly, the lifted action ``with $g^{-1}$'', 
corresponding to eq. \ref{defineleft}, is: $\Psi_{A}({\bf q},{\bf p}) := 
T^*(\Phi_{A^{-1}})({\bf q},{\bf p}) = (A{\bf q},A{\bf p})$.

Again, one readily checks that these lifted actions preserve both $\o = dq^i 
\wedge dp^i$ (another exercise in manipulating the exterior derivative!) and $H 
:= p^2/2m.$ So $SO(3)$ is a Hamiltonian symmetry group.\\
\indent Like the original action of $SO(3)$ on $Q$, these actions are faithful. 
But they are not transitive: the orbits are labelled by the radii of two 
spheres centred at the origins of $\mathR^3$ and $(\mathR^3)^*$. And they are 
not free: suppose $\bf q$ and $\bf p$ are parallel and on the axis of rotation 
of $A$. 

(viii): Now we consider the pivoted rigid body. But unlike examples (vi) and 
(vii), we will consider only kinematics, not dynamics: even for a free body. 
That is, we will say nothing about the definitions of, and invariance of, $\o$ 
and $H$; for details of these, cf. e.g. Abraham and Marsden (1978: Sections 4.4 
and 4.6) and the other references given in (3) of Section \ref{rednprospectus}. 
We will in any case consider the dynamics  of this example in more general 
terms (using momentum maps) in Sections \ref{Egscotgtliftedmms} and 
\ref{redn}.\\
\indent We recall from the discussion at the end of Section \ref{rotgrp} that 
the configuration space of the pivoted rigid body is $SO(3) =: G$. We also saw 
there that the space and body representations of the angular velocity $v = 
{\dot g} \in T_gG$ are given by right and left translation. Thus eq. 
\ref{callitrtrn} and \ref{callitltrn} give:
\be
v^S \equiv {\dot g}^S := T_g R_{g^{-1}}({\dot g}) \;\;\;{\rm{and}} \;\;\;
v^B \equiv {\dot g}^B := T_g L_{g^{-1}}({\dot g}) \;\; .
\label{introSBsuperscript}
\ee   
But we are now concerned with the {\em cotangent} lift of left (or right)  
translation. So let $SO(3)$ act on itself by left translation: $\Phi_g h \equiv 
L_g h = gh$. Let us lift this action ``with $g$'', i.e. eq. \ref{defineright}, 
to get a right action on $T^*G$. So let $\al \in T^*_h G$ and $(TL_g)(h, {\dot 
h}) = (gh, g{\dot h})$. Then eq. \ref{defineT^*f} implies that: first
\be
(T^*L_g)(\al) \in T^*_{g^{-1}h}G \; ,
\ee
and second that for all $v \in T_{g^{-1}h}G$
\be
< \; T^*(L_g)(\al) ; v \; > \; = \; < \; \al ; gv \; > \; .
\label{dedwithgcotgtliftSO3}
\ee
In other words, on analogy with eq. \ref{deducedwithgaction} and 
\ref{deducedwithgactionsecond}: for eq. \ref{dedwithgcotgtliftSO3} to hold for 
all $v \in T_{g^{-1}h}G$ requires that with $gv \in T_h G$:
\be
T^*(L_g)(\al) : v \in T_{g^{-1}h}G \mapsto  \al(gv) \; . 
\ee 
Similarly, the lifted action ``with $g^{-1}$'' corresponding to eq. 
\ref{defineleft}, i.e. the left action on $T^*G$, is
\be
< \; T^*(L_{g^{-1}})(\al) ; v \; > \; = \; < \; \al ; g^{-1}v \; > \; , 
\forall \al \in T^*_h G , v \in T_{gh}G
\label{dedwithg-1cotgtliftSO3}
\ee
We will continue this example in Section \ref{kinicslgs}, after developing more 
of the theory of Lie group actions.

Finally, let us sketch another mechanical example: the case of $N$ particles 
with configuration space $Q := \mathR^{3N}$ interacting by Newtonian 
gravity---discussed in Section \ref{spacesactions}. This will combine and 
generalize examples (vi) and (vii); and lead on to the next Sections' 
discussions of orbits and quotients.

(ix): As I mentioned above (before eq. \ref{trslnrotnActonR3}), the 
cotangent-lifted actions of translations and rotations on a single particle  
carry over straightforwardly to the case of $N$ particles: the action defined 
on a single particle is just repeated, component by component, for each of the 
$N$ particles to give an action on $T^*Q \cong \mathR^{3N} \times 
(\mathR^{3N})^*$.\\
\indent  Furthermore, the groups of translations and rotations are subgroups of 
a single group, the Euclidean group $E$. I shall not define $E$ exactly. Here, 
let it suffice to say that:\\
\indent (a): $E$'s component-wise action on the configuration space $Q := 
\mathR^{3N}$ has a cotangent lift, which is of course also component by 
component.\\
\indent (b): $E$'s cotangent-lifted action is not transitive, nor free; but it 
is faithful.\\
\indent (c): If we take as the Hamiltonian function the $H$ of eq. 
\ref{HNNewtonianpps}, describing the particles as interacting by Newtonian 
gravity, then $E$ is a Hamiltonian symmetry group. In fact, the kinetic and 
potential energies are separately invariant, essentially because the particles' 
interaction depends only on the inter-particle distances, not on their 
positions or orientations; cf. the  discussion in Section \ref{spacesactions}.

A final comment about  example (ix), which points towards the following 
Sections:--- \\
\indent  Recall that in Sections \ref{relationproced} and 
\ref{reductionproced}, we used  this example as a springboard to discussing 
Relationist and Reductionist procedures, which quotiented the configuration 
space or phase space. But in order for the quotient spaces (and orbits) to be 
manifolds, and in particular  for dimensions to add or subtract in a simple 
way, we needed to excise two classes of ``special'' points, before quotienting. 
These were: the class of symmetric configurations or states (i.e. those fixed 
by some element of $E$), and  the class of collision configurations or states. 
For the quotienting of phase space advocated by Reductionism, the classes of 
states were $\delta \subset T^*\mathR^{3N}$ and $\Delta \subset 
T^*\mathR^{3N}$; (cf. Section \ref{reductionproced} for definitions.)\\
\indent With examples (vi) to (ix) in hand, we can now see that:\\
\indent \indent  (a):  $\delta$ and $\Delta$ are each closed under the 
cotangent-lifted action of $E$ on $T^*\mathR^{3N}$; i.e., each is a union of 
orbits. So $E$ acts on $M : = T^*\mathR^{3N} - (\delta \cup \Delta)$.\\
\indent \indent (b): $E$ acts freely on $M$.\\
We will see in the sequel (especially in Sections \ref{332Aproper}.B and 
\ref{quotpm}) that an action being free is one half (one conjunct) of an 
important sufficient condition for orbits and quotient spaces to be manifolds. 
The other conjunct will be the notion of an action being {\em proper}: which we 
will define in Section \ref{332Aproper}.

\subsection{Quotient structures from group actions}\label{quotstruc}
In finite dimensions, any orbit ${\rm {Orb}}(x)$ is an immersed submanifold of 
$M$. This can be proved directly (Abraham and Marsden (1978: Ex. 1.6F(b), p. 
51, and 4.1.22 p. 265)). But for our purposes, this is best seen as a corollary 
of some  conditions under which quotient structures are manifolds; as follows.

The relation, $x \cong y$ if there is a $g \in G$ such that $g \cdot x = y$, is 
an equivalence relation, with the orbits as equivalence classes. We denote the 
quotient space, i.e. the set of orbits, by $M/G$ (sometimes called the {\em 
orbit space}). We write the canonical projection as
\be
\pi: M \rightarrow M/G, \;\;\; x \mapsto {\rm {Orb}}(x) \; ;
\ee
and we give $M/G$ the quotient topology by defining $U \subset M/G$ to be open 
iff $\pi^{-1}(U)$ is open in $M$.

Simple examples (e.g. (ii) of Section \ref{basicaction}.A) show that this 
quotient topology need not be Hausdorff. However, it is easy to show that if 
the set
\be
R : = \{(x,\Phi_g x) \in M \times M : \; (g,x) \in G \times M \}
\ee
is a closed subset of $M \times M$, then the quotient topology on $M/G$ is 
Hausdorff.\\
\indent But to ensure that $M/G$ has a manifold structure, further conditions 
are required. The main one (and a much harder theorem) is:
\begin{quote}
$R$ is a closed submanifold  of $M \times M$ iff $M/G$ is a manifold with $\pi: 
M \rightarrow M/G$ a submersion.
\end{quote}

\indent This theorem has two Corollaries which are important for us.\\
\indent (1): A map $h: M/G \rightarrow N$, from the manifold $M/G$, for which 
$\pi: M \rightarrow M/G$ is a submersion, to the manifold $N$, is smooth iff $h 
\circ \pi: M \rightarrow N$ is smooth.\\
\indent This corollary has a useful implication, called {\em passage to the 
quotients}, about the notion of {\em equivariance}---which will be important in 
symplectic reduction.\\
\indent A smooth map $f: M \rightarrow N$ is called {\em equivariant} if it 
respects the action of a Lie group $G$ on the manifolds. That is: Let $G$ act 
on $M$ and $N$ by $\Phi_g: M \rightarrow M$ and $\Psi_g: N \rightarrow N$ 
respectively. $f: M \rightarrow N$ is called  equivariant  with respect to 
these actions if for all $g \in G$
\be
f \circ \Phi_g = \Psi_g \circ f.
\label{equivarcefinite}
\ee
That is, $f$ is equivariant iff for all $g$, the following diagram commutes:
 \begin{equation}
    \bundle{M}{\Phi_g}{M}
    \bundlemap{f}{f}
    \bundle{N}{\Psi_g}{N}    
	\label{basiceqvarcediag}
\end{equation}
Equivariance immediately implies that $f$ naturally induces a map, ${\hat f}$ 
say, on the quotients. That is: the map
\be
{\hat f}: {\rm{Orb}}(x) \in M/G \mapsto {\rm{Orb}}(f(x)) \in N/G
\label{hatfforpasstoquots}
\ee
is well-defined, i.e. independent of the chosen representative $x$ for the 
orbit.\\
\indent Applying the corollary we have: If $f: M \rightarrow N$ is equivariant, 
and the quotients $M/G$ and $M/N$ are manifolds with the canonical projections 
both submersions, then $f$ being smooth implies that ${\hat f}$ is smooth. This 
is called {\em passage to the quotients}.

\indent (2): Let $H$ be a closed subgroup of the Lie group $G$. (By (2) of 
Section \ref{examsubsub}, this is equivalent to $H$ being a subgroup that is a 
submanifold of $G$.) Let $H$ act on $G$ by left translation: $(h,g) \in H 
\times G \mapsto hg \in G$, so that the orbits are the right cosets $Hg$. Then 
$G/H$ is a manifold and $\pi: G \rightarrow G/H$ is a submersion.

\subsection{Proper actions}\label{332Aproper}
By adding to the Section \ref{quotstruc}'s main theorem (i.e.,  $R$ is a closed 
submanifold  of $M \times M$ iff $M/G$ is a manifold with $\pi: M \rightarrow 
M/G$ a submersion), the notion of a proper action we can give useful sufficient 
conditions for:\\
\indent\indent (A):  orbits to be submanifolds;\\
\indent\indent (B):  $M/G$ to be a manifold.

\indent An action $\Phi: G \times M \rightarrow M$ is called {\em proper} if 
the map
\be
{\tilde{\Phi}}: (g,x) \in G \times M \mapsto (x,\Phi(g,x)) \in M \times M
\label{defineproper}
\ee  
is proper. By this we mean that if $\{x_n \}$ is a convergent sequence in $M$, 
and $\{ \Phi_{g_n}(x_n) \}$ is a convergent sequence in $M$, then $\{g_n \}$ 
has a convergent subsequence in $G$. In finite dimensions, this means that 
compact sets have compact inverse images; i.e. if $K \subset M \times M$ is 
compact, then ${\tilde{\Phi}}^{-1}(K)$ is compact.\\
\indent If $G$ is compact, this condition is automatically satisfied. Also, the 
action of a group on itself by left (or by right)  translation (Example (iv) of 
Section \ref{basicaction}.A) is always proper. Furthermore, the cotangent lift 
of left (or right) translation ((2) and Example (viii) of Section 
\ref{basicaction}.A) is always proper. We shall not prove this, but it will be 
important in the sequel. 
 
\paragraph{4.3.A Isotropy groups; orbits as manifolds}\label{653AIsortopyGrps}
For $x \in M$ the {\em isotropy} (or {\em stabilizer} or {\em symmetry}) group 
of $\Phi$ at $x$ is
\be
G_x := \{g \in G: \Phi_g(x) \equiv \Phi(g,x) = x \} \subset G.
\label{defineisotropy}
\ee
(So an action is free iff for all $x \in M$, $G_x = \{ e \}$.)

So if we define 
\be
\Phi^x: G \rightarrow M \;\; : \;\; \Phi^x(g) := \Phi(g,x)
\label{definesuperscripaction}
\ee
we have: $G_x = (\Phi^x)^{-1}(x)$. (The notation $\Phi^x$ is a ``cousin'' of 
the notation $\Phi_g$ defined in eq. \ref{definesubscripaction}.)\\
\indent So since $\Phi^x$ is continuous, $G_x$ is a closed subgroup of $G$. So, 
by the result in (2) of Section \ref{examsubsub} (i.e. the result before eq. 
\ref{characliesubalg}), $G_x$ is a submanifold (as well as Lie subgroup) of 
$G$. And if the action is proper, $G_x$ is compact.\\
\indent Furthermore, the fact that for all $h \in G_x$ we have  $\Phi^x(gh) = 
\Phi_g \circ \Phi_h (x) = \Phi_g (x)$, implies that $\Phi^x$ naturally induces 
a map 
\be
{\tilde{\Phi}}^x: [g] = gG_x \in G/G_x \mapsto \Phi_g x \in {\rm{Orb}}(x) 
\subset M \; .
\label{definePhitildex}
\ee
That is, this map is well-defined. ${\tilde{\Phi}}^x$ is injective because if 
$\Phi_g x = \Phi_h x$ then $g^{-1}h \in G_x$, so that $gG_x = hG_x$.

It follows from Section \ref{quotstruc}'s main theorem (i.e.,  $R$ is a closed 
submanifold  of $M \times M$ iff $M/G$ is a manifold with $\pi: M \rightarrow 
M/G$ a submersion) that:\\
\indent (a): If $\Phi: G \times M \rightarrow M$ is an action and $x \in M$, 
then ${\tilde{\Phi}}^x$ defined by eq. \ref{definePhitildex} is an injective 
immersion.\\
\indent Here we recall from Section \ref{323ASubmanifolds}  that injective 
immersions need not be embeddings. But:---\\
\indent (b): If also $\Phi$ is proper, the orbit ${\rm{Orb}}(x)$ is a closed 
submanifold of $M$ and ${\tilde{\Phi}}^x$ is a diffeomorphism. In other words: 
the manifold structure of ${\rm {Orb}}(x)$ is given by  the bijective map $[g] 
\in G/G_x \mapsto g \cdot x \in {\rm {Orb}}(x)$ being a diffeomorphism.

{\em Examples}:--- \\
(We  use the numbering of corresponding examples in Section 
\ref{basicaction}.A):---\\
\indent (i): $G = SO(3)$ acts on $M = \mathR^3$ by $(A,x) \mapsto Ax$. Since 
${\rm {Orb}}(x)$ is a sphere centred at the origin of radius $\parallel x 
\parallel$, $M/G \cong \mathR^+$: which is not a manifold. But results (a) and 
(b) are illustrated: the isotropy group $G_x$ at $x$ is the group of rotations 
with $x$ on the axis; the action is proper (for $G$ is compact); the orbit 
${\rm {Orb}}(x)$ is a closed manifold of $M$; and the isotropy group's cosets 
$[g] \in G/G_x$ are mapped diffeomorphically by ${\tilde{\Phi}}^x$ to points on 
the sphere ${\rm {Orb}}(x)$.\\
\indent (iii): Let $X$ be the constant vector field $\pl_x$ on $M = \mathR^3$. 
$X$ is complete. The action of $\mathR$ on $M$ has as orbit through the point 
${\bf x} = (x,y,z) \in \mathR^3$, the line $y =$ constant, $z =$ constant. The 
action is free, and therefore faithful and the isotropy groups are trivial. So 
$G/G_x \equiv G$. The action is proper. Again results (a) and (b) are 
illustrated: the orbits ${\rm{Orb}}({\bf x})$ are closed submanifolds of $M$, 
viz.  copies of the real line $\mathR = G \equiv G/G_x$ that are diffeomorphic 
to $\mathR$ by ${\tilde{\Phi}}^{\bf x}$.

\paragraph{4.3.B A sufficient condition for the orbit space $M/G$ to be a 
manifold}\label{653BSufftForQuotToBeMfd}
With result (b) from the end of Section \ref{332Aproper}.A,, we can prove that:
\begin{quote}
If $\Phi: G \times M \rightarrow M$ is a proper free action, then the orbit 
space $M/G$ is a manifold with $\pi: M \rightarrow M/G$ a submersion.
\end{quote}

{\em Examples}: (again using the numbering in  Section \ref{basicaction}.A):--- 
\\
\indent (i): $G = SO(3)$ acts on $M = \mathR^3$ by $(A,x) \mapsto Ax$. Since 
${\rm {Orb}}(x)$ is a sphere centred at the origin of radius $\parallel x 
\parallel$, $M/G \cong \mathR^+$: which is not a manifold, and indeed the 
action is not free.\\
\indent (iii): Let $X$ be the constant vector field $\pl_x$ on $M = \mathR^3$. 
$X$ is complete, and the action of $\mathR$ on $M$ has as orbits the lines $y 
=$ constant, $z =$ constant. The action is faithful, free and proper, so that 
the orbit space $M/G$ is a manifold: $M/G \cong \mathR^2$.\\
\indent (iv): Left (or right) translation is obviously a free action of a group 
$G$ on itself, and we noted above that it is proper. But since it is 
transitive, the orbit space $G/G$ is the trivial 0-dimensional manifold (the 
singleton set of $G$).\\
\indent (viii): The cotangent lift of left (or right) translation by $SO(3)$, 
or more generally, by a Lie group $G$. This action is proper (noted after eq. 
\ref{defineproper}), and obviously free. \\
\indent (ix): The Euclidean group $E$ acts freely on $M : = T^*\mathR^{3N} - 
(\delta \cup \Delta)$. This action is also proper: a (harder!) exercise for the 
reader.

\subsection{Infinitesimal generators of actions}\label{inflgenor}
We now connect this Subsection's topic, group actions, with the Lie algebra of 
the Lie group concerned, i.e. with the topic  of Section \ref{lg}, especially 
\ref{LieOfLie}.

Let $\Phi: G \times M \rightarrow M$ be a (left) action by the Lie group $G$ on 
a manifold $M$. Then each $\xi \in \mg$ defines an action of $\mathR$ on $M$, 
which we write as $\Phi^\xi$, in the following way.\\
\indent  We can think either in terms of exponentiation of $\xi$'s 
corresponding left-invariant vector field $X_{\xi}$ (cf. eq. \ref{expnlintrodd} 
and \ref{expnlintroddagain}); or in terms of of exponentiating $\xi$ itself 
(cf. eq. \ref{explofLiealg} and \ref{relateexpxitoexpXsubxi}):
\be
\Phi^{\xi}: \mathR \times M \rightarrow M \;\; :\;\;
\Phi^{\xi}(\tau,x) := \Phi(\exp(\tau X_{\xi}),x) \equiv \Phi(\exp(\tau \xi),x).
\label{ladefineRealaction}
\ee 
That is, in terms of our subscript notation for the original action $\Phi$ (cf. 
eq. \ref{definesubscripaction}): $\Phi_{\exp(\tau X_{\xi})} \equiv 
\Phi_{\exp(\tau \xi)}: M \rightarrow M$ is a flow on $M$.

That the flow is complete, i.e. that an action of all of $\mathR$ is defined, 
follows from (2) {\em Exponentiation again} of Section \ref{LieOfLie}, 
especially after eq. \ref{oneparasubgrp}. Cf. also example (iii) of Section 
\ref{basicaction}.  
 
We say that the corresponding vector field on $M$, written $\xi_M$, i.e. the 
vector field defined at $x \in M$ by
\be
\xi_M (x) := \frac{d}{d \tau}\mid_{\tau = 0} \Phi_{\exp(\tau X_{\xi})}(x) 
\equiv \frac{d}{d \tau}\mid_{\tau = 0} \Phi_{\exp(\tau \xi)}(x)
\label{defineinflgenor}
\ee
is the {\em infinitesimal generator} of the action corresponding to $\xi$.\\
\indent In terms of the map $\Phi^x$ defined in eq. 
\ref{definesuperscripaction}, we have that for all $\xi \in \mg$
\be
\xi_M (x) = (T_e \Phi^x)(\xi) \; .
\label{inflgenorintermsofPhisuperx}
\ee

So NB: the words `infinitesimal generator' are used in different, though 
related, ways. In Remark (2) at the end of Section \ref{LieOfLie}, a vector 
field on the group $G$, or an element $\xi \in \mg$, was called an 
`infinitesimal generator'. Here the infinitesimal generator is a vector field 
on the action-space $M$. Similarly, beware the notation: $\xi_M$ is a vector 
field on $M$, while $X_{\xi}$ is a vector field on $G$.

As an example, we again take the rotation group $SO(3)$ acting on $\mathR^3$: 
$(A, {\bf x}) \in SO(3) \times \mathR \mapsto A{\bf x}$. One readily checks 
that with $\o \in \mathR^3$, so that $\Theta(\o) \in \mso(3)$, the 
infinitesimal generator of the action corresponding to $\xi \equiv \Theta(\o)$ 
is the vector field on $\mathR^3$
\be
\xi_{\mathR^3}({\bf x}) \equiv (\Theta(\o))_{\mathR^3}({\bf x}) = \o \wedge 
{\bf x} \; .
\ee
In particular, the vector field on $\mathR^3$ representing infinitesimal 
anti-clockwise rotation about the $x$-axis is $e_1 := y{\pl_z} - z{\pl_y}$ (cf. 
eq. \ref{so3genors}). Similarly, the infinitesimal generators of the action of 
rotating about the $y$ axis and about the $z$-axis are, respectively: $e_2 := 
z{\pl_x} - x{\pl_z}$ and
$e_3 := x{\pl_y} - y{\pl_x}$.  The Lie brackets are given by:
\be
[e_1,e_2] = -e_3 \;\;\; [e_3,e_1] = -e_2 \;\;\; [e_2,e_3] = -e_1.
\ee
The minus signs here are a general feature of the transition $\xi \in \mg 
\mapsto \xi_M \in {\cal X}(M)$; cf. result (4) below.

As another example, we take the infinitesimal generator of left and right 
translation on the group $G$. (We will need this example for our theorems about 
symplectic reduction; cf. Sections \ref{Egscotgtliftedmms}, \ref{rednthm} and 
\ref{rednmommfn}.) NB: There will be a ``left-right flip'' here, which 
continues the discussion in (4) of Section \ref{LieOfLie}.B, comparing using 
left-invariant vs. right-invariant vector fields to define the Lie algebra of a 
Lie group.\\
\indent For left translation $\Phi(g,h) \equiv L_g h := gh$, we have for all 
$\xi \in \mg$:
\be
\Phi^{\xi}(\tau, h) = (\exp \tau \xi)h = R_h (\exp \tau \xi) \; ;
\ee
so that the infinitesimal generator  is
\be
\xi_G(g) = (T_e R_g)\xi \; .
\label{xi_Gisri}
\ee
So $\xi_G$ is a {\em right}-invariant vector field; and unless $G$ is abelian, 
it is {\em not} equal to the left-invariant vector field $g \mapsto X_{\xi}(g) 
:= (T_e L_g)\xi$; cf. eq. \ref{derivsleftrighttrans} and \ref{defalbeta}.\\
\indent Similarly, for right translation (which is a {\em right} action, cf. 
(1) (iv) in Section \ref{basicaction}.A), the infinitesimal generator is the 
left-invariant vector field
\be
g \mapsto X_{\xi}(g) :=  (T_e L_g)\xi \; .
\label{theotherxi_Gisli}
\ee   

Three straightforward results connect the notion of an infinitesimal generator  
with previous ideas. I will not give proofs, but will present them in the order 
of the previous ideas.

(1): Recall the correspondence between Lie subgroups and Lie subalgebras, at 
the end of Section \ref{examsubsub}; eq. \ref{characliesubalg}. This implies 
that the Lie algebra of the isotropy group $G_x, x \in M$ (called the {\em 
isotropy algebra}), is
\be
\mg_x = \{\xi \in \mg : \xi_M(x) = 0 \} \; .
\ee 

(2): Infinitesimal generators $\xi_M$ give a differential version of the notion 
of equivariance, discussed in (1) of Section \ref{quotstruc}: a version called 
{\em infinitesimal equivariance}.\\
\indent In eq. \ref{equivarcefinite}, we set $g = \exp(\tau \xi)$ and 
differentiate with respect to $\tau$ at $\tau = 0$. This gives $Tf \circ \xi_M 
= \xi_N \circ f$. That is: $\xi_M$ and $\xi_N$ are $f$-related. In terms of the 
pullback $f^*$ of $f$, we have: $f^*\xi_N = \xi_M$.

\indent (3): Suppose the action $\Phi$ is proper, so that by result (b) at the 
end of Section \ref{332Aproper}.A: the orbit Orb($x$) of any point $x \in M$ is 
a (closed) submanifold of $M$. Then the tangent space to Orb($x$) at a point 
$y$ in Orb($x$) is
\be
T{\rm{Orb}}(x)_y = \{\xi_M(y) : \xi \in \mg \} \; .
\label{tgtspaceatorbitintermsofinflgenors}
\ee

Finally, there is a fourth result relating infinitesimal generators $\xi_M$ to 
previous ideas; as follows. (But it is less straightforward than the previous 
(1)-(3): its proof requires the notion of the adjoint representation, described 
in the next Section.) 

(4): The infinitesimal generator map $\xi \mapsto \xi_M$ establishes a Lie 
algebra {\em anti}-homomorphism between $\mg$ and the Lie algebra ${\cal X}_M$ 
of all vector fields on $M$. (Contrast the Lie algebra isomorphism between 
$\mg$ and the set ${\cal X}_L(G)$ of left-invariant vector fields on the group  
$G$; Section \ref{LieOfLie} especially eq. \ref{defalbeta}.) That is:
\be
(a\xi + b\eta)_M = a\xi_M + b\eta_M \;\; ; \;\; [\xi_M,\eta_M] = - [\xi,\eta]_M 
\;\; \forall \xi,\eta \in \mg, \; {\rm{and}} \; a,b \in \mathR.
\label{lalgactionantihomo}
\ee
Incidentally, returning to (4) of Section \ref{LieOfLie}.B, which considered 
defining the Lie algebra of a Lie group in terms of right-invariant vector 
fields, instead of left-invariant vector fields: had we done so, the 
corresponding map $\xi \mapsto \xi_M$ would have been a Lie algebra 
homomorphism.

\subsection{The adjoint and co-adjoint representations}\label{example:adcoad}
A leading idea of later Sections (especially Sections \ref{sympcoad}, 
\ref{eqvarcemm} and \ref{redn}) will be that there is a natural symplectic 
structure in the orbits of a certain natural representation of any Lie group: 
namely a representation of the group on the dual of its own Lie algebra, called 
the {\em co-adjoint representation}. Here we introduce this representation. But 
we lead up to it by first describing the {\em adjoint representation} of a Lie 
group on its own Lie algebra. Even apart from symplectic structure (and so 
applications in mechanics), both representations illustrate the ideas of 
previous Subsections. I will again use $SO(3)$ and $\mso(3)$ as examples.
 
\subsubsection{The adjoint representation}\label{adj;was6.5.5.A}
We proceed in four stages. We first define the representation, then discuss 
infinitesimal generators, then discuss matrix Lie groups, and finally discuss 
the rotation group.
  
(1): {\em The representation defined}:---\\
Let $G$ be a Lie group and $\mg$ its Lie algebra, i.e. the tangent space to the 
group at the identity $e \in G$, equipped with the commutator bracket operation 
$[,]$.

Recall (e.g. from the beginning of Section \ref{LieOfLie}) that $G$ acts on 
itself by left and right translation: each $g \in G$ defines diffeomorphisms of 
$G$ onto itself by
\be
L_g: h \in G \mapsto gh \in G \;\; ; \;\; 
R_g: h \in G \mapsto hg \in G .
\ee 
The induced maps of the tangent spaces are, for each $h \in G$:
\be
L_{g*}:TG_h \rightarrow TG_{gh} \;\; \mbox{ and} \;\; R_{g*}:TG_h \rightarrow 
TG_{hg} .
\ee
The diffeomorphism $K_g := R_{g^{-1}} \circ L_g$ (i.e. conjugation by $g, K_g: 
h \mapsto ghg^{-1}$) is an inner automorphism of $G$. (Cf. example (v) at the 
end of Section \ref{basicaction}.) Its derivative at the identity $e \in G$ is 
a linear map from the Lie algebra $\mg$ to itself, which is denoted:
\be
  Ad_g := (R_{g^{-1}} \circ L_g)_{*e} : \mg \rightarrow \mg.
\label{defineAd_g}
\ee 
So letting $g$ vary through $G$, the map $Ad: g \mapsto Ad_g$ assigns to each 
$g$ a member of End($\mg$), the space of linear maps on (endomorphisms of) 
$\mg$. The chain rule implies that $Ad_{gh} = Ad_g Ad_h$. So 
\be
Ad: g \mapsto Ad_g
\ee
is a left action, a representation, of $G$ on $\mg$: $G \times \mg \rightarrow 
\mg$. It is called the {\em adjoint representation}.

Three useful results about $Ad$ follow from our results (1) and (3) in Section 
\ref{LieOfLie}.B (cf. eq. \ref{homorespect}: {\em Homomorphisms respect 
exponentiation}):\\
\indent [1]: If $\xi \in \mg$ generates the one-parameter subgroup $H = 
\{\exp(\tau X_{\xi}): \tau \in \mathR \}$, then $Ad_g(\xi)$ generates the 
conjugate subgroup $K_g(H) = gHg^{-1}$.
\be
 \exp(Ad_g(\xi)) = K_g(\exp \xi) := g(\exp \xi)g^{-1}.
\label{Adggeneratesconjuggrp}
\ee
Incidentally, eq. \ref{Adggeneratesconjuggrp} has a many-parameter 
generalization. Let $H$ and $H'$ be two connected $r$-dimensional Lie subgroups 
of the Lie group $G$, with corresponding Lie subalgebras $\mh$ and $\mh'$  of 
the Lie algebra $\mg = \mg(G)$. Then $H$ and $H'$ are conjugate subgroups, $H' 
= gHg^{-1}$, iff $\mh$ and $\mh'$  are corresponding conjugate subalgebras, 
i.e. $\mh' = Ad_g(\mh)$.

[2]: Eq. \ref{Adggeneratesconjuggrp} also implies another result which will be 
needed for a crucial result about symplectic reduction, in Section 
\ref{mmsforcotgtlifts}. (The many-parameter generalization just mentioned will 
not be needed.) It relates $Ad$ to the pullback of an arbitrary action 
$\Phi$.\\
\indent Thus let $\Phi$ be a left action of $G$ on $M$. Then for every $g \in 
G$ and $\xi \in \mg$
\be
(Ad_g \xi)_M = \Phi^*_{g^{-1}} \xi_M \; ,
\label{AdandPullbackifnlgenor}
\ee
where $\Phi^*$ indicates pullback of the vector field. For we have:
\begin{eqnarray}
(Ad_g \xi)_M (x) := \frac{d}{d \tau}\mid_{\tau = 0} \Phi (\exp(\tau Ad_g 
\xi),x) \\
= \frac{d}{d \tau}\mid_{\tau = 0} \Phi (g(\exp \tau \xi)g^{-1},x) \;\;
{\rm{\; by \; eq. \; \ref{Adggeneratesconjuggrp}}} \\
= \frac{d}{d \tau}\mid_{\tau = 0} (\Phi_g \circ \Phi_{\exp \tau \xi} \circ 
\Phi_{g^{-1}}(x)) \\
= T_{\Phi_{g^{-1}}(x)} \Phi_g (\xi_M (\Phi_{g^{-1}} (x)))  \;\;
{\rm{\; by \; the \; chain \; rule \; and \; eq. \; \ref{defineinflgenor}}} \\
= \left(\Phi^*_{g^{-1}} \xi_M \right) (x) 
\;\;
{\rm{\; by \; the \; definition \; of \; pullback.}}
\label{proveAdandPullbackifnlgenor}
\end{eqnarray}
Not only is this result needed later. Also, incidentally: it is the main part 
of the proof of result (4) at the end of Section \ref{inflgenor}, that $\xi 
\mapsto \xi_M$ is a Lie algebra anti-homomorphism. 

\indent [3]: $Ad_g$ is an algebra homomorphism, i.e.
\be
Ad_g[\xi,\eta] = [Ad_g \xi, Ad_g \eta] \;\; , \;\; \xi,\eta \in {\mg}.
\ee 

(2): {\em Infinitesimal generators: the map $ad$}:---\\
The map $Ad$ is differentiable. Its derivative at $e \in G$ is a linear map 
from the Lie algebra $\mg$ to the space of linear maps on $\mg$. This map is 
called $ad$, and its value for argument $\xi \in {\mg}$ is written $ad_{\xi}$. 
That is:
\be
ad := Ad_{*e}: {\mg} \rightarrow \mbox{End} {\mg} \;\;\; ; \;\;\; ad_{\xi} =
\frac{d}{d \tau}\mid_{\tau =0} Ad_{\exp (\tau \xi)} \; 
\label{sumupAddiffble}
\ee 
where $\exp (\tau \xi)$ is the one-parameter subgroup with tangent  vector  
$\xi$ at the identity. But if we apply the definition eq. \ref{defineinflgenor} 
of the infinitesimal generator of an action, to the adjoint action $Ad$, we get 
that for each $\xi \in \mg$, the generator $\xi_{\mg}$, i.e. a vector field on 
$\mg$, is
\be
\xi_{\mg}: \eta \in \mg \mapsto \xi_{\mg}(\eta) \in {\mg} {\rm{\;\;\; with 
\;\;\; }} 
\xi_{\mg}(\eta) := \frac{d}{d {\tau}}\mid_{\tau = 0} Ad_{\exp (\tau 
\xi)}(\eta).
\label{defineinfgenoradjaction}
\ee
Comparing eq. \ref{sumupAddiffble}, we see that $ad_{\xi}$ is just the 
infinitesimal generator $\xi_{\mg}$ of the adjoint action corresponding to 
$\xi$:
\be
ad_{\xi} = \xi_{\mg} \; . 
\label{adxi=xisubg}
\ee

We now compute the infinitesimal  generators of the adjoint action. It will be 
crucial to later developments (especially Section \ref{sympcoad}) that these 
are given by the Lie bracket in $\mg$.

We begin by considering the function $Ad_{\exp (\tau \xi)}(\eta)$ to be 
differentiated. By eq. \ref{defineAd_g}, we have
\begin{eqnarray}
Ad_{\exp (\tau \xi)}(\eta) = 
T_e(R_{\exp (-\tau \xi)} \circ L_{\exp (\tau \xi)})(\eta) \\ \nonumber
= (T_{\exp (\tau \xi)}(R_{\exp (-\tau \xi)}) \circ T_e L_{\exp (\tau 
\xi)})(\eta) \\ \nonumber
= (T_{\exp (\tau \xi)}(R_{\exp (-\tau \xi)}) \cdot X_{\eta}(\exp (\tau \xi)) 
\end{eqnarray}
where the second line follows by the chain rule, and the third by definition of 
left-invariant vector field. Writing the flow of $X_{\xi}$ as $\phi_{\tau}(g) = 
g \exp \tau \xi = R_{\exp(\tau \xi)}g$, and applying the definition of the Lie 
derivative (eq. \ref{Liederivgenldefn}), we then have
\begin{eqnarray}
\xi_{\mg}(\eta) := \frac{d}{d {\tau}}\mid_{\tau = 0} Ad_{\exp (\tau \xi)}(\eta)  
= \frac{d}{d {\tau}}\left[ 
T_{\phi_{\tau}(e)}\phi^{-1}_{\tau} \cdot X_{\eta}(\phi_{\tau}(e))
\right]\mid_{\tau = 0} \\ \nonumber
= [X_{\xi}, X_{\eta}](e) = [\xi,\eta]. 
\end{eqnarray}
where the final equation is the definition eq. \ref{defineLA'sbracket}  of the 
Lie bracket in the Lie algebra.

\indent So for the adjoint action, the infinitesimal generator corresponding to 
$\xi$ is taking the Lie bracket: $\eta \mapsto [\xi,\eta]$. To sum up: eq. 
\ref{sumupAddiffble} and \ref{defineinfgenoradjaction} now become 
\be
ad = Ad_{*e}: {\mg} \rightarrow \mbox{End} {\mg} \;\;\; ; \;\;\; ad_{\xi} =
\frac{d}{d \tau}\mid_{\tau =0} Ad_{\exp (\tau \xi)} = \xi_{\mg}: \eta \in \mg 
\mapsto [\xi,\eta] \in {\mg}.
\label{sumupad}
\ee 

(3): {\em Example: matrix Lie groups}:---\\
In the case where $G \subset GL(n,\mathR)$ is a matrix Lie group with Lie 
algebra $\mg \subset \mgl(n)$, these results are easy to verify. Writing $n 
\times n$ matrices as $A,B \in G$, conjugation is $K_A(B) = ABA^{-1}$, and the 
adjoint map $Ad$ is also given by conjugation
\be
Ad_A(X) = AXA^{-1}, \;\;\; A \in G, X \in \mg.
\ee 
So with $A(\tau) = \exp(\tau X)$, so that $A(0) = I$ and $A'(0) = X$, we have 
with $Y \in \mg$
\begin{eqnarray}
\frac{d}{d \tau}\mid_{\tau = 0} Ad_{\exp \tau X} Y = 
\frac{d}{d \tau}\mid_{\tau = 0} \left[ A(\tau)YA(\tau)^{-1}
\right] \\ \nonumber
= A'(0)YA^{-1}(0) + A(0)YA^{-1 '}(0).
\label{adforln} 
\end{eqnarray}
But differentiating $A(\tau)A^{-1}(\tau) = I$ yields
\be
\frac{d}{d \tau} (A^{-1}(\tau)) = -A^{-1}(\tau)A'(\tau)A^{-1}(\tau), 
\;\;\;{\rm{and\;so}}\;\; A^{-1 '}(0) = -A'(0) = -X
\ee
so that indeed we have
\be
\frac{d}{d \tau}\mid_{\tau = 0} Ad_{\exp \tau X} Y = XY - YX = [X,Y].
\ee

(4): {\em Example: the rotation group}:--- \\
It is worth giving details for the case of $G = SO(3)$, $\mg = \mso(3)$. We saw 
in Section \ref{rotgrp} (eq. \ref{so3genors}) that the three matrices   
\be
A^x = \left( \begin{array}{ccc}
0 & 0 & 0 \\
0 & 0 & -1 \\
0 & 1 & 0 
\end{array}\right), \;\;
A^y = \left( \begin{array}{ccc}
0 & 0 & 1 \\
0 & 0 & 0 \\
-1 & 0 & 0 
\end{array}
\right), \;\;
A^z = \left( \begin{array}{ccc}
0 & -1 & 0 \\
1 & 0 & 0 \\
0 & 0 & 0 
\end{array}
\right)
\label{so3genorsrepeat}
\ee
span $\mso(3)$, and generate the one-parameter subgroups 
\be
R^x_{\theta} = \left( \begin{array}{ccc}
1 & 0 & 0 \\
0 & \cos \theta & -\sin \theta \\
0 & \sin \theta & \cos \theta 
\end{array}\right), \;\;
R^y_{\theta} = \left( \begin{array}{ccc}
\cos \theta & 0 & \sin \theta \\
0 & 1 & 0 \\
-\sin \theta & 0 & \cos \theta 
\end{array}
\right), \;\;
R^z_{\theta} = \left( \begin{array}{ccc}
\cos \theta & -\sin \theta & 0 \\
\sin \theta & \cos \theta & 0 \\
0 & 0 & 1 
\end{array}
\right)
\label{so3subgrpsrepeat}
\ee
representing anticlockwise rotation around the respective coordinate axes in 
the physical space $\mathR^3$. To calculate the adjoint action of 
$R^x_{\theta}$ on the generator $A^y$, we differentiate the product 
$R^x_{\theta}R^y_{\tau}R^x_{-\theta}$ with respect to $\tau$ and set $\tau = 
0$. That is, we find
\be
Ad_{R^x_{\theta}}(A^y) = R^x_{\theta}(A^y)R^x_{\theta} = 
 \left( \begin{array}{ccc}
0 & -\sin \theta & \cos \theta \\
\sin \theta & 0 & 0 \\
-\cos \theta & 0 & 0 
\end{array}
\right) = \cos \theta \cdot A^y + \sin \theta \cdot A^z.
\label{so3subgrps}
\ee
We similarly find
\be
Ad_{R^x_{\theta}}(A^x) = A^x , \;\; 
Ad_{R^x_{\theta}}(A^z) = -\sin \theta \cdot A^y + \cos \theta \cdot A^z.
\ee
So the adjoint action of the subgroup $R^x_{\theta}$ representing rotations 
around the $x$-axis of physical space is given by rotations around the 
$A^x$-axis in the Lie algebra space $\mso(3)$. Similarly for the other 
subgroups representing rotations around the $y$ or $z$-axis. And so for any 
rotation matrix $R \in SO(3)$, relative to given axes $x,y,z$ for $\mathR^3$, 
its adjoint map $Ad_R$ acting on $\mso(3) \cong \mathR^3$ has the same matrix 
representation  relative to the induced basis $\{A^x,A^y,A^z\}$ of $\mso(3)$. 
(NB: This agreement between $SO(3)$'s adjoint  representation  and  its natural 
physical interpretation is   special to $SO(3)$: it does not hold for other 
matrix Lie groups.)

Finally, the infinitesimal generators of the adjoint action are given by 
differentiation. For example, using eq. \ref{so3subgrps}, we find that
\be
ad_{A^x}(A^y) := 
\frac{d}{d \theta}\mid_{\theta =0} Ad_{R^x_{\theta}} A^y = A^z \;\; ;
\ee
which agrees with the commutator: $A^z = [A^x,A^y]$. 

\subsubsection{The co-adjoint representation}\label{coadj;was6.5.5.B}
Again we proceed in stages. We first define the representation, then discuss 
infinitesimal generators, and then take the rotation group as an example.

(1): {\em The representation defined}:---\\
We recall that a linear map $A:V \rightarrow W$ induces (basis-independently) a 
{\em transpose} (dual) map, written $A^*$  (or $\tilde{A}$ or $A^T$), $A^*: W^* 
\rightarrow V^*$ on the dual spaces, $V^* := \{ \al: V \rightarrow \mathR \mid 
\al \; {\rm{linear}}\; \}$ and similarly for $W^*$;  by
\be
\forall \al \in W^*, \forall v \in V: \; \; {A^*}(\al)(v) \; \equiv \; 
<{A^*}(\al) \; ; \; v> \; := \; \al(A(v)) \; \equiv \;(\al \circ A)(v) \; .
\label{deftranspose}
\ee 
\indent So any representation, ${\cal R}$ say, of a group $G$ on a vector space 
$V$, ${\cal R}: G \rightarrow {\rm{End}}(V)$, induces a representation ${\cal 
R}^*$ of $G$ on the dual space  $V^*$, by taking the transpose. We shall call 
${\cal R}^*$ the {\em dual} or {\em transpose} of ${\cal R}$; it is also 
sometimes called a `contragredient representation'. That is: for ${\cal R}(g): 
V \rightarrow V$, we define ${\cal R}^*(g): V^* \rightarrow V^*$ by
\be
{\cal R}^*(g): \al \in V^* \; \mapsto \;\; {\cal R}^*(g)(\al) := \al({\cal 
R}(g)) \in V^* \;\; .
\label{definecontragredrepn}
\ee

Thus the adjoint representation of $G$ on $\mg$ induces a {\em co-adjoint  
representation} of $G$ on the dual ${\mg}^*$ of its Lie algebra $\mg$, i.e. on 
the cotangent space to the group $G$ at the identity, ${\mg}^* = T^*_eG$. The 
co-adjoint representation will play a central role in symplectic reduction 
(starting in Section \ref{sympcoad}).

\indent So let $Ad^*_g: {\mg}^* \rightarrow {\mg}^*$ be the dual (aka: 
transpose) of $Ad_g$, defined by
\be
\forall \al \in {\mg}^*, \xi \in \mg: \;\;\; 
< Ad^*_g \al ; \xi > \; := \; <\al ; Ad_g \xi >.
\label{defineAd^*}
\ee 
Since $Ad: g \mapsto Ad_g$ is a left action ($Ad_{gh} = Ad_g Ad_h$), the 
assignment $g \mapsto Ad^*_g$ is a right action. So to define a left action, we 
use the inverse $g^{-1}$; cf. eq. \ref{rightfromleftbyinverse} and 
\ref{defineleft}. Namely, we define the left action
\be
(g, \al) \in G \times {\mg}^* \mapsto Ad^*_{g^{-1}} \al \in {\mg}^* \; ;
\label{defineAd^*_g-1}
\ee
called the {\em co-adjoint action} of $G$ on ${\mg}^*$. And the corresponding 
{\em co-adjoint representation} of $G$ on ${\mg}^*$ is denoted by
\be
Ad^*: G \rightarrow {\rm{End}}({\mg}^*), \;\;\; 
Ad^*_{g^{-1}} = (T_e(R_g \circ L_{g^{-1}}))^* \; .
\label{defineAd^*_g-1;repn}
\ee

(2): {\em The map $ad^*$; infinitesimal generators}:---\\ 
The map $Ad^*$ is differentiable. Its derivative at $e \in G$ is a linear map 
from the Lie algebra $\mg$ to the space of linear maps on ${\mg}^*$. This map 
is called $ad^*$, and its value for argument $\xi \in {\mg}$ is written 
$ad^*_{\xi}$. Thus $ad^*_{\xi}$ is an endomorphism of ${\mg}^*$, and we have
\be
ad^* = Ad^*_{*e}: \xi \in {\mg} \rightarrow ad^*_{\xi} \in \mbox{End} {\mg}^* 
\; . 
\ee 
Now recall our deduction from eq. \ref{sumupAddiffble} and 
\ref{defineinfgenoradjaction} that $ad_{\xi} = \xi_{\mg}$, i.e. eq. 
\ref{adxi=xisubg}. In the same way we here deduce an equality to the 
infinitesimal generator of the co-adjoint action:
\be
ad^*_{\xi} = \xi_{\mg^*} \; .
\ee 

In fact, $ad^*_{\xi}$ is, {\em modulo} a minus sign, the {\em adjoint} of 
$ad_{\xi}$, in the usual sense of the natural pairing of a vector space with 
its dual: as we now show. (So the notation $ad^*$ is justified, {\em modulo} a 
minus sign.)

Let us compute for this action, the value of the infinitesimal generator 
$\xi_{\mg^*}$ (a vector field on $\mg^*$, induced by $\xi \in \mg$)  at the 
point $\al \in \mg^*$. That is, we will compute the value $\xi_{\mg^*}(\al)$. 
As usual, we identify the tangent space $(T\mg^*)_{\al}$ in which this value 
lives, with $\mg^*$ itself; and similarly for $\mg$. So, with $\xi_{\mg^*}$ 
acting on $\eta \in \mg$, we compute:
\begin{eqnarray}
<ad^*_{\xi}(\al) ; \eta> \;\; \equiv \;\; <\xi_{\mg^*}(\al) ; \eta > \;\; = 
\;\; \left< 
\frac{d}{d \tau}\mid_{\tau = 0} Ad^*_{\exp -\tau \xi}(\al) 
; \eta \right> \\ 
= \;\; 
\frac{d}{d \tau}\mid_{\tau = 0} \left<Ad^*_{\exp -\tau \xi}(\al) 
; \eta \right> \;\;
= \;\; \frac{d}{d \tau}\mid_{\tau = 0} \left<\al 
; Ad_{\exp -\tau \xi} \eta \right> \\ 
= \;\; \left<\al ; \frac{d}{d \tau}\mid_{\tau = 0}  Ad_{\exp -\tau \xi} \eta 
\right> \;\;
= \;\; < \al ; - [\xi,\eta]> \;\; = \;\; - < \al ; ad_{\xi}(\eta)>.
\label{computead*}
\end{eqnarray}
So $ad^*_{\xi}$, defined as the derivative of $Ad^*$ is, up to a sign, the 
adjoint of $ad_{\xi}$.

(3): {\em Example: the rotation group}:---\\
 Let us now write the elementary vector product in $\mathR^3$ as $\wedge$, and 
identify $\mso(3) \cong (\mathR^3, \wedge)$ and $\mso(3)^* \cong \mathR^{3^*}$. 
And let us have the natural pairing given by the elementary euclidean inner 
product $\cdot$. Then the result just obtained (now with $\bullet$ marking the 
argument-place)
\be
<\xi_{\mg^*}(\al) ; \bullet > = - < \al ; [\xi,\bullet]>
\ee
becomes for $\al \in \mso(3)^*$ and $\xi \in so(3)$
\be
\xi_{\mso(3)^*}(\al) \cdot \bullet = - \al \cdot (\xi \wedge \bullet) \; .
\ee
So for $\eta \in \mso(3)$, we have 
\be
<\xi_{\mso(3)^*}(\al) ; \eta > \; = \xi_{\mso(3)^*}(\al) \cdot \eta = - \al 
\cdot (\xi \wedge \eta) = - (\al \wedge \xi) \cdot  \eta = -
< \al \wedge \xi ; \eta >.
\ee
In short: 
\be
\xi_{\mso(3)^*}(\al) = - \al \wedge \xi = \xi \wedge \al.
\label{sumxial}
\ee

 Now since $SO(3)$ is compact, we know that the co-adjoint action is proper; so  
${\rm{Orb}}(\al)$ is a closed submanifold of $\mso(3)^*$, and eq. 
\ref{tgtspaceatorbitintermsofinflgenors} of Section \ref{inflgenor} applies. So  
if we fix $\al$, and let $\xi$ vary through $\mso(3) \cong \mathR^3$, we get 
all of the tangent space $T_{\al}{\rm{Orb}}(\al)$ to the orbit passing through 
$\al$. Then eq. \ref{sumxial} implies that the tangent space is the plane 
normal to $\al$, and passing through $\al$'s end-point. Letting $\al$ vary 
through $\mso(3)^*$, we conclude that the {\em co-adjoint orbits are the 
spheres centred on the origin}.

In the following Sections, we will see that the orbits of the co-adjoint 
representation of {\em any} Lie group $G$ have a natural symplectic structure.  
So the orbits are always even-dimensional; and by considering all Lie groups 
and all possible orbits, we can get a series of examples of symplectic 
manifolds.\\
\indent Besides, this fact will play a central role in our generalized  
formulation of Hamiltonian mechanics, and in symplectic reduction. And we will 
(mercifully!) get a good understanding of that role, already in Section 
\ref{pmspreamble}. To prepare for that, it is worth gathering some threads 
about our recurrent example, $SO(3)$; and generalizing them to other Lie groups 
...

\subsection{Kinematics on Lie groups}\label{kinicslgs}
To summarize some aspects of this Section, and to make our later discussion of 
reduction clearer, it is worth collecting and generalizing some of our results 
about $SO(3)$ and the description it provides of the rigid body. More 
precisely, we will now combine:\\
\indent (i): the description of space and body coordinates in terms of left and 
right translation, at the end of Section \ref{rotgrp}; \\
\indent (ii): the cotangent lift of translation (example (viii) of Section 
\ref{basicaction}.A);\\
\indent (iii): the adjoint and co-adjoint representations of $SO(3)$ (as in (4) 
of Section \ref{adj;was6.5.5.A}, and (3) of Section \ref{coadj;was6.5.5.B}.

We will also generalize: namely, we will consider (i) to (iii) for an arbitrary 
Lie group $G$, not just for $SO(3)$. (The point of doing so will become clear 
in (3) of Section \ref{pmspreamble}.) This will occur already in Section 
\ref{spbodyonG;was6.5.6.A}. Then in Section \ref{Fundldiffms;was6.5.6.B}, we 
will show how this material yields natural diffeomorphisms 
\be
TG \rightarrow G \times \mg \;\;\; {\rm{and}} \;\;\; T^*G \rightarrow G \times 
\mg^* \;\; ;
\label{announce2ndimldiffs}
\ee 
(so if ${\rm{dim}} G = n$, then all four manifolds are $2n$-dimensional). We 
will also see that by applying Section \ref{quotstruc}'s notion of 
equivariance, we can ``pass to the quotients'', and get from eq. 
\ref{announce2ndimldiffs}, the natural diffeomorphisms 
\be
TG/G \rightarrow \mg \;\;\; {\rm{and}} \;\;\; T^*G/G \rightarrow  \mg^* \;\; ;
\label{announcendimldiffs}
\ee 
where the quotients on the left hand sides (the domains) is by the action of 
left translation; (to be precise: by the action of its derivative for $TG$, and 
its cotangent lift for $T^*G$).

\subsubsection {Space and body coordinates generalized to 
$G$}\label{spbodyonG;was6.5.6.A}
So let a (finite-dimensional) Lie group $G$ act on itself by left and right 
translation, $L_g$ and $R_g$. For any $g \in G$, we define
\be
\la_g : T_g G \rightarrow \mg \;\; {\rm{by}} \;\; v \in T_g G \mapsto 
(T_eL_g)^{-1}(v) \equiv (T_g L_{g^{-1}})(v) \in \mg \; .
\label{definelambdag}
\ee
We similarly define
\be
\rho_g : v \in T_g G \mapsto (T_eR_g)^{-1}(v) \equiv (T_g R_{g^{-1}})(v) \in 
\mg \; .
\label{definerhog}
\ee
On analogy with the case of the pivoted rigid body (cf. eq. \ref{callitrtrn} 
and \ref{callitltrn}, or eq. \ref{introSBsuperscript}), we say that $\la_g$ 
represents $v \in T_g G$ in {\em body coordinates}, and $\rho_g$ represents $v$ 
in {\em space coordinates}. We also speak of body and space {\em 
representations}. The transition from body to space coordinates is then an 
isomorphism of $\mg$; viz. by eq. \ref{defineAd_g}
\be
\forall \xi \in \mg, \;\; (\rho_g \circ \la^{-1}_g)(\xi) = \rho_g(T_eL_g(\xi)) 
\equiv Ad_g \xi \; .
\label{bodytospTGgetAd}
\ee  
So we can combine the $S$ and $B$ superscript notation of eq. 
\ref{introSBsuperscript} with Section \ref{adj;was6.5.5.A}'s notion of the 
adjoint representation, and write
\be
v^S = Ad_g v^B \; .
\ee

In a similar way, the cotangent lifts of left and right translation provide 
isomorphisms between the dual spaces $T^*_g G, g \in G$ and $\mg^*$. Thus for 
any $g \in G $, we define
\be
{\bar \la}_g : T^*_g G \rightarrow \mg^* \;\; {\rm{by}} \;\; \al \in T^*_g G 
\mapsto \al \circ T_eL_g \equiv (T_eL_g)^*(\al) \equiv (T^*_eL_g)(\al) \in 
\mg^* \; ;
\label{definebarlambdag}
\ee
and similarly 
\be
{\bar \rho}_g : \al \in T^*_g G \mapsto \al \circ T_e R_g \equiv (T^*_e 
R_g)(\al)  \in \mg^* \; .
\label{definebarrhog}
\ee
And we again use the $S$ and $B$ superscript notation of eq. 
\ref{introSBsuperscript}, and define for $\al \in T^*_g G$
\be
\al^S := (T^*_e R_g)(\al) \equiv {\bar \rho}_g (\al) \;\;\; {\rm{and}} \;\;\;
\al^B := (T^*_e L_g)(\al) \equiv {\bar \la}_g (\al) \; ,
\ee
which are called the {\em space} (or `spatial') and {\em body} representations, 
respectively, of $\al$. The transition from body to space representations  is 
now an isomorphism of $\mg^*$; viz. 
\be
\forall \al \in \mg^*, \;\; ({\bar \rho}_g \circ {\bar \la}^{-1}_g)(\al) =  
Ad^*_{g^{-1}}(\al) \; , \;\;\; {\rm{i.e.}} \;\;\; \al^S = Ad^*_{g^{-1}}(\al^B) 
\; .
\label{problemcotgtisom}
\ee  

\subsubsection{Passage to the quotients}\label{Fundldiffms;was6.5.6.B}
For later purposes, we need to develop the details of how the element  $g \in 
G$ ``carries along throughout'' in eq. \ref{definelambdag} to 
\ref{problemcotgtisom}. More precisely, we have two isomorphisms: 
\be
TG \cong G \times \mg \;\;\; {\rm{and}} \;\;\; T^*G \cong G \times \mg^* \; .
\ee
These are isomorphisms of vector bundles; but we shall not develop the language   
of fibre bundles. What matters for us is that once we exhibit these 
isomorphisms, we will see that we have equivariant maps relating two group 
actions, in the sense of eq. \ref{equivarcefinite} and \ref{basiceqvarcediag}. 
And this will mean that we can pass to the quotients to infer that $TG/G$ is 
diffeomorphic to $\mg$, and correspondingly that $T^*G/G$ is diffeomorphic to 
$\mg^*$.\\
\indent This last diffeomorphism will form the first part of Section 
\ref{redn}'s main theorem, the Lie-Poisson reduction theorem, which says that 
$T^*G/G$ and $\mg^*$ are isomorphic as Poisson manifolds. In Section \ref{pms} 
onwards, we will develop the notion of a Poisson manifold, and the significance 
of this isomorphism  for the reduction  of mechanical problems.\\
\indent I should note here that there is a parallel story about the first 
diffeomorphism, i.e. about $TG/G$ being diffeomorphic to $\mg$. It forms the 
first part of another reduction theorem,  which is the Lagrangian analogue of 
Section \ref{redn}'s Lie-Poisson theorem. But since this Chapter has adopted 
the Hamiltonian approach, I will not go into details. They can be found in 
Marsden and Ratiu (1999: Sections 1.2, 13.5, 13.6), under the title 
`Euler-Poincar\'{e} reduction'.

Thus corresponding to eq. \ref{definelambdag}, we define the isomorphism
\be
\la: TG \rightarrow G \times \mg \;\;\; {\rm{by}} \;\;\;
\la(v) := (g, (T_eL_g)^{-1}(v))  \equiv (g, (T_g L_{g^{-1}})(v)) 
\ee
with $v \in T_g G$, i.e. $g = \pi_G(v)$ and $\pi_G: TG \rightarrow G$ the 
canonical projection. (As mentioned concerning eq. \ref{defineT^*f}, it is  
harmless to (follow many presentations and) conflate a point in $TG$, i.e. 
strictly speaking a pair $(g, v), g \in G, v \in T_gG$, with its vector  $v$.) 
And corresponding to eq. \ref{definerhog}, we define the isomorphism
\be
\rho: TG \rightarrow G \times \mg \;\;\; {\rm{by}} \;\;\;
\rho(v) := (g, (T_eR_g)^{-1}(v))  \equiv (g, (T_g R_{g^{-1}})(v)) \; . 
\label{definerho}
\ee
The transition from body to space representations given by eq. 
\ref{bodytospTGgetAd} now implies
\be
(\rho \circ \la^{-1})(g,\xi) = \rho(g, T_eL_g(\xi)) = (g, (T_eR_g)^{-1} \circ 
T_eL_g(\xi)) = (g, Ad_g \xi).
\ee

In a similar way, the cotangent bundle $T^*G$ is isomorphic in two ways to $G 
\times \mg^*$: namely by
\be
{\bar \la}(\al) := (g, \al \circ T_eL_g) \equiv (g, (T^*_eL_g)\al) \in G \times 
\mg^*  \; , 
\label{definebarlambda}
\ee
and by
\be
{\bar \rho}(\al) := (g, \al \circ T_eR_g) \equiv (g, (T^*_eR_g)\al) \in G 
\times \mg^*
\label{definebarrho}
\ee
where $\al \in T^*_g G$, i.e. $g = \pi^*_G(\al)$ with $\pi^*_G: T^*G 
\rightarrow G$ the canonical  projection. (Again, we harmlessly conflate a 
point $(g, \al)$ in $T^*G$ with its form $\al \in T^*_gG$.)

Let us now compute in the body representation, the actions of: (i) the 
(derivative of the) left translation map, $TL_g$, and (ii) the corresponding 
cotangent lift $T^*L_g$. This will show that $\la$ and $\bar \la$ are 
equivariant maps for certain group actions. 

(i): We compute:
\begin{eqnarray}
(\la \circ TL_g \circ \la^{-1})(h, \xi) = (\la \circ TL_g)(h, TL_h(\xi)) = 
\la(gh, (TL_g \circ TL_h)(\xi)) \\ 
= (gh, ((TL_{gh})^{-1} \circ TL_{gh})(\xi)) = (gh, \xi) .
\label{bodyrepnlefttrn}
\end{eqnarray}
So in the body representation, left translation does not act on the vector 
component. (That is intuitive, in that the vector $\xi$ is ``attached to the 
body'' and so should not vary relative to coordinates fixed in it.) Eq. 
\ref{bodyrepnlefttrn} means that $\la$ is an equivariant map relating left 
translation $TL_g$ on $TG$ to the $G$-action on $G \times \mg$ given just by 
left translation on the first component:
\be
\Phi_g((h, \xi)) \equiv g \cdot (h,\xi) := (gh, \xi) \; .
\label{defineleftonlyfirstcpt}
\ee

Equivariance means that $\la$ induces a map $\hat \la$ on the quotients. That 
is: as in eq. \ref{hatfforpasstoquots}, the map
\be
{\hat \la}: TG/G \rightarrow (G \times \mg)/G 
\label{definehatlambda}
\ee 
defined as mapping, for any $g$, the orbit of any $v \in T_g G$ to the orbit of 
$\la(v)$, i.e.
\begin{eqnarray}
{\hat \la}: {\rm{Orb}}(v) \equiv \{ u \in TG \mid T_g L_h(v) = u, \; 
{\rm{some}} \; h \in G \} \mapsto {\rm{Orb}}(\la(v)) \\
\equiv \{ (hg, (T_e L_g)^{-1}(v)) \mid 
{\rm{some}} \; h \in G \}
\label{definehatlambdadetails}
\end{eqnarray}
is well-defined, i.e. independent of the chosen representative $v$ of the 
orbit.

Besides, since the canonical  projections, $v \in TG \mapsto {\rm{Orb}}(v) \in 
TG/G$ and $(g, \xi) \mapsto {\rm{Orb}}((g, \xi)) \in (G \times \mg)/G$, are 
submersions, we can apply result (1) of Section \ref{quotstruc} and conclude 
that $\hat \la$ is smooth.\\
\indent Finally, we notice that since the action of left translation is 
transitive, we can identify each orbit of the $\Phi$ of eq. 
\ref{defineleftonlyfirstcpt} with its right component $\xi \in \mg$; and so we 
can identify the set of orbits $(G \times \mg)/G$ with $\mg$.

To sum up: we have shown that $TG/G$ and $(G \times \mg)/G$, i.e. in effect 
$\mg$, are diffeomorphic:
\be
{\hat \la}: TG/G \; \rightarrow \; (G \times \mg)/G \equiv \mg \; .
\ee

(ii): The results for the cotangent bundle are similar to those in (i). On 
analogy with eq. \ref{bodyrepnlefttrn}, the action of the cotangent lift of 
left translation $T^*L_g$ is given in body representation by applying eq. 
\ref{definebarlambda} to get
\be
({\bar \la} \circ (T^*L_g) \circ {\bar \la}^{-1})(h, \al) = (g^{-1}h, \al) \; ;
\label{bodyrepnlefttrncotgt}
\ee
or equivalently, now taking the cotangent  lift of left translation to define a 
left action (cf. eq. \ref{defineleft}),
\be
({\bar \la} \circ (T^*L_{g^{-1}}) \circ {\bar \la}^{-1})(h, \al) = (gh, \al) \; 
.
\label{barlambdaeqvart}
\ee
So in body representation, left translation does not act on the covector 
component; (again, an intuitive result in so far as $\al$ is ``attached to the 
body''). So eq. \ref{barlambdaeqvart} means that $\bar \la$ is an equivariant 
map relating the cotangent lifted left action of left translation on $T^*G$ to 
the $G$-action on $G \times \mg^*$ given just by left translation on the first 
component:
\be
\Phi_g((h, \al)) \equiv g \cdot (h,\al) := (gh, \al) \; .
\label{defineleftonlyfirstcptforcotgtlift}
\ee
So, on analogy with eq. \ref{definehatlambda} and \ref{definehatlambdadetails}, 
we can pass to the quotients, defining a map
\be
{\hat {\bar \la}}: T^*G/G \rightarrow (G \times \mg^*)/G
\ee
by requiring that for $\al \in T^*_gG$, so that $T^*L_{h^{-1}} \al \in 
T^*_{hg}G$:
\begin{eqnarray}
{\hat {\bar \la}}: {\rm{Orb}}(\al) \equiv \{ \bb \in T^*G \mid \bb = 
T^*L_{h^{-1}}(\al) , \; {\rm{some}} \; h \in G \} \mapsto \;\;\;\;\;\;\; \\
{\rm{Orb}}({\bar \la}(\al)) 
\equiv \{ (hg, (T^*_e L_g)(\al)) \mid 
{\rm{some}} \; h \in G \} \equiv 
\{ (h, (T^*_e L_g)\al ) \mid {\rm{some}} \; h \in G \} \; .
\label{definehatbarlambdadetails}
\end{eqnarray}
And finally, we identify the set of orbits $(G \times \mg^*)/G$ with $\mg^*$, 
so that we conclude that $T^*G/G$ and $\mg^*$ are diffeomorphic. That is, we 
think of the diffeomorphism ${\hat {\bar \la}}$ as mapping $T^*G/G$ to $\mg^*$: 
\be
{\hat {\bar \la}}: {\rm{Orb}}(\al) \equiv \{ \bb \in T^*G \mid \bb = 
T^*L_{h^{-1}}(\al) , \; {\rm{some}} \; h \in G \} \in T^*G/G \mapsto (T^*_e 
L_g)(\al) \in \mg^*.
\label{statehatbarlambdaforLPredndiffm}
\ee 
As I said above, this diffeomorphism is the crucial first part of Section 
\ref{redn}'s main reduction theorem. But we will see its role there, already in 
(3) of Section \ref{pmspreamble}.  

Finally, a result which will {\em not} be needed later. To calculate the 
derivatives and cotangent lifts of left translation in {\em space} 
representation, we replace $\la$ and $\bar \la$ by $\rho$ and $\bar \rho$ as 
defined by eq. \ref{definerho} and \ref{definebarrho}. We get as the analogues 
of eq. \ref{bodyrepnlefttrn} and \ref{bodyrepnlefttrncotgt} respectively:
\be
(\rho \circ TL_g \circ \rho^{-1})(h, \xi) = (gh, Ad_g (\xi)) \; ,
\label{spacerepnlefttrn}
\ee
and
\be
({\bar \rho} \circ T^*L_g \circ {\bar \rho}^{-1})(h, \al) = (g^{-1}h, Ad^*_g 
(\al)) \; .
\label{spacerepnlefttrncotgt}
\ee
Though these results are not needed later, they are also analogues of some 
later results, eq. \ref{JLisTeRg} and \ref{JRisTeLg}, which we will need. (Note 
that, in accordance with the discussion between eq. \ref{defineAd^*} and 
\ref{defineAd^*_g-1}, eq. \ref{spacerepnlefttrncotgt} involves right actions.)

\section{Poisson manifolds}\label{pms}
\subsection{Preamble: three reasons for Poisson manifolds}\label{pmspreamble}
Now that we are equipped with Sections \ref{tool} and \ref{actionlg}'s toolbox 
of modern geometry, we can develop, in this Section and the two to follow, the 
theory of symplectic reduction. This Section develops the general theory of 
Poisson manifolds, as a framework for a generalized Hamiltonian mechanics. Its 
main results concern the foliation, and quotienting, of Poisson manifolds. Then 
Section \ref{symyredn} returns us to symmetries and conserved quantities:  
topics which are familiar from Section \ref{noetcomplete}, but which Section 
\ref{symyredn} will discuss in the generalized framework using the notion of a 
momentum map. Finally, in Section \ref{redn} all the pieces of our jigsaw 
puzzle will come together, in our symplectic reduction theorem. 

We already glimpsed in (1) of Section \ref{rednprospectus} the idea of a 
Poisson manifold as a generalization of a symplectic manifold, that provides 
the appropriate framework for a  generalized Hamiltonian mechanics.  It is a 
manifold equipped with a bracket, called a `Poisson bracket', that has 
essentially the same formal defining properties as in symplectic geometry 
except that it can be ``degenerate''. In particular, the dimension $m$ of a 
Poisson manifold $M$  can be even or odd. As we will see, Hamiltonian mechanics 
can be set up on Poisson manifolds, in a natural generalization of the usual 
formalism: there are $m$ first-order ordinary differential equations for the 
time evolution of local coordinates $x^1,...,x^m$, and the time-derivative of 
any dynamical variable (scalar function on the Poisson manifold  $M$) is given 
by its Poisson bracket with the Hamiltonian. Besides, this generalization 
reduces to the usual formalism in the following sense. Any Poisson manifold $M$ 
is foliated into symplectic manifolds, and any Hamiltonian mechanics of our 
generalized kind defined on $M$ restricts on each symplectic leaf to a 
conventional Hamiltonian mechanics using the induced symplectic form.

This last point, the invariance of the symplectic leaves under the dynamics,  
prompts the question `why bother with the Poisson manifold, since the dynamics 
can be written down on each leaf?'. There are three reasons. I will just 
mention the first; the rest of  Section \ref{pms} will develop the second; and 
the two subsequent Sections will develop the third.

(1): {\em Parameters and stability}:---\\
The first two reasons concern the fact that for many problems in Hamiltonian 
mechanics, it is natural to consider an odd-dimensional state-space. One 
principal way this happens is if the system is characterized by  some  odd 
number, say $s$ (maybe $s=1$), of parameters that are constant in time. Then 
even though for a fixed value of the parameter(s), there is a Hamiltonian 
mechanics on a symplectic manifold, of dimension $2n$ say, it is useful to 
envisage the $2n+s$ dimensional space in order to keep track of how the 
behaviour of systems depends on the parameters.
 For example, this is very useful for analysing stability, especially if one 
can somehow control the value of the parameters. Stability theory (and related 
fields such as bifurcation theory) are crucially important, and vast, 
topics---which I will not go into.\footnote{Except to note a broad 
philosophical point. These parameters illustrate the modal or counterfactual 
involvements of mechanics. The $s$ dimensions of the state-space, and the 
mathematical constructions built on them, show how rich and structured these 
involvement are. For a detailed discussion of the modal involvements of 
mechanics, cf. Butterfield (2004).}

(2): {\em Odd-dimensional spaces: the rigid body again}:---\\
Secondly, even in the absence of such controllable parameters, there are 
mechanical systems whose description leads naturally to an odd-dimensional 
state-space. The paradigm elementary example is the rigid body pivoted at a 
point (mentioned in (3) of Section \ref{rednprospectus}). An elementary 
analysis, repeated in every textbook, leads to a description of the body by the 
three components of the  angular momentum (relative to body coordinates, i.e. 
coordinates fixed in the body): these components evolve according to the three 
first-order Euler equations.\\
\indent This situation prompts two foundational questions; (which of course 
most textbooks ignore!). First, we note that a configuration of the body is 
given by three real numbers: viz. to specify the rotation required to rotate 
the body into the given configuration, from a fiducial configuration. So a 
conventional Hamiltonian description of the rigid body would use six 
first-order equations. (Indeed, similarly for a Lagrangian description, if we 
treat the three ${\dot q}$s as variables.) So how is  the description by 
Euler's equations  related to a six-dimensional Hamiltonian (or indeed 
Lagrangian) description?\\
\indent Second, can the description by the Euler equations be somehow regarded 
as itself Hamiltonian, or Lagrangian?

\indent This Chapter will not pursue these questions about the rigid body; for 
details, cf. the references at the end of (3) of Section \ref{rednprospectus}.  
For us, the important point is that the theory of symplectic reduction shows 
that the answer to the second question is Yes. Indeed, a ``resounding Yes''. 
For we will see very soon (in Section \ref{LPso(3)}.A) that the 
three-dimensional space of the components, in body coordinates, of the angular 
momentum is our prototype example of a Poisson manifold; and the evolution by 
Euler's equations is the Hamiltonian mechanics on each  symplectic leaf of this 
manifold. In short: in our generalized framework, Euler's equations are {\em 
already} in Hamiltonian form. 

\indent Furthermore, this Poisson manifold is already familiar: it is 
$\mso(3)^*$, the dual of the Lie algebra of the rotation group. Here we connect 
with several previous discussions (and this Chapter's second  motto).\\
\indent First: we connect with the discussion of rotation in Relationist and 
Reductionist  mechanics (Sections \ref{relationproced} to \ref{compareproced}). 
In particular, cf. comment (iii) about $\gamma$, the three variables encoding 
the total angular momentum of the system, at the end of Section 
\ref{reductionproced}. (So as regards (1)'s idea of labelling the symplectic 
leaves by parameters constant in time: in this example, it is the magnitude $L$ 
of the total body angular momentum which is the parameter.)

\indent Second: we connect with Section \ref{rotgrp}'s discussion of $\mso(3)$,  
with Section \ref{coadj;was6.5.5.B}'s discussion of the co-adjoint 
representation on $\mso(3)^*$, and with Section \ref{kinicslgs}'s discussion of 
kinematics on an arbitrary Lie group. As regards the rigid body, the main 
physical idea is that the action of $SO(3)$ on itself by left translation is 
interpreted in terms of the coordinate transformation, i.e. rotation, between 
the space and body coordinate systems.\\
\indent But setting aside the rigid body: recall that in Section 
\ref{coadj;was6.5.5.B} we saw that for $\mso(3)^*$, the co-adjoint orbits are 
the spheres centred on the origin. I also announced that they have a natural 
symplectic structure---and that this was true for the orbits of the co-adjoint 
representation of {\em any} Lie group. Now that we have the notion of a Poisson 
manifold,  we can say a bit more, though of course the proofs are yet to 
come:---
\begin{quote} 
For any Lie group $G$, the dual of its Lie algebra $\mg^*$  is a Poisson 
manifold; and $G$ has on $\mg^*$ a co-adjoint representation, whose orbits are 
the symplectic leaves of $\mg^*$ as a Poisson manifold.
\end{quote}

In particular, we remark that the theory of the rigid body  just sketched is 
independent of the dimension of physical space being three: it carries over to 
$\mso(n)^*$ for any $n$. So we can readily do the Hamiltonian mechanics of the 
rigid body in arbitrary dimensions. That sounds somewhat academic! But it leads 
to a more general point, which is obviously of vast practical importance.\\
\indent In engineering we often need to analyse or design bodies consisting of 
two or more rigid bodies jointed together, e.g. at a universal joint. Often the 
configuration space of such a jointed body can be given by a sequence of 
rotations (in particular about the joints) and-or translations from a fiducial 
configuration; so that we can take an appropriate Lie group $G$ as the body's 
configuration space. If so, we can try to mimic our strategy for the rigid 
body, i.e. to apply the result just announced. And indeed, for such bodies, the 
action of left translation, and so the adjoint and co-adjoint representations 
of $G$ on $\mg$ and $\mg^*$, can often be physically significant.

But leaving engineering aside, let us sum up this second reason for Poisson 
manifolds as follows. For some mechanical systems the natural state-space for a 
Hamiltonian mechanics is a Poisson manifold. And in the paradigm case of the 
rigid body, there is a striking interpretation of the Poisson manifold's leaves 
as the orbits of the co-adjoint representation of the rotation group $SO(3)$.

(3): {\em Reduction}:---\\
My first two reasons have not mentioned reduction. But unsurprisingly, they 
have several connections with the notion. Here I shall state just one main 
connection, which links Section \ref{kinicslgs}'s kinematics on Lie groups to 
our main reduction theorem: this will be my third motivation for studying 
Poisson manifolds.

In short, the connection is that:---\\
\indent (i): For various systems, the configuration space is naturally taken to 
be a Lie group $G$; (as we have just illustrated with the rigid body).\\
\indent (ii): So it is natural to set up an orthodox Hamiltonian mechanics of 
the system on the cotangent bundle $T^*G$. But (as in the Reductionist 
procedure of Section \ref{reductionproced}) it is also natural to quotient by 
the lift to the cotangent bundle of $G$'s action on itself by left 
translation.\\
\indent (iii): When we do this, the resulting reduced phase space $T^*G/G$ is a 
Poisson manifold. Indeed it is an isomorphic copy of $\mg^*$. That is, we have 
an isomorphism of Poisson manifolds: $\mg^* \cong T^*G/G$. This is the 
Lie-Poisson reduction theorem.\\
\indent I shall give a bit more detail about each of (i)-(iii).

(i): For various systems, any configuration can be obtained by acting with an 
element of the Lie group $G$ on some reference configuration which can itself 
be labelled by an element of $G$, say the identity $e \in G$. So we take the 
Lie group  $G$ to be the configuration space. As mentioned in (3) of Section 
\ref{rednprospectus}, there is even an infinite-dimensional example of this: 
the ideal fluid.

(ii): So $T^*G$ is the conventional Hamiltonian phase space of the system. But  
$G$ acts on itself by left translation. We can then consider the quotient of  
$T^*G$ by the cotangent lift  of left translation. Intuitively, this is a 
matter of ``rubbing out'' the way that $T^*G$ encodes  (i)'s choice of 
reference configuration. By passing to the quotients as in Section 
\ref{kinicslgs}, we infer that $T^*G/G$ is a manifold. But of course it is in 
general {\em not}  even-dimensional. For its dimension is $\frac{1}{2}{\rm 
{dim}}(T^*G) \equiv {\rm {dim}}(G)$. So consider any odd-dimensional $G$: for 
example, our old friend, the three-dimensional rotation group $SO(3)$.

(iii):  But $T^*G/G$ {\em is} always a Poisson manifold. And it is always  
isomorphic as a Poisson manifold to $\mg^*$, with its symplectic leaves being 
the co-adjoint orbits of $\mg^*$: $\mg^* \cong T^*G/G$.

I end this third reason for studying Poisson manifolds  with two remarks about 
examples.\\
\indent The first remark echoes the end of  Section \ref{coadj;was6.5.5.B}, 
where I said that by considering all possible Lie groups and all the orbits of 
their co-adjoint representations, we get a series of examples of symplectic 
manifolds. We can now put this together with the notion of a Poisson manifold, 
and with the comment  at the end of Section \ref{examsubsub}, that every 
(finite-dimensional) Lie algebra is the Lie algebra of a Lie group. In short:
we get a series of examples of Poisson manifolds, in either of two equivalent 
ways: from the dual $\mg^*$ of any (finite-dimensional) Lie algebra $\mg$; or 
from the quotient $T^*G/G$ of the cotangent lift of left translation. In either 
case, the example is the co-adjoint representation. \\
\indent The second remark is that there are yet other examples of Poisson 
manifolds and reductions. Indeed, we noted one in Section 
\ref{reductionproced}: viz. the Reductionist's reduced phase space ${\bar M} := 
M/E$, obtained by quotienting the phase space $M := T^*\mathR^{3N} - (\delta 
\cup \Delta)$ by the (cotangent lift) of the action of the euclidean group $E$ 
on $\mathR^{3N}$. But I shall not go into further details about this example; 
(for which cf. the Belot papers listed in Section \ref{compare2quots}, and 
references therein). Here it suffices to note that this example is not of the 
above form: $\mathR^{3N}$ is not $E$, and the action of $E$ on $\mathR^{3N}$ is 
not left translation. This of course echoes my remarks at the end of Section 
\ref{introprospectus} that the theory of symplectic reduction is too large and 
intricate for this Chapter to be more than an ``appetizer''.

So much by way of motivating Poisson manifolds. The rest of this Section will 
cover reasons (1) and (2); but reason
(3), about reduction,  is postponed to Sections \ref{symyredn} and \ref{redn}.
 We give some basics about Poisson manifolds, largely in coordinate-dependent 
language, in Section \ref{basics}. 
  In Section \ref{sympstruc}, we move to a more coordinate-independent language 
and show that Poisson manifolds are foliated into symplectic manifolds. In 
Section \ref{sympcoad}, we  show that the leaves of the foliation of a 
finite-dimensional Lie algebra $\mg^*$ are the orbits of the co-adjoint 
representation of $G$ on $\mg^*$. Finally in Section \ref{quotpm}, we prove a  
general theorem about quotienting a Poisson manifold by the action of Lie 
group, which will be important  for Section \ref{redn}'s main theorem.

\subsection{Basics}\label{basics}
In Sections \ref{pbs} to \ref{sfs}, we develop some basic definitions and 
results about Poisson manifolds. This leads up to Section \ref{LPso(3)}, where 
we see that the dual of any finite-dimensional Lie algebra has a natural (i.e. 
basis-independent) Poisson manifold structure. Throughout, there will be some 
obvious echoes of previous discussions of anti-symmetric  forms, Poisson 
brackets, Hamiltonian vector fields and Lie brackets (Sections \ref{july05} and 
\ref{lgla}). But I will for the most part {\em not} articulate these echoes. 
 
\subsubsection{Poisson brackets}\label{pbs}
A manifold $M$ is called a {\em Poisson manifold} if it is equipped with a {\em 
Poisson bracket} (also known as: {\em Poisson structure}).  A {\em Poisson 
bracket} is an assignment to each pair of smooth real-valued functions $F,H: M 
\rightarrow \mathR$, of another such function, denoted by $\{F,H\}$, subject to 
the following four conditions:---\\
(a) Bilinearity:\\
\be
\{aF + bG, H\} = a\{F, H\} + b\{G, H\} \; ; \; \{F, aG + bH\} = a\{F, G\} + 
b\{F, H\} \;\; \forall a,b \in \mathR.
\ee
(b) Anti-symmetry:\\
\be
\{F, H\} = - \{H, F\} \; .
\ee
(c) Jacobi identity:\\
\be
\{\{F, H\}, G\} + \{\{G, F\}, H\} + \{\{H, G\}, F\}= 0 \; .
\ee
(d) Leibniz' rule:\\
\be
\{F, H \cdot G\} = \{F, H\}\cdot G + H \cdot \{F, G\} \; .
\label{LeibPoim}
\ee
In other words: $M$ is a Poisson manifold iff both: (i) the set ${\cal F}(M)$ 
of smooth scalar functions on $M$, equipped with the bracket $\{,\}$, is a Lie 
algebra; and (ii) the bracket $\{,\}$ is a derivation in each factor. 

Any symplectic manifold is a Poisson manifold. The Poisson bracket is defined 
by the manifold's symplectic form; cf.  eq. \ref{sympPBgeomic}.   

``Canonical'' Example:---\\
 Let $M = \mathR^m, m = 2n + l$, with standard coordinates $(q,p,z) = 
(q^1,...,q^n,p^1,...,p^n,z^1,...,z^l)$. Define the Poisson bracket of any two 
functions $F(q,p,z)$, $H(q,p,z)$ by
\be
\{F, H\} := \Sigma^n_i \left(\frac{\pl F}{\pl q^i}\frac{\pl H}{\pl p^i} - 
\frac{\pl F}{\pl p^i}\frac{\pl H}{\pl q^i}
\right).
\label{canlbr}
\ee
Thus this bracket ignores the $z$ coordinates; and if $l$ were equal to zero, 
it would be the standard Poisson bracket  for $\mathR^{2n}$ as a symplectic 
manifold. We can immediately deduce the Poisson brackets for the coordinate 
functions. Those for the $q$s and $p$s are as for the usual symplectic case:
\be
\{q^i, q^j\} = 0 \;\;\;\; \{p^i, p^j\} = 0 \;\;\;\; \{q^i, p^j\} = \delta_{ij}.
\label{tradPB}
\ee
On the other hand, all those involving the $z$s vanish:
\be
\{q^i, z^j\} =  \{p^i, z^j\} =  \{z^i, z^j\} \equiv 0.
\label{newPBforz}
\ee 
Besides, any function $F$ depending only on the $z$'s, $F \equiv F(z)$ will 
have vanishing Poisson brackets with all functions $H: \{F, H\} = 0.$ 

\indent This example seems special in that $M$ is foliated into 
$2n$-dimensional symplectic manifolds, each labelled by $l$ constant values of 
the $z$s. But Section \ref{DarbouxPoiss} will give a generalization for Poisson 
manifolds of Darboux's theorem (mentioned at the end of Section 
\ref{cotgtblesymp}): a generalization saying, roughly speaking, that every 
Poisson manifold ``looks locally like this''.

For any Poisson manifold, we say that a function $F: M \rightarrow \mathR$ is 
{\em distinguished} or {\em Casimir} if its Poisson bracket with all smooth 
functions $H: M \rightarrow \mathR$ vanishes identically: $\{F, H\} = 0.$ 

\subsubsection{Hamiltonian vector fields}\label{hvfs}
Given a smooth function $H: M \rightarrow \mathR$, consider the map on smooth 
functions: $F \mapsto \{F, H\}$. The fact that the Poisson bracket is bilinear 
and obeys Leibniz's rule implies that this map $F \mapsto \{F, H\}$ is a 
derivation on the space of smooth functions, and so  determines a vector field 
on $M$; (cf. (ii) of Section \ref{fieldtoflow}.B). We call this vector field 
the {\em Hamiltonian vector field} associated with (also known as: generated 
by) $H$, and denote it by $X_H$.\\
\indent But independently of the Poisson structure, the action of any vector 
field  $X_H$ on a smooth function $F$, $X_H(F)$, also equals $L_{X_H}(F) \equiv 
dF(X_H)$; (cf. eq. \ref{defineLiederiv}).  So we have for all smooth $F$
\be
L_{X_H}(F) \equiv dF(X_H) \equiv X_H(F) = \{F, H\} \; .
\label{defhamvf}
\ee
The equations describing the flow of $X_H$ are called {\em Hamilton's 
equations}, for the choice of $H$ as ``Hamiltonian''.

In the previous example with $M = \mathR^{2n+l}$, we have
\be
X_H = \Sigma^n_i \left(\frac{\pl H}{\pl p^i}\frac{\pl }{\pl q^i} - \frac{\pl 
H}{\pl q^i}\frac{\pl }{\pl p^i}
\right),
\ee
and the flow is given by the ordinary differential equations 
\be
\frac{d q^i}{dt} = \frac{\pl H}{\pl p^i} \;\;\;\;
\frac{d p^i}{dt} = - \frac{\pl H}{\pl q^i} \;\;\;\;
\frac{d z^j}{dt} = 0. \;\;\;\; i = 1,...,n ;  \;\; j = 1,...,l.
\ee
Again, the $z$s, and any function $F(z)$ solely of them, are distinguished and 
have a vanishing Hamiltonian vector field.  On the other hand, the coordinate 
functions $q^i$ and $p^i$ generate the Hamiltonian vector fields $-\frac{\pl 
}{\pl p^i}$ and $\frac{\pl }{\pl q^i}$ respectively.

Two further remarks about eq. \ref{defhamvf}:---\\
\indent (1): It follows that a function $H$ is distinguished (i.e. has 
vanishing Poisson brackets with all functions) iff its Hamiltonian vector field 
$X_H$ vanishes everywhere. And since the Poisson bracket is antisymmetric, this 
is so iff $H$ is constant along the flow of all Hamiltonian vector fields.\\
\indent (2): This equation is the beginning of the theory of constants of the 
motion (first integrals), and of Noether's theorem, for Poisson manifolds; just 
as the corresponding equation was the beginning for the symplectic case. This 
will be developed in Section \ref{symyredn}. 

{\em Poisson brackets and Lie brackets}:---\\
With the definition eq. \ref{defhamvf} in hand, we can readily establish our 
first important connection between Poisson manifolds and Section \ref{tool}'s 
Lie structures. Namely: result (2) at the end of Section \ref{lavecfields}, eq.  
\ref{HamvfsclosedunderLie;symp}, is also valid for Poisson manifolds.\\
\indent That is: the Hamiltonian vector field of the Poisson bracket of scalars 
$F,H$ on a Poisson manifold $M$ is, upto a sign, the Lie bracket of the 
Hamiltonian vector fields, $X_F$ and $X_H$, of $F$ and $H$:
\be
X_{\{F, H\}} = - [X_F, X_H] = [X_H, X_F] \; \; .
\label{HamvfsclosedunderLie}
\ee 
The proof is exactly as for eq. \ref{HamvfsclosedunderLie;symp}.\\
\indent So the Hamiltonian vector fields, with the Poisson bracket, form a Lie 
subalgebra of the Lie algebra ${\cal X}_M$ of all vector fields on the Poisson 
manifold $M$. This result will be important in Section \ref{Pomaps}'s 
 proof that every Poisson manifold is a disjoint union of symplectic manifolds.

\subsubsection{Structure functions}\label{sfs}
We show that to compute the Poisson bracket of any two functions given in some 
local coordinates ${\bf x} = x^1,...,x^m$, it suffices to know the Poisson 
brackets of the coordinates. For any function $H: M \rightarrow \mathR$, let 
the components of its Hamiltonian vector field in the coordinate system $\bf x$ 
be written as $h^i(x)$. So $X_H = \Sigma^m_i \; h^i(x)\frac{\pl }{\pl x^i}$. 
Then for any other function  $F$, we have
\be
\{F, H \} = X_H(F) = \Sigma^m_i \; h^i(x)\frac{\pl F}{\pl x^i}.
\label{Tostrucfns}
\ee
Taking $x^i$ as the function $F$, we get: $\{x^i, H \} = X_H(x^i) = h^i(x)$. So 
eq. \ref{Tostrucfns} becomes
\be
\{F, H \}  = \Sigma^m_i \; \{x^i, H \} \frac{\pl F}{\pl x^i}.
\label{Tostrucfns2}
\ee
If we now put $x^i$ for $H$ and $H$ for $F$ in eq. \ref{Tostrucfns2}, we get
\be
\{x^i, H \}  = - \{H, x^i \} = - X_{x^i}(H) = - \Sigma^m_j \; \{x^j, x^i \} 
\frac{\pl H}{\pl x^j}.
\label{Tostrucfns3}.
\ee
Combining eq.s \ref{Tostrucfns2} and \ref{Tostrucfns3}, we get the basic 
formula for the Poisson bracket of any two functions in terms of the Poisson 
bracket of local coordinates:
\be
\{F, H \}  = \Sigma^m_i \; \Sigma^m_j \; \{x^i, x^j \} \frac{\pl F}{\pl 
x^i}\frac{\pl H}{\pl x^j}.
\label{Tostrucfns4}
\ee
We assemble these basic brackets, which we call the {\em structure functions} 
of the Poisson manifold, 
\be
J^{ij}(x) := \{x^i, x^j \} \;\;\; i,j = 1,...,m
\label{JdefinedPBscoords}
\ee
into a $m \times m$ anti-symmetric matrix of functions, $J(x)$, called the {\em 
structure matrix} of $M$. More precisely, it is the structure matrix for $M$ 
relative to our coordinate system $\bf x$. Of course, the transformation  of 
$J$ under a coordinate change ${\bf }x'^i := x'^i(x^1,...,x^m)$ is  determined 
by setting $F := x'^i, H := x'^j$ in the basic formula eq. \ref{Tostrucfns4}.

 Then, writing $\nabla H$ for the (column) gradient vector of $H$, eq. 
\ref{Tostrucfns4} becomes
\be
\{F, H \}  = \nabla F \cdot J \nabla H.
\label{PBfromJnaive}
\ee 
For example, the canonical bracket on $\mathR^{2n+l}$, eq.\ref{canlbr}, written 
in the $(q,p,z)$ coordinates, has the simple form
\be
J = \left( \begin{array}{ccc}
0 & I & 0 \\
-I & 0 & 0 \\
0 & 0 & 0 
\end{array}
\right).
\label{Jmatrix}
\ee 
where $I$ is the $n \times n$ identity matrix.

We can write the Hamiltonian vector field, and the Hamilton's equations, 
associated with the function $H$ in terms of $J$. Since 
\be
\{x^i, H \}   =  \Sigma^m_j \; \{x^i, x^j \} \frac{\pl H}{\pl x^j}
\ee
we get:
\be
X_H = \Sigma^m_i \; \left( \Sigma^m_j \; J^{ij}(x) \frac{\pl H}{\pl x^j} 
\frac{\pl }{\pl x^i} \right),
\label{X_HfromJ}
\ee
or in matrix notation: $X_H = (J \nabla H) \cdot {\pl}_x$. Similarly, 
Hamilton's equations
\be
\frac{dx^i}{dt} = \{x^i, H \}
\label{hampois} 
\ee
get the matrix form
\be
\frac{dx}{dt} = J(x) \nabla H(x) \;\; ; \;\;\; {\rm{i.e.}} \;\;\;
\frac{dx^i}{dt} = \Sigma^m_j \; J^{ij}(x) \frac{\pl H}{\pl x^j} \; .
\label{HamMatrix}
\ee
To summarize how we have generalized from the usual form of Hamilton's 
equations: compare eq. \ref{HamMatrix}, \ref{PBfromJnaive} and \ref{Jmatrix}  
respectively with eq. \ref{HEgeomic1}, \ref{sympPBgeomic} and 
\ref{eq;gramschmidt}.

 Note that not every $m \times m$ anti-symmetric matrix of functions on an 
$m$-dimensional manifold (or even: on an open subset of $\mathR^m$)  is the 
structure matrix of a Poisson manifold: for the Jacobi identity constrains the 
functions. In fact it is readily shown that the Jacobi identity corresponds to 
the following $m^3$ partial differential equations governing the $J^{ij}(x)$, 
which are in general non-linear. Writing as usual ${\pl}_l$ for ${\pl}/{\pl 
x^l}$: 
\be
\Sigma^m_{l=1} \left(J^{il}{\pl}_l J^{jk} + J^{kl}{\pl}_l J^{ij} + 
J^{jl}{\pl}_l J^{ki} \right) = 0 
\;\;\;\; i,j,k, = 1,...,m; \forall x \in M.
\label{JacobiForJ}
\ee
In particular, any constant anti-symmetric matrix $J$ defines a Poisson 
structure.

\subsubsection{The Poisson structure on $\mg^*$}\label{LPso(3)}
We can now show that any $m$-dimensional Lie algebra $\mg$ defines a Poisson 
structure, often called the {\em Lie-Poisson bracket}, on any $m$-dimensional 
vector space $V$. We proceed in two stages.

\indent (1): We first present the definition in a way that seems to depend on a 
choice of  bases, both in $\mg$ (where the definition  makes a choice of 
structure constants) and in the space $V$.\\
\indent (2): Then we will see that choosing $V$ to be $\mg^*$, the definition 
is in fact basis-independent.\\
\indent \indent This Poisson structure on $\mg^*$ will be of central importance 
from now on. As Marsden and Ratiu write: `Besides the Poisson structure on a 
symplectic manifold, the Lie-Poisson bracket on $\mg^*$, the dual of a Lie 
algebra, is perhaps the most fundamental example of a Poisson structure' (1999: 
415). Here we return to our motivating discussion of Poisson manifolds, 
especially reasons (2) and (3) of Section \ref{pmspreamble}: which concerned 
the rigid body and reduction, respectively. Indeed,  we will see already in the 
Example at the end of this Subsection (Section \ref{LPso(3)}.A) how the 
Lie-Poisson bracket on the special case $\mg^* := \mso(3)^*$ clarifies the 
theory of the rigid body. And we will see in  Sections \ref{rednthm} and 
\ref{rednmommfn} how for any $\mg$, the Lie-Poisson bracket on $\mg^*$ is 
induced by reduction, from the canonical Poisson (viz. symplectic) structure on 
the cotangent bundle $T^*G$. This will be our reduction theorem, that $T^*G/G 
\cong \mg^*$.

After (2), we will see that the Lie-Poisson bracket on $\mg^*$ implies that 
Hamilton's equations on $\mg^*$ can be expressed using $ad^*$: a form that will 
be needed later. This will be (3) below. Then we will turn in Section 
\ref{LPso(3)}.A to the example $\mg^* := \mso(3)^*$.

(1): {\em A Poisson bracket on any vector space $V$}:--- \\
 Take a basis, say $e_1,...,e_m$, in $\mg$, and so structure constants 
$c^k_{ij}$ (cf. eq. \ref{defstrucconst}).  Consider the space $V$ as a 
manifold, and coordinatize it by taking a basis, $\epsilon_1,...,\epsilon_m$ 
say, determining  coordinates $x^1,...,x^m$. We now define the Poisson bracket 
(in this case, often called the {\em Lie-Poisson bracket}) between two smooth 
functions $F,H: V \rightarrow \mathR$ to be
\be
\{F, H \} := \Sigma^m_{i,j,k = 1} \;\; c^k_{ij}x^k \frac{\pl F}{\pl 
x^i}\frac{\pl H}{\pl x^j}.
\label{defLPbr}
\ee  
This takes the form of eq. \ref{Tostrucfns4}, with linear structure functions 
$J^{ij}(x) =  \Sigma^m_k \;\; c^k_{ij}x^k$. One easily checks that 
anti-symmetry, and the Jacobi identity, for the structure constants,
eq. \ref{propstrucconst}, implies that these $J^{ij}$ are anti-symmetric and 
obey their Jacobi identity  eq. \ref{JacobiForJ}. So eq. \ref{defLPbr} defines 
a Poisson bracket on $V$.\\
\indent In particular, the associated Hamiltonian equations, eq.s \ref{hampois} 
and \ref{HamMatrix}, take the form
\be
\frac{dx^i}{dt} = \Sigma^m_{j,k = 1} \;\; c^k_{ij}x^k \frac{\pl H}{\pl x^j}.
\label{HamLiePois}
\ee 

(2): {\em The Lie-Poisson bracket on $\mg^*$}:--- \\ 
To give a basis-independent characterization of the Lie-Poisson bracket, we 
first recall that:\\
\indent (i): the gradient $\nabla F(x)$ of $F: V \rightarrow \mathR$ at any 
point $x \in V$ is in the dual space $V^*$ of (continuous) linear functionals 
on $V$;:\\
\indent (ii): any finite-dimensional vector space is canonically, i.e. 
basis-independently, isomorphic to its double dual:  $(V^*)^* \cong V$.\\
\indent Then writing $< \; ; \; >$ for the natural pairing between $V$ and 
$V^*$, we have, for any $y \in V$
\be
< \nabla F(x) ; y > := {\rm {lim}}_{\tau \rightarrow 0}
\frac{F(x + \tau y) - F(x)}{\tau}.
\ee

  Now let us take $V$ in our definition of the Lie-Poisson bracket to be 
$\mg^*$. So we will show that $\mg$ makes $\mg^*$ a Poisson manifold, in a 
basis-independent way. And let the basis $\epsilon_1,...,\epsilon_m$  be dual 
to the basis $e_1,...,e_m$ of $\mg$. If $F:\mg^* \rightarrow \mathR$ is any 
smooth function, its gradient $\nabla F(x)$ at any point $x \in \mg^*$ is an 
element of $(\mg^*)^* \cong \mg$. One now checks that the Lie-Poisson bracket 
defined by eq. \ref{defLPbr} has the basis-independent expression
\be
\{F, H \}(x) \; = \; < x; [\nabla F(x), \nabla H(x)]> \;\; , \;\; x \in \mg^*
\label{intrLPbr}  
\ee
where $[,]$ is the ordinary Lie bracket on the Lie algebra $\mg$ itself.

(3): {\em Hamilton's equations on $\mg^*$}:--- \\ 
We can also give a basis-independent  expression of the Hamilton's equations 
eq. \ref{HamLiePois}: viz. by expressing the Lie bracket in eq. \ref{intrLPbr} 
in terms of $ad$, as indicated by eq. \ref{sumupad}.

Thus let $F \in {\cal F}(\mg^*)$ be an arbitrary smooth scalar function on 
$\mg^*$. By the chain rule
\be
\frac{dF}{dt} = {\bf D}F (x) \cdot {\dot x} = < {\dot x} ; \nabla F(x) > \; .
\ee   
But applying eq.s \ref{sumupad} and \ref{computead*} to eq. \ref{intrLPbr} 
implies:
\be
\{F, H \}(x) = < x; [\nabla F(x), \nabla H(x)]> = - < x ; ad_{\nabla 
H(x)}(\nabla F(x)) > = < ad^*_{\nabla H(x)}(x) ; \nabla F(x) > \; .
\label{intrLPbrandad}  
\ee
Since $F$ is arbitrary and the pairing is non-degenerate, we deduce that 
Hamilton's equations take the form
\be
\frac{dx}{dt} = ad^*_{\nabla H(x)}(x) \; .
\label{HamEqWithad^*}
\ee

\paragraph{7.2.4.A Example: $\mso(3)$ and 
$\mso(3)^*$}\label{724Aso(3)andso(3)^*} As an example of the dual of a Lie 
algebra as a Poisson manifold, let us consider again our standard example 
$\mso(3)^*$. We will thereby make good our promise in (2) of Section 
\ref{pmspreamble}, to show that Euler's equations for a rigid body are {\em 
already} in Hamiltonian form---in our generalized sense. We will also see why 
in the Chapter's second motto, Arnold mentions the three  dual spaces, 
$\mathR^{3*}, \mso(3)^*$ and $T^*(SO(3))_g$; (cf. the discussion at the end of 
Section \ref{rotgrp}).

The Lie algebra   $\mso(3)$ of $SO(3)$ has a basis $e_1,e_2,e_3$ representing 
infinitesimal rotations around the $x$-, $y$- and $z$-axes of $\mathR^3$. As we 
have seen, we can think of these basis elements: as vectors in $\mathR^3$ with  
$[,]$ as elementary  vector multiplication; or as anti-symmetric matrices with 
$[,]$ as the matrix commutator; or as left-invariant vector fields on $SO(3)$ 
with $[,]$ as the vector field commutator (i.e. Lie bracket).

\indent Let $\epsilon_1,\epsilon_2,\epsilon_3$ be a dual basis for $\mso(3)^*$, 
with $x = x^1\epsilon_1 + x^2\epsilon_2 + x^3\epsilon_3$ a typical point 
therein. If $F: \mso(3)^* \rightarrow \mathR$, its gradient at $x$ is the 
vector
\be
\nabla F = \frac{\pl F}{\pl x^1}e_1 + \frac{\pl F}{\pl x^2}e_2 +
\frac{\pl F}{\pl x^3}e_3 \;\; \in \mso(3).
\ee
Then eq. \ref{intrLPbr} tells us that, if we write $\mso(3)$ as $\mathR^3$ with 
$\times$ for elementary vector multiplication, the Lie-Poisson bracket on 
$\mso(3)^*$ is 
\begin{eqnarray}
\{F, H \}(x) = x^1 \left(\frac{\pl F}{\pl x^3}\frac{\pl H}{\pl x^2} - \frac{\pl 
F}{\pl x^2}\frac{\pl H}{\pl x^3}\right) + ... 
+ x^3 \left(\frac{\pl F}{\pl x^2}\frac{\pl H}{\pl x^1} - \frac{\pl F}{\pl 
x^1}\frac{\pl H}{\pl x^2}\right) \\ 
= - x \cdot (\nabla F \times \nabla H).
\label{LPBmso(3)}
\end{eqnarray}
So the structure matrix $J(x)$ is
\be
J(x) = 
\left( \begin{array}{ccc}
0 & -x_3 & x_2 \\
x_3 & 0 & -x_1 \\
-x_2 & x_1 & 0 
\end{array}
\right), \;\;\;
x \in \mso(3)^*.
\label{Jforso(3)}
\ee 
Hamilton's equations corresponding to the Hamiltonian function $H(x)$ are 
therefore
\be
\frac{dx}{dt} = x \times \nabla H(x) \; .
\label{HamrbFromLPbr}
\ee

Now consider the Hamiltonian representing the kinetic energy of a free pivoted 
rigid body
\be
H(x) = \frac{1}{2} \left(\frac{(x^1)^2}{I_1} + \frac{(x^2)^2}{I_2} + 
\frac{(x^3)^2}{I_2} \right),
\label{HamrbForso(3)*}
\ee
in which the $I_i$ are the moments of inertia about the three coordinate axes, 
and the $x^i$ are the corresponding components of the {\em body} angular 
momentum. For this Hamiltonian, Hamilton's equations eq. \ref{HamrbFromLPbr} 
become
\be
\frac{dx^1}{dt} = \frac{I_2 - I_3}{I_2 I_3}x^2x^3 \;\; , \;\;
\frac{dx^2}{dt} = \frac{I_3 - I_1}{I_3 I_1}x^3x^1 \;\; , \;\;
\frac{dx^3}{dt} = \frac{I_1 - I_2}{I_1 I_2}x^1x^2 \;\; , \;\;
\label{EulerFromLPbr}
\ee
Indeed, these are the Euler equations for a free pivoted rigid body. I shall 
not go into details about the rigid body. I only note that:\\
\indent (i): In the elementary theory of such a body, the magnitude $L$ of the 
angular momentum is conserved, and eq. \ref{EulerFromLPbr} describes the motion 
of the $x^i$ on a sphere of radius $L$ centred at the origin.\\
\indent (ii): In Section \ref{sympcoad}, we will return to seeing these spheres 
as the orbits of the co-adjoint representation of $SO(3)$ on $\mso(3)^*$ (cf. 
Section \ref{coadj;was6.5.5.B}).\\
\indent (iii): Let us sum up this theme by saying, with Marsden and Ratiu 
(1999, p.11) that here we see: `a simple and beautiful Hamiltonian structure 
for the rigid body equations'.

\subsection{The symplectic foliation of Poisson manifolds}\label{sympstruc}
We first reformulate some ideas of Section \ref{basics} in more 
coordinate-independent language, starting with Section \ref{sfs}'s idea of the 
structure matrix $J(x)$ (Section \ref{rank}). Then we discuss canonical 
transformations on a Poisson manifold (Section \ref{Pomaps1}). This will lead 
up  to showing that any Poisson manifold is foliated by symplectic leaves 
(Section \ref{Pomaps}). Finally, we state a generalization of Darboux's 
theorem; and again take $\mso(3)$ as an example (Section \ref{DarbouxPoiss}).

\subsubsection{The Poisson structure and its rank}\label{rank}
We now pass from the structure matrix $J$, eq. \ref{JdefinedPBscoords},  to a 
coordinate-independent  object, the {\em Poisson structure} (also known as: 
co-symplectic structure), written  $\mathsf{B}$. Whereas $J$ multiplied naive 
gradient vectors, as in eq. \ref{PBfromJnaive} and \ref{HamMatrix}, 
$\mathsf{B}$ is to map the 1-form $dH$ into its Hamiltonian vector field; as 
follows.

At each point $x$ in a Poisson manifold $M$, there is a unique linear map 
$\mathsf{B}_x$, which we will also write as $\mathsf{B}$
\be
\mathsf{B} \equiv \mathsf{B}_x : T^*_xM \rightarrow T_xM
\label{defineB}
\ee
such that 
\be
\mathsf{B}_x(dH(x)) = X_H(x).
\label{requireB}
\ee
For the requirement eq. \ref{requireB} implies, by eq. \ref{X_HfromJ}, that for 
each $j = 1,...,m$
\be
\mathsf{B}_x(dx^j) = \Sigma_i J^{ij}(x) \frac{\pl }{\pl x^i}\mid_x
\ee
Since the differentials $dx^i$ span $T^*_xM$, this fixes $\mathsf{B}_x$, by 
linearity. $\mathsf{B}_x$'s action on any one-form $\al = \Sigma a_j dx^j$ is: 
\be
\mathsf{B}_x(\al) = \Sigma_{i,j}  J^{ij}(x)a_j \frac{\pl }{\pl x^i}\mid_x
\label{BessentialJ}
\ee
so that $\mathsf{B}_x$ is essentially matrix multiplication by $J(x)$. Here, 
compare again eq. \ref{hampois} and \ref{HamMatrix}.

Here we recall that any linear map between (real finite-dimensional) vector 
spaces, $B: V \rightarrow W^*$, has an associated bilinear form  $B^{\sh}$ on 
$V \times W^{**} \cong V \times W$ given  by
\be
B^{\sh}(v,w) \;\; := \;\; < B(v) \; ; \; w > \; .
\label{definemap'sassocbil}
\ee
Accordingly, some authors introduce the Poisson structure as a bilinear form 
$\mathsf{B}^{\sh}_x : T^*_xM \times T^*_xM \rightarrow \mathR$, often called 
the {\em Poisson tensor}. Thus  eq. \ref{definemap'sassocbil} gives, for $\al, 
\bb \in T^*_xM$
\be
\mathsf{B}^{\sh}_x (\al,\bb) \; := \; <  \mathsf{B}(\al), \bb  > \; .
\label{Poistsraltive}
\ee
$\mathsf{B}^{\sh}_x$ is antisymmetric, since the matrix $J(x)$ is. So, if we 
now let $x$ vary over $M$, we can sum up in the traditional terminology of 
tensor analysis: $\mathsf{B}^{\sh}$ is an antisymmetric contravariant 
two-tensor field.  

Example:--- Consider our first example, $M = \mathR^{2n+l}$ with the ``usual 
bracket'' eq. \ref{canlbr}, from the start of Section \ref{pbs}. For any 
one-form 
\be 
\al = \Sigma^n_{i=1} (a_i dq^i + b_i dp^i) + \Sigma^l_{j=1} c_j dz^j
\ee 
we have 
\be
\mathsf{B}(\al) = \Sigma^n_{i=1} \left(b_i\frac{\pl }{\pl q^i} - a_i\frac{\pl 
}{\pl p^i} \right).
\ee
In this example the form of $\mathsf{B}$ is the same from point to point. In 
particular, the kernel of $\mathsf{B}$ has everywhere the same dimension, viz. 
$l$, the number of distinguished coordinates.

We now define the {\em rank} at $x$ of a Poisson manifold $M$ to be the rank  
of its Poisson structure $\mathsf{B}$ at $x$, i.e. the dimension of the range 
of $\mathsf{B}_x$. This range is also the span of all the Hamiltonian vector 
fields on $M$ at $x$:
\be 
{\rm {ran}}(\mathsf{B}_x) := \{ X \in T_xM  : \; X = \mathsf{B}_x(\al), \; {\rm 
{some}}\;\; \al \in T^*_xM\} \; = \; \{X_H(x)  : \; H:M \rightarrow \mathR \; 
{\rm{smooth}} \; \} \; .
\label{rangeB}
\ee
So the rank of $M$ at $x$ is also  equal to the dimension of $\mathsf{B}_x$'s 
domain, i.e. dim($T^*_xM$)=dim($M$), minus the dimension of the kernel, 
${\rm{dim}}(\mathsf{B}_x)$.\\
\indent Since in local coordinates, $\mathsf{B}_x$ is given by multiplication 
by the structure matrix $J(x)$, the rank of $M$ at $x$ is the rank (the same in 
any coordinates) of the matrix $J(x)$. That $J(x)$ is anti-symmetric implies 
that the rank of $M$ is {\em even}: cf. again the normal form of antisymmetric 
bilinear forms, eq. \ref{omegawiths} and \ref{eq;gramschmidt}.

\indent The manifold  $M$ being symplectic corresponds, of course, to the rank 
of $\mathsf{B}$ being everywhere maximal, i.e. equal to dim($M$).\\
\indent In this case, the kernel of $\mathsf{B}$ is trivial, and any 
distinguished function $H$ is constant on $M$. For $H$ is distinguished iff 
$X_H = 0$; and if the rank is maximal, then $dH = 0$, so that $H$ is 
constant.\\
\indent Besides, each of the Poisson structure and symplectic form on  $M$ 
determine the other. In particular, the Poisson tensor $\mathsf{B}^{\sh}$ of 
eq. \ref{Poistsraltive} is, up to a sign, the ``contravariant cousin'' of $M$'s 
symplectic form $\o$. For recall: (i) the relation between a symplectic 
manifold's Poisson bracket and its form, eq. \ref{sympPBgeomic}, viz.
\be
\{F, H \} = dF(X_H) = \o(X_F,X_H) \; ;
\ee
and (ii) eq. \ref{defhamvf} for Hamiltonian vector fields on a Poisson 
manifold, viz. 
\be
X_H(F) = \{F, H \} \; .
\ee
Applying these equations yields, if we start from eq. \ref{Poistsraltive} and 
eq. \ref{requireB}:
\be
\mathsf{B}^{\sh}(dH,dF) \; := \; < B(dH), dF > = dF(X_H) = X_H(F) = \{F, H \} = 
\o(X_F, X_H) \; .
\label{PBandSFforasympPoM}
\ee

\indent  We have also seen examples where the Poisson structure $\mathsf{B}$ is 
of non-maximal rank:\\
\indent \indent (i): In our opening ``canonical'' example, the Poisson bracket 
eq. \ref{canlbr} on $M = \mathR^{2n+l}$ has rank $2n$ everywhere. \\
\indent \indent (ii): In the Lie-Poisson structure on $\mso(3)^*$, the rank 
varies across the manifold: it is 2 everywhere, except at the origin $x=0$ 
where it is 0. (Cf. the rank of the matrix $J$ in eq. \ref{Jforso(3)}.) 

\subsubsection{Poisson maps}\label{Pomaps1}
Already at the beginning of our development of Poisson manifolds, we saw that a 
scalar function $H:M \rightarrow \mathR$ defines  equations of motion, with $H$ 
as ``Hamiltonian'', for all other functions $F:M \rightarrow \mathR$, of the 
familiar Poisson bracket type:
\be
{\dot F} = \{F, H \}.
\ee
(Cf. Section \ref{hvfs}, especially the remarks around eq. \ref{defhamvf}.) We 
now develop the generalization for Poisson manifolds of some related notions 
and results.

We say that a smooth map $f:M_1 \rightarrow M_2$ between Poisson manifolds 
$(M_1, \{, \}_1)$ and $(M_2, \{, \}_2)$ is {\em Poisson} or {\em canonical} iff 
it preserves the Poisson bracket. To be precise: we first need the idea of the 
{\em pullback} of a function; cf. Section \ref{fieldtoflow}.A. In this context, 
the pullback  $f^*$ of a function $F: M_2 \rightarrow \mathR$ is given  by
\be
 f^*F := F \circ f; \;\;\;\; {\rm{i.e.}} \;\; f^*F: x \in M_1 \mapsto F(f(x)) 
\in \mathR.
\label{pullback} 
\ee
Then we say that $f:M_1 \rightarrow M_2$ is {\em Poisson} iff for all smooth 
functions $F, G: M_2 \rightarrow \mathR$ ($F,G \in {\cal F}(M_2)$)
\be
f^*\{F, G\}_2  = \{f^*F, f^*G \}_1 \;\; ;
\label{definePoissmaponPoissmfd}
\ee
where by the definition eq. \ref{pullback}, the lhs $\equiv \{F, G\}_2 \circ 
f$, and the rhs $\equiv \{F \circ f, G \circ g \}_1$.

\indent We note the special case where $M_1 = M_2 =: M$ and $M$ is symplectic; 
i.e. the Poisson bracket is of maximal rank, and so defines a symplectic form 
on $M$, as in eq. \ref{PBandSFforasympPoM}. In this case, we return to the 
equivalence in Section \ref{Hammechs}'s usual formulation of Hamiltonian 
mechanics, between preserving the Poisson bracket and preserving the symplectic 
form. That is: a map $f:M \rightarrow M$ on a symplectic manifold $M$ is 
Poisson iff it is symplectic.\\
\indent Besides, we already have for symplectic manifolds an infinitesimal  
version of the idea of a Poisson or symplectic map: viz. the idea of a locally 
Hamiltonian vector field; cf. Section \ref{noetcomplete}. Similarly for Poisson 
manifolds, we will need the corresponding infinitesimal version of a Poisson 
map; but not till Section \ref{canlactioninflgenors}.

One can show (using in particular the Jacobi identity) that the flows of a 
Hamiltonian vector field are Poisson. (Here of course, $(M_1, \{, \}_1) = (M_2, 
\{, \}_2)$.) That is: if $\phi_{\tau}$ is the flow of $X_H$ (i.e. $\phi_{\tau} 
= \exp(\tau X_H)$), then 
\be
\phi^*_{\tau} \{F, G \} =  \{\phi^*_{\tau} F, \phi^*_{\tau} G \}
\;\;\;\; {\rm{i.e.}} \;\; 
\{F, G \} \circ \phi_{\tau} =  \{F \circ \phi_{\tau}, G \circ \phi_{\tau} \} \; 
.
\label{HamFlowisPois}
\ee
Similarly, one can readily show the equivalent proposition, that along the flow 
of a Hamiltonian vector field the Lie derivative of the Poisson tensor 
$\mathsf{B}^{\sh}$ vanishes. That is: for any smooth function $H:M \rightarrow 
\mathR$, we have:
\be
{\cal L}_{X_H} \mathsf{B}^{\sh} = 0 \; .
\label{hamvfLiederivesPoisstsr}
\ee 

Since preserving the Poisson bracket implies in particular preserving its rank, 
it follows from eq. \ref{HamFlowisPois} (or from eq. 
\ref{hamvfLiederivesPoisstsr}) that:\\
\indent If $X_H$ is a Hamiltonian vector field on a Poisson manifold $M$, then 
for any $\tau \in \mathR$ and $x \in M$, the rank of $M$ at $\exp(\tau X_H)(x)$ 
is the same as the rank at $x$. In other words: Hamiltonian vector fields are 
rank-invariant in the sense used in the general form of Frobenius' theorem 
(Section \ref{323BFrobTheorem}).\\
\indent This result will be important for the foliation theorem for Poisson 
manifolds.

We will also need the result (also readily shown) that Poisson maps push 
Hamiltonian flows forward to Hamiltonian flows. More precisely: let $f: M_1 
\rightarrow M_2$ be a Poisson map; so that at each $x \in M_1$, we have the 
derivative map on the tangent space, $T f: (T M_1)_x \rightarrow (T 
M_2)_{f(x)}$. And  let $H: M_2 \rightarrow \mathR$ be a smooth function. If 
$\phi_{\tau}$ is the flow of $X_H$ and $\psi_{\tau}$ is the flow (on $M_1$) of 
$X_{H \circ f}$, then:
\be
\phi_{\tau} \circ f = f \circ \psi_{\tau} \;\; {\rm{and}}\;\; 
Tf \circ X_{H \circ f} = X_H \circ f \;\; .
\label{PoisspushHamtoHamn}
\ee
In particular, this square commutes:
 \begin{equation}
    \bundle{M_1}{\psi_{\tau}}{M_1}
    \bundlemap{f}{f}
    \bundle{M_2}{\phi_{\tau}}{M_2}    
	\label{poispushHamtoHam}
\end{equation}

\subsubsection{Poisson submanifolds: the foliation theorem}\label{Pomaps}
To state the foliation theorem for Poisson manifolds, we need the idea of a 
{\em Poisson immersion}, which leads to the closely related idea of a {\em 
Poisson submanifold}. In effect, these ideas combine the idea of a Poisson map 
with the ideas about injective immersions in (2) of Section 
\ref{323ASubmanifolds}. We recall from that discussion that for an injective 
immersion, $f: N \rightarrow M$, the range $f(N)$ is not necessarily a 
submanifold of $M$: but $f(N)$ is nevertheless called an `injectively immersed 
submanifold' of $M$.  (But as mentioned in Section \ref{323BFrobTheorem}, many 
treatments ignore this point: they in effect assume that an injective immersion 
$f$ is also an embedding, i.e. a homeomorphism between $N$ and $f(N)$, so that 
$f(N)$ is indeed a submanifold of $M$ and $f$ is a diffeomorphism.) 

An injective immersion $f: N \rightarrow M$, with $M$ a Poisson manifold, is 
called a {\em Poisson immersion} if any Hamiltonian vector field defined on an 
open subset of $M$ containing $f(N)$ is in the range of the derivative map of 
$f$ at $y \in N$, i.e. ran($T_y f$), at all points $f(y)$ for $y \in N$.

Being a Poisson immersion is equivalent to the following rather technical 
condition.
\begin{quote}
{\bf {Characterization of Poisson immersions}} An injective immersion $f: N 
\rightarrow M$, with $M$ a Poisson manifold, is a Poisson immersion iff:\\
if $F,G: V \subset N \rightarrow \mathR$, where $V$ is open in $N$, and if 
$\bar{F}, \bar{G}: U \rightarrow \mathR$ are extensions of $F \circ f^{-1}, G 
\circ f^{-1}: f(V) \rightarrow \mathR$ to an open neighbourhood $U$ of $f(V)$ 
in $M$, then $\{\bar{F}, \bar{G}\}\mid_{f(V)}$ is well-defined and independent 
of the extensions.
\end{quote}
The main point of this equivalence is that it ensures that if $f: N \rightarrow 
M$ is a Poisson immersion, then $N$ has a Poisson structure, and $f: N 
\rightarrow M$ is a Poisson map. It is worth seeing how this comes about---by 
proving the equivalence.

\indent {\em Proof}: Let $f:N \ra M$ be a Poisson immersion, and let $F,G:V 
\subset N \ra \mathR$ and let $\bar{F}, \bar{G}: U \supset f(V)  \rightarrow 
\mathR$ be extensions of $F \circ f^{-1}, G \circ f^{-1}: f(V) \rightarrow 
\mathR$. Then for $y \in V$, there is a unique vector $v \in TN_y$ such that 
\be
X_{\bar{G}}(f(y)) = (T_y f)(v) \; .
\ee 
So evaluating the Poisson bracket  of $\bar{F}$ and $\bar{G}$ at $f(y)$ yields, 
by eq. \ref{defhamvf},
\be
\{{\bar F}, {\bar G} \}(f(y)) = d{\bar F}(f(y)) \cdot X_{\bar G}(f(y)) =
d{\bar F}(f(y)) \cdot (T_y f)(v) = d({\bar F} \circ f)(y) \cdot v \equiv dF(y) 
\cdot v \; .
\label{indpdtextensn}
\ee
So $\{{\bar F}, {\bar G} \}(f(y))$ is independent of the extension $\bar F$ of 
$F \circ f^{-1}$. Since the Poisson bracket is antisymmetric, it is also 
independent of the extension $\bar G$ of $G \circ f^{-1}$. So we can define a 
Poisson structure on $N$ by defining for any $y$ in an open $V \subset N$
\be
\{F, G \}_N (y) := \{{\bar F}, {\bar G} \}_M (f(y)) \; .
\label{PoissOnN}
\ee
This makes $f:N \ra M$ a Poisson map, since for any ${\bar F}, {\bar G}$ on $M$ 
and any $y \in N$, we have that
\be
[f^* \{{\bar F}, {\bar G} \}_M](y) \equiv [\{{\bar F}, {\bar G} \}_M \circ 
f](y) =
\{F, G \}_N (y) \equiv \{f^*{\bar F}, f^*{\bar G} \}_N (y) \; ;
\ee
where the middle equality uses eq. \ref{PoissOnN}.\\
\indent For the converse implication, assume that eq. \ref{indpdtextensn} 
holds, and let $H:U \ra \mathR$ be a Hamiltonian defined on an open subset $U$ 
of $M$ that intersects $f(N)$.  Then as we have just seen, $N$ is a Poisson 
manifold and $f:N \ra M$ is a Poisson map. Because $f$ is Poisson, it pushes 
$X_{H \circ f}$ to $X_H$. That is: eq. \ref{PoisspushHamtoHamn} implies that if 
$y \in N$ is such that $f(y) \in U$, then
\be
X_H (f(y)) = (T_y f)(X_{H \circ f}(y)) \; .
\ee 
So $X_H (f(y))$ is in the range of $T_y f$; so $f:N \ra M$ is a Poisson 
immersion. QED.

Now suppose that the inclusion $id: N \rightarrow M$ is a Poisson immersion. 
Then we call $N$ a {\em Poisson submanifold} of $M$. We emphasise, in line with 
the warning we recalled from (2) of Section \ref{323ASubmanifolds}, that  $N$ 
need not be a submanifold of $M$; but it is nevertheless called an `injectively 
immersed submanifold' of $M$.

 From the definition of a Poisson immersion, it follows that any Hamiltonian 
vector field must be tangent to a Poisson submanifold. In other words: writing  
$\cal X$ for the system of Hamiltonian vector fields on $M$, and ${\cal X} 
\mid_x$ for their values at $x \in M$, we have: if $N$ is a Poisson submanifold 
of $M$, and $x \in N$, ${\cal X} \mid_x \subset TN_x$.\\
\indent For the special case where $M$ is a symplectic manifold, we have ${\cal 
X} \mid_x = T_xM$, and the only Poisson submanifolds of $M$ are its open sets.

Finally, we define the following equivalence relation on a Poisson manifold 
$M$. Two points $x_1, x_2 \in M$  are {\em on the same symplectic leaf} if 
there is a piecewise smooth curve in $M$ joining them, each segment of which is 
an integral curve of a locally defined Hamiltonian vector field. An equivalence 
class of this equivalence relation is a {\em symplectic leaf}.

We can now state and prove that Poisson manifolds are foliated.

\paragraph{7.3.3.A Foliation theorem for Poisson 
manifolds}\label{733APomFoliate} The result is:---
\begin{quote}
A Poisson manifold $M$ is the disjoint union of its symplectic leaves. Each 
symplectic leaf is an injectively immersed Poisson submanifold, and the induced 
Poisson structure on the leaf is symplectic. The leaf through the point $x$, 
$N_x$ say, has dimension equal to the rank of the Poisson structure at $x$; and 
the tangent space to the leaf at $x$ equals
\begin{eqnarray}
TN_x = {\rm {ran}}(\mathsf{B}_x) := \{ X \in T_xM \;\; : \;\;  X = 
\mathsf{B}_x(\al), \;\; {\rm {some}}\;\; \al \in T^*_xM\} \\ 
= \{X_H(x) : H \in {\cal F}(U), \;\; U \; {\rm {open}}\;{\rm {in}}\;\;M\;\; \}
\label{PoisFolia}
\end{eqnarray}
\end{quote}

{\em Proof}: We apply the general form of Frobenius' theorem (Section 
\ref{323BFrobTheorem}) to the system $\cal X$ of Hamiltonian vector fields on 
$M$. We know from eq. \ref{HamvfsclosedunderLie} (Section \ref{hvfs}) that 
$\cal X$ is involutive, and from eq. \ref{HamFlowisPois} above that it is 
rank-invariant. So by Frobenius' theorem, $\cal X$ is integrable. The integral 
submanifolds are by definition given by the rhs of eq. \ref{PoisFolia}. QED.

One also readily shows that:\\
\indent (i): One can evaluate the Poisson bracket of $F,G:M \rightarrow \mathR$ 
at $x \in M$ by restricting $F$ and $G$ to the symplectic leaf $N_x$ through 
$x$, and evaluating the Poisson bracket that is defined by the symplectic form 
on the leaf $N_x$; (i.e. the Poisson bracket defined in eq. 
\ref{sympPBgeomic}).\\
\indent (ii): A distinguished function is constant on any symplectic leaf $N_x$ 
of $M$.

We end with two remarks. The first is a mathematical warning; the second 
concerns physical interpretation.\\
\indent (1): Recall our warning that symplectic leaves need not be 
submanifolds. This also means that all the distinguished functions being 
constants does {\em not} imply that the Poisson structure is non-degenerate. 
Indeed, one can readily construct an example in which the symplectic leaves are 
not manifolds, all distinguished functions  are constants, and the Poisson 
structure is degenerate. Namely, one adapts an example mentioned before, in 
Section \ref{examsubsub}: the flows on the torus $\mathsf{T}^2$ that wind 
densely around it. (For more details about this example, cf. Arnold (1973: 
160-167) or Arnold (1989: 72-74) or Butterfield (2004a: Section 2.1.3.B); for 
how to adapt it, cf. Marsden and Ratiu (1999: 347).\\
\indent (2): As we have seen, any integral curve of any Hamiltonian vector 
field $X_H$ is confined to one of the symplectic leaves. So if we are 
interested only in  the behaviour of a single solution through a point $x \in 
M$, we can restrict our attention to the symplectic leaf $N_x$ through $x$: for 
the solution will always remain in $N_x$. But as stressed in Section 
\ref{pmspreamble}, there are at least three good reasons not to ignore the more 
general Poisson structure!

\subsubsection{Darboux's theorem}\label{DarbouxPoiss}
At the end of Section \ref{cotgtblesymp}, we mentioned Darboux's theorem: it  
said that any symplectic manifold ``looks locally like'' a cotangent bundle. 
The generalization for Poisson manifolds says that any Poisson manifold ``looks 
locally like'' our canonical example on $\mathR^m, m = 2n + l$, given at the 
start of Section \ref{pbs}. More precisely, we have:
\begin{quote}
Let $M$ be an $m$-dimensional Poisson manifold, and let $x \in M$ be a point 
with an open neighbourhood $U \subset M$ throughout which the rank is a 
constant $2n \leq m$. Then defining $l := m - 2n$, there is a possibly smaller 
neighbourhood $U' \subset U$ of $x$, on which there exist local coordinates 
$(q,p,z) = (q^1,...,q^n,p^1,...,p^n,z^1,...,z^l)$, for which the Poisson 
bracket takes the form
\be
\{F, H\} := \Sigma^n_i \left(\frac{\pl F}{\pl q^i}\frac{\pl H}{\pl p^i} - 
\frac{\pl F}{\pl p^i}\frac{\pl H}{\pl q^i}
\right) \; .
\label{canlbrDarbouxGnlzd}
\ee
(So the Poisson brackets for the coordinate functions take the now-familiar 
form given by eq. \ref{tradPB} and \ref{newPBforz}.) The symplectic leaves of 
$M$ intersect the coordinate  chart in the slices $\{z^1 = c_1,\dots,z^l = c_l 
\}$ given by constant values of the distinguished coordinates $z$.
\end{quote}

We shall not give the proof. Suffice it to say that:\\
\indent (i): Like Darboux's theorem for symplectic manifolds: it proceeds by 
induction on the ``half-rank'' $n$; and it begins by taking any function $F$ as 
the ``momentum'' $p^1$ and constructing the canonically conjugate coordinate 
$q^1$ such that $\{q^1, p^1\} = 1$.\\
\indent (ii): The induction step invokes a version of Frobenius' theorem in 
which the fact that the rank $2n$ is constant  throughout $U$ secures a 
coordinate system in which the $2n$-dimensional  integral manifolds are given 
by slices defined by constant values of the remaining $l$ coordinates. The 
Poisson structure then secures that these remaining coordinates are 
distinguished.

\paragraph{7.3.4.A Example: $\mso(3)^*$ yet again}\label{mso(3)}
We illustrate (1) the foliation theorem and (2) Darboux's theorem, with 
$\mso(3)^*$; whose Lie-Poisson structure we described in Section 
\ref{LPso(3)}.A.

\indent (1): At $x \in \mso(3)^*$, the subspace ${\cal X}\mid_x := \{X_H(x) : H 
\in {\cal F}(U), \;\; U \; {\rm {open}}\;{\rm {in}}\;\;M\; \}$ of values of 
locally Hamiltonian vector fields is spanned by $e_1 := y{\pl_z} - z{\pl_y}$ 
representing infinitesimal rotation about the $x$-axis (cf. eq. 
\ref{infmlrotnx}); $e_2 := z{\pl_x} - x{\pl_z}$ for rotation about the 
$y$-axis; and $e_3 := x{\pl_y} - y{\pl_x}$ for rotation about the $z$-axis. If 
$x \neq 0$, these vectors span a two-dimensional subspace of $T \mso(3)^*_x$: 
viz. the tangent plane to the sphere $S_{\mid x \mid}$ of radius $\mid x \mid$ 
centred at the origin. So the foliation theorem implies that $\mso(3)^*$'s 
symplectic leaves are these spheres; and the origin.\\
\indent We can compute the Poisson bracket of $F,G: S_{\mid x \mid} \ra \mathR$ 
by extending $F$ and $G$ to a neighbourhood of $S_{\mid x \mid}$; cf. eq. 
\ref{PoissOnN}. That is: we can consider extensions ${\bar F}, {\bar G}: U 
\supset S_{\mid x \mid} \ra \mathR$, and calculate the Poisson bracket in 
$\mso(3)^*$, whose Poisson structure we already computed in eq. 
\ref{LPBmso(3)}.\\
\indent Adopting spherical polar coordinates with $r = \mid x \mid$, i.e. $x^1 
= r \cos \theta \sin \phi, x^2 = r \sin \theta \sin \phi, x^3 = r \cos \phi$, 
we can define ${\bar F}, {\bar G}$ merely by ${\bar F}(r,\theta,\phi) := 
F(\theta,\phi), {\bar G}(r,\theta,\phi) := G(\theta,\phi)$; so that the partial 
derivatives with respect to the spherical angles $\theta, \phi$ are equal, i.e. 
${\bar F}_{\theta} = F_{\theta}, {\bar F}_{\phi} = F_{\phi}, {\bar G}_{\theta} 
= G_{\theta}, {\bar G}_{\phi} = G_{\phi}$.\\
\indent Besides, eq. \ref{Tostrucfns4} implies that we need only calculate  the 
Poisson bracket  in $\mso(3)^*$ of the spherical angles $\theta$ and $\phi$. So 
eq. \ref{LPBmso(3)} gives 
\be
\{\theta, \phi \} = - x \cdot (\nabla \theta \times \nabla \phi) = \frac{-1}{r 
\sin \phi} \; ;
\ee 
and eq. \ref{PoissOnN} and \ref{Tostrucfns4} give
\be
\{F, G \}= \{{\bar F}, {\bar G} \} = \frac{-1}{r \sin \phi} (F_{\theta}G_{\phi} 
- F_{\phi}G_{\theta})\; . 
\ee

(2): $z := x^3$ defines the Hamiltonian  vector field $X_z = x^2 \pl_{x^1} - 
x^1 \pl_{x^2}$ that generates clockwise rotation about the $z \equiv x^3$-axis. 
So away from the origin the polar angle $\theta := \arctan (x^2/x^1)$ has a 
Poisson bracket with $z$ equal to: $\{\theta, z \} = X_z (\theta) = -1$. 
Exprssing $F,H: \mso(3)^* \ra \mathR$ in terms of the coordinates $z, \theta$ 
and $r := \mid x \mid$, we find that the Lie-Poisson bracket is: $\{F, H \} = 
F_z H_{\theta} - F_{\theta} H_z$. So $(z,\theta, r)$ are canonical coordinates. 

\subsection{The symplectic structure of the co-adjoint 
representation}\label{sympcoad}
Section \ref{LPso(3)} described how the dual $\mg^*$ of a finite-dimensional 
Lie algebra  of a Lie group $G$ has the structure of a Poisson manifold. In 
this case, the foliation established in the previous Subsection has an 
especially neat interpretation. Namely: the leaves are the orbits of the 
co-adjoint representation of $G$ on $\mg^*$.

This symplectic structure in the co-adjoint representation sums up themes from 
Sections \ref{example:adcoad} (especially \ref{coadj;was6.5.5.B}), and 
\ref{LPso(3)} and \ref{sympstruc}. In particular, it connects two properties of  
the  Lie bracket in $\mg$, which we have already seen: viz.\\
\indent (i): The  Lie bracket in $\mg$ gives the infinitesimal generators of 
the adjoint action; cf. eq. \ref{sumupad}.\\
\indent (ii): The  Lie bracket in $\mg$ defines (in a basis-independent way) a 
Lie-Poisson bracket on $\mg^*$, thus making $\mg^*$ a Poisson manifold. (Cf. 
the definition in eq. \ref{defLPbr}, shown to be basis-independent by eq. 
\ref{intrLPbr}.)

In fact, there is a wealth of instructive results and examples about the 
structure of the co-adjoint representation: we will only scratch the 
surface---as in other Sections! We will give a proof, under a simplifying 
assumption, of one main result; and then make a few remarks about other 
results.

The result is:
\begin{quote}
{\bf{The orbits of the co-adjoint representation are $\mg^*$'s leaves}} \\
Let $G$ be a Lie group, with its co-adjoint representation $Ad^*$ on $\mg^*$. 
That is, recalling eq. \ref{defineAd^*_g-1;repn}, we have:
\be
Ad^*: G \rightarrow {\rm{End}}({\mg}^*), \;\;\; 
Ad^*_{g^{-1}} = (T_e(R_g \circ L_{g^{-1}}))^* \; .
\label{defineAd^*_g-1;representation;again}
\ee
The orbits of this representation are the symplectic leaves of $\mg^*$, taken 
as equipped with its natural Poisson structure, i.e. the Lie-Poisson bracket 
eq. \ref{intrLPbr}.   
\end{quote}
{\em Proof}:--- We shall prove this under the simplifying assumption that the 
co-adjoint action of $G$ on $\mg^*$ is proper. (We recall from the definition 
of proper actions, eq. \ref{defineproper}, that for any compact Lie group, such 
as $SO(3)$, this condition is automatically satisfied.) Then we know from 
result (3) and eq. \ref{tgtspaceatorbitintermsofinflgenors}, at the end of 
Section \ref{inflgenor}, that this implies that the co-adjoint orbit 
${\rm{Orb}}(\al)$ of any $\al \in \mg^*$ is a closed submanifold of $\mg^*$, 
and that the tangent space to ${\rm{Orb}}(\al)$ at a point $\bb \in 
{\rm{Orb}}(\al)$ is
\be
T{\rm{Orb}}(\al)_{\bb} \; = \;  \{ \; \xi_{\mg^*}(\bb) \; : \; \xi \in \mg \; 
\} \; . 
\label{TOrbsimpleforcoad}
\ee 
We will see shortly how this assumption implies that $\mg^*$'s symplectic 
leaves are submanifolds.\footnote{To verify that our condition is indeed 
simplifying---i.e. that in general the co-adjoint orbits in $\mg^*$ are not 
submanifolds---consider the example in Marsden and Ratiu (1999: 14.1.(f), p. 
449); taken from Kirillov (1976: 293).}

We now argue as follows. For $\xi \in \mg$, consider the scalar function on 
$\mg^*$, $K_{\xi}: \al \in \mg^* \mapsto K_{\xi}(\al) \; := \; <\al ; \xi > \; 
\in \mathR$; and its Hamiltonian vector field $X_{K_{\xi}}$. At each $\al \in 
\mg^*$, the gradient $\nabla K_{\xi}(\al) \equiv d K_{\xi}(\al)$, considered as 
an element of $(T^*{\mg}^*)_{\al} \cong \mg$, is just $\xi$ itself. Now we will 
compute $X_{K_{\xi}}(F)(\al)$ for any $F:\mg^* \rightarrow \mathR$ and any $\al 
\in \mg^*$, using in order:\\
\indent (i): the intrinsic definition of the Lie-Poisson bracket on $\mg^*$, 
eq. \ref{intrLPbr};\\
\indent (ii): the fact that the infinitesimal generator of the adjoint action 
is the Lie bracket in $\mg$, eq. \ref{sumupad};\\
\indent (iii): the fact that the derivative $ad^*$ of the co-adjoint action 
$Ad^*$ is, up to a sign, the adjoint of $ad_{\xi}$; eq. \ref{computead*}.

Thus we get, for all $F:\mg^* \ra \mathR$ and $\al \in \mg^*$:
\begin{eqnarray}
X_{K_{\xi}}(F)(\al) \;\; \equiv \;\; \{F, K_{\xi}\}(\al) \;\; = \;\; < \; \al 
\; ; \; [\nabla F(\al), \nabla K_{\xi}(\al)] \;  > \\ 
= \;\; < \; \al \; ; \; [\nabla F(\al), \xi ] \; > \;\;  = \;\; - \; < \; \al 
\; ; \; [\xi, \nabla F(\al) ] \; >\\ 
= \;\; - \; < \; \al \; ; \; ad_{\xi} (\nabla F(\al))  \; >\\ 
= \;\; < \;  ad^*_{\xi}(\al) \; ; \; \nabla F(\al)  \; >.
\label{computecoadsympleaves}
\end{eqnarray}
But on the other hand, the vector field $X_{K_{\xi}}$ is uniquely determined by 
its action on all such functions $F$ at all $\al \in \mg^*$:
\be
X_{K_{\xi}}(F)(\al) \;\; \equiv \;\; < \; X_{K_{\xi}}(\al) \; ; \nabla F(\al) 
\; > \; .
\ee
So we conclude that at each $\al \in \mg^*$:
\be
X_{K_{\xi}} = ad^*_{\xi} \; .
\ee
But the subspace ${\cal X}\mid_{\al}$ of values at $\al$ of Hamiltonian vector 
fields is spanned by the $X_{K_{\xi}}(\al)$, with $\xi$ varying through $\mg$. 
And as $\xi$ varies through $\mg$, $ad^*_{\xi}(\al)$ is the tangent space 
$T{\rm{Orb}}(\al)_{\al}$ to the co-adjoint orbit ${\rm{Orb}}(\al)$ of $G$ 
through $\al$. So
\be
{\cal X}\mid_{\al} = T{\rm{Orb}}(\al)_{\al} \; .
\ee  
So the integral submanifolds of the system $\cal X$ of Hamiltonian vector 
fields, which are the symplectic leaves of $\mg^*$ by Section \ref{Pomaps}.A's 
foliation theorem, are the co-adjoint orbits. QED.

For the illustration of this theorem by our standard example, $\mso(3)^*$, cf. 
our previous discussions of it: in Section \ref{coadj;was6.5.5.B} for its 
co-adjoint structure; in Section \ref{LPso(3)}.A for its Lie-Poisson structure; 
and in Section \ref{DarbouxPoiss}.A for its symplectic leaf structure.

We end this Subsection by stating  two other results. They are not needed 
later, but they are enticing hints of how rich is the theory of co-adjoint 
orbits.\\
\indent (1): For each $g \in G$, the co-adjoint map $Ad^*_g : \mg^* \ra \mg^*$ 
is a Poisson map that preserves the symplectic leaves of $\mg^*$.\\
\indent (2): A close cousin of the theorem just proven is that the Lie bracket 
on $\mg$ defines (via its definition of the Lie-Poisson bracket on $\mg^*$, eq. 
\ref{intrLPbr}) a symplectic form, i.e. a non-degenerate closed two-form, on 
each co-adjoint orbit, by:
\be
\o (\al)(ad^*_{\xi}(\al), ad^*_{\eta}(\al)) \;\; := \;\;
< \; \al \; ; [\xi, \eta]_{\mg} \; > \;\; , \;\; \forall \al \in \mg^*, \; 
\forall \xi, \eta \in \mg \; . 
\ee 
This theorem is proven in detail (without our simplifying assumption that $G$'s 
action is proper) by Marsden and Ratiu (1999: Thm 14.3.1, pp. 453-456); and 
much more briefly by Arnold (1989: 321, 376-377, 457); and rather differently 
(even without using the notion of a Poisson manifold!) in Abraham and Marsden 
(1978: 302-303).

\subsection{Quotients of Poisson manifolds}\label{quotpm}
We now end Section \ref{pms} with the simplest general theorem about 
quotienting a Lie group action on a Poisson manifold, so as to get a quotient  
space (set of orbits) that is itself a Poisson manifold. So this theorem 
combines themes from Sections \ref{actionlg}---in particular,  the idea from 
Section \ref{332Aproper}.B that for a free and proper group action, the orbits 
and quotient space are manifolds---with material about Poisson manifolds from 
Section \ref{basics}. (The material in Sections \ref{sympstruc} and 
\ref{sympcoad} will not be needed.) This theorem will be important in Section 
\ref{redn}. We call this result the 
\begin{quote}
{\bf Poisson reduction  theorem}:
Suppose the Lie group $G$ acts on Poisson manifold $M$ is such a way that each 
$\Phi_g: M \rightarrow M$ is a Poisson map. Suppose also that the quotient 
space $M/G$ is a manifold and the projection $\pi: M \rightarrow M/G$ is a 
smooth submersion (say because $G$'s action on $M$ is free and proper, cf. 
Section \ref{332Aproper}.B). Then there is a unique Poisson structure on $M/G$ 
such that $\pi$ is a Poisson map. The Poisson bracket on $M/G$ is called the 
{\em reduced} Poisson bracket.
\end{quote}
{\em Proof}: Let us first assume that $M/G$ is a Poisson manifold and that 
$\pi$ is a Poisson map; and show uniqueness. We first note that for any $f: M/G 
\rightarrow \mathR$, the function ${\bar f} := f \circ \pi: M \rightarrow 
\mathR$ is obviously the unique $G$-invariant function on $M$ that projects by 
$\pi$ to $f$. That is:  if $[x] \equiv {\rm{Orb}}(x) \equiv G \cdot x$ is the 
orbit of $x \in M$, then ${\bar f}$ assigns the same value $f([x])$ to all 
elements of the orbit $[x]$. Besides, in terms of pullbacks (eq. 
\ref{pullback}), ${\bar f} = \pi^* f$.\\
\indent Then the condition that $\pi$ be Poisson, eq. 
\ref{definePoissmaponPoissmfd}, is that for any two smooth scalars $f,h: M/G 
\rightarrow \mathR$, we have an equation  of smooth scalars on $M$:
\be
\{f,h\}_{M/G} \; \circ \pi = \{ f \circ \pi, h \circ \pi \}_M = \{ {\bar f}, 
{\bar h} \}_M \; 
\label{quotpoistruc}
\ee
where the subscripts indicate on which space the Poisson bracket is defined. 
Since $\pi$ is surjective, eq. \ref{quotpoistruc} determines the value 
$\{f,h\}_{M/G}$ uniquely.\\
\indent But eq. \ref{quotpoistruc} also {\em defines} $\{f,h\}_{M/G}$ as a 
Poisson bracket; in two stages. (1): The facts that $\Phi_g$ is Poisson, and 
${\bar f}$ and ${\bar h}$ are constant on orbits imply that
\be
\{{\bar f}, {\bar h} \}(g\cdot x) \; = \; ( \{{\bar f}, {\bar h} \}\circ 
\Phi_g)(x)  \; = \; 
\{{\bar f}\circ \Phi_g, {\bar h}\circ \Phi_g \} (x)
  \; = \; \{{\bar f}, {\bar h} \}(x).
\ee
That is: $\{{\bar f}, {\bar h} \}$ is also constant on orbits, and so defines 
$\{f, h\}$ uniquely.\\
\indent(2): We show that $\{f, h\}$, as thus defined, is a Poisson structure on 
$M/G$, by checking that the required properties, such as the Jacobi identity, 
follow from the Poisson structure $\{\;, \;\}_M$ on $M$. QED. 

This theorem is a ``prototype'' for material to come. We spell this out in two 
brief remarks, which look forward to the following two Sections.

(1): {\em Other theorems}:--- This theorem is one of many that yield new 
Poisson manifolds and symplectic manifolds from old ones by quotienting. In 
particular, as we will see in detail in Section \ref{redn}, this theorem is 
exemplified by the case where $M = T^*G$ (so here $M$ is symplectic, since it 
is a cotangent bundle), and $G$ acts on itself by left translations, and so 
acts on $T^*G$ by a cotangent lift. In this case, we will have $M/G \cong 
\mg^*$; and the reduced Poisson bracket just defined, by eq. 
\ref{quotpoistruc}, will be the Lie-Poisson bracket we have already met in 
Section \ref{LPso(3)}. 

(2): {\em Reduction of dynamics}:--- Using this theorem, we can already fill 
out a little what is involved in reduced dynamics; which we only glimpsed in 
our introductory discussions, in Section \ref{relnalmechsexample} and 
\ref{pmspreamble}. We can make two basic points, as follows.

\indent (A): If $H$ is a $G$-invariant Hamiltonian function on $M$, it defines 
a corresponding function $h$ on $M/G$ by $H = h \circ \pi$. The fact that 
Poisson maps push Hamiltonian flows forward to Hamiltonian flows (eq. 
\ref{PoisspushHamtoHamn}) implies,  since $\pi$ is Poisson, that $\pi$ 
transforms $X_H$ on $M$ to $X_h$ on $M/G$. That is: 
\be
T\pi \circ X_H = X_h \circ \pi \; \; ;
\label{hamflowspirelated}
\ee
i.e. $X_H$ and $X_h$ are $\pi$-related. Accordingly, we say that the 
Hamiltonian system $X_H$ on $M$ {\em reduces} to that on $M/G$.

\indent (B): We shall see in Section \ref{mommmapsnoet} that $G$-invariance of 
$H$ is associated  with a family of conserved quantities (constants of the 
motion,  first integrals), viz. a constant of the motion  $J(\xi): M 
\rightarrow \mathR$ for each $\xi \in \mg$. Here, $J$ being conserved means  
$\{J, H\} = 0$; just as in our discussion of Noether's theorem in ordinary 
Hamiltonian mechanics (Section \ref{noetcomplete}). Besides, if $J$ is also 
$G$-invariant, then the corresponding function $j$ on $M/G$ is conserved by 
$X_h$ since
\be
\{j, h\} \circ \pi = \{J, H\} = 0 \;\; {\rm{implies}}\;\; \{j, h\} = 0 \; .
\ee

\section{Symmetry and conservation revisited: momentum maps}\label{symyredn}
We now develop the topics of symmetry and conserved quantities (and so 
Noether's theorem) in the context of Poisson manifolds. At the centre of these 
topics lies the idea of a {\em momentum map} of a Lie group action on a Poisson 
manifold; which we introduce in Section \ref{mommmapsbasics}. This is the 
modern geometric generalization of a conserved quantity, such as linear or 
angular momentum for the Euclidean group---hence the name. Formally, it will be 
a map $\bf J$ from the Poisson manifold $M$ to the dual $\mg^*$ of the Lie 
algebra of the symmetry  group $G$. Since its values lie in a vector space, it 
has components. So our description of conserved quantities will no longer be 
``one-dimensional'', i.e. focussed on a single vector field in the state space, 
as it was in Sections \ref{NoetherLag} and \ref{Hammechs}. The map $\bf J$ will 
be associated with a linear map $J$ from $\mg$ to ${\cal F}(M)$, the scalar 
functions on the manifold $M$. That is: for each $\xi \in \mg$, $J(\xi)$ will 
be a conserved quantity if the Hamiltonian $H$ is invariant under the 
infinitesimal generator $\xi_M$, i.e. if $\xi_M(H) = 0$. 

The conservation of momentum maps will be expressed by the Poisson manifold  
version of Noether's theorem (Section \ref{mommmapsnoet}), and  illustrated by 
the familiar examples of linear and angular momentum (Section 
\ref{mommmapsnoetexample}). Then we discuss the equivariance of momentum maps, 
with respect to the co-adjoint representation of $G$ on $\mg^*$; Section 
\ref{eqvarcemm}. 
Finally in Section \ref{mommmapcotgt}, we discuss the crucial special case of  
momentum maps on cotangent bundles, again with examples.

\subsection{Canonical actions and momentum maps}\label{mommmapsbasics}
We first apply the definition of Poisson maps (from Section \ref{Pomaps1}) to 
group actions (Section \ref{canlactioninflgenors}). This will lead to  the idea 
of the momentum map (Section \ref{momentummapintro}).

\subsubsection{Canonical actions and infinitesimal 
generators}\label{canlactioninflgenors}
Let $G$ be a Lie group acting on a Poisson manifold $M$ by a smooth left action 
$\Phi: G \times M \rightarrow M$; so that as usual we write $\Phi_g: x \in M 
\mapsto \Phi_g(x):= g \cdot x \in M$. As in the definition of a Poisson map 
(eq. \ref{definePoissmaponPoissmfd}), we say the action is {\em canonical} if 
\be
\Phi^*_g \{F_1, F_2\} = \{\Phi^*_g F_1, \Phi^*_g F_2\}
\label{defcanlaction}  
\ee  
for any $F_1, F_2 \in {\cal F}(M)$ and any $g \in G$. If $M$ is symplectic with 
symplectic form $\o$, then the action is canonical iff it is symplectic, i.e. 
$\Phi^*_g \o = \o$ for all $g \in G$.
 
We will be especially interested in the infinitesimal version of this notion; 
and so with infinitesimal generators of actions. We recall from eq. 
\ref{defineinflgenor} that the infinitesimal generator of the action 
corresponding to a Lie algebra element $\xi \in \mg$ is the vector field 
$\xi_M$ on $M$ obtained by differentiating the action with respect to $g$ at 
the identity in the direction $\xi$:
\be
\xi_M(x) = \frac{d}{d \tau} [\exp(\tau \xi) \cdot x] \mid_{\tau = 0}.
\ee
So we differentiate eq. \ref{defcanlaction} with respect to $g$ in the 
direction $\xi$, to give:
\be
\xi_M(\{F_1, F_2 \}) =  \{\xi_M(F_1), F_2 \} +  \{F_1, \xi_M(F_2) \} \; .
\label{defineinfPoissautom}
\ee
Such a vector field $\xi_M$ is called an {\em infinitesimal Poisson 
automorphism}.

Side-remark:--- We will shortly see that it is the universal quantification 
over $g \in G$ in eq. \ref{defcanlaction}, and correspondingly in eq. 
\ref{defineinfPoissautom} and \ref{requirexiMgbllyhamn} below, that means our 
description of conserved quantities is no longer focussed on a single vector 
field; and in particular, that a momentum map representing a conserved quantity 
has components.

In the symplectic case, differentiating $\Phi^*_g \o = \o$ implies that the Lie 
derivative ${\cal L}_{\xi_M}\o$ of $\o$ with respect to $\xi$ vanishes: ${\cal 
L}_{\xi_M}\o = 0$. We saw in Section \ref{noetcomplete} that this is equivalent 
to $\xi_M$ being locally Hamiltonian, i.e. there being a local scalar $J: U 
\subset M \rightarrow \mathR$ such that $\xi_M = X_J$.  This was how Section 
\ref{noetcomplete} vindicated eq. \ref{naivenoetham}'s ``one-liner'' approach 
to Noether's theorem: because the vector field $X_f$ is locally Hamiltonian, it 
preserves the symplectic structure, i.e. Lie-derives the symplectic form ${\cal 
L}_{X_f}\o = 0$---as a symmetry should.\\
\indent We also saw in result (2) at the end of Section \ref{lavecfields} that 
the ``meshing'', up to a sign, of the Poisson bracket on scalars with the Lie 
bracket on vector fields implied that the locally Hamiltonian vector fields 
form a Lie subalgebra of the Lie algebra ${\cal X}(M)$ of all vector fields.

Turning to the context of Poisson manifolds, we need to note two points. The 
first is a similarity with the symplectic case; the second is a contrast.\\
\indent (1): One readily checks, just by applying eq. 
\ref{defineinfPoissautom}, that the infinitesimal Poisson automorphisms are 
closed under the Lie bracket. So we write the Lie algebra of these vector 
fields as ${\cal P}(M)$: ${\cal P}(M) \subset {\cal X}(M)$. 

\indent (2): On the other hand, Section \ref{noetcomplete}'s equivalence 
between a vector field being locally Hamiltonian and preserving the geometric 
structure of the state-space breaks down.\\
\indent Agreed, the first implies the second: a locally Hamiltonian vector 
field preserves the Poisson bracket. We noted this already in Section 
\ref{Pomaps1}. The differential statement was that such a field $X_H$ 
Lie-derives the Poisson tensor: ${\cal L}_{X_H} \mathsf{B}^{\sh} = 0$ (eq. 
\ref{hamvfLiederivesPoisstsr}). The finite statement was that the flows of such 
a field are Poisson maps: $\phi^*_{\tau} \{F, G \} =  \{\phi^*_{\tau} F, 
\phi^*_{\tau} G \}$ (eq. \ref{HamFlowisPois}).\\
\indent But the converse implication fails: an infinitesimal Poisson 
automorphism on a Poisson manifold need {\em not} be locally Hamiltonian. For 
example, make $\mathR^2$ a Poisson manifold by defining the Poisson structure 
\be
\{F, H \} = x \left(\frac{\pl F}{\pl x}\frac{\pl H}{\pl y} - \frac{\pl H}{\pl 
x}\frac{\pl F}{\pl y} \right) \; ;
\ee
then the vector field $X = {\pl }/{\pl y}$ in a neighbourhood of a point on the 
$y$-axis is a non-Hamiltonian infinitesimal Poisson automorphism.\\
\indent This point will affect the formulation of Noether's theorem for Poisson 
manifolds, in Section \ref{mommmapsnoet}. 

Nevertheless, we shall from now on be interested in cases where for all $\xi$, 
$\xi_M$ is globally Hamiltonian. This means there is a map $J: \mg \ra {\cal 
F}(M)$ such that
\be
X_{J(\xi)} = \xi_M
\label{requirexiMgbllyhamn}
\ee
for all $\xi \in \mg$. There are three points we need to note about this 
condition.

(1): Since the right hand side of eq. \ref{requirexiMgbllyhamn} is linear in 
$\xi$, we can require such a $J$ to be a linear map. For given any  $J$ obeying 
eq. \ref{requirexiMgbllyhamn}, we can take a basis $e_1,\dots,e_m$ of $\mg$ and 
define a new linear ${\bar J}$ by setting, for any $\xi = \xi^i e_i$, ${\bar 
J}(\xi) := \xi^i J(e_i)$.

(2): Eq. \ref{requirexiMgbllyhamn} does not determine $J(\xi)$. For by the 
linearity of the map $\mathsf{B}: d J(\xi) \mapsto X_{J(\xi)}$, we can add to  
such a $J(\xi)$ any distinguished function, i.e. an $F: M \rightarrow \mathR$ 
such that $X_F = 0$. That is: $X_{J(\xi) + F} \equiv X_{J(\xi)}$. (Of course, 
in the symplectic case, the only distinguished functions are constants.)   

(3): It is worth expressing eq. \ref{requirexiMgbllyhamn} in terms of Poisson 
brackets. Recalling that for any $F, H \in {\cal F}(M)$, we have $X_H (F) = 
\{F, H \}$, this equation becomes
\be
\{F, J(\xi) \} = \xi_M (F) \;\; , \; \forall F \in {\cal F}(M), \;\; \forall 
\xi \in \mg \; .
\label{requirexiMgbllyhamn;repeat;PBversion}
\ee

We will also need the following result:
\be
X_{J([\xi, \eta])} = X_{\{J(\xi), J(\eta) \}_M} \; .
\label{mmmeshPBstruc}
\ee
To prove this, we just apply two previous results, each giving a Lie algebra 
anti-homomorphism.\\
\indent (i): Result (4) at the end of Section \ref{inflgenor}: for any left 
action of Lie group $G$ on any manifold $M$, the map $\xi \mapsto \xi_M$ is a 
Lie algebra anti-homomorphism between $\mg$ and the Lie algebra ${\cal X}_M$ of 
all vector fields on $M$:
\be
(a\xi + b\eta)_M = a\xi_M + b\eta_M \;\; ; \;\; [\xi_M,\eta_M] = - [\xi,\eta]_M 
\;\; \forall \xi,\eta \in \mg, \; {\rm{and}} \; a,b \in \mathR.
\label{lalgactionantihomo;repeat}
\ee
\indent (ii): The ``meshing'' up to a sign, just as in the symplectic case, of 
the Poisson bracket on scalars with the Lie bracket on vector fields, as in eq. 
\ref{HamvfsclosedunderLie} at the end of Section \ref{hvfs}:
\be
X_{\{F, H\}} = - [X_F, X_H] = [X_H, X_F] \; \; .
\label{HamvfsclosedunderLie;repeat}
\ee 
So for a Poisson manifold $M$, the map $F \in {\cal F}(M) \mapsto X_F \in {\cal 
X}(M)$ is a Lie algebra anti-homomorphism.

\indent Applying (i) and (ii), we deduce eq. \ref{mmmeshPBstruc} by:
\be
X_{J([\xi, \eta])} = [\xi, \eta]_M = - [\xi_M, \eta_M] = - [X_{J(\xi)}, 
X_{J(\eta)}] =  X_{\{J(\xi), J(\eta) \}_M} \; .
\label{mmmeshPBstruc;proven}
\ee

\subsubsection{Momentum maps introduced}\label{momentummapintro}
So suppose that there is a canonical left action of $G$ on a Poisson manifold 
$M$. And suppose there is a linear map $J: \mg \rightarrow {\cal F}(M)$ such 
that
\be
X_{J(\xi)} = \xi_M
\label{requirexiMgbllyhamn;repeat}
\ee
for all $\xi \in \mg$.\\
\indent  The two requirements---that the action be infinitesimally canonical 
(i.e. each  $\xi_M \in {\cal P}(M)$) and that each $\xi_M$ be globally 
Hamiltonian---can be expressed as requiring that there be a $J: \mg \rightarrow 
{\cal F}(M)$ such that there is a commutative diagram. Namely, the map $\xi \in 
\mg \mapsto \xi_M \in {\cal P}(M)$ is to equal the composed map: 
\be
\mg \stackrel{J}{\longrightarrow} {\cal F}(M) \stackrel{F \mapsto 
X_F}{\longrightarrow} {\cal P}(M) \; .
\label{inflgenorfromtriangle}
\ee

Then the map ${\bf J}:M \rightarrow \mg^*$ defined by
\be
< \; {\bf J}(x) \; ; \; \xi \; > \;\; := \;\;  J(\xi)(x)
\label{definemm}
\ee
for all $\xi \in \mg$ and $x \in M$, is called the {\em momentum map} of the 
action.

Another way to state this definition is as follows. Any smooth function  ${\bf 
J}:M \rightarrow \mg^*$ defines at each $\xi \in \mg$ a scalar $J(\xi): x \in M 
\mapsto ({\bf J}(x))(\xi) \in \mathR$. By taking $J(\xi)$ as a Hamiltonian 
function, one defines a Hamiltonian vector field $X_{J(\xi)}$. But since $G$ 
acts on $M$, each $\xi \in \mg$ defines a vector field on $M$, viz. $\xi_M$. So 
we say that $\bf J$ is a {\em momentum map} for the action if for each $\xi \in 
\mg$, these two vector fields are identical: $X_{J(\xi)} = \xi_M$.

Three further remarks by way of illustrating this definition:---\\
\indent (1): {\em An isomorphism}:--- One readily checks that eq. 
\ref{definemm} defines an isomorphism between the space of smooth maps $\bf J$ 
from $M$ to $\mg^*$, and the space of linear maps $J$ from $\mg$ to scalar 
functions ${\cal F}(M)$. We can take $J$ to define $\bf J$ by saying that at 
each $x \in M$, ${\bf J}(x): \xi \in \mg \mapsto {\bf J}(x)(\xi) \in \mathR$ is 
to be given by the composed map
\be
\mg \stackrel{J}{\longrightarrow} {\cal F}(M) 
\stackrel{\mid_x}{\longrightarrow} \mathR \; ,
\ee
where $\mid_x$ means evaluation at $x \in M$. Or we can take $\bf J$ to define 
$J$ by saying that at each $\xi \in \mg$, $J(\xi): x \in M \mapsto J(\xi)(x) 
\in \mathR$ is to be given by the composed map
\be
M \stackrel{{\bf J}}{\longrightarrow} \mg^* 
\stackrel{\mid_{\xi}}{\longrightarrow} \mathR \; ,
\ee
where $\mid_{\xi}$ means evaluation at $\xi  \in \mg$.

(2): {\em Differential equations for the momentum map}:--- Using Hamilton's 
equations, we can readily express the definition of momentum map as a set of  
differential equations. Recall that on a Poisson manifold, Hamilton's equations 
are determined by eq. \ref{requireB}, which was that at each $x \in M$
\be
\mathsf{B}_x(dH(x)) = X_H(x) \;\; ;
\label{requireBrepeat?}
\ee
or in local coordinates $x^i, i = 1,\dots, m \equiv {\rm{dim}}(M)$, with 
$J^{ij}(x) \equiv \{x^i, x^j \}$ the structure matrix,
\be
\mathsf{B}_x(\frac{\pl H}{\pl x^j} d x^j) = \Sigma_{i,j}  J^{ij}(x) 
\frac{\pl H}{\pl x^j} \frac{\pl }{\pl x^i}\mid_x \; ;
\label{BessentialJ;repeat}
\ee
(cf. eq. \ref{BessentialJ}). So in local coordinates, Hamilton's equations are 
given by eq. \ref{HamMatrix}, which was:
\be
\frac{dx^i}{dt} = \Sigma^m_j \; J^{ij}(x) \frac{\pl H}{\pl x^j} \; .
\label{HamMatrix;repeat}
\ee 
So the condition for a momentum map $X_{J(\xi)} = \xi_M$ is that for all $\xi 
\in \mg$ and all $x \in M$
\be
\mathsf{B}_x(d(J(\xi))(x)) = \xi_M(x) \;\; .
\label{requireBmmap}
\ee
In coordinates, this is the requirement that for all $i = 1,\dots, m$
\be
\Sigma^m_j \; J^{ij}(x) \frac{\pl J(\xi)}{\pl x^j} \; = \; (\xi_M)^i(x) \; ,
\label{requireBmmapMatrix}
\ee
where---apologies!---the two $J$s on the left hand side  have very different 
meanings. 

In the symplectic case, ${\rm{dim}}(M) \equiv m = 2n$ and we have Hamilton's 
equations as eq. \ref{HEgeomic4}, viz. 
\be
{\bf i}_{X_H} \o :=  \o(X_H, \cdot) = dH(\cdot) \;\; .
\label{HEgeomic4;repeat}
\ee 
So the condition  for a momentum  map is that for all $\xi$
\be
 \o(\xi_M, \cdot) = d(J(\xi))(\cdot) \;\; .
\label{requireommap}
\ee 
In Hamiltonian mechanics, it is common to write the $2n$ local coordinates 
$q,p$ as $\xi$, i.e. to write
\be
\xi^{\al} := q^{\al}, \;\; {\al} = 1,...,n \;\;\; ; \;\;\; \xi^{\al} := 
p_{{\al}- n}, \;\; {\al} = n+1,...,2n \;\; .
\label{definexisupal} 
\ee
So in order to express eq. \ref{requireommap} in local coordinates, let us 
temporarily write $\eta$ for the arbitrary element of $\mg$. Then writing 
$\eta_M = (\eta_M)^{\al}\frac{\pl}{\pl \xi^{\al}}$ and $\o_{\al \bb} := 
\o(\frac{\pl}{\pl \xi^{\al}}, \frac{\pl}{\pl \xi^{\bb}})$, eq. 
\ref{requireommap} becomes
\be
\o_{\al \bb}(\eta_M)^{\al} = \frac{\pl J(\eta)}{\pl \xi^{\bb}} \;\; .
\label{requireommapinlocalxicoords}
\ee

(3): {\em Components: an example}:--- As discussed after eq. 
\ref{defineinfPoissautom}, we think of the collection of functions $J(\xi)$, as 
$\xi$ varies through $\mg$, as the {\em components} of $\bf J$.\\
\indent To take our standard example: the angular momentum of a particle in 
Euclidean space, in a state $x = ({\bf q},{\bf p})$ is ${\bf J}(x) := {\bf q} 
\wedge {\bf p}$. Identifying $\mso(3)^*$ with $\mathR^3$ so that the natural 
pairing is given by the dot product (cf. (3) at the end of Section 
\ref{coadj;was6.5.5.B}), we get that the component of ${\bf J}(x)$ around the 
axis $\xi \in \mathR^3$ is $< \; {\bf J}(x) \; ; \; \xi \; > = \xi \cdot ({\bf 
q} \wedge {\bf p})$. The Hamiltonian vector field determined by this 
Hamiltonian function  $x = ({\bf q},{\bf p}) \mapsto \xi \cdot ({\bf q} \wedge 
{\bf p})$ is of course the infinitesimal generator of rotations about the 
$\xi$-axis. In Section \ref{mommmapsnoetexample}, we will see more examples of 
momentum maps.

\subsection{Conservation of momentum maps: Noether's theorem 
}\label{mommmapsnoet}
In ordinary Hamiltonian mechanics, we saw that Noether's theorem had a simple  
expression as a ``one-liner'' based on the antisymmetry of the Poisson bracket: 
namely, in eq. \ref{naivenoetham}, which was that for any scalar functions $F, 
H$ 
\be
X_F(H) = \{H, F \} = 0 \;\;\;\; {\mbox{ iff }} \;\; \;\;
0 = \{F, H \} = X_H(F)  \; \; .
\label{naivenoetham;repeat} 
\ee
In words: the Hamiltonian $H$ is constant under the flow induced by $F$  iff 
$F$ is a constant of the motion under the dynamical flow $X_H$.\\
\indent More precisely, Section  \ref{noetcomplete} vindicated this one-liner 
as expressing Noether's theorem. For the one-liner respected the requirement 
that a symmetry should preserve the symplectic form (equivalently, the Poisson 
bracket), and not just (as in the left hand side of eq. 
\ref{naivenoetham;repeat}) the Hamiltonian function $H$; for, by Cartan's magic 
formula, a vector field's preserving the symplectic form was equivalent to its 
being locally Hamiltonian.

For Poisson manifolds, the equivalence corresponding to this last statement 
fails. That is, as we noted in (2) of Section \ref{canlactioninflgenors}: an 
infinitesimal Poisson automorphism need not be locally Hamiltonian.\\
\indent Nevertheless, most of the ``one-liner'' approach to Noether's theorem 
carries over to the framework of Poisson manifolds. In effect, we just restrict 
discussion to cases where the relevant Hamiltonian vector fields exist: recall 
our saying after (2) of Section \ref{canlactioninflgenors} that we would 
concentrate on cases where all the $\xi_M$ are globally Hamiltonian.

 Thus, it is straightforward to show that for a Poisson manifold $M$, just as 
for symplectic manifolds: if $F, H \in {\cal F}(M)$, $H$ is constant along the 
integral curves of $X_F$ iff $\{H, F \} = 0$ iff $F$ is constant along the 
integral curves of $X_H$. (We could have proved this already in Section 
\ref{hvfs}; but postponed it till now, when it will be used.)   

With this result as a lemma, one immediately gets
\begin{quote}
{\bf Noether's theorem for Poisson manifolds} Suppose that $G$ acts canonically 
on a Poisson manifold $M$ and has a momentum map ${\bf J}: M \rightarrow 
\mg^*$; and that  $H$ is invariant under $\xi_M$ for all $\xi \in \mg$, i.e. 
$\{H, J(\xi) \} = \xi_M (H) = 0, \; \forall \xi \in \mg$;
(cf. eq. \ref{requirexiMgbllyhamn;repeat;PBversion}). Then $\bf J$ is a 
constant of the motion determined by $H$. That is:
\be
{\bf J} \circ \phi_{\tau} = {\bf J} 
\ee 
where $\phi_{\tau}$ is the flow of $X_H$.
\end{quote}
{\em Proof}: By the lemma, the fact that $\{H, J(\xi) \} = \xi_M (H) = 0$ 
implies that $J(\xi)$ is constant along the flow of $X_H$. So by the definition 
of momentum map, eq. \ref{definemm}, the corresponding $\mg^*$-valued map $\bf 
J$ is also a constant of the motion. QED.

It follows immediately that $H$ itself, and any distinguished function, is a 
constant of the motion. Besides, as remarked in (2) at the end of Section 
\ref{canlactioninflgenors}: a constant of the motion $J(\xi)$ is determined 
only up to an arbitrary choice of a distinguished function. Indeed, though this 
Chapter has set aside (ever since (iii) of Section \ref{introprospectus}) 
time-dependent functions: if one considers them, then there is here an 
arbitrary choice of a time-dependent distinguished function.

\subsection{Examples}\label{mommmapsnoetexample}
We give two familiar examples; and then, as a glimpse of the general power of 
the theory, two abstract examples (which will {\em not} be needed later on).

(1): {\em Total linear momentum of $N$ particles} :---\\
In (3) at the end of Section \ref{basicaction}.A, we showed that the {\em left} 
cotangent lift of the action of the translation group $\mathR^3$ on $Q = 
\mathR^{3N}$ to $M = T^*\mathR^{3N}$, i.e. the left action corresponding to eq. 
\ref{defineleft}, is
\be
\Psi_{\bf x}({\bf q}_i,{\bf p}^i) := T^*(\Phi_{- {\bf x}})({\bf q}_i,{\bf p}^i) 
= ({\bf q}_i + {\bf x},{\bf p}^i) \; , \;\; i = 1,..., N \; .
\label{leftcotgtliftlinmomm3N}
\ee
(Here we combine the discussions of examples (vi) and (ix) in Section 
\ref{basicaction}.A.)\\
\indent To find the momentum  map, we: (a) compute the infinitesimal generator 
$\xi_M$ for an arbitrary element $\xi$ of $\mg = \mathR^3$; and then (b) solve 
eq. \ref{requireommap}, or in coordinates eq. 
\ref{requireommapinlocalxicoords}.\\
\indent (a): We differentiate eq. \ref{leftcotgtliftlinmomm3N} with respect to 
${\bf x}$ in the direction $\xi$, getting
\be
\xi_M ({\bf q}_i,{\bf p}^i) = (\xi,..., \xi, {\bf 0},...,{\bf 0}) \; .
\ee 
\indent (b): Any function $J(\xi)$ has Hamiltonian vector field 
\be
X_{J(\xi)} ({\bf q}_i,{\bf p}^i) = \left(\frac{\pl J(\xi)}{\pl {\bf p}^i}, - 
\frac{\pl J(\xi)}{\pl {\bf q}_i} \right) \; ;
\ee
so that the desired $J(\xi)$ with $X_{J(\xi)} = \xi_M$ solves
\be
\frac{\pl J(\xi)}{\pl {\bf p}^i} = \xi \;\;{\rm{\; and \;}} \;\; 
\frac{\pl J(\xi)}{\pl {\bf q}_i} = {\bf 0} \; , \; 1 \leq i \leq N \; .
\ee
Choosing constants so that $J$ is linear, the solution is 
\be
J(\xi) ({\bf q}_i,{\bf p}^i) = \left( \Sigma^N_{i = 1} {\bf p}^i \right) \cdot 
\xi \; , \;\;{\rm{\; i.e. \;}} \;\;
{\bf J}({\bf q}_i,{\bf p}^i) = \Sigma^N_{i = 1} {\bf p}^i \; ;
\label{linmommasmmap;fromdiffeqn}
\ee
i.e. the familiar total linear momentum.

(2): {\em Angular momentum of a single particle} :---\\
$SO(3)$ acts on $Q = \mathR^3$ by $\Phi_{{A}}({\bf q}) = A{\bf q}$. So the 
tangent (derivative) map is
\be
T_{\bf q}\Phi_{{A}}: ({\bf q},{\bf v}) \in T \mathR^3_{\bf q} \mapsto
({A}{\bf q},{A}{\bf v}) \in T \mathR^3_{{A}{\bf q}} \; .
\ee
As we saw in example (vii) of Section \ref{basicaction}.A, the {\em left} 
cotangent lift of the action to $M = T^*\mathR^3$ (the lifted action ``with 
$g^{-1}$'', corresponding to eq. \ref{defineleft}) is: 
\be
T^*_{A{\bf q}}(\Phi_{A^{-1}})({\bf q},{\bf p}) = (A{\bf q},A{\bf p}) \;.
\label{leftcotgtliftangmommsingle}
\ee
To find the momentum map, we proceed in two stages, (a) and (b), as in example 
(1).\\
\indent (a): We differentiate eq. \ref{leftcotgtliftangmommsingle} with respect 
to $A$ in the direction $\xi = \Theta(\o) \in \mso(3)$, where $\o \in \mathR^3$ 
and $\Theta$ is as in eq. \ref{3dcorrspce} and \ref{theta3dcorrspce}. We get
\be
\xi_M({\bf q},{\bf p}) = (\xi {\bf q}, \xi{\bf p}) = (\o \wedge {\bf q}, \o 
\wedge {\bf v}) \; .
\ee
\indent (b): So the desired $J(\xi)$ is the solution linear in $\xi$ to the 
Hamilton's equations
\be
\frac{\pl J(\xi)}{\pl {\bf p}} = \xi {\bf q} \;\;{\rm{\; and \;}} \;\; 
\frac{\pl J(\xi)}{\pl {\bf q}} = - \xi {\bf p} \; .
\ee
So a solution is given by
\be
J(\xi)({\bf q},{\bf p}) = (\xi{\bf q}) \cdot {\bf p} = (\o \wedge {\bf q}) 
\cdot {\bf p} = ({\bf q} \wedge {\bf p}) \cdot \o \; ,
\ee
so that 
\be
{\bf J}({\bf q},{\bf p}) = {\bf q} \wedge {\bf p} \; ,
\label{angularmommasmmap;fromdiffeqn}
\ee
i.e. the familiar angular momentum.

(3): {\em Dual of a Lie algebra homomorphism} :---\\
We begin by stating a {\em Lemma}, which we will not prove; for details cf. 
Marsden and Ratiu (1999: 10.7.2, p. 372). Namely: let $G,H$ be Lie groups and 
let $\al: \mg \rightarrow \mh$ be a linear map between their Lie algebras. Then 
$\al$ is a Lie algebra homomorphism iff its dual $\al^*:\mh^* \rightarrow 
\mg^*$ is a (linear) Poisson map (where $\mh^*, \mg^*$ are equipped with their 
natural Lie-Poisson brackets as in Section \ref{LPso(3)}).\\
\indent Now let $G, H$ be Lie groups, let $A:H \rightarrow G$ be a Lie group 
homomorphism, and let $\al:\mh \rightarrow \mg$ be the induced Lie algebra 
homomorphism; so that by the Lemma, $\al^*: \mg^* \rightarrow \mh^*$ is a 
Poisson map. We will prove that $\al^*$ is also a momentum map for the action 
of $H$ on $\mg^*$ given by, with $h \in H, x \in \mg^*$: 
\be
\Phi(h,x) \equiv h \cdot x := Ad^*_{A(h)^{-1}} x \; .
\label{Honmg^*}
\ee 
\indent {\em Proof}: We first recall the adjoint and co-adjoint actions $Ad_g : 
\mg \rightarrow \mg$ and $Ad^*_g : \mg^* \rightarrow \mg^*$; in particular, eq. 
\ref{defineAd^*}. So the action in eq. \ref{Honmg^*} is:
\be
\forall x \in \mg^*, \forall \xi \in \mg \; : \;\;
< h \cdot x ; \xi > \; = \; < x; Ad_{A(h)^{-1}} \xi > \; .
\label{Honmg^*2}
\ee
As usual, we compute for $\eta \in \mh$, the infinitesimal generator 
$\eta_{\mg^*}$ at $x \in \mg^*$ by differentiating eq. \ref{Honmg^*2} with 
respect to $h$ at $e$ in the direction $\eta \in \mh$. We get (cf. eq. 
\ref{computead*}):
\be
< \eta_{\mg^*}(x) ; \xi > \; = \; - < x ; ad_{\al(\eta)} \xi > \; = \; 
<ad^*_{\al(\eta)}(x) ; \xi > \; .
\label{computeetageestar}
\ee
We define ${\bf J}(x) := \al^*(x)$: that is,
\be
J(\eta)(x) \; \equiv \; < {\bf J}(x) ; \eta > \; := \; < \al^*(x) ; \eta > \; 
\equiv \; < x ; \al(\eta) > \; ;
\ee
which implies 
\be
\nabla_x J(\eta) \; = \; \al (\eta) \; .
\ee
Now we recall that Hamilton's equations for $J(\eta)$ as the Hamiltonian are 
(cf. eq. \ref{HamEqWithad^*})
\be
{\dot x} \equiv X_{J(\eta)}(x) = ad^*_{\nabla_x J(\eta)} (x) \; .
\label{HamEqWithad^*forJ(eta)}
\ee
Combining eq. \ref{computeetageestar} to eq. \ref{HamEqWithad^*forJ(eta)}, we 
get: 
\be
X_{J(\eta)}(x) = ad^*_{\al(\eta)} (x) = \eta_{\mg^*} (x) \; ;
\ee
proving that ${\bf J}(x) := \al^*(x)$ is a momentum map. QED.

(4): {\em Momentum maps for subgroups} :---\\
Assume that ${\bf J}: M \rightarrow \mg^*$ is a momentum map for a canonical 
left action of $G$ on $M$; and let $H < G$ be a subgroup of $G$. Then $H$ also 
acts canonically on $M$, and this action has as a momentum map the restriction 
of ${\bf J}$'s values to $\mh \subset \mg$. That is: the map 
\be
{\bf J}_H: M \rightarrow \mh^* \;\; {\rm{\; given \; by \;}} \;\;
{\bf J}_H (x) := {\bf J}(x)\mid_{\mh} \; .
\ee
For the canonical action of $G$ ensures that if $\eta \in \mh \subset \mg$, 
then $\eta_M = X_{J(\eta)}$. Then $J_H(\eta) := J(\eta) \forall \eta \in \mh$ 
defines a momentum map for $H$'s action. That is
\be
\forall x \in M, \; \forall \eta \in \mh : \;\;\; < {\bf J}_H (x) ; \eta > \; = 
\; < {\bf J} (x) ; \eta > \; .
\ee

\subsection{Equivariance of momentum maps}\label{eqvarcemm}
In (1) of Section \ref{quotstruc}, we defined the general notion of an 
equivariant map $f:M \rightarrow N$ between manifolds  as one that respects the 
actions of a group $G$ on $M$ and on $N$: eq. \ref{equivarcefinite}. We now 
develop an especially important case of this notion: the equivariance of 
momentum maps ${\bf J}: M \rightarrow \mg^*$, where the action on $\mg^*$ is 
the co-adjoint action, eq. \ref{defineAd^*_g-1}.

\indent For us, this notion will have two main significances:--- \\
\indent (i): many momentum maps that occur in examples are equivariant in this 
sense; \\
\indent (ii): equivariance has various theoretical consequences: in particular, 
momentum maps for cotangent lifted actions are always equivariant (Section 
\ref{mommmapcotgt}), and equivariance is crucial in theorems about reduction 
(Section \ref{redn}).

In this Section, we will glimpse these points by:\\
\indent (i): defining the notion, and remarking on a weakened differential 
version of the notion (Section \ref{earceinflearce});\\
\indent (ii): proving that equivariant momentum maps are Poisson (Section 
\ref{eqvartmmsarePoisson}).

\subsubsection{Equivariance and infinitesimal 
equivariance}\label{earceinflearce}
Let $\Phi$ be a canonical left action of $G$ on $M$, and let ${\bf J}: M 
\rightarrow \mg^*$ be a momentum map for it. We say $\bf J$ is {\em 
equivariant} if for all $g \in G$
\be
{\bf J} \circ \Phi_g \; = \; Ad^*_{g^{-1}} \circ {\bf J}  \; ;
\label{eqvarce2nddefn}
\ee
cf. eq. \ref{equivarcefinite} and the definition of co-adjoint action, eq. 
\ref{defineAd^*_g-1;repn}:
 \begin{equation}
    \bundle{M}{\Phi_g}{M}
    \bundlemap{{\bf J}}{{\bf J}}
    \bundle{\mg^*}{Ad^*_{g^{-1}}}{\mg^*}    
	\label{eqvardiag}
\end{equation}
 
An equivalent formulation arises by considering that we can add to the 
commutative square in eq. \ref{eqvardiag} the two commutative triangles:
\be
M \stackrel{J(\xi)}{\longrightarrow} \mathR \;\; {\rm{\; is \;}} \;\;
M \stackrel{{\bf J}}{\longrightarrow} \mg^* 
\stackrel{\mid_{\xi}}{\longrightarrow} \mathR \; ;
\label{firsttriangle}
\ee
representing the fact that $J(\xi)(x) = {\bf J}(x)(\xi)$;
and
\be
\mg^* \stackrel{\mid_{\xi}}{\longrightarrow} \mathR \;\; {\rm{\; is \;}} \;\;
\mg^* \stackrel{Ad^*_{g^{-1}}}{\longrightarrow} \mg^* 
\stackrel{\mid_{Ad_{g}\xi}}{\longrightarrow} \mathR \; ;
\label{secondtriangle}
\ee
representing the fact that for all $\eta \in g^*$
\be
< Ad^*_{g^{-1}}(\eta) ;  Ad_g(\xi) > \;  = \;  < \eta ; Ad_{g^{-1}}Ad_g(\xi) > 
\; \equiv \; < \eta ; \xi > \; .
\ee
Eq.s \ref{firsttriangle} and \ref{secondtriangle} imply that an equivalent 
formulation of equivariance is that for all $x \in M, g \in G$ and $\xi \in 
\mg$ (and with $g \cdot x \equiv \Phi_g(x)$)
\be
{\bf J} (g \cdot x)(Ad_g \xi) \equiv J(Ad_g \xi)(g \cdot x) = J(\xi)(x) \equiv 
{\bf J}(x)(\xi) \; .
\label{eqvarcereformltn}
\ee

In (2) of Section \ref{inflgenor}, we differentiated the general notion of an  
equivariant map, and got the weaker differential notion that the infinitesimal 
generators $\xi_M$ and $\xi_N$ of the actions of $G$ on $M$ and on $N$ are 
$f$-related.\\
\indent Here also we can differentiate equivariance, and get the notion of {\em 
infinitesimal equivariance}. But I will not go into details since:\\
\indent (i): we will not need the notion, not least because (as mentioned 
above), many momentum maps are equivariant;\\
\indent (i): under certain common conditions (e.g. the group $G$ is compact, or 
is connected) an infinitesimally equivariant momentum map can always be 
replaced by an equivariant one.

So let it suffice to say that infinitesimal equivariance {\em is} theoretically 
important. In particular, the result eq. \ref{mmmeshPBstruc}, viz.
\be
X_{J([\xi, \eta])} = X_{\{J(\xi), J(\eta) \}_M} \; 
\label{mmmeshPBstrucrepeat}
\ee
implies that
\be
\Sigma (\xi, \eta) :=  J([\xi, \eta]) - \{J(\xi), J(\eta) \}_M 
\label{definesigmadisting}
\ee
is a distinguished function on the Poisson manifold  $M$, and so constant on 
every symplectic leaf.\\
\indent This makes it natural to ask when $\Sigma \equiv 0$. After all, cf. eq. 
\ref{inflgenorfromtriangle}. Both  $\xi \mapsto \xi_M$ and $F \mapsto X_F$ are 
Lie algebra anti-homomorphisms. So it is natural to ask whether $J$ is a Lie 
algebra homomorphism, i.e. whether $\Sigma = 0$.  And it turns out that 
infinitesimal equivariance is equivalent to $\Sigma = 0$.

\subsubsection{Equivariant momentum maps are 
Poisson}\label{eqvartmmsarePoisson}
The following result is important, both as a general method of finding 
canonical  maps between Poisson manifolds, and for the Lie-Poisson reduction 
theorem of Section \ref{redn}.
\begin{quote}
{\bf{Equivariant momentum maps are Poisson}}
Let $\bf J: M \rightarrow \mg^*$ be an equivariant momentum map for a canonical 
left action of $G$ on a Poisson manifold $M$. Then $\bf J$ is a Poisson map: 
for all $F_1, F_2 \in {\cal F}(\mg^*)$,
\be
{\bf J}^* \{F_1, F_2\}_{\mg^*} = \{{\bf J}^*  F_1, {\bf J}^*  F_2 \}_M
\;\; {\rm{; \; i.e. \;}} \;\; 
\{F_1, F_2\}_{\mg^*} \circ {\bf J} = \{  F_1 \circ {\bf J},  F_2 \circ {\bf J} 
\}_M \; .
\label{eqvartispoisson}  
\ee  
\end{quote}

{\em Proof}:--- We will relate (i) the left hand side, then (ii) the right hand 
side of eq. \ref{eqvartispoisson} to $J$; and finally we will use the fact that 
the Poisson bracket on $M$ depends only on the values of the first 
derivatives.\\
\indent (i): Let $x \in M, \al = {\bf J}(x) \in \mg^*$; and let $\xi = \nabla 
F_1$ and $\eta = \nabla F_2$ evaluated at $\al$, so that $\xi, \eta \in 
\mg^{**} = \mg$. Then
\be
\{F_1, F_2\}_{\mg^*} ({\bf J}(x)) \equiv < \al ; [\nabla F_1, \nabla F_2] > = 
< \al ; [\xi, \eta] > = J([\xi, \eta])(x) = \{ J(\xi), J(\eta) \} (x) \; ;
\ee
where the third equation just applies the definition of $\bf J$, eq. 
\ref{definemm}, and the fourth equation uses (infinitesimal) equivariance. \\
\indent (ii): We show that $(F_1 \circ {\bf J})(x)$ and $J(\xi)(x)$ have equal 
$x$-derivatives. For any $x \in M$ and $v_x \in T_xM$
\be
{\bf d}(F_1 \circ {\bf J})(x) \cdot v_x = {\bf d}F_1 (\al) \cdot T_x {\bf 
J}(v_x) = < T_x {\bf J}(v_x) ; \nabla F_1 > = {\bf d} J(\xi)(x) \cdot v_x \; ;
\label{rerhs}
\ee 
where the first equation uses the chain rule, and the last uses the definition 
of $\bf J$, eq. \ref{definemm} and the fact that $\xi = \nabla F_1$.\\
\indent Finally, since the Poisson bracket on $M$ depends only on the values of 
the first derivatives, we infer from eq. \ref{rerhs} that
\be
\{  F_1 \circ {\bf J},  F_2 \circ {\bf J} \} (x) =  \{ J(\xi), J(\eta) \} (x) 
\; .
\ee
Combining this with (i), the result follows. QED.

\subsection{Momentum maps on cotangent bundles}\label{mommmapcotgt}
Let a Lie group $G$ act on a manifold (``configuration space'') $Q$. We saw in 
Section \ref{basicaction}.A that this action can be {\em lifted} to the 
cotangent bundle $T^*Q$; cf. eq.s \ref{defineT^*f}, \ref{defineright} and 
\ref{defineleft}. In this Section, we focus on momentum maps for such {\em 
cotangent lift} actions. We shall see that any such action has an equivariant 
momentum map, for which there is an explicit general formula. The general 
theory (Sections \ref{mommfns}, \ref{mmsforcotgtlifts}) will need just one main 
new notion, the {\em momentum function}. We end with some examples (Section 
\ref{Egscotgtliftedmms}).

\subsubsection{Momentum functions}\label{mommfns}
Given a manifold $Q$ and its vector fields ${\cal X}(Q)$, we define the map
\be
{\cal P}: {\cal X}(Q) \rightarrow {\cal F}(T^*Q) \;\; {\rm{\; by: \;}} \;\;
({\cal P}(X))(\al_q) \; := \; < \al_q ; X(q) > \; 
\label{definemommfn}
\ee
for $q \in Q, X \in {\cal X}(Q)$ and $\al_q \in T^*_q Q$. Here, $\al_q$ is, 
strictly speaking, a point in the cotangent bundle above the base-point $q \in 
Q$: so $\al_q$ can be written as $(q, \al)$ with $\al$ a covector at $q$, i.e. 
$\al \in T^*_q Q$. But as we mentioned just before defining cotangent lifts 
(eq. \ref{defineT^*f}): it is harmless to (follow many presentations and) 
conflate a point in $T^*Q$, i.e.  a pair $(q, \al), q \in Q, \al \in T^*_{q}Q$, 
with its form $\al$, provided we keep track of the $q$ by writing the form as 
$\al_q$.

 ${\cal P}(X)$, as defined by eq. \ref{definemommfn}, is called the {\em 
momentum function} of $X$. In coordinates,  ${\cal P}(X)$ is given by
\be
{\cal P}(X)(q^i,p_i) = X^j(q^i)p_j
\label{Pdefinedincoords}
\ee
where we sum on $j = 1,..., n := {\rm{dim}} \; Q$. (So NB: This ${\cal P}$ is 
different from that in ${\cal P}(M)$, the infinitesimal Poisson automorphisms 
of $M$, discussed in Section \ref{canlactioninflgenors}.)

We also denote by ${\cal L}(T^*Q)$ the space of smooth functions $F:T^*Q 
\rightarrow \mathR$ that are linear on fibres of $T^*Q$: i.e. writing the 
bundle points $\al_q, \bb_q \in T^*_q Q$ as $(q, \al)$ and $(q, \bb)$, we have 
for $\lambda, \mu \in \mathR$
\be
F(q, (\lambda \al + \mu \bb )) = \lambda F((q, \al )) +  \mu F((q, \bb )) \; .
\label{definelinonfibres}
\ee  
So functions $F, H$ that are in ${\cal L}(T^*Q)$ can be written in coordinates 
as (summing on $i = 1,...,n$)
\be
F(q,p) = X^i(q)p_i \;\; {\rm{\; and \;}} \;\; H(q,p) = Y^i(q)p_i
\label{coordexpressnlinfibers}
\ee
for functions $X^i$ and $Y^i$; and so any momentum function ${\cal P}(X)$ is in 
${\cal L}(T^*Q)$.

One readily checks that the standard Poisson bracket (from $T^*Q$'s symplectic 
structure, Section \ref{cotgtblesymp}) of such an $F$ and $H$ is also linear on 
the fibres of $T^*Q$. In fact, eq. \ref{coordexpressnlinfibers} implies
\be
\{F, H \}(q,p) := \frac{\pl F}{\pl q^j}\frac{\pl H}{\pl p_j} - \frac{\pl H}{\pl 
q^j}\frac{\pl F}{\pl p_j} = \left(\frac{\pl X^i}{\pl q^j} Y^j - \frac{\pl 
Y^i}{\pl q^j}X^j \right) \; .
\label{PBalsolinfibres} 
\ee 
So ${\cal L}(T^*Q)$ is a Lie subalgebra of ${\cal F}(T^*Q)$.

The next result summarizes how momentum functions relate ${\cal X}(Q)$ and 
Hamiltonian vector fields on $T^*Q$ to ${\cal L}(T^*Q)$.
\begin{quote}
{\bf {Three (anti)-isomorphic Lie algebras}} The two Lie algebras \\
 (i) $({\cal X}(Q), [, ] )$ of vector fields on $Q$;\\
 (ii) Hamiltonian vector fields $X_F$ on $T^*Q$ with $F \in {\cal L}(T^*Q)$\\
 are isomorphic. And each is anti-isomorphic to \\
  (iii) (${\cal L}(T^*Q),  \{, \}$). \\
  In particular, the map $\cal P$ is an anti-isomorphism from (i) to (iii), so 
that we have
\be
\{{\cal P}(X), {\cal P}(Y) \}_{T^*Q} = - {\cal P}([X, Y]) \; .
\label{calPantiisom}
\ee 
\end{quote}  
{\em Proof}: Since ${\cal P}(X): T^*Q \rightarrow \mathR$ is linear  on fibres, 
$\cal P$ maps ${\cal X}(Q)$ into ${\cal L}(T^*Q)$. $\cal P$ is also onto ${\cal 
L}(T^*Q)$: given $F \in {\cal L}(T^*Q)$, we can define $X(F) \in {\cal X}(Q)$ 
by
\be
< \al_q ; X(F)(q) > \; := \; F(\al_q) \;\;\;  \forall \al_q \in T^*_qQ
\ee
so that ${\cal P}(X(F)) = F$. $\cal P$ is linear and ${\cal P}(X) = 0$ implies 
that $X = 0$. Also, eq. \ref{calPantiisom} follows immediately by comparing eq. 
\ref{PBalsolinfibres} with the Lie bracket of $X, Y \in {\cal X}(Q)$; cf. eq. 
\ref{Liebr1storder}. So $\cal P$ is an anti-isomorphism from $({\cal X}Q, [, ] 
)$ to (${\cal L}(T^*Q),  \{, \}$).\\
\indent The map
\be
F \in ({\cal L}(T^*Q),  \{, \}) \mapsto X_F \in \left( \{ X_F \mid F \in {\cal 
L}(T^*Q) \}, [, ] \right)
\ee   
is surjective by definition. It is a Lie algebra anti-homomorphism, by eq. 
\ref{HamvfsclosedunderLie;symp} (i.e. result (2) in Section \ref{lavecfields}). 
And if $X_F = 0$, then $F$ is constant on $T^*Q$; and hence $F \equiv 0$ since 
$F$ is linear on the fibres (cf. eq. \ref{definelinonfibres}). QED.

\subsubsection{Momentum maps for cotangent lifted 
actions}\label{mmsforcotgtlifts}
We begin this Subsection with a result relating the Hamiltonian flow on $T^*Q$ 
induced by the momentum function ${\cal P}(X)$ to the Hamiltonian flow on $X$ 
induced by $X$. From this result, our main result---the guarantee of an 
equivariant  momentum map for a cotangent lifted action, and an explicit 
formula for it---will follow directly.

\begin{quote}
{\bf {The Hamiltonian flow of a momentum function}}
Let $X \in {\cal X}(Q)$ have flow $\phi_{\tau}$ on $Q$; cf. Section 
\ref{fieldtoflow}.B. Then the flow of $X_{{\cal P}(X)}$ on $T^*Q$ is $T^* 
\phi_{- \tau}$. That is: the flow of $X_{{\cal P}(X)}$ is the cotangent lift 
(Section \ref{basicaction}.A) of $\phi_{- \tau}$, as given by the diagram, with 
$\pi_Q$ the canonical projection:
 \begin{equation}
    \bundle{T^*Q}{\pi_Q}{Q}
    \bundlemap{T^* \phi_{- \tau}}{\phi_{\tau}}
    \bundle{T^*Q}{\pi_Q}{Q}    
	\label{hamflowmomfndiag}
\end{equation}
\end{quote}
{\em Proof}: We differentiate the relation in eq. \ref{hamflowmomfndiag}, i.e.
\be
\pi_Q \circ T^* \phi_{- \tau} = \phi_{\tau} \circ \pi_Q
\ee
at $\tau = 0$ to get
\be
T \pi_Q \circ Y = X \circ \pi_Q \;\; {\rm{ \; with \;}} \;\;
\forall \al_q \in T^*_q Q, \; Y(\al_q) = \frac{d}{d \tau} \mid_{\tau = 0} T^* 
\phi_{- \tau} (\al_q) \; ;
\label{defineYforflow} 
\ee
i.e. $T^* \phi_{- \tau}$ is the flow of $Y$.\\
\indent Now we will show that $Y = X_{{\cal P}(X)}$, using eq. 
\ref{defineYforflow} and the geometrical formulation of Hamiltonian mechanics 
of Section \ref{july05}, especially Cartan's magic formula, eq. \ref{magic}, 
applied to the canonical one-form $\theta \equiv \theta_H$ (defined by eq. 
\ref{derivprojectionpi} and \ref{definethetaH}).\\
\indent We reported (at the start of (2) of Section \ref{basicaction}.A) that 
the cotangent lift $T^* \phi_{- \tau}$ preserves $\theta \equiv \theta_H$ on 
$T^*Q$. So ${\cal L}_Y \theta = 0$. Then the definition of $\o$ as the negative 
exterior derivative of $\theta$, and Cartan's magic formula, eq. \ref{magic}, 
yields
\be
{\bf i}_Y \o = - {\bf i}_Y {\bf d} \theta = {\bf d}{\bf i}_Y \theta \; .
\label{magicapplied}
\ee
On the other hand, we also have
\be
{\bf i}_Y \theta (\al_q) \equiv < \theta (\al_q) ; Y(\al_q) > = 
< \al_q ; T \pi_Q (Y(\al_q)) > = < \al_q ; X(q) > = {\cal P}(X)(\al_q)
\label{lovely}
\ee
where the second equation applies the definition of the canonical one-form (eq. 
\ref{derivprojectionpi}), the third applies eq. \ref{defineYforflow}, and the 
fourth applies the definition eq. \ref{definemommfn} of momentum functions. \\
\indent Combining eq. \ref{magicapplied} and \ref{lovely}, we have:
\be
{\bf i}_Y \o = {\bf d}{\cal P}(X)
\ee
which is Hamilton's equations (eq. \ref{HEgeomic4}) telling us that $Y = 
X_{{\cal P}(X)}$. QED.

Accordingly the Hamiltonian vector field $X_{{\cal P}(X)}$ on $T^*Q$ is called 
the {\em cotangent lift} of $X \in {\cal Q}$ to $T^*Q$. In local coordinates, 
we can write, by combining eq. \ref{HamDelta0} and \ref{Pdefinedincoords}
\be
X_{{\cal P}(X)} = \frac{\pl {\cal P}(X)}{\pl p_i}\frac{\pl }{\pl q^i} - 
\frac{\pl {\cal P}(X)}{\pl q^i}\frac{\pl }{\pl p_i} =
X^i \frac{\pl }{\pl q^i} - \frac{\pl X^i}{\pl q^j} p_i \frac{\pl }{\pl p_j} \; 
.
\ee
Note in particular that, combining the usual sign-change between Lie algebras 
and Poisson brackets (eq. \ref{HamvfsclosedunderLie;symp}) with the sign-change 
for momentum functions (eq. \ref{calPantiisom}), we have
\be
[X_{{\cal P}(X)}, X_{{\cal P}(Y)}] = - X_{\{ {\cal P}(X), {\cal P}(Y) \}} =
- X_{- {\cal P}([X, Y])} = X_{{\cal P}([X, Y])} \; .
\ee

We can now readily prove our main result guaranteeing, and giving a formula 
for, equivariant momentum maps.
\begin{quote}
{\bf {Equivariant momentum maps}}
Let $G$ act on the left on $Q$ and so by cotangent lift on $T^*Q$. The 
cotangent lifted action has an equivariant momentum map ${\bf J}: T^*Q 
\rightarrow \mg^*$ given by 
\be
< {\bf J}(\al_q) ; \xi > \; = \; < \al_q ; \xi_Q(q) > \; \equiv \; {\cal 
P}(\xi_Q)(\al_q) \; .
\label{guarmmap}
\ee
In coordinates $q^i, p_i$ on $T^*Q$ and $\xi^a$ on $\mg$, and with $\xi^i_Q = 
\xi^a A^i_a$ the components of $\xi_Q$, this reads
\be
J_a \xi^a = p_i \xi^i_Q = p_i A^i_a \xi^a
\ee
so that $J_a(q,p) = p_i A^i_a(q)$.
\end{quote}
{\em Proof}: The preceding result tells us that for any $\xi \in \mg$, the 
infinitesimal generator of the cotangent lifted action on $T^*Q$ is $\xi_{T^*Q} 
\equiv X_{{\cal P}(\xi_Q)}$. So a momentum map for this action is given by 
\be
J(\xi) = {\cal P}(\xi_Q) \; .
\ee
This gives eq. \ref{guarmmap}, just by applying the definitions of the momentum 
map $\bf J$ (eq. \ref{definemm}) and of momentum function (eq. 
\ref{definemommfn}).\\
\indent To prove equivariance, we argue as follows:
\begin{eqnarray}
< {\bf J}(g \cdot \al_q) ; \xi > \; = \; < (g \cdot \al_q) ; \xi_Q(g \cdot q) > 
\\
= \; < \al_q ; (T \Phi_{g^{-1}}) \xi_Q (g \cdot q) >  \; \equiv \; 
< \al_q ; (T_{g \cdot q} \Phi_{g^{-1}} \circ \xi_Q \circ \Phi_g)(q) . \\
= \; < \al_q ; (\Phi^*_g \xi_Q)(q) > \\
= \; < \al_q ; (Ad_{g^{-1}} \xi)_Q (q) > \\
= \; < {\bf J}(\al_q) ; Ad_{g^{-1}} \xi > \; = \; <  Ad^*_{g^{-1}}({\bf 
J}(\al_q)) ; \xi > \; .
\end{eqnarray}
Here we have  applied in succession:
(i) eq. \ref{guarmmap}; (ii) the fact that $g \cdot \al_q$ is short for 
$T^*(\Phi_{g^{-1}})(\al_q)$, cf. eq. \ref{defineleft} and \ref{defineT^*f}; 
(iii) the definition of pullback, cf. eq. \ref{proveAdandPullbackifnlgenor}; 
(iv) result [2],  eq. \ref{AdandPullbackifnlgenor}, of Section 
\ref{adj;was6.5.5.A}; (v) eq. \ref{guarmmap} again; and finally, 
 (vi) the fact that $Ad^*$ is the adjoint of $Ad$, cf. eq. \ref{defineAd^*}. 
QED.

\subsubsection{Examples}\label{Egscotgtliftedmms}
We discuss first our familiar examples, linear and angular momentum i.e. (1) 
and (2) from Section \ref{mommmapsnoetexample}; and then the cotangent lift of 
left and right translations on $G$---an example motivated by Section 
\ref{kinicslgs}'s description of kinematics on a Lie group $G$. 

(1): {\em Total linear momentum of $N$ particles}:--- \\
Since the translation group $\mathR^3$ acts on $Q := \mathR^{3N}$ by $\Phi({\bf 
x}, ({\bf q}_i)) = ({\bf q}_i + {\bf x})$, the infinitesimal generator on $Q$ 
is
\be
\xi_{\mathR^{3N}}({\bf q}_i) = (\xi, \dots, \xi) (\xi \; N \;{\rm{\; times}})
\ee
Applying eq. \ref{guarmmap}, the equivariant momentum map is given by
\be
J(\xi) ({\bf q}_i,{\bf p}^i) = \left( \Sigma^N_{i = 1} {\bf p}^i \right) \cdot 
\xi \; , \;\;{\rm{\; i.e. \;}} \;\;
{\bf J}({\bf q}_i,{\bf p}^i) = \Sigma^N_{i = 1} {\bf p}^i \; ;
\label{linmommasmmap;fromformula}
\ee
agreeing with our previous solution, eq. \ref{linmommasmmap;fromdiffeqn}, based 
on the differential equation eq. \ref{requireommapinlocalxicoords}.

(2): {\em Angular momentum of a single particle}:--- \\
$SO(3)$ acts on $\mathR^3$ by $\Phi(A, {\bf q}) = A{\bf {q}}$. Writing $\xi \in 
\mso(3)$ as $\xi = \Theta \o$ (cf. eq. \ref{ansym}, \ref{theta3dcorrspce} and 
\ref{ansymxi}), the infinitesimal generator is
\be
\xi_{\mathR^3}({\bf q}) = \xi {\bf q} = \o \wedge {\bf q} \; .
\ee
So applying eq. \ref{guarmmap}, the equivariant momentum map ${\bf J}: 
T^*\mathR^3 \rightarrow \mso(3) \cong \mathR^3$ is given by
\be
< {\bf J}({\bf q}, {\bf p}) ; \o > = < {\bf p} ; \o \wedge {\bf q} > = 
{\bf p} \cdot (\o \wedge {\bf q}) = \o \cdot ({\bf q} \wedge {\bf p}) 
\; , \;\;{\rm{\; i.e. \;}} \;\;
{\bf J}({\bf q},{\bf p}) = {\bf q} \wedge {\bf p} \; ;
\ee
agreeing with our previous solution, eq. \ref{angularmommasmmap;fromdiffeqn}, 
based on the differential equation eq. \ref{requireommapinlocalxicoords}.

(3): {\em The cotangent lift of left and right translations on $G$}:--- \\
Recalling eq. \ref{xi_Gisri}, viz. that the infinitesimal generator  of left 
translation is
\be
\xi_G(g) = (T_e R_g)\xi \; ,
\label{xi_Gisri;repeat}
\ee 
a {\em right}-invariant vector field, and applying eq. \ref{guarmmap}, we see 
that the momentum map ${\bf J}_L: T^*G \rightarrow \mg^*$ for the cotangent 
lift of left translation is given by
\be
< {\bf J}_L (\al_g) ; \xi > \; = \; < \al_g ; \xi_G(g) > \; = \; < \al_g ; (T_e 
R_g)\xi > \; = \; < (T^*_e R_g)(\al_g) ; \xi >
\ee
where the last equation applies the definition of the cotangent lift eq. 
\ref{defineT^*f}. That is: the equivariant  momentum map is
\be
{\bf J}_L (\al_g) = T^*_e R_g (\al_g) \; .
\label{JLisTeRg}
\ee 
In words: the momentum map ${\bf J}_L$ of the cotangent lift of  left 
translation is the cotangent lift of right translation.

In a similar way, we could consider right translation: $R_g: h \mapsto hg$. 
Right translation  defines a right action on $G$,  has $\xi_G(g) = (T_e 
L_g)\xi$ as its infinitesimal generator, and so has 
\be
{\bf J}_R: T^*G \rightarrow \mg^* \;\; ; \;\; {\bf J}_R (\al_g) := T^*_e L_g 
(\al_g)
\label{JRisTeLg}
\ee
as the momentum map of its cotangent lift. Note that this momentum map is 
equivariant with respect to $Ad^*_g$: which, as discussed after eq. 
\ref{defineAd^*}, is a {\em right} action.

\section{Reduction}\label{redn}
\subsection{Preamble}\label{preamblerednthm}
In this final Section, the themes of Section \ref{rednintro} onwards come 
together---at last! As announced in Section \ref{pmspreamble}, we will 
concentrate on proving what is nowadays called the {\em Lie-Poisson reduction 
theorem}: that is, the isomorphism  of Poisson manifolds
\be
  T^*G/G \cong \mg^* \; .
  \ee 
Here the quotient of $T^*G$ is by the cotangent lift of $G$'s action on itself 
by left translation.

\indent As it happens, this Chapter's main sources (i.e. Abraham and Marsden 
(1978), Arnold (1989), Olver (2000) and Marsden and Ratiu (1999)) do not 
contain what is surely the most direct proof of this result. So we give it in 
Section \ref{rednthm}. The result will follow directly from four previous main 
results, one from Section \ref{pms} and three from Section \ref{symyredn}.\\
\indent `Directly', but for one wrinkle! This relates to ``flipping'' between 
left and right translation, and their various lifts. In short: the four 
previous results show that $T^*G/G$ is isomorphic as a Poisson manifold, {\em 
not} to $\mg^*$ with the Lie-Poisson bracket familiar since eq. \ref{defLPbr} 
and \ref{intrLPbr}, but instead to $\mg^*$ equipped with this bracket's {\em 
negative}, i.e. equipped with
\be
\{F, H \}_- (x) \; := \; - < x; [\nabla F(x), \nabla H(x)]> \;\; , \;\; x \in 
\mg^* \;.
\label{intrLPbrminus}
\ee   
But we shall (mercifully!) not reproduce, with minus signs appropriately added, 
our entire discussion of the Lie-Poisson bracket that ensued after eq. 
\ref{defLPbr}; (exercise for the reader!).\\
\indent To avoid ambiguity, we shall sometimes write $\mg^*_+$ for $\mg^*$ 
equipped  with the positive Lie-Poisson bracket of eq. \ref{intrLPbr}; and 
$\mg^*_-$ for $\mg^*$ equipped  with the negative Lie-Poisson bracket of eq. 
\ref{intrLPbrminus}.

In fact, it will be clearest from now on,  to treat right actions on a par with 
left actions; despite our previous emphasis on the latter. This will mean that 
we will also treat right-invariant vector fields (and another notion of 
right-invariance defined in Section \ref{lirifns}) on a par with left-invariant 
vector fields (and Section \ref{lirifns}'s corresponding new notion of 
left-invariance). Indeed, we have already glimpsed this would be necessary 
in:\\
\indent (i):  Section \ref{inflgenor}'s result that the infinitesimal generator  
of left translation is a right-invariant vector field, and vice versa (eq. 
\ref{xi_Gisri}, \ref{theotherxi_Gisli}); and its corollaries in Example (3) of 
Section \ref{Egscotgtliftedmms}, that\\
\indent (ii): the momentum map ${\bf J}_L$ of the cotangent lift of  left 
translation is the cotangent lift of right translation; (eq. \ref{JLisTeRg}); 
and \\
\indent (iii): the momentum map ${\bf J}_R$ of the cotangent lift of  right 
translation is the cotangent lift of left translation; (eq. \ref{JRisTeLg}).

So by the end of Section \ref{rednthm}, we will have a short proof of the 
Lie-Poisson reduction theorem. But (as often happens), the most direct proof 
does not give very much information about the situation. So in Section 
\ref{invartT^*G} we give more information (following Marsden and Ratiu (1999)). 
Then in Section \ref{redndyns}, we discuss the reduction of dynamics (as 
against Poisson structure) from $T^*G$ to $\mg^*$. 

 Finally, in Section \ref{soon} we state another reduction theorem, which is 
cast in terms of symplectic, not Poisson, manifolds---but which uses several 
notions from  Section \ref{tool}, such as free and proper actions, and isotropy 
groups.  But we do not prove this theorem: we include it mostly in order to 
emphasize our previous remark, that (despite its length!) this Chapter just 
scratches the surface of the subject. We also discuss the relation between it 
and the Lie-Poisson reduction theorem.

\subsection{The Lie-Poisson Reduction Theorem}\label{rednthm}
First we recall from the end of Section \ref{Fundldiffms;was6.5.6.B} (eq. 
\ref{barlambdaeqvart}) that 
${\bar \la}: T^*G \rightarrow G \times \mg^*$ is an equivariant map relating 
the cotangent lifted left action of left translation on $T^*G$ to the 
$G$-action on $G \times \mg^*$ given just by left translation on the first 
component. So we  passed to the quotients, and defined ${\hat {\bar \la}}: 
T^*G/G \rightarrow (G \times \mg^*)/G$ by eq. \ref{definehatbarlambdadetails}, 
viz.
\begin{eqnarray}
{\hat {\bar \la}}: {\rm{Orb}}(\al) \equiv \{ \bb \in T^*G \mid \bb = 
T^*L_{h^{-1}}(\al) , \; {\rm{some}} \; h \in G \}  \mapsto \;\;\;\;\;\;\; \\
{\rm{Orb}}({\bar \la}(\al)) 
\equiv \{ (hg, (T^*_e L_g)(\al)) \mid 
{\rm{some}} \; h \in G \} \equiv 
\{ (h, (T^*_e L_g)\al ) \mid {\rm{some}} \; h \in G \} \; .
\label{definehatbarlambdadetailsrepeat}
\end{eqnarray}
where $\al \in T^*_gG$, so that $T^*L_{h^{-1}} \al \in T^*_{hg}G$.
Finally, we identified $(G \times \mg^*)/G$ with $\mg^*$, so that the 
diffeomorphism ${\hat {\bar \la}}$ maps $T^*G/G$ to $\mg^*$, as in eq. 
\ref{statehatbarlambdaforLPredndiffm}: 
\be
{\hat {\bar \la}}: {\rm{Orb}}(\al) \equiv \{ \bb \in T^*G \mid \bb = 
T^*L_{h^{-1}}(\al) , \; {\rm{some}} \; h \in G \} \in T^*G/G \mapsto (T^*_e 
L_g)(\al) \in \mg^*.
\label{stateLPreddiffm}
\ee 

So now, we are to show that the diffeomorphism ${\hat {\bar \la}}: T^*G/G 
\rightarrow \mg^*$ is a Poisson map, in the sense of eq. 
\ref{definePoissmaponPoissmfd} (Section \ref{Pomaps1}). So we need to show:\\
\indent (i): $T^*G/G$ is a Poisson manifold;\\
\indent (ii): ${\hat {\bar \la}}$ maps (i)'s Poisson structure on $T^*G/G$ to 
that of $\mg^*$. In fact, as announced in Section \ref{preamblerednthm}, ${\hat 
{\bar \la}}$ maps on to the Poisson structure of $\mg^*_-$, i.e. as given by  
eq. \ref{intrLPbrminus}.

{\em Prima facie}, there could be a judicious choice to be made about (i), i.e. 
about how to define the Poisson structure on $T^*G/G$, so as to secure (ii), 
i.e. so that ${\hat {\bar \la}}$ respects the Poisson structure. But in fact 
our previous work gives a pre-eminently obvious choice---which works. Namely: 
we use the Poisson structure induced on $T^*G/G$ by the Poisson  reduction 
theorem  of Section \ref{quotpm}. The result follows directly by combining with 
this theorem, three results from Section \ref{symyredn}:\\
\indent (i): that equivariant   momentum maps   are Poisson; eq. 
\ref{eqvartispoisson} in Section \ref{eqvartmmsarePoisson};\\
\indent (ii): that a cotangent lifted left action has an equivariant momentum 
map; eq. \ref{guarmmap} in Section \ref{mmsforcotgtlifts}; \\
\indent (iii): that the momentum maps of the cotangent lifts of left and right 
translation on $G$ are ${\bf J}_L = T^*_e R_g$ and ${\bf J}_R = T^*_e L_g$; eq. 
\ref{JLisTeRg} and \ref{JRisTeLg} in Section \ref{Egscotgtliftedmms}.\\
\indent In particular, combining (i)-(iii): one deduces (exercise!) that ${\bf 
J}_R = T^*_e L_g$ is equivariant with respect to $Ad^*_g$, and so Poisson with 
respect to the  negative Lie-Poisson bracket (eq. \ref{intrLPbrminus}'s 
bracket)  on $\mg^*$. That is: it is Poisson with the codomain $\mg^*_-$.

Thus we have the
\begin{quote}
{\bf Lie-Poisson reduction theorem} The diffeomorphism ${\hat {\bar \la}}: 
T^*G/G \rightarrow \mg^*$:
\be
{\hat {\bar \la}}: {\rm{Orb}}(\al) \equiv \{ \bb \in T^*G \mid \bb = 
T^*L_{h^{-1}}(\al) , \; {\rm{some}} \; h \in G \} \in T^*G/G \mapsto (T^*_e 
L_g)(\al) \in \mg^* 
\label{stateLPreddiffm;repeat}
\ee
is Poisson.
\end{quote} 
{\em Proof}: First, eq. \ref{stateLPreddiffm;repeat} means we have a 
commutative triangle.  For with $\pi: T^*G \rightarrow T^*G/G$ the canonical 
projection,  the momentum  map ${\bf J}_R: T^*G \rightarrow \mg^*, \al_g 
\mapsto (T^*_e L_g)\al_g$ is  equal to ${\hat {\bar \la}} \circ \pi$: 
\be
T^*G \stackrel{\pi}{\longrightarrow} T^*G/G \stackrel{{\hat {\bar 
\la}}}{\longrightarrow} \mg^* \; .
\label{commtriangleforlprt}
\ee
Since left translation is a diffeomorphism of $G$, and the cotangent lift of 
any diffeomorphism of a manifold to its cotangent bundle is symplectic (cf. 
after eq. \ref{trslnrotnActonR3} in Section \ref{basicaction}.A), the Poisson  
reduction theorem  of Section \ref{quotpm} applies. That is, there is a unique 
Poisson structure on $T^*G/G$ such that $\pi$ is Poisson. We also know from eq. 
\ref{guarmmap}, \ref{eqvartispoisson} and \ref{JRisTeLg} that ${\bf J}_R = 
T^*_e L_g$ is Poisson with respect to eq. \ref{intrLPbrminus}'s bracket  on 
$\mg^*$.\\
\indent We can now deduce that ${\hat {\bar \la}}$ is Poisson, i.e. that for 
all $x \in T^*G/G$ and all $F, H \in {\cal F}(\mg^*_-)$
\be
(\{ F, H \}_{\mg^*_-} \circ {\hat {\bar \la}}) \; (x) \; = \;
\{ F \circ {\hat {\bar \la}}, H \circ {\hat {\bar \la}} \}_{T^*G/G} \; (x) \; .
\label{hatbarlaispoisson}
\ee
We just use (in order) the facts that:\\
\indent (i): $\pi$ is surjective, so that for all $x \in T^*G/G$ there is an 
$\al_g \in T^*G$ with $x = \pi(\al_g) \equiv {\rm{Orb}}(\al_g)$;\\
\indent (ii): ${\bf J}_R = {\hat {\bar \la}} \circ \pi$; \\
\indent (iii): ${\bf J}_R$ is Poisson; and \\
\indent (iv): $\pi$ is Poisson:
\begin{eqnarray}
(\{ F, H \}_{\mg^*_-} \circ {\hat {\bar \la}}) \; (x) \; = \; 
\{ F, H \}_{\mg^*_-} \circ ({\hat {\bar \la}} \circ \pi) \; (\al_g) \\
 = \;
\{ F, H \}_{\mg^*_-} \circ  {\bf J}_R \; (\al_g) \; = \; 
\{ F \circ  {\bf J}_R, H \circ  {\bf J}_R \}_{T^*G} \; (\al_g) \\
 = \;
\{ F \circ  {\hat {\bar \la}}, H \circ  {\hat {\bar \la}} \}_{T^*G/G} \; 
(\pi(\al_g)) \; \equiv \; \{ F \circ {\hat {\bar \la}}, H \circ {\hat {\bar 
\la}} \}_{T^*G/G} \; (x) \; . \;\; {\rm{QED}}.
\label{lprtmaindeduce}
\end{eqnarray}

\subsection{Meshing with the symplectic structure on $T^*G$: invariant 
functions}\label{invartT^*G}
We turn to giving more information about the situation described by the 
Lie-Poisson reduction theorem.  The general idea will be that the Lie-Poisson 
bracket on $\mg^*$ meshes  with the canonical symplectic structure on $T^*G$. 
This will be made precise in two ways: the first is discussed in the first two 
Subsections, the second is discussed in the third Subsection.

\indent  The first discussion will have three stages:\\
\indent (i): we show that scalars on $\mg^*$, $F \in {\cal F}(\mg^*)$, are in 
one-one correspondence with scalars on $T^*G$ that are constant on the orbits 
of the cotangent lift of left translation, which will be called {\em 
left-invariant} functions; and similarly, for the cotangent  lift of right 
translation (a correspondence with {\em right-invariant} functions);\\
\indent (ii): we take the usual canonical Poisson bracket in $T^*G$ of these 
left-invariant or right-invariant scalars; and restrict this bracket to $\mg^*$ 
regarded as the cotangent space $T^*_eG$ at the identity $e \in G$; and then \\
\indent (iii): we show that this restriction is the Lie-Poisson bracket on 
$\mg^*$: the familiar positive one for {\em right}-invariant functions, and the 
new negative one of eq. \ref{intrLPbrminus} for the {\em left}-invariant 
functions.\\
\indent We do stages (i) and (ii) in Section \ref{lirifns}. These stages will 
not involve the choice between the positive and negative Lie-Poisson brackets.  
But stage (iii), in Section \ref{rednthmbrief}, will involve this choice. It 
will be a one-liner corollary of Section \ref{eqvartmmsarePoisson}'s result 
that equivariant momentum maps are Poisson maps, eq. \ref{eqvartispoisson}; 
(unsurprisingly, in that we also used this result in Section \ref{rednthm}'s 
proof of the reduction theorem).
 
In the third Subsection, we use invariant functions to show a different sense 
in which the Lie-Poisson bracket on $\mg^*$ meshes  with the symplectic 
structure on $T^*G$. Namely, we derive the Lie-Poisson bracket on $\mg^*$ from 
the Poisson  reduction theorem  of Section \ref{quotpm}, by using the ideas of 
invariant functions and momentum functions. 

\subsubsection{Left-invariant and right-invariant functions on 
$T^*G$}\label{lirifns}
We say that a function $F: T^*G \rightarrow \mathR$ is {\em left-invariant} if 
for all $g \in G$, and all $\al_g \in T^*_g G$
\be
(F \circ T^*L_g)(\al_g) = F(\al_g)
\ee
where $T^*L_g$ is the cotangent lift of $L_g: G \rightarrow G$. Similarly, $F: 
T^*G \rightarrow \mathR$ is called {\em right-invariant} if for all $g \in G$
\be
(F \circ T^*R_g) = F \; .
\ee
So if $F: T^*G \rightarrow \mathR$ is left-invariant or right-invariant, it is 
determined by its values for arguments in $T^*_eG = \mg^*$.\\
\indent Since any $\al \in \mg^*$ is mapped by $T^*L_{g^{-1}} \equiv 
(T^*L_g)^{-1}$ to an element of $T^*_g G$, a function is left-invariant iff it 
is constant on the orbits of the various $T^*L_g$ for $g \in G$, i.e. constant 
on the orbits of the cotangent  lift of left translation.  Similarly, a 
function is right-invariant iff it is constant on the orbits of the cotangent 
lift of right translation.\\
\indent So left-invariant functions induce well-defined functions on the 
quotient space $T^*G/G$; and so, by Section \ref{rednthm}, on its diffeomorphic 
(indeed Poisson manifold) copy $\mg^*$. Similarly for right-invariant 
functions. 

But let us for the moment consider the  smooth left-invariant (or 
right-invariant) functions on $T^*G$, rather than the induced maps on the 
quotient space.  We will denote the space of all smooth left-invariant 
functions on $T^*G$ by ${\cal F}_L(T^*G)$, and similarly the space of smooth 
right-invariant functions by ${\cal F}_R(T^*G)$.\\
\indent Recalling (from the discussion after eq. \ref{trslnrotnActonR3}) that 
cotangent lifts are symplectic maps, i.e. $T^*L_g$ and $T^*R_g$ are symplectic 
maps on $T^*G$, it follows immmediately that ${\cal F}_L(T^*G)$ and ${\cal 
F}_R(T^*G)$ are each closed under the canonical Poisson bracket on $T^*G$. So 
they are each a Lie algebra with this bracket. 

Now we can use the momentum maps ${\bf J}_L$ and ${\bf J}_R$ of Example (3) of 
Section \ref{Egscotgtliftedmms} to extend any scalar $F:\mg^* \rightarrow 
\mathR$, i.e. $F \in {\cal F}(\mg^*)$, to a left-invariant, or right-invariant, 
scalar on $T^*G$.\\
\indent Thus, given $F:\mg^* \rightarrow \mathR$ and $\al_g \in T^*_gG$, we 
define $F_L \in {\cal F}_L(T^*G)$ by
\be
F_L(\al_g) := (F \circ {\bf J}_R)(\al_g) \equiv (F \circ T^*_e L_g)(\al_g) \; .
\ee
So $F_L$ is by construction left-invariant, and is called the {\em 
left-invariant extension} of $F$ from $\mg^*$ to $T^*G$.\\
\indent One similarly defines the {\em right-invariant extension} $F_R \in 
{\cal F}_R(T^*G)$ of any $F \in {\cal F}(\mg^*)$ by
\be
F_R(\al_g) := (F \circ {\bf J}_L)(\al_g) \equiv (F \circ T^*_e R_g)(\al_g) \; .
\ee
Then the maps
\be
F \in {\cal F}(\mg^*) \mapsto F_L \in {\cal F}_L(T^*G) \;\; {\rm{and}} \;\;
F \in {\cal F}(\mg^*) \mapsto F_R \in {\cal F}_R(T^*G)
\label{LPRTisoms}
\ee
are vector space isomorphisms (exercise for the reader!) whose inverse is just 
restriction to the fiber $T^*_eG = \mg^*$.\\
\indent This completes what we called `stages (i) and (ii)': describing a 
correspondence between scalars on $\mg^*$ and scalars on $T^*G$ that are 
constant on the orbits of the cotangent lifts of left and right translation; 
and considering the canonical Poisson bracket (on $T^*G$) of these scalars, 
i.e. the Lie algebras ${\cal F}_L(T^*G)$ and ${\cal F}_R(T^*G)$.

\subsubsection{Recovering the Lie-Poisson bracket}\label{rednthmbrief}
We now do stage (iii): we show that the restriction of the canonical Poisson 
bracket on $T^*G$ of the right/left invariant functions, to $\mg^*$ regarded as 
the cotangent space $T^*_eG$ at the identity $e \in G$, is the 
positive/negative Lie-Poisson bracket.

 Since the inverses of the maps eq. \ref{LPRTisoms} are just restriction to the 
fiber $T^*_eG = \mg^*$, it suffices to show that the maps eq. \ref{LPRTisoms} 
are Lie algebra isomorphisms. More precisely:
\begin{quote}
{\bf {Recovery of the Lie-Poisson bracket}} Using the positive Lie-Poisson 
bracket on $\mg^*$ (we write $\mg^*_+$): $F \mapsto F_R$ is a Lie algebra 
isomorphism.\\
Similarly: using the negative Lie-Poisson bracket on $\mg^*$ (we write 
$\mg^*_-$):\\
 $F \mapsto F_L$ is a Lie algebra isomorphism.

That is: for all $F, H \in {\cal F}(\mg^*)$
\be
\{F, H \}_+ = \{F_R, H_R \}_{T^*G} \mid_{\mg^*} \;\; ; \;\; 
\{F, H \}_- = \{F_L, H_L \}_{T^*G} \mid_{\mg^*} 
\label{lprt2aspbequality}
\ee
\end{quote}

{\em Proof}:  Consider ${\bf J}_L: T^*G \rightarrow \mg^* \equiv \mg^*_+$, 
${\bf J}_L = T^*_eR_g$. ${\bf J}_L$ is an equivariant momentum map. So, by the 
result eq. \ref{eqvartispoisson} of Section \ref{eqvartmmsarePoisson}, it is 
Poisson. That is:
\be
\{F, H \}_+ \circ {\bf J}_L = \{F \circ {\bf J}_L, H \circ {\bf J}_L \}_{T^*G} 
= \{F_R, H_R \}_{T^*G} \; .
\label{JLisPoisson}
\ee
Restricting eq. \ref{JLisPoisson} to $\mg^*$ gives the first equation of eq. 
\ref{lprt2aspbequality}.\\
\indent Similarly, one proves the second equation by using the fact that ${\bf 
J}_R: T^*G \rightarrow \mg^* \equiv \mg^*_-, {\bf J}_R = T^*_eL_g$ is an 
equivariant momentum map and so is Poisson. That is:
\be
\{F, H \}_- \circ {\bf J}_R = \{F \circ {\bf J}_R, H \circ {\bf J}_R \}_{T^*G} 
= \{F_L, H_L \}_{T^*G} \; . 
\label{JRisPoisson}
\ee
We then restrict eq. \ref{JRisPoisson} to $\mg^*$. QED.

\subsubsection{Deriving the Lie-Poisson bracket}\label{rednmommfn}
Our discussion so far, in both Section \ref{rednthm} and the two previous 
Subsections, has taken the Lie-Poisson bracket (whether positive or negative) 
as given. We now show, using  invariant functions and Section \ref{mommfns}'s 
idea of momentum functions, how to {\em derive} the Lie-Poisson bracket on 
$\mg^*$.\\
\indent So this derivation will amount to another, more ``constructive'', proof 
of the Lie-Poisson reduction theorem. As in Section \ref{rednthm}'s proof, two 
main ingredients will be:\\
\indent \indent (a): the diffeomorphism ${\hat {\bar \la}}$ between $T^*G/G$ 
and  $\mg^*$ (eq. \ref{statehatbarlambdaforLPredndiffm} or 
\ref{stateLPreddiffm} or \ref{stateLPreddiffm;repeat}), and \\
\indent \indent (b): the Poisson  reduction theorem  of Section \ref{quotpm}, 
applied to $G$'s action on $T^*G$. \\
But instead of Section \ref{rednthm}'s proof's using the facts that (i) the 
momentum maps ${\bf J}_R \equiv T^*_e L_g$ and ${\bf J}_L \equiv T^*_e R_g$ are 
equivariant and (ii) equivariant momentum maps are Poisson, we will now use the 
ideas of invariant functions and momentum functions. 

We begin by recalling that (since left translation is a diffeomorphism of $G$, 
and the cotangent lift of any diffeomorphism of a manifold to its cotangent 
bundle is symplectic), the Poisson  reduction theorem  implies that there is a 
unique Poisson structure on $T^*G/G$ such that $\pi: T^*G \rightarrow T^*G/G$ 
is Poisson. We now use the diffeomorphism  ${\hat {\bar \la}}: T^*G/G 
\rightarrow \mg^*$ to transfer this Poisson structure to $\mg^*$. Let us call 
the result $\{ , \}_-$. Though this is not to be read (yet!) as the negative 
Lie-Poisson bracket,  our aim now is to calculate that it is in fact this 
bracket.

Notice first that since the momentum  map ${\bf J}_R: T^*G \rightarrow \mg^*, 
\al_g \mapsto (T^*_e L_g)\al_g$ is equal to ${\hat {\bar \la}} \circ \pi$ (eq. 
\ref{commtriangleforlprt}), we know that ${\bf J}_R$ is Poisson with respect to 
this induced bracket on $\mg^*$. That is
\be
\{ F, H \}_- \circ  {\bf J}_R \; (\al_g) \; = \; 
\{ F \circ  {\bf J}_R, H \circ  {\bf J}_R \}_{T^*G} \; (\al_g) \;
= \; \{ F_L, H_L \}_{T^*G} \; (\al_g) \; .
\label{JRPoisswrtmysterybr}
\ee 
To calculate the right hand side, we will apply the ideas of invariant 
functions and momentum functions to each argument of the bracket; in particular 
to the first:
\be
F_L (\al_g) = F(T^*_e L_g \cdot \al_g) \; .
\label{FLready}
\ee 

We observe that since a Poisson bracket depends only on the values of first 
derivatives, we can replace $F \in {\cal F}(\mg^*)$ by its linearization. That 
is, we can assume $F$ is linear, so that at any point $\al \in \mg^*$, $F(\al) 
= <  \al ; \nabla F  >$, where $\nabla F$ is a constant in $\mg \equiv 
\mg^{**}$. Applying this, and the definition of a momentum function eq. 
\ref{definemommfn}, to eq. \ref{FLready}, we get:
\be
F(T^*_e L_g \cdot \al_g) \; = \; < T^*_e L_g \cdot \al_g ; \nabla F > \;
= \; <  \al_g ; T_e L_g \cdot \nabla F > \; = \; {\cal P}(X_{\nabla F})(\al_g) 
\; ,
\label{FLP}
\ee
where the last equation applies the definition of a momentum function to the 
left-invariant vector field on $G$, $X_{\xi}(g) \equiv T_e L_g (\xi)$, for the 
case $\xi = \nabla F$.

Now we apply to eq. \ref{FLP}, in order: eq. \ref{calPantiisom}, the definition 
of the Lie algebra bracket (cf. eq. \ref{LIcommuteLiebr}),  eq. 
\ref{definemommfn} again, and the definition of left-invariant vector fields. 
We get:
\begin{eqnarray}
\{ F_L, H_L \}_{T^*G} \; (\al_g) \; = \; \{ {\cal P}(X_{\nabla F}), {\cal 
P}(X_{\nabla H}) \}_{T^*G} \; (\al_g) \; = \; - {\cal P}([X_{\nabla F}, 
X_{\nabla H}]) (\al_g) \\
= \; -  {\cal P}(X_{[{\nabla F},{\nabla H}]}) (\al_g) \; = \; 
- < \al_g ; X_{[{\nabla F},{\nabla H}]} > \\ 
= \; 
- < \al_g ; T_e L_g([{\nabla F},{\nabla H}]) > \; = \; 
- < T^*_e L_g(\al_g) ; [{\nabla F},{\nabla H}] > \; .
\label{deductiondone}
\end{eqnarray}
Combining eq. \ref{JRPoisswrtmysterybr} and eq. \ref{deductiondone}, and 
writing $\al \in \mg^*$ for $(T^*_e L_g) \al_g \equiv {\bf J}_R (\al_g)$, we 
have our result:
\be
\{ F, H \}_- (\al) = - < \al ; [{\nabla F},{\nabla H}] > \; .
\ee
One similarly derives the positive Lie-Poisson bracket by considering 
right-invariant extensions of linear functions. The minus sign coming from eq. 
\ref{calPantiisom} is cancelled by the sign reversal in the Lie bracket of 
right-invariant vector fields. That is, it is cancelled by a minus sign coming 
from  eq. \ref{RILBminusUsualLB}.  

\subsection{Reduction of dynamics}\label{redndyns}
We end our account of the Lie-Poisson reduction theorem by discussing the 
reduction of dynamics from $T^*G$ to $\mg^*$.\\ 
\indent We can be brief since we have already stated the main idea, when 
discussing the Poisson reduction theorem; cf. (2)(A) in Section \ref{quotpm}. 
Thus recall that (under the conditions of the theorem) a $G$-invariant 
Hamiltonian function on a Poisson manifold $M$, $H: M \rightarrow \mathR$, 
defines a corresponding function $h$ on $M/G$ by $H = h \circ \pi$, where $\pi$ 
is the projection $\pi: M \rightarrow M/G$; and since $\pi$ is Poisson, and so 
pushes Hamiltonian flows forward to Hamiltonian flows, $\pi$ pushes $X_H$ on 
$M$ to $X_h$ on $M/G$: 
\be
T\pi \circ X_H = X_h \circ \pi \; \; .
\label{hamflowspirelated;repeat}
\ee
Applying this, in particular eq. \ref{hamflowspirelated;repeat}, to the 
Lie-Poisson reduction theorem, we get
\begin{quote} 
{\bf{Reduction of dynamics}} Let $H: T^*G \rightarrow \mathR$ be 
left-invariant. That is: the function $H^- := H\mid_{\mg^*}$ on $\mg^*$ 
satisfies 
\be
H(\al_g) = H^-({\bf J}_R (\al_g)) \equiv H^-(T^*_eL_g \cdot \al_g) \; , \;\;\;
\al_g \in T^*_gG \; .
\ee 
Then ${\bf J}_R$ pushes $X_H$ forward to $X_{H^-}$. Or in terms of the flows 
$\phi(t)$ and $\phi^-(t)$ of $X_H$ and $X_{H^-}$ respectively:
\be
{\bf J}_R(\phi(t)(\al_g)) \; = \; \phi^-(t)({\bf J}_R(\al_g)) \; .
\ee 
Similar statements hold for a right-invariant function $H: T^*G \rightarrow 
\mathR$, its restriction $H^+ := H\mid_{\mg^*}$ and ${\bf J}_L \equiv 
T^*_eR_g$.
\end{quote} 

Besides, we already know the vector field of $H^-$ on $\mg^*$. For eq. 
\ref{HamEqWithad^*} in (3) of Section \ref{LPso(3)} gave a basis-independent 
expression of Hamilton's equations on $\mg^*$ in terms of $ad^*$. We just need 
to note that since we are now using the negative Lie-Poisson bracket on 
$\mg^*$, all terms in the deduction (eq. \ref{intrLPbrandad}) apart from the 
left hand side, get a minus sign. So writing $\al \in \mg^*$, eq. 
\ref{HamEqWithad^*} for the vector field  $X_{H^-}$ becomes:
\be
\frac{d \al}{dt} \; = \; - \; ad^*_{\nabla H^-(\al)}(\al) \; .
\label{HamEqWithad^*;repeat}
\ee

On the other hand, we can go in the other direction, reconstructing the 
dynamics on $T^*G$ from eq. \ref{HamEqWithad^*;repeat} on $\mg^*$. The 
statement of the main result, below, is  intuitive, in that the 
``reconstruction equation'' for $g(t) \in T^*G$ is
\be
g^{-1}{\dot g} =  \nabla H^-  \;\; .
\ee
This is intuitive since it returns us to the basic idea of mechanics  on $\mg$ 
and $\mg^*$, viz. that the map
\be
\la_g: {\dot g} \in T_gG \mapsto \la_g({\dot g}) := (T_gL_{g^{-1}}){\dot g} \in 
\mg
\ee
maps the generalized velocity to its body representation; cf. eq. 
\ref{definelambdag}. However, the  proof of this result is involved (Marsden 
and Ratiu (1999: theorems 13.4.3, 13.4.4, p. 423-426); so we only state the 
result. It is:---
\begin{quote} 
{\bf{Reconstruction of dynamics}} Suppose given a Lie group $G$, a 
left-invariant $H: T^*G \rightarrow \mathR$, its restriction $H^- := 
H\mid_{\mg^*}$, and an integral curve $\al(t)$ of the Lie-Poisson Hamilton's 
equations eq. \ref{HamEqWithad^*;repeat} on $\mg^*$, with the initial condition 
$\al(0) = T^*_eL_{g_{0}}(\al_{g_{0}})$. Then the integral curve in $T^*G$ of 
$X_H$ is given by
\be
T^*_{g(t)}L_{g(t)^{-1}}(\al(t)) \;\; ;
\ee
where $g(t)$ is the solution of the reconstruction equation
\be
g^{-1}{\dot g} =  \nabla H^-
\ee
with initial condition $g(0) = g_0$.
\end{quote}

\subsection{Envoi: the Marsden-Weinstein-Meyer theorem}\label{soon}
I emphasize that our discussion of reduction has only scratched the surface: 
after all this Section has been relatively short! But now that the reader is 
armed with the long and leisurely exposition from Section \ref{tool} onwards, 
they are well placed to pursue the topic of reduction; e.g. through this 
Chapter's main sources, Abraham and Marsden (1978), Arnold (1989), Olver (2000) 
and Marsden and Ratiu (1999). 

In particular, the reader can now relate the Lie-Poisson reduction theorem to 
another main theorem about symplectic reduction, usually called the 
Marsden-Weinstein-Meyer or Marsden-Weinstein theorem (after these authors' 
papers in 1973 and 1974).\\
\indent This theorem concerns a symplectic action of a Lie group $G$ on a 
symplectic manifold $(M, \o)$. For the sake of completeness, and to orient the 
reader to Landsman's discussion of this theorem (this vol., ch. 5, especially 
Section 4.5), it is worth stating it (as usual, for the finite-dimensional case 
only), together with the lemma used to prove it, and the ensuing reduction of 
dynamics. These statements will also round off our discussion by illustrating 
how some notions expounded from Section \ref{tool} onwards, but not used in 
this Section, are nevertheless  useful---e.g. in stating the hypotheses of this 
theorem.

So suppose the Lie group $G$ acts symplectically (eq. \ref{defcanlaction}) on 
the symplectic manifold $(M, \o)$; and that ${\bf J}: M \rightarrow \mg^*$ is 
an $Ad^*$-equivariant momentum map for this action (eq. \ref{eqvarce2nddefn} 
and \ref{eqvarcereformltn}). Assume also that $\al \in \mg^*$ is a regular 
value of $\bf J$, i.e. that at every point $x \in {\bf J}^{-1}(\al)$, $T_x{\bf 
J}$ is surjective. So the submersion theorem of (1) of Section 
\ref{323ASubmanifolds} applies; in particular, ${\bf J}^{-1}(\al)$ is a 
sub-manifold of $M$ with dimension dim($M$) - dim($\mg^*$) $\equiv$ dim($M$) - 
dim($G$).\\
\indent Let $G_{\al}$ be the isotropy group (eq. \ref{defineisotropy}) of $\al$ 
under the co-adjoint action, i.e. 
\be
G_{\al} := \{ g \in G \mid Ad^*_{g^{-1}} \al = \al \} .
\ee  
So since $\bf J$ is $Ad^*$-equivariant under $G_{\al}$, the quotient space 
$M_{\al} := {\bf J}^{-1}(\al)/G_{\al}$ is well-defined.\\
\indent Now assume that $G_{\al}$ acts freely and properly on ${\bf 
J}^{-1}(\al)$, so that (Section \ref{332Aproper}.B) the quotient space  
$M_{\al} = {\bf J}^{-1}(\al)/G_{\al}$ is a manifold. $M_{\al}$ is the {\em 
reduced phase space} (corresponding to the momentum value $\al$).\\
\indent Now we assert:

\begin{quote} 
{\bf{Marsden-Weinstein-Meyer theorem}}
$M_{\al}$ has a natural symplectic form $\o_{\al}$ induced from $(M, \o)$ as 
follows. Let $u, v$ be two vectors tangent to $M_{\al}$ at some point $p \in 
M_{\al}$: so $p$ is an orbit of $G_{\al}$'s action on ${\bf J}^{-1}(\al)$, and 
$u, v \in T_p M_{\al}$. Then $u$ and $v$ are  obtained, respectively, from some 
vectors $u'$ and $v'$ tangent to ${\bf J}^{-1}(\al)$ at some point $x \in {\bf 
J}^{-1}(\al)$ of the orbit $p$, by the projection $\pi_{\al}: {\bf J}^{-1}(\al) 
\rightarrow M_{\al}$. That is: 
\be
T \pi_{\al}(u') = u \;\; ; \;\; T \pi_{\al}(v') = v \; .
\ee 
It turns out that the value assigned by $M$'s symplectic form $\o$ is the same 
whatever choice of $x, u', v'$ is made. So we define the symplectic form 
$\o_{\al}$ on $M_{\al}$ as assigning this value. In other words: writing 
$\pi_{\al}$ for the projection, $i_{\al}:  {\bf J}^{-1}(\al) \rightarrow M$ for 
the inclusion, and $^*$ for pullback:
\be
\pi^*_{\al} \o_{\al} \; = \; i^*_{\al} \o \; .
\ee 
\end{quote} 

The proof of this theorem uses the following Lemma. Let us write $G \cdot x$ 
for the orbit Orb($x$) of $x$ under the action of all of $G$, and similarly 
$G_{\al} \cdot x$ for the orbit under $G_{\al}$, i.e. $\{\Phi(g,x) \mid g \in 
G_{\al} \}$. Then the Lemma states:
\begin{quote} For any $x \in {\bf J}^{-1}(\al)$:--- \\
(i): $T_x(G_{\al} \cdot x) = T_x(G \cdot x) \cap T_x({\bf J}^{-1}(\al))$; and 
\\
(ii): $T_x(G \cdot x)$ and $T_x({\bf J}^{-1}(\al))$ are $\o$-orthogonal 
complements of one another in $TM$. That is: for all $u' \in T_x M$:\\
$u' \in T_x({\bf J}^{-1}(\al))$ iff $\o(u', v') = 0$ for all $v' \in T_x(G 
\cdot x)$.
\end{quote} 
Both the Lemma and the theorem are each proven in some dozen lines. For 
details, cf. Abraham and Marsden (1978: Theorems 4.3.1-2, p. 299-300), or 
Arnold (1989: Appendix 5.B, p. 374-376).

Two final remarks. (1): The reduction of dynamics secured by the 
Marsden-Weinstein-Meyer theorem is similar to what we have seen before, for 
both the Poisson reduction theorem ((2) of Section \ref{quotpm}), and the 
Lie-Poisson reduction theorem (Section \ref{redndyns}). One proves, again in a 
few lines (Abraham and Marsden (1978: Theorems 4.3.5, p. 304):
\begin{quote} 
{\bf{Marsden-Weinstein-Meyer reduction of dynamics}} Let $H: M \rightarrow 
\mathR$ be invariant under the action of $G$ on $M$, so that by Noether's 
theorem for momentum maps (Section \ref{mommmapsnoet}) $\bf J$ is conserved, 
i.e. ${\bf J}^{-1}(\al)$ is invariant under the flow $\phi(t)$ of $X_H$ on $M$. 
Then $\phi(t)$ commutes with the action of $G_{\al}$ on ${\bf J}^{-1}(\al)$ 
(i.e. $\phi(t) \circ \Phi_g = \Phi_g \circ \phi(t)$ for $g \in G_{\al}$), and 
so defines a flow ${\hat \phi}(t)$ on $M_{\al}$ such that $\pi_{\al} \circ 
\phi(t) = {\hat \phi}(t) \circ \pi_{\al}$, i.e.
\be
\bundle{{\bf J}^{-1}(\al)}{\phi(t)}{{\bf J}^{-1}(\al)}
    \bundlemap{\pi_{\al}}{\pi_{\al}}
    \bundle{M_{\al}}{{\hat \phi}(t)}{M_{\al}} 
\ee
The flow ${\hat \phi}(t)$ is Hamiltonian with the Hamiltonian $H_{\al}$ defined 
by $H_{\al} \circ \pi_{\al} = H \circ i_{\al}$. 
\end{quote} 

(2): I said at the start of this Subsection that the reader can now relate the 
Lie-Poisson reduction theorem to the Marsden-Weinstein-Meyer theorem. It is not 
hard to show that the former is an example of the latter. As the symplectic 
manifold $M$ one takes $T^*G$, acted on symplectically by the cotangent lift of 
left translation. So we know (from (3) of Section \ref{Egscotgtliftedmms}) that 
${\bf J}_L := T^*_e R_g$ is an $Ad^*$-equivariant momentum map ... and so on: I 
leave this as an exercise for the reader! The answer is supplied at Arnold 
(1989: 377, 321) and Abraham and Marsden (1978: 302). (Abraham and Marsden call 
it the `Kirillov-Kostant-Souriau theorem'.)\\
\indent Suffice it to say here that this exercise gives another illustration of 
one of our central themes, that $\mg^*$'s symplectic leaves are the orbits of 
the co-adjoint representation. For the reduced phase space $M_{\al}$ is 
naturally identifiable with the co-adjoint orbit Orb($\al$) of $\al \in \mg^*$, 
with the symplectic forms also naturally identified; (cf. also  result (2) at 
the end of Section \ref{sympcoad}).

{\em Acknowledgements}:--- I am grateful to audiences in Irvine, Oxford, 
Princeton and Santa Barbara; to several colleagues for encouragement; and to 
Gordon Belot, Klaas Landsman, David Wallace, and especially Graeme Segal, for 
very helpful, and patient!, conversations and correspondence.

\section{References}
R. Abraham and J. Marsden (1978), {\em Foundations of Mechanics}, second 
edition: Addison-Wesley.

V. Arnold (1973), {\em Ordinary Differential Equations}, MIT Press.

V. Arnold (1989), {\em Mathematical Methods of Classical Mechanics}, Springer, 
(second edition).

J. Barbour and B. Bertotti (1982), `Mach's principle and the structure of 
dynamical theories', {\em Proceedings of the Royal Society of London} {\bf A 
382}, p. 295-306.

G. Belot (1999), `Relationism rehabilitated', {\em International Studies in 
Philosophy of Science} {\bf 13}, p. 35-52.

G. Belot (2000), `Geometry and motion', {\em British Journal for the Philosophy 
of Science} {\bf 51}, p. 561-596.  

G. Belot (2001), `The principle of sufficient reason',  {\em Journal of 
Philosophy} {\bf 98}, p. 55-74. 

G. Belot (2003), `Notes on symmetries', in Brading and Castellani (ed.s) 
(2003), pp. 393-412.

G. Belot (2003a), `Symmetry and gauge freedom',  {\em Studies in the History 
and Philosophy of Modern Physics} {\bf 34}, p. 189-225. 

G. Belot (this volume).

G. Belot and J. Earman (2001), `Pre-Socratic quantum gravity', in C. Callender 
and N. Huggett (ed.s) {\em Physics meets Philosophy at the Planck Scale}, 
Cambridge University Press, pp. 213-255. 

R. Bishop and S. Goldberg (1980), {\em Tensor Analysis on Manifolds}, New York: 
Dover.

J. Boudri (2002), {\em What was Mechanical  about Mechanics: the Concept of 
Force between Metaphysics and Mechanics from Newton to Lagrange}, Dordrecht: 
Kluwer Academic.

K. Brading and E. Castellani (ed.s) (2003), {\em Symmetry in Physics}, 
Cambridge University Press.

K. Brading and E. Castellani (this volume).

J. Butterfield (2004), `Some Aspects of Modality in Analytical mechanics',  in 
{\em Formal Teleology and Causality}, ed. M. St\"{o}ltzner, P. Weingartner, 
Paderborn: Mentis. \\
Available at Los Alamos arXive: http://arxiv.org/abs/physics/0210081 ;
 and at Pittsburgh archive: http://philsci-archive.pitt.edu/archive/00001192.

J. Butterfield (2004a), `Between Laws and Models: Some Philosophical Morals of 
Lagrangian Mechanics'; available at Los Alamos arXive: 
http://arxiv.org/abs/physics/0409030 ;
 and at Pittsburgh archive: http://philsci-archive.pitt.edu/archive/00001937/.
 
J. Butterfield (2005), `Between Laws and Models: Some Philosophical Morals of 
Hamiltonian Mechanics', in preparation.

J. Butterfield (2006), `On Symmetry and Conserved Quantities in Classical 
Mechanics', forthcoming in a {\em Festschrift} for Jeffrey Bub, ed. W.  
Demopoulos and I. Pitowsky, Kluwer: University of Western Ontario Series in 
Philosophy of Science. 
available at Los Alamos arXive: http://arxiv.org/abs/physics/ ;
 and at Pittsburgh archive: http://philsci-archive.pitt.edu/archive/00002362/

R. Darling (1994), {\em Differential Forms and Connections}, Cambridge 
University Press.

M. Dickson (this volume).

E. Desloge (1982), {\em Classical Mechanics}, John Wiley.

J. Earman (2003), `Tracking down gauge: an ode to the constrained Hamiltonian 
formalism', in Brading and Castellani (ed.s) (2003), pp. 140-162.

H. Goldstein (1950), {\em Classical Mechanics}, Addison-Wesley; (1966 third 
printing).

I. Grattan-Guinness (2000), `A sideways look at Hilbert's twenty-three problems 
of 1900', {\em Notices of the American Mathematical Society} {\bf 47}, p. 
752-757.

I. Grattan-Guinness (2006), `Classical mechanics as a formal(ised) science', in 
B. Loewe (ed.), {\em Foundations of the Formal Sciences}, Kluwer, to appear.  

T. Hawkins (2000), {\em Emergence of the Theory of Lie Groups: an essay in the 
history of mathematics 1869-1926}, New York: Springer. 

M. Jammer (1957), {\em Concepts of Force}, Harvard University Press.

M. Jammer (1961), {\em Concepts of Mass in Classical and Modern Physics}, 
Harvard University Press; republished by Dover in 1997.

M. Jammer (2000), {\em Concepts of Mass in Contemporary Physics and 
Philosophy}, Princeton University Press.

O. Johns (2005), {\em Analytical Mechanics for Relativity and Quantum 
Mechanics}, Oxford University Press, forthcoming.

A. Kirillov (1976), {\em Elements of the Theory of Representations}, Grunlehren 
Math. Wiss., Springer-Verlag.

H. Kragh (1999), {\em Quantum Generations}, Princeton University Press.

N. Landsman (this volume), `Between Classical and Quantum'.

S. Lie (1890). {\em Theorie der Transformationsgruppen: zweiter abschnitt}, 
Leipzig: B.G.Teubner.  

J. Lutzen (1995), {\em Denouncing Forces; Geometrizing Mechanics: Hertz's 
Principles of Mechanics}, Copenhagen University Mathematical Institute Preprint 
Series No 22. 

J. Lutzen (2003), `Between rigor and applications: developments in the concept 
of function in mathematical analysis', in {\em Cambridge History of Science, 
vol. 5: The modern physical and mathematical sciences},  ed. M.J. Nye, p. 
468-487.

J. Lutzen (2005) {\em Mechanistic Images in Geometric Form:
Heinrich Hertz's 'Principles of Mechanics'}, Oxford University Press.

J. Marsden and T. Hughes (1982), {\em Mathematical Foundations of Elasticity}, 
Prentice-Hall; Dover 1994.

J. Marsden and T. Ratiu (1999), {\em Introduction to Mechanics and Symmetry}, 
second edition: Springer-Verlag.

E. McMullin (1978), {\em Newton on Matter and Activity}, University of Notre 
Dame Press.

I. Newton (1687), {\em Principia: Mathematical Principles of Natural 
Philosophy}, trans. I.B. Cohen and A. Whitman, Cambridge University Press 
(1999). 

P. Olver (2000), {\em Applications of Lie Groups to Differential Equations}, 
second edition: Springer-Verlag.

C. Rovelli (this volume), `Quantum Gravity'.

T. Ryckman (2005), {\em The Reign of Relativity: Philosophy in Physics 
1915-1925} Oxford University Press.

S. Singer (2001), {\em Symmetry in Mechanics: a Gentle Modern Introduction}, 
Boston: Birkhauser.

E. Slovik (2002), {\em Cartesian Spacetime: Descartes' Physics and the 
Relational Theory of Space and Motion}, Dordrecht: Kluwer Academic.

M. St\"{o}ltzner  (2003), `The Principle of Least Action as the Logical 
Empiricist's Shibboleth',  {\em Studies in History and Philosophy of Modern 
Physics} {\bf 34B}, p. 285-318.

D. Wallace (2003), `Time-dependent Symmetries: the link between gauge 
symmetries and indeterminism', in Brading and Castellani (ed.s) (2003), pp. 
163-173.

\end{document}